%% file: Flavor_TASI2016.tex
\documentclass[12pt]{article}

\usepackage{amsmath}
\usepackage{amssymb}
\usepackage{amsfonts}
\usepackage{graphicx}
\usepackage[utf8]{inputenc}	
\usepackage{amsthm}				

\usepackage{slashed}       
\usepackage{mathrsfs}      
\usepackage{pifont}        
\usepackage{bbm}           
\usepackage{cancel}

\usepackage[dvipsnames]{xcolor}
\usepackage{fancyhdr}		
\usepackage{lipsum}         
\usepackage{framed}        
\usepackage{cite}          
\usepackage{xspace}			

\usepackage{booktabs}      
\usepackage{nicefrac}      

\usepackage[font=small]{caption} 
\usepackage{float}         

\usepackage{answers}			
	\Newassociation{sol}{Solution}{answers}
	\newtheorem{prob}{Problem}[section]

	\Opensolutionfile{answers}[answers1]

\usepackage{tikzfeynman}   

\usepackage[margin=2cm]{geometry}   
\graphicspath{{figures/}}			
\numberwithin{equation}{section}    

\usepackage{multicol}
\usepackage{etoolbox}
\usepackage{relsize}
\patchcmd{\thebibliography}
  {\list}
  {\begin{multicols}{2}\smaller\list}
  {}
  {}
\appto{\endthebibliography}{\end{multicols}}
\let\oldenumerate\enumerate
\renewcommand{\enumerate}{
  \oldenumerate
  \setlength{\itemsep}{1pt}
  \setlength{\parskip}{0pt}
  \setlength{\parsep}{0pt}
}

\let\olditemize\itemize
\renewcommand{\itemize}{
  \olditemize
  \setlength{\itemsep}{1pt}
  \setlength{\parskip}{0pt}
  \setlength{\parsep}{0pt}
}

\newcommand{\acro}[1]{\textsc{\MakeLowercase{#1}}}    
\newcommand{\PDG}{\acro{PDG}\xspace}
\newcommand{\SM}{\acro{SM}\xspace}
\newcommand{\UV}{\acro{UV}\xspace}
\newcommand{\CKM}{\acro{CKM}\xspace}
\newcommand{\FCNC}{\acro{FCNC}\xspace}
\newcommand{\FCNCs}{\acro{FCNCs}\xspace}
\newcommand{\GIM}{\acro{GIM}\xspace}
\newcommand{\QCD}{\acro{QCD}\xspace}
\newcommand{\QED}{\acro{QED}\xspace}
\newcommand{\CP}{\acro{CP}\xspace}
\newcommand{\CPV}{\acro{CPV}\xspace}

\renewcommand{\tilde}{\widetilde}   
\renewcommand{\vec}[1]{\mathbf{#1}} 

\newcommand{\email}[1]{\href{mailto:#1}{#1}}
\newenvironment{institutions}[1][2em]{\begin{list}{}{\setlength\leftmargin{#1}\setlength\rightmargin{#1}}\item[]}{\end{list}}

\newtheorem{theorem}{Theorem}[section]

\theoremstyle{definition}						
\newtheorem{definition}[theorem]{Definition}	
	
\newtheorem{eg}[theorem]{Example}

\usepackage[
	colorlinks=true,
	citecolor=green!50!black,
	linkcolor=NavyBlue!75!black,
	urlcolor=green!50!black,
	hypertexnames=false]{hyperref}

\fancypagestyle{firststyle}{
	\rhead{\footnotesize%
	\texttt{UCR-TR-2017-FLIP-K-2SO}%
	}}

\usepackage{etoc}

\begin{document}

\thispagestyle{firststyle} 	

\begin{center}

    {{\huge \bf Just a Taste\\[5pt]}
    {\large Lectures on Flavor Physics}
    }

    \vskip .7cm

    { 
    	\textbf{Yuval Grossman}$^{a}$
    	and 
    	\textbf{Philip Tanedo}$^{b}$ 
    	} 
    \\ 
    \vspace{-.2em}
    { \tt \footnotesize
	    \email{yg73@cornell.edu},
	    \email{flip.tanedo@ucr.edu}
    }
	
    \vspace{-.2cm}

    \begin{institutions}[2.25cm]
    \footnotesize
    $^{a}$ 
    {\it 
        Department of Physics, \acro{LEPP}, Cornell University, Ithaca, \acro{NY} 14853
        }    
	\\ 
	\vspace*{0.05cm}   
	$^{b}$ 
	{\it 
	    Department of Physics \& Astronomy, 
	    University of  California, Riverside, 
	    \acro{CA} 92521	    
	    }
    \end{institutions}

\end{center}


\begin{abstract}
\noindent 
We review the flavor structure of the Standard Model and the ways in which the flavor parameters are measured. 
This is an extended writeup of the \acro{TASI} 2016 lectures on flavor physics. Earlier versions of these notes were presented at pre-\acro{SUSY} 2015 and Cornell University's Physics 7661 course in 2010.
\end{abstract}

\small
\setcounter{tocdepth}{2}
\tableofcontents
\normalsize



\section{Introduction}

In this set of lectures, we introduce basics of flavor physics, that is, the part of Nature where the differences between the quarks plays a role. While this writeup includes more material than presented at the lectures, this write up is still just a taste of the entire field; for more in-depth reading we refer to other recent \acro{TASI} lectures~\cite{Ligeti:2015kwa, Grinstein:2015nya, Gedalia:2010rj}, reviews~\cite{Kamenik:2017znu,Blanke:2017ohr,Lee:2015bka,Isidori:2013ez,Nir:2010jr,Grossman:2010gw,Gedalia:2010rj,Isidori:2010gz,Buras:2009if,Nierste:2009wg,Artuso:2009jw,Nir:2007xn,Hocker:2006xb,Neubert:2005mu,Nir:2005js,Buras:2005xt,Ligeti:2003fi} and books~\cite{Bigi:2000yz,Branco:1999fs} on the subject.

To start off, here's a list of branching ratios collected from the \PDG.%
\footnote{%
The \textit{Review of Particle Physics} is prepared by the Particle
Data Group and is often referred to as 'the
\PDG'~\cite{Olive:2016xmw}. It just about contains everything you ever
wanted to know about particle physics.}
\begin{align}
\text{Br}(B \to X \mu \nu) &= 0.1086 (16) \\
\text{Br}(B \to X e \nu) &= 0.1086 (16) \\
\text{Br}(B \to X_s \gamma) &= 3.49(19) \times 10^{-4} \\
\text{Br}(B_s \to \mu^+ \mu^-) &= 2.4(8) \times 10^{-9} \\
\text{Br}(B^+\to \bar D^0 \ell^+ \nu) &= 2.27(11) \times 10^{-2} \\
\text{Br}(B^-\to \pi^0 \ell^- \bar \nu) &= 7.80(27) \times 10^{-5} \\
\text{Br}(K_L \to \mu^+ \mu^-) &= 6.84(11) \times 10^{-9}\\
\text{Br}(K^+ \to \mu^+ \nu) &= 0.6356(11) \\
\text{Br}(\psi \to \mu^+\mu^-) &= 5.961(33) \times 10^{-2} \\
\text{Br}(D \to \mu^+ \mu^-) & < 6.2 \times 10^{-9} \ .
\end{align}
Stare at these for a moment---do you see a pattern? If you were
trapped on a desert island without your smart phone and only the
\acro{PDG}, some of the observations from these branching ratios that
you may come up with are:
\begin{enumerate}
\item
\textbf{Lepton universality}. Swapping one generation of leptons with another does not appear to affect the branching ratios of these transitions.
\item
\textbf{Flavor-changing neutral currents are small}.
On the other hand, processes that change flavor are suppressed for charge-neutral transitions compared to transitions between hadrons of different charge.
\item
\textbf{Generation hierarchy}.
Decays between third and first generation are suppressed  
compared to that of third to second generation.
\end{enumerate}

In these lectures we uncover why these properties and others exist in the Standard Model (\acro{SM}) of particle physics. We elucidate that these features are, in fact, predicted once we specify the particle content and electroweak charges of the \SM. In contrast, other features of the theory are particular to specific parameters of this effective theory.
In the second part of these lectures, we tackle the question of how
these parameters are actually measured in low-energy systems where
\acro{QCD} confines the quarks into hadrons.

\begin{prob}\textbf{Using the PDG.}
Use the \acro{PDG} to answer the following questions:
\begin{enumerate}
\item What are the component quarks of the $D^+$ meson? What is its mass?
\item What are the component quarks of the $\Lambda$ baryon? What is its spin?
\item What is Br$(\tau \to \mu \nu \bar\nu)$?
\item What is the width (in eV) of the $B^+$ meson?
\item What is the average distance a $B^+$ meson will travel if $\gamma=4$?
\end{enumerate}
\begin{sol}
These are trivial to look up in the \acro{PDG}. The paper version is
slightly more satisfying to thumb through, while the online version is
kinder to the environment.
\begin{enumerate}
\item The easiest place to look is the Meson Summary Table. The $D^+$
is a $c\bar d$ bound state with mass $m_{D^+}\approx 1870$ MeV.
\item The easiest place to look is the Baryon Summary Table. The
$\Lambda$ is a $uds$ bound state with spin $1/2$.
\item The easiest place to look is the Lepton Summary Table. (Do you
see a theme?) The $\tau$ will decay to $\mu \nu\bar\nu$ 17\% of the
time.
\item The $B^+$ meson has a lifetime of $\tau_{B^\pm} = 1.6 \times
10^{-12}$ s. This translates to a width of
\begin{align}
\Gamma_{B^+} = (1.6 \times 10^{-12} \text{ s})^{-1}\times \frac{1
\text{ GeV}}{1.5 \times 10^{24} \text{ s}^{-1}} = 4.4 \times 10^{-13}
\text{ GeV} = 4.4 \times 10^{-4} \text{ eV}.
\end{align}
\item The \acro{PDG} tells us that $c\tau = 491$ $\mu$m. With a boost
factor $\gamma = 4$ this gives us $\gamma c\tau = 1.9 $ mm.
\end{enumerate}
\end{sol}
\end{prob}


\section{Model building}\label{sec:model:building}

\begin{quote}
	\textit{A theorist excitedly runs into an experimentalist's office one day, exclaiming that not only has the theorist uncovered an elegant model that solves the biggest outstanding questions in particle physics, but that it makes a robust experimental prediction. The excitement is contagious, and the experimentalist applies for grants, hires postdocs, builds new laboratory equipment, and starts to take data. Many years after the initial encounter, the experimentalist's entire research group---now exclusively focused on proving this one, elegant model---is ready to unblind its data. That afternoon, the experimentalist sulks despondently into the theorist's office. ``Prof.~Theorist? I'm really sorry to break this to you, but the data rules out the model.'' The theorist looks up at the experimentalist, considers the streaks of white hair and their many years of collaboration on this project, and finally says, ``What a shame! Did you know that I spent two whole weeks of my life developing that model?''} --Tim M.P.\ Tait 
\end{quote}

The goal of high-energy physics is to fill in the right-hand side of the following equation:
\begin{align}
	\mathcal L \;=\; ?
\end{align}
Our job is to find the effective Lagrangian of nature and experimentally determine its parameters. 

In order to do this, we build models: these are our hypotheses. In fact, it is perhaps more accurate to say that a theorist's job is not model \textit{building}, but rather model \textit{designing}. In order to design our Lagrangian, we need three ingredients:
\begin{enumerate}
	\item The \textbf{gauge group} of the model,
	\item The \textbf{representations} of the fields under this gauge group,
	\item The pattern of \textbf{spontaneous symmetry breaking}.
\end{enumerate}
The last point is typically represented by a sign, such as the sign of the Higgs mass-squared parameter at the unstable vacuum ($\mu^2<0$). 

Once we have specified these ingredients, the next step is to write the \textit{most general renormalizable Lagrangian} that is invariant under the gauge symmetry and provides the required spontaneous symmetry breaking pattern. This is far from a trivial statement. The `most general' qualifier tells us that \textit{all} terms that satisfy the above conditions \textit{must} be present in the Lagrangian, even the terms that may be phenomenologically problematic. For example, even though we might not want to include a term that induces proton decay, we cannot simply omit it from our model without some symmetry principle that forbids it.

On the other hand, renormalizability strongly constrains the form of a
Lagrangian and limits us to only a finite number of terms. This
condition comes to us from the principles of effective field theory
and Wilsonian renormalization. We assume that the more fundamental
ultraviolet (\UV) theory generates all possible operators---including
non-renormalizable terms---at the \UV scale. By dimensional analysis,
the non-renormalizable operators depend on negative powers of the \UV
scale $\Lambda$. Thus at the low energies $\mu \ll \Lambda$ where the
theory
is valid, we expect that these operators are suppressed by powers of $\mu/\Lambda \ll 1$. Thus such non-renormalizable operators exist in principle but they should come with small coefficients. The effect of these terms on low-energy observables is something we want to understand, but we expect them to be subdominant to phenomenology induced by renormalizable terms. For more background on this subject, see reviews by Hollowood \cite{Hollowood:2009eh}, Manohar \cite{Manohar:1996cq}, and Georgi \cite{Georgi:2009vn}.

There are a few other rules for designing models that we have thus far left implicit:
\begin{enumerate}
	\item We use the language of quantum field theory. 
	\item We impose Poincar\'e invariance. This can be relaxed, for example in the case of non-relativistic QCD which breaks Poincar\'e invariance at the \textit{effective} level, but not the \textit{fundamental} level.
	\item We do not \emph{impose} global symmetries. Global symmetries are an \emph{output} once we specify the gauge group, the matter content, and write down the most general Lagrangian. In fact, continuous global symmetries may be problematic on a theoretical level: we expect quantum gravity to break global symmetries---black holes can become electrically charged if they eat a proton, but they do not pick up baryon number~\cite{Barr:1992qq,Kamionkowski:1992mf,Holman:1992us,Ghigna:1992iv,Kallosh:1995hi}. This, of course, does not prevent the structure of a theory from having approximate global symmetries that were not imposed by hand.
	\item The basic fermion representation are two-component chiral (Weyl) spinors. Four component fields end up becoming useful for some calculations, but four dimensional theories are generally chiral. An encyclopedic reference for Weyl spinors is \cite{Dreiner:2008tw}. 
	\item A model is not a description of nature until one
          experimentally measures the finite number of physical
          parameters associated with it. Below we clarify what makes a
          parameter \emph{physical}. The fact that there are finite number of such parameters comes from renormalizability. If a Lagrangian contains $k$ physical parameters, then one must first perform $k$ measurements to be able to predict the value of any further measurements. 
\end{enumerate}
In the last point, the point is that the number of parameters required to define a theory is independent of how they are parameterized. In fact, many not-so-obvious ideas in field theory such as the renormalization group become obvious once one notices that they are reparameterizations of these physical parameters.

In practice, if you make $n>k$ measurements of a theory, you don't first do $k$ \emph{parameter measurements} and then do $(n-k)$ \emph{observations} of the theory. Instead, you take all $n$ measurements and do a statistical fit for the $k$ parameters to check for self-consistency; see e.g.~\cite{Wells:2005vk, Wells:statusofSM} for a theorist's perspective. Philosophically, however, it is useful and important to remember that a model---a Lagrangian---must come with measurements of its parameters.

\subsection{The Standard Model}


Following the model building rules above, it is useful to distinguish between
\begin{align*}
	\text{\emph{a} standard model}&
	&
	&\text{and}
	&
	\text{\emph{the} Standard Model} \ .
\end{align*}
A standard model is an effective theory with the same $\text{SU}(3)_c\times\text{SU}(2)_L\times\text{U}(1)_Y$ gauge group and the same representations of matter fields that we know and love. One can write out the Lagrangian for \emph{a} standard model as a series of all renormalizable, gauge-invariant operators built out of the specified matter fields and with unknown coefficients. In contrast, \textbf{\emph{the} Standard Model} is all of the above plus \emph{specific values} for each of the \emph{physical} combinations of operator coefficients.
\emph{The} Standard Model is predictive and determines which of the gauge symmetries are spontaneously broken. For example, it may be clear that in \emph{a} standard model, quarks will mix, whereas in \emph{the} Standard Model, we already know the mixing angles. In these lectures, we identify features of flavor phenomenology that are generic to \emph{a} standard model and those that are specific to \emph{The} Standard Model and its particular physical parameters.

The three ingredients of a standard model are:
\begin{enumerate}
	\item \textsc{Gauge group}: $\text{SU}(3)_\text{c}\times \text{SU}(2)_L \times \text{U}(1)_Y$.
	\item \textsc{Matter representations}: a useful mnemonic for the fermion fields of a standard model is $QUDLE$ (``cuddly''), which is short for the left-handed quark doublet $Q$, right-handed up $U$, right-handed down $D$, left-handed lepton doublet $L$, and the right-handed charged lepton $E$. Using the convention where all fields are written in terms of left-handed Weyl spinors (taking $CP$ conjugations as necessary), we may write the gauge representations using the notation $(c,L)_Y$:
	\begin{align*}
		Q_i &\quad (\mathbf{3}, \mathbf{2})_{1/6}\\
		U_i^c &\quad (\bar{\mathbf{3}}, \mathbf{1})_{-2/3}\\
		D_i^c &\quad (\bar{\mathbf{3}}, \mathbf{1})_{1/3}\\
		L_i &\quad (\mathbf{1}, \mathbf{2})_{-1/2}\\
		E_i &\quad (\mathbf{1}, \mathbf{1})_{1},
	\end{align*}
	we've written the generation index $i=1,2,3$. Note that this index does \textit{not} imply a global symmetry. In addition to the fermions we also have a complex scalar Higgs field, $H$, with the representation $(\mathbf{1},\mathbf{2})_{1/2}$. $SU(2)$ is pseudoreal, so the $\epsilon$ tensor allows us to convert between the fundamental $\mathbf{2}$ and antifundamental $\bar{\mathbf{2}}$ representations. This allows us to write $\tilde H_a = \epsilon_{ab}H^b$.
	\item \textsc{Spontaneous symmetry breaking}: $SU(2)_\text{L}\times U(1)_\text{Y} \to U(1)_{\text{EM}}$. We can see this from the sign of the Higgs mass term, $\mathcal L \supset \mu^2|H|^2$, so that the potential contains a term $-\mu^2 |H|^2$.
\end{enumerate}
The most general renormalizable Lagrangian containing the aforementioned matter fields and subject to the \SM gauge group can be divided into three parts,
\begin{align}
	\mathcal L_{\text{SM}} &= \mathcal L_{\text{kin}} + \mathcal L_{\text{Higgs}} + \mathcal L_{\text{Yuk}}.
\end{align}
The kinetic Lagrangian $\mathcal L_{\text{kin}}$ includes the gauge interactions through the covariant derivative and non-Abelian field strengths. The Higgs Lagrangian $\mathcal L_\text{Higgs}$ provides the Mexican hat potential
\begin{align}
	\mathcal L_{\text{Higgs}} = \mu^2 |H|^2 - \lambda |H|^4.
\end{align}
We have assumed the sign of $\mu^2$ to induce electroweak symmetry breaking and have further chosen the sign of $\lambda$ to ensure stability of the vacuum.
The Yukawa part of the Lagrangian $\mathcal L_{\text{Yuk}}$ contains
\begin{align}
	\mathcal L_{\text{Yuk}} = y^e_{ij}\bar{L}^iHE^j  + y_{ij}^d \bar{Q}^iH D^j + y_{ij}^u \bar{Q}^j\tilde H U^j + \text{h.c.} \ .\label{eq:SM:L:yukawa}
\end{align}
We have left the flavor indices, $i$ and $j$, explicit.
In a standard model, fermions only get masses through these terms. In a general field theory you can also have bare masses when gauge and spacetime symmetries permit it: for example, vector-like masses if fermion gauge representations are not chiral.

\subsection{Global, accidental, and approximate symmetries}

One might wonder why we talk about ideas like baryon number
conservation if our model buildings rules state that we may not impose
any global symmetries. The global symmetries of a standard model are \textit{outputs} of the theory rather than input constraints. They come from the restriction of renormalizability and gauge invariance. Global symmetries that appear only because non-renormalizable terms aren't considered are called \textbf{accidental symmetries}. These are broken explicitly by non-renormalizable terms, but since these terms are small one can often make use of these symmetries. When people ask why the proton doesn't decay in the Standard Model, baryon number conservation is only part of the answer; the `fundamental' reason is that baryon-number-violating operators in the Standard Model are non-renormalizable. 

In addition to accidental symmetries there are other \textbf{approximate symmetries} which are parametrically small in the sense that they become exact global symmetries when a parameter (or set of parameters) is set to zero. 
An example is \textbf{isospin symmetry}, which is an approximate global symmetry of the Standard Model that rotates the up and down quarks. It is broken by the quark masses and by \acro{QED}. Are these parametrically small? `Small' only means something when considering something dimensionless. It turns out that the relevant parameter is the ratio of the quark mass splittings to the \acro{QCD} confinement scale,
\begin{align}
	\frac{m_u - m_d}{\Lambda_{\text{QCD}}} \ .
\end{align}
Isospin symmetry is additionally broken by `the most famous small parameter in the world,' $\mathcal \alpha_{\text{EM}}$ since the charge of the $u$ is different from the charge of the $d$. 

\begin{eg}
\noindent\textbf{Symmetries in \emph{a} standard model and in
  \emph{the} Standard Model}. The distinction between accidental and
approximate symmetries highlight a distinction between a general
theory and a theory with specific values for its
parameters. Accidental symmetries are independent of the values of
parameters, these symmetries show up in \emph{any} `standard model'
with the same gauge group and matter representations of our known
universe. Approximate symmetries, on the other hand, depend explicitly
on the smallness of specific parameters, such as the smallness of the
up and down quark mass splitting. The approximate symmetries in these
lectures are thus exclusive properties of \emph{the} Standard
Model\footnote{Borrowing from string theory, one could imagine a
  \emph{standard model multiverse} with infinite copies of possible
  universes that all have the same gauge group and the same matter
  representations. These universes will be very different, with
  diffeent masses and couplings. They will all have the same accidental symmetries. They will not have the same approximate symmetries.}.
\end{eg}

There's one more small parameter that you might be familiar with. The
Higgs potential obeys a \textbf{custodial symmetry}. It is broken at
one-loop by the Yukawa couplings and electromagnetism. While one might
argue that the $\mathcal O(1)$ top Yukawa coupling is not small, since the breaking only occurs at loop-level, the relevant parameter is $y^2/16\pi^2$ which \textit{is} small. For more information about custodial symmetry and a nice overview of the symmetries of the Standard Model, see Scott Willenbrock's \acro{TASI} 2004 lectures \cite{Willenbrock:2004hu}. 

These accidental and approximate symmetries can be very useful, but we should remember that they are not as fundamental as the gauge symmetries that are a part of our definition of the theory.

\subsection{How to count physical parameters}

A model will never graduate to a `description of Nature' until its physical parameters are measured. But what do we mean by \textit{physical} parameters? Let us begin by reviewing some \textit{unphysical} parameters which cannot be measured. In Newtonian mechanics, the absolute value of the energy is not physical---only differences are measurable. In quantum mechanics, the overall phase of a system's wave function is not physical---only the magnitude. 

We know that the Standard Model has 18 physical
parameters,\footnote{At this point we do not consider the 19th parameter that is
  related to the strong \acro{CP} problem. We return to this in Section~\ref{sec:CP}.} see \cite{Cahn:1996ag} for a friendly reminder. It doesn't matter \textit{how} we write these 18 parameters. We may parameterize the Standard Model in different ways---but the number of physical parameters does not change. 

To learn how to count these parameters, let's consider the leptonic Yukawa term,
\begin{align}
	y^e_{ij}\bar L^i H E^j.
\end{align}
How many parameters does the matrix $y^e$ have? Since it is a $3\times 3$ complex matrix, it contains 18 real parameters. We know, however, that the \SM lepton sector  only has three physical parameters: the masses of the charged leptons. Thus there are 15 unphysical parameters hidden in (and obfuscating) the $y^e$ matrix! Since unphysical parameters are useless for predictions, it's always nicer to choose a parameterization where everything is physical.

How could we have figured out that there were 3 physical parameters in this sector of the Standard Model? If $y^e=0$, then the Lagrangian would enjoy a larger global symmetry: the global symmetry of the kinetic term, $\mathcal L_{\text{kin}}\supset \bar E_i \slashed{D} E_i + \bar L_i\slashed{D}L_i$. This is a $\text{U}(3)_E\times \text{U}(3)_L$ global symmetry associated with the rotation between the three generations of $E$ and $L$. The Yukawa term breaks the global symmetry down to $\text{U}(1)^3$, corresponding to the phases of the mass eigenstate fields. We may use the now-broken $\text{U}(3)_{E}\times \text{U}(3)_{L}$ symmetry  to rotate the symmetry-breaking order parameter $y^e$ to a convenient direction. Clearly this rotation will use only the broken generators of $\text{U}(3)_E\times \text{U}(3)_L$. We call $y^e$ a \textbf{spurion} of the broken  $\text{U}(3)_{E}\times \text{U}(3)_{L}$ symmetry. A particularly convenient direction is to rotate $y^e$ to a real diagonal matrix proportional to the charged lepton masses. In this form, it is manifest that there are three physical parameters. To recap, we used the \textit{broken} generators of the symmetry group to remove \textit{unphysical} parameters by orienting the spurion $y^e$ to convenient direction in symmetry space. It is tautological that the unbroken generators would not affect the spurion direction.

From this simple example one can see that the number of physical parameters is related to the generators of the unbroken symmetry by
\begin{align}
	\text{\# physical parameters} = \text{\# parameters} - \text{\# broken generators}.
\end{align}
In the case of the lepton Yukawa, U(3)$_E\times$U(3)$_L\to$U(1)$^3$ yields 15 broken generators. Given that $y^e$ has 18 parameters,  we learn that there are $18-15=3$ physical parameters in this sector. Make sure you understand why this formula makes sense. Unlike unbroken generators which represent symmetries of the system, the generator of a \textit{broken} symmetry will change a system and so allow us to align the system to match our parameterization and, by doing so, eliminate unphysical parameters. This rather abstract statement can be made very concrete with the following example.

\begin{framed}
\noindent\textbf{A very simple example}. Recall the Zeeman effect in quantum mechanics: the energy levels of the hydrogen atom are split in the presence of a magnetic field. In the absence of such a field the system enjoys an $\text{SO}(3)$ rotational symmetry and so has degenerate states. The field breaks the symmetry and splits the degeneracy by breaking $\text{SO}(3)$ down to $\text{SO}(2)$, which is the remaining rotational symmetry perpendicular to the direction of the field. There are two broken generators. The physical parameter in the system is the magnitude of the magnetic field. In general, the field points in an arbitrary direction $\mathbf{B}=B_x {\mathbf{\hat x}}+ B_y{\mathbf{\hat y}}+ B_z {\mathbf{\hat z}}$. We can use the two broken generators to rotate the coordinates so that $B_x=B_y=0$. These were unphysical parameters. This leaves a single physical parameter, $B_z$, which in these coordinates is precisely the magnitude of $\mathbf{B}$.
\end{framed}

Much of this language should sound familiar from your study of spontaneous symmetry breaking. The spurions, however, are manifestly \emph{explicit} symmetry breaking---they are reminders that the global symmetry is not sacrosanct. The spurion is useful because renormalizability and gauge invariance limit the number of operators that can break the global symmetry, sometimes to just a single possible term. We identify the symmetry by taking the spurion$\to 0$ limit; we may even colloquially say that the `symmetry is restored' in this limit: but let us be clear that unlike spontaneous symmetry breaking, the symmetry never existed to begin with\footnote{%
We note that one is free to construct theories where spurions are actually the remnants of a spontaneous breaking. This can be useful for organizing a hierarchy of parameters~\cite{Froggatt:1978nt}. However, this typically turns your theory into a different theory: if the global symmetry is spontaneously broken, then you must include massless Goldstones in your theory which changes the matter content. If you want to remove the Goldstones by feeding them to gauge bosons, then you must gauge the broken symmetry, which changes the gauge symmetries of the theory. Either way, you have changed the defining characteristics of your theory.}%
. The power of the spurion is that we can still use the not-really-there symmetry to organize our theory to understand the parametric dependence of observables on the spurions.

\begin{framed}
\noindent\textbf{Mass-dependence of dipole moments}. The leading contribution to a lepton dipole moment in the \SM comes from a loop-level diagram that generates the dipole operator, $\bar e_L\sigma^{\mu\nu} e_R F_{\mu\nu}+\text{h.c.}$. Observe that the operator involves one left-chiral lepton and one right-chiral lepton. This operator violates chiral symmetry. At low energies where the lepton sector of the \SM is basically \acro{QED}, the only renormalizable term that violates chiral symmetry is the lepton mass term. The lepton mass is a spurion for chiral symmetry breaking in \acro{QED}. Thus we deduce that the loop diagram that generates the dipole must have an amplitude proportional to the lepton mass. One could have equivalently noticed this from electroweak gauge invariance since $e_L$ must be promoted to an electroweak doublet which, in turn, implies the presence of a Higgs vacuum expectation value (vev). In this case, the Higgs vev is a spurion that happens to come from symmetry breaking.
\end{framed}

\subsection{Counting parameters in the Standard Model}\label{sec:paramter:counting:SM}

Where do the Standard Model's 18 parameters come from? 
\begin{itemize}
	\item The kinetic Lagrangian $\mathcal L_{\text{kin}}$ contains three physical parameters, the couplings $g',g,g_3$ which determine the strengths of each force.
	\item The Higgs Lagrangian $\mathcal L_{\text{Higgs}}$ contains two physical parameters which we take to be the vacuum expectation value of its neutral component $v$ and the quartic interaction strength $\lambda$. (One could also have chosen the unstable mass $-\mu^2$ or the physical mass $m_H^2$.)
	\item Finally, the Yukawa Lagrangian contains the three charged lepton masses which we discussed above plus ten physical parameters associated with the quark Yukawa sector.
\end{itemize}
To explain the last ten parameters, recall that the quark Yukawa sector takes the form
\begin{align}
	\mathcal L_\text{Yuk} \sim y^u \bar Q\tilde H U + y^d \bar Q H D.
\end{align}
There are 36 total real parameters between the $y^u$ and $y^d$. The generational symmetry of the kinetic term is $\text{U}(3)^3$ which gives 27 total generators. How many of these survive after turning on the $y^u$ and $y^d$ interactions? Baryon number, $\text{U}(1)_B$ survives with one generator, and so we have 26 broken generators. The number of physical parameters is $36-26=10$. 

In fact, we can separate the  real physical parameters from the physical phases. A generic Yukawa matrix has 9 real parameters and 9 phases, a pair for each element: $y_{ij} = r_{ij}e^{i\theta_{ij}}$. How many real parameters and phases are there in a unitary matrix? We know that:
\begin{enumerate}
	\item A unitary $\text{U}(N)$ matrix has $N^2$ total parameters: $N^2$ complex elements with the constraint equation $U_{ik}(U^\dag)_{kj}=\delta_{ij}$ for each element.
	\item An $\text{O}(N)$ \textit{orthogonal} matrix has $\frac 12 N(N-1)$ parameters which are all real.
\end{enumerate}
Since an orthogonal matrix is just a unitary matrix ``without the phases,'' we may subtract the latter from the former to find that an $N\times N$ unitary matrix has $\frac 12 N(N+1)$ phases and $\frac 12 N(N-1)$ real parameters. Thus our $\text{U}(N)^3$ kinetic term symmetry gives 9 real parameters and 18 phases. The Yukawas break this to $\text{U}(1)_B$ so that there are 9 broken real generators and 17 broken complex phases. Thus,
the number of real physical parameters is
\begin{align}
	2\text{ Yukawas }\times 9 \text{ $\mathbbm{R}$ param. each} - 9 \text{ broken $\mathbbm{R}$ generators } = 9 \text{ physical $\mathbbm{R}$ parameters}.
\end{align}
Similarly for the complex phase,
\begin{align}
	2\text{ Yukawas }\times 9 \text{ phases each} - 17 \text{ broken phase generators } = 1 \text{ physical phase}.
\end{align}
We show below that these physical parameters are the six quark masses,
the three mixing angles, and the one \acro{CP}-violating phase. 

The above counting came from a classical analysis. Indeed, careful students will argue that this classical analysis is too cavalier---after all, anomalies break some of these classical symmetries. In the quantum picture, the $\text{U}(1)_A$ symmetry is anomalous and so should not be counted as having ever been a symmetry of the kinetic term. This gives one less broken generator, but introduces a new term in the Lagrangian, the $\Theta_\text{QCD}$ term which encodes the same strong \acro{CP}-violating phase\footnote{Skeptical students may want to argue that such a term should be included in the classical Lagrangian, and so our classical counting was short by one parameter. However, at the \emph{classical} level, the strong \acro{CP} phase $\Theta_\text{QCD}G\tilde G$ is a total derivative and vanishes upon integration in the action.}.

\begin{prob}\label{prob:extra:generations}
	\textbf{Extra generations in the Standard Model}. Count the number of physical flavor parameters in the quark sector of the Standard Model with $N$ generations. Show that the quark sector of such a model has $N(N+3)/2$ real parameters and $(N-1)(N-2)/2$ phases. Determine the number of mixing angles. 
	\begin{sol}
		Let us review our counting for real parameters and phases. Each Yukawa matrix has $N^2$ elements which can each be written as $y_{ij} = r_{ij}e^{i\theta_{ij}}$. Thus each Yukawa matrix contains $N^2$ real parameters and $N^2$ phases. There are two Yukawas (up and down), so we have a total of $2N^2$ real parameters and $2N^2$ phases. 
		
		We would now like to identify the number of broken generators. An element of $\text U(N)$ is an $N\times N$ complex matrix ($N^2$ real parameters and $N^2$ phases) satisfying
		\begin{align}
			UU^\dag = \mathbbm{1}.
		\end{align}
		This gives one equation of constraint for each complex component for a total of $N^2$ parameters per unitary rotation.  How many of these parameters are real? This is precisely the number of parameters of an element of $\text O(N)$, namely $\frac 12 N(N-1)$. The remaining $\frac 12 N(N+1)$ parameters are phases. 
				
		In the quark sector of our $N$ generation Standard Model, we have $2N^2$ real parameters and $2N^2$ phases coming from the $y^u$ and $y^d$ matrices. How many of these are physical? We subtract the number of broken generators. For generic Yukawa matrices with no underlying structure, the $\text U(N)^3$ flavor symmetry of the $Q, u_R, d_R$ is broken down to $\text U(1)_B$.  Thus we have $3\times \frac{1}{2}N(N-1)$ broken generators associated with real parameters and $3\times \frac 12 N(N+1) -1$ broken generators associated with phases. This gives a total of
		\begin{align}
			2N^2 - \frac 32 N(N-1) &= \frac 12 N(N+3) \text{\; physical real parameters}\\
			2N^2 - \frac 32 N(N+1) + 1 &= \frac 12 (N-2)(N-1) \text{\; physical phases}.
		\end{align}
		Of the real parameters, $2N$ of them are Dirac masses and so $\frac 12 N(N-1)$ must be mixing angles---this is consistent with our counting of O($N$) generators. Let us remark that for the full $N$-generation Standard Model we must also add $N$ real parameters for the lepton masses, the 3 real parameters for the gauge couplings, and the two real parameters governing the Higgs sector. 
	\end{sol}
\end{prob}

\begin{prob} \textbf{Exotic light quarks, part I.}\label{prob:exotic:quarks}
	Consider a copy of the Standard Model with a modified quark sector without the $c$, $b$, and $t$ quarks. The remaining quark representations are modified to $Q_L = (u_L, d_L)$ forming an $\text{SU}(2)_L$ doublet and $s_L$ an $\text{SU}(2)_L$ singlet. All the right-handed quarks are singlets ($u_R, d_R, s_R$). All color and electric charges are the same as in the Standard Model. Electroweak symmetry follows as in the usual Standard Model. How many physical parameters are in the model? How many of them are real and how many are phases?
	\begin{sol}
		The problem already defined the three main ingredients of a model. It's up to us to now write the most general Lagrangian and identify the number of physical parameters. We will only consider the quark sector since all other sectors are the same as the Standard Model. There are five matter fields in the quark sector ($Q_L$, $u_R$, $d_R$, $s_L$, $s_R$). The only difference from the Standard Model is that the $s_L$ is an $\text{SU}(2)_L$ singlet so that it must have $Y=-1/3$ to obtain the correct electric charge. The kinetic terms have a $\text U(2)\times \text U(1)^3$ symmetry where the $\text U(2)$ corresponds to mixing the $d_R$ and $s_R$ fields. Note that the $s_L$ cannot mix with these fields even though it seems to have the same quantum numbers as the $s_R$; the $s_L$ and $s_R$ are different Lorentz representations. In other words, $s_L$ is a left-chiral spinor $\chi_\alpha$ while $s_R$ is a right-chiral spinor $\bar\psi^{\dot\alpha}$. As one can see from the indices, these transform differently under rotations. Using the counting introduced in the previous problem for the number of real parameters and phases in unitary matrices, we see that the kinetic terms have a symmetry group with 1 real parameter generator and 6 phase generators. 
		
		Let us now identify how many total parameters are in the Lagrangian and how many of of the above symmetries are broken. The most general quark Yukawa sector is
		\begin{align}
			\mathcal L_{\text{Yuk.,} Q} &= y_u \bar{Q}_LH u_R + y_d \bar{Q}_L\tilde H d_R + y_s \bar{Q}_L\tilde H s_R + \text{h.c.},\label{eq:sol:exotic:quarks:Yuk}
		\end{align}
		where the $y_i$ are complex numbers so that we have 3 real parameters and 3 phases. In the Standard Model the Yukawa sector was the only source of additional quark interactions. In our exotic model, however, the fact that the $s_L$ is a singlet allows us to write additional bare mass terms,
		\begin{align}
			\mathcal L_{\text{mass}} &= m_s \bar{s}_L s_R + m_d \bar{s}_L d_R + \text{h.c.}\label{eq:sol:exotic:quarks:mass}
		\end{align}
		Each $m_a$ is a complex number so that this gives 2 additional real parameters and 2 additional phases. We now have a total of 5 real parameters and 5 phases. What symmetries are preserved by these Lagrangian terms? We are left with an analog of $U(1)_B$ where the left-handed and right-handed fields all transform with the same phase.  This means that of the symmetries in the kinetic sector, the one real parameter and 5 of the 6 phases are broken. The number of physical parameters is equal to the number of total parameters minus broken generators, so this gives us a total of four real parameters and no phases. 
		
		Does this make sense? Three real parameters are associated to the Dirac masses of the particles. We have one left over real parameter, but we can see that when the Higgs gets its electroweak symmetry-breaking vev there is a mass term mixing the $s$ and $d$ quarks so that the mass term after EWSB looks like
		\begin{align}
			\mathcal L \supset 
			\begin{pmatrix}
				\bar d_L & \bar s_L
			\end{pmatrix}
			\begin{pmatrix}
				y_d \frac{v}{\sqrt{2}} & y_s \frac{v}{\sqrt{2}}\\
				m_d & m_s
			\end{pmatrix}
			\begin{pmatrix}
				d_R\\
				s_R
			\end{pmatrix}
			+ \text{h.c.}\label{eq:sol:exotic:quarks:matrix}
		\end{align}
		so that we can see that the left over real parameter gives the $d-s$ mixing angle.
	\end{sol}
\end{prob}

\begin{prob}
	\textbf{Parameters of the MSSM}. Show that there are 110 physical parameters in the flavor sector of the minimal supersymmetric Standard Model (\acro{MSSM}). Don't forget to include general soft \acro{SUSY}-breaking terms. How do these divide into masses, mixing angles, and phases?
	\begin{sol}
		A careful counting by Dimopoulos and Sutter gives 110 parameters: 30 masses, 29 mixing angles, and 41 phases \cite{Dimopoulos:1995ju}. You are referred to their paper for a very nice analysis. See also a nice presentation by Haber \cite{Haber:2000jh} for some variants of the MSSM.
	\end{sol}
\end{prob}

\begin{prob}
	\textbf{The $\Theta_\text{Weak}$ angle for $\text{SU}(2)_L$}. We discussed how the counting of parameters pointed to the appearance of $\Theta_\text{YM}$ as a physical parameter when taking into account quantum anomalies. Use a similar argument to show that such an angle is \textit{not} physical for $\text{SU}(2)_L$. \textbf{Hint:} one could also ask why the anomalous $\text{U}(1)_{B+L}$ doesn't affect our Standard Model counting.
	\begin{sol}
		The conventional explanation for why the $\text{SU}(2)_L$ $\Theta$ angle is not physical comes from the use of the anomalous $\text{U}(1)_{B+L}$ symmetry to rotate it away. (One cannot do this in \acro{QCD} with $\text{U}(1)_A$ since this would give the quark masses an imaginary component.) A cute way to check this is to count the number of physical parameters in the $\text{SU}(2)_L$ sector at a \textit{quantum} level and check that they are all accounted for \textit{without} having to introduce a $\Theta_{\text{L}}$ term. We can cast this argument in a different way in light of our parameter counting. We saw that the $\text{U}(1)_A$ anomaly of \acro{QCD} led to the appearance of another physical parameter at the quantum level, $\Theta_{\text{QCD}}$. If we look at the $\text{SU}(2)_L$ sector of the Standard Model, we might also expect an additional parameter to appear, $\Theta{\text{weak}}$, because of the $\text{U}(1)_{B+L}$ anomaly. However, the key point is that $\text{U}(1)_{B+L}$ isn't broken by the Yukawas, and so it was never counted as a `broken symmetry' in our tree-level analysis. Thus the $\text{U}(1)_{B+L}$ anomaly doesn't lead to one less broken symmetry and so does not introduce an additional physical parameter. In other words, $\Theta_{\text{Weak}}$ is not physical.  By the way, a useful way to remember which global symmetries are broken anomalies is to look at the instanton that induces the appropriate $\Theta$ term. In this case the sphaleron violates $B+L$ while preserving $B-L$.
	\end{sol}
\end{prob}

\section{The flavor structure of the Standard Model}

From our parameter counting, we know that the Standard Model has then physical flavor parameters, six quark masses and three mixing angles, and a phase. The experimental goal of flavor physics is to measure these four parameters in as many ways as possible to check for consistency and hope for a signal of new physics. Let's see how these parameters show up phenomenologically.

\subsection{The CKM matrix}

It should be no surprised that we are interested in the \textbf{CKM matrix}, named after Cabbibo, Kobayashi, and Maskawa. Cabbibo, unfortunately, passed away in 2010 before he had a chance to compete for a Nobel prize after Kobayashi and Maskawa each took part of the 2008 prize. We should hold a special regard for Cabbibo since he was the one to diagonalize the relevant $2\times 2$ matrix between the $d$ and $s$ quarks\footnote{High-minded students may feel like this was \emph{just} diagonalizing a $2\times 2$ matrix. Experienced physicists know that there are many phenomena that reduce to diagonalizing $2\times 2$ matrices, and \emph{clever} physicists know how to identify the \emph{right} $2\times 2$ matrix to diagonalize.}. 

At its heart, the \CKM matrix comes from a basis rotation and the fact that one cannot simultaneously diagonalize all of the flavor matrices in the Standard Model. The \CKM matrix is the `leftover' when moving form one flavor basis to another and is the source of flavor mixing. The particular matrices that cannot be simultaneously diagonalized are the $W$-boson couplings, the masses of the $d$-type quarks, and the masses of the $u$-type quarks. At most we can diagonalize two of these. We chose to diagonalize the masses and leave the $W$ coupling non-diagonal. It is important to emphasize over and over again that this is a \textit{choice} that we are free to make. In neutrino physics, for example, we choose the neutrino masses to be non-diagonal. In principle we could have even chosen a different basis where none of the matrices are diagonal.

From our principle of writing the most general renormalizable Lagrangian, the only non-trivial---that is, not from the kinetic terms---quark interactions are Yukawa couplings (\ref{eq:SM:L:yukawa}),
\begin{align}
	\mathcal L_{\text{Yuk}} = y_{ij}^d \bar{Q}^iH D^j + y_{ij}^u \bar{Q}^j\tilde H U^j + (\text{lepton term}) + \text{h.c.}
\end{align}
Upon electroweak symmetry breaking, the Higgs vacuum expectation value
(`vev') turns these into mass matrices. We choose the Higgs vev to be real and in the down component of $H$ (and so the up component of $\tilde H$).
Writing the doublet fields out explicitly,
\begin{align}
	Q_L^i = \begin{pmatrix}
		u_L^i\\d_L^i
	\end{pmatrix},
\end{align}
we see that the Higgs vev gives us mass terms of the form
\begin{align}
	\mathcal L_{\text{mass,}q} = m_{ij}^d \bar d_L^i d_{R}^j + m_{ij}^u \bar u_L^i u_R^j,\label{eq:SM:quark:mass:terms}
\end{align}
where the mass matrices are related to the Yukawas and the Higgs vev $\langle H\rangle_2 = v/\sqrt{2}$ by
\begin{align}
	m^q_{ij} = \frac{v}{\sqrt{2}}y^q_{ij}.
\end{align}
In general these $m^q$s are not diagonal and contain many unphysical parameters. To remove these, we move to the mass basis. Again we recall that there's nothing \textit{holy} about the mass basis! This is a \textit{choice} that we are free to make, we could have chosen a different basis without affecting the physics. 

\begin{framed}
	\noindent\textbf{Is the mass basis special}? When we learn \acro{QFT} we always start out in the mass basis. We should, however, remember our roots in relativistic quantum mechanics where the `mass basis' is just the energy basis and is just another basis that one is free to choose for one's states. In \acro{QM} we're used to working with states that aren't in the energy basis, so we should be willing to transfer this intuition to \acro{QFT}. \textbf{\textit{However}}, there is an important counter point about why the mass basis is special in \acro{QFT}. In the mass basis, one can completely solve the quadratic part of the path integral without any of the problems associated with the infinite-dimensionality of \acro{QFT}. This means that it is a natural basis to expand about in perturbation theory. Of course, we are free \textit{not} to use the mass basis and in the limit where the masses are perturbative relative to the energy scales of a process one can work with two-point mass insertion Feynman rules. So it is perfectly valid to do QFT in bases other than the mass basis, though there are good reasons why the mass basis is a default in calculations.
\end{framed} 

Let's diagonalize $m$. At this stage, $m$ is an \emph{arbitrary} complex matrix. In order to diagonalize such a matrix, we perform a bi-unitary transformation: a unitary transformation acting on the left and another unitary transformation acting on the right. By comparison, a Hermitian matrix can be diagonalized with the \emph{same} rotation acting from the left and right. In general, 
\begin{align}
	\hat{m}^q_{ij} = \left(V^q_{L}\right)_{ik} m^q_{k\ell} \left(V^{q\dag}_{R}\right)_{\ell j}\ .
\end{align}
We use the notation that a matrix with a hat, $\hat m$, is diagonal. In principle these $L,R$ indices have nothing to do with left- and right- chiral fermions, but one can see from (\ref{eq:SM:quark:mass:terms}) that the $i$ index is associated with the left-chiral quark while the $j$ index is associated with the right-chiral quark. Diagonalizing $m$ evidently rotates the left-chiral and right-chiral fields by $V_{L}$ and $V_R$ accordingly:
\begin{align}
	q^i_{L} &= \left(V^q_{L}\right)_{ij}q'^j_L\\
	q^i_{R} &=\left(V^q_{R}\right)_{ij}q'^j_R.
\end{align}
On the left-hand side we've written the interaction-basis field and on the right-hand side we've written the corresponding linear combination of mass-basis fields. From here on we will assume the mass basis and drop the prime. (Don't you hate it when textbooks do this?)

How does our Lagrangian look in this mass basis? The Yukawa interactions are now diagonal since the mass matrices are proportional to the Yukawa matrices, but now the coupling of the $W$ is contains off-diagonal terms. Going from the interaction to the mass basis, we find
\begin{align}
	\mathcal{L}_{Wqq} =  \frac{g}{\sqrt{2}}\bar{u}_L i\gamma_\mu d_L W^\mu \to \frac{g}{\sqrt{2}}\bar u_L i\gamma_\mu (V_{uL}V_{dL}^\dag)d_L W^\mu,
\end{align}
where the factor of $1/\sqrt{2}$ came from the normalization of the $W^\pm$ states relative to the $W^{1,2}$ states. We now identify the famous \textbf{CKM} matrix,
\begin{align}
	V\equiv V_{uL}V_{dL}^\dag.
\end{align}
The \CKM matrix has some important properties:
\begin{itemize}
	\item The \CKM matrix is unitary. $V^\dag V = V^\dag V = \mathbbm{1}$. We will see, however, that at times it is convenient to approximate it with a non-unitary matrix that captures all of the relevant physics. 
	\item The number of physical parameters is always four (and always three mixing angles and a complex phase), but we there are many convenient parameterizations. Choosing a convenient parameterization is independent of physics.
	\item The index structure of the \CKM matrix is
	\begin{align}
		V=\begin{pmatrix}
			V_{ud} & V_{us} & V_{ub}\\
			V_{cd} & V_{cs} & V_{cb}\\
			V_{td} & V_{ts} & V_{tb}
		\end{pmatrix}.\label{eq:VCKM:in:elements}
	\end{align}
	We have \textit{chosen} to order the quarks by increasing masses. We have also \textit{chosen} to name the mass eigenstates. 
\end{itemize}

Before moving on, let us re-emphasize once again for the millionth time that we are making an arbitrary (but convenient) choice about how we choose the physical parameters of $V$. We are free to choose any parameterization and the physics must be parameterization-invariant. Any quantity which is parameterization dependent \textit{cannot} be physical. 
\begin{eg}
	$V_{ub}$ is parameterization dependent. However, we know that since the \CKM matrix is unitary, $VV^\dag$ is the unit matrix and so its elements are parameterization \emph{in}dependent. We make use of this in Section~\ref{sec:unitarity:triangle} to define unitarity triangles.
\end{eg}
In this sense, parameterization-invariance is like gauge invariance. Anything physical must be parameterization invariant, and any `physical' quantity which you calculate that ends up being parameterization-dependent must be a mistake.

\subsection{Parameterizations of the CKM matrix}

Our parameter counting taught us that the quark sector of the Standard Model contains three mixing angles and one complex phase. 
%
%
Given that, we the understand that the general unitary matrix we wrote in (\ref{eq:VCKM:in:elements}) must contain many unphysical parameters since each element $V_{ij}$ is a complex number. 

\begin{prob}\textbf{Simplest unitary parameterization of the CKM}. Given that we know that the physical content of the \CKM matrix must boil down to three mixing angles and one phase, 
show that the most general \CKM matrix can be parameterized by at most 4 real parameters that cannot all be in the same row. That is, that we must have a minimum of five complex elements.
	\begin{sol}
		See \cite{Silva:CPviolation}; one of the authors of that book attended the lecture where this problem was posed and nodded approvingly when the book was referenced.
	\end{sol}
\end{prob}

In light of the above mathematical result, it is customary to choose $V_{ud}, V_{us}$, $V_{cb}$, and $V_{tb}$ to be purely real. The remaining elements are complex.  One standard parameterization of the mixing angles is: $\theta_{12}, \theta_{13}, \theta_{23}$. In the limit of two generations (e.g.\ $d$ and $s$), $\theta_{12}$ is the usual \textbf{Cabbibo angle}. The two indices tell us which plane we're rotating about. The phase $\delta_{KM}$ is typically named after Kobayashi and Maskawa. In terms of these parameters, the \CKM matrix takes the form (writing $c_{12} = \cos\theta_{12}$, etc.)
\begin{align}
	V_{\text{CKM}} =
	\begin{pmatrix}
		c_{12}c_{13} & s_{12}c_{13} & s_{13}e^{-i\delta}\\
		-s_{12}c_{23}-c_{12}s_{23}s_{13}e^{i\delta} & c_{12}c_{23}-s_{12}s_{23}s_{13}e^{i\delta} & s_{23}c_{13}\\
		s_{12}s_{23}-c_{12}c_{23}s_{13}e^{i\delta} & -c_{12}s_{23}-s_{12}c_{23}s_{13}e^{i\delta} & c_{23}c_{13}
	\end{pmatrix}.\label{eq:CKM:standard:parameterization}
\end{align}
%
From the \acro{UTFIT} website,\footnote{\url{http://www.utfit.org/UTfit/}} we find that the data give
\begin{align}
\sin\theta_{12}	&= 0.22497 \pm 0.00069 \\
 \sin\theta_{23} &=0.04229 \pm 0.00057	\\
 \sin\theta_{13} &= 0.00368 \pm 0.00010	\\
 \delta [^{\circ}] &= 65.9 \pm 2.0 \ .
\end{align}
Notice that all mixing angles are small! This is an important case where \emph{the} Standard Model is not a generic standard model. Thus it would be nice to have an approximation that captures the essential physics in a way that makes it more transparent. This is a bit of a shift in paradigm, so let's take a moment to discuss some philosophy. The reason why we use approximations in physics is because we often can't solve things exactly. However, here we have an \textit{exact} parameterization of the \CKM matrix, but we want to move away from it to an approximation. This sounds extremely stupid! It will seem even more stupid when you realize that the approximation that we make is not even unitary---we \textit{lose} one of the fundamental properties of the matrix! However, part of being a physicist means knowing what you can neglect. 
\begin{quote}
	\textit{We tell the undergrads at Cornell that it is not by mistake that our department is in the College of Arts and Sciences. Physics is about the \emph{art} of making the right approximation. Making the right approximation can teach you a lot.}
\end{quote}
The first person to make such an approximation was Wolfenstein---you might be familiar with him from the \acro{MSW} effect in neutrino physics. His key insight was that the orders of magnitude of the \CKM matrix seem to follow a particular pattern:
\begin{align}
	|V| \sim 
	\begin{pmatrix}
		1 & \lambda & \lambda^3\\
		\lambda &1 & \lambda^2\\
		\lambda^3 & \lambda^2 & 1
	\end{pmatrix}\ ,\label{eq:CKM:order:of:mag}
\end{align}
where $\lambda \approx 0.2$. Motivated by this, he defined four
\textit{different} parameters to describe the physical content of the
\CKM matrix: $\lambda, A, \rho, \eta$. One takes $\lambda$ to be a
small parameter worthy of expanding about and the others are formally $\mathcal O(1)$. These parameters are defined relative to the standard parameterization by
\begin{align}
	s_{12} &= \lambda = \frac{|V_{us}|}{\sqrt{|V_{ud}|^2+|V_{us}|^2}}\\
	s_{13} &= A\lambda^2 = \lambda\left|\frac{V_{cb}}{V_{us}}\right|
\end{align}
and finally 
\begin{align}
s_{13}e^{i\delta} &=  A\lambda^3(\rho+i\eta) = V^*_{ub} \ .
\end{align}
One is free to plug in these relations into (\ref{eq:CKM:standard:parameterization}) to get a perfectly unitary and even uglier representation of the \CKM matrix. The beauty of the \textbf{Wolfesntein parameterization}, however, is that we may use it to write the CKM matrix as a Taylor expansion in $\lambda$, so that
\begin{align}
	V = 
	\begin{pmatrix}
		1-\lambda^2/2 & \lambda & A\lambda^3(\rho-i\eta)\\
		-\lambda & 1-\lambda^2/2 & A\lambda^2\\
		A\lambda^3(1-\rho-i\eta) & -A\lambda^2 & 1
	\end{pmatrix} + \mathcal O(\lambda^4). \label{eq:CKM:Wolfenstein}
\end{align}
Looking at the factors of $\lambda$ we see that the diagonals are order one, and we get the structure in (\ref{eq:CKM:order:of:mag}) that Wolfenstein observed. Now look at the upper left $2\times 2$. Do you recognize this? (\emph{If you don't see it, then there's something wrong with your physics education!}) The elements are just the expansion for sine and cosine. In fact, it is the $2\times 2$ Cabbibo mixing matrix! To first order in $\lambda$, the first two generations don't know about the third. There's one more feature: complex numbers only show up in the 1--3 and 3--1 mixing elements. Let us make a brief detour to highlight the physical significance of this complex number.

\subsection{CP violation}

Let's remind ourselves of some properties of discrete symmetries. We have the usual suspects: $C$, $P$, $CP$, and $CPT$. Any local Lorentz-invariant field theory preserves $CPT$. So far everything we've observed agrees with $CPT$, so we assume that this is respected by nature at a fundamental level---though there \textit{are} interesting cases where \emph{effective} $CPT$-violating theories are useful. What about the other discrete symmetries? None of them \textit{needs} to be conserved by a theory. Experimentally we know that \acro{QCD} and \acro{QED} each conserve both $C$ and $P$ separately, and so also conserve $CP$. The weak interaction is a different story.

Electroweak theory is, by construction, parity violating. This is because it is a chiral theory: it treats left- and right-handed fields differently. Clearly once you write down such a theory, interchanging left and right brings you to a different theory. Any chiral theory tautologically violates parity. What about charge conjugation? Applying $C$ to a left-handed field transforms it into a right-handed field, and so it is also violated in by the weak interaction. The particular transformation isn't important to us here, but is explained thoroughly in, e.g.\ \cite{Ticciati:QFT,Silva:CPviolation}. Thus a chiral theory violates \textit{both} $P$ and $C$. To repeat things over again but with more refined language, we may say
\begin{quote}
	\acro{QED} and \acro{QCD} are vectorial and so preserve $P$ and $C$ separately, but electroweak theory is chiral and so violates both $P$ and $C$.
\end{quote}

What about $CP$? Since both $P$ and $C$ transformations take left- and right-handed fields into one another, $\chi \leftrightarrow \bar\psi$, at first glance $CP$ isn't \textit{necessarily} violated. This is indeed true at a model-building level. Following our prescription for designing a model, we may specify the key ingredients of a model---gauge group, representations, breaking pattern---and then write down the most general renormalizable Lagrangian. This is a model, but it's not actually a ``predictive description of nature'' until its physical parameters are determined explicitly. \emph{A} `standard model' written with undetermined parameters may or may not violate $CP$. It is only once the parameters of the Standard Model were measured\footnote{This is an anachronistic statement since the Standard Model was developed theoretically hand-in-hand with the experiments that probed its parameters.} that we found that indeed, the values of some of the parameters violates $CP$. 

How does a parameter violate $CP$? $CP$-violation reduces to the presence of a non-vanishing physical phase: this is precisely what we called $\delta_{KM}$ above. You should read these as synonyms: 
\begin{center}
	``physical non-zero phase'' = ``this theory is $CP$ violating.''
\end{center}
We won't go into details here, though we've already done most of the work in our parameter counting. For details see \cite{Burgess:StandardModel, Silva:CPviolation}. Instead, we provide a hand-wavy argument and focus on the Yukawa terms. Let us consider the up-type Yukawa term and---for this one time only---explicitly write out its Hermitian conjugate,
\begin{align}
\mathcal L_{\text{Yuk,} u} =	y_{ij}\chi_Q^i \tilde H \psi_U^j + y^*_{ij}\bar\chi_Q^i \tilde H^\dag \bar\psi_U^i,
\end{align}
where we've explicitly written out Weyl spinors, $\Psi_\text{Dirac} = (\chi,\ \bar\psi)^T$ to emphasize that our fields are chiral\footnote{In our notation, the bar distinguishes the $(1/2,0)$ and $(0,1/2)$ spin representations. These carry different indices, $\chi_\alpha$ and $\bar\psi^{\dot \alpha}$. See~\cite{Dreiner:2008tw} for an encyclopedic treatment of this formalism.}.
What happens when we apply $CP$ to this? The key point is that the fermion bilinear $\bar\Psi \Psi$ in Dirac notation\footnote{It is arguable whether or not the discussion is simpler using Dirac spinors. On the one hand the transformation properties are straightforward, but on the other hand one should technically also write out explicit chiral projection operators which end up not mattering. The original lecture was given in Dirac spinors. The choice to use Weyl spinors here was a source of heated debate between the authors.} is invariant under $CP$. In terms of Weyl spinors,
\begin{align}
	\chi_Q^i \tilde H \psi_U^j \stackrel{CP}{\longleftrightarrow } 
	\bar\chi_Q^i \tilde H^\dag \bar\psi_U^j \ .
\end{align}
The $CP$ conjugated Yukawa terms thus look like
\begin{align}
	(CP)\mathcal L_{\text{Yuk,} u} = y_{ij}\bar\chi_Q^i \tilde H^\dag \bar\psi_U^j + y^*_{ij}\chi_Q^i\tilde H\psi_U^j\ .
\end{align}
Note that we did \textit{not} act on the coefficient, which is just a number and does not transform. We conclude that the Lagrangian is $CP$-invaraint \textit{only} when $y=y^*$, i.e.\ when the Yukawa matrix is real. We could have \emph{forced} $y$ to be real by re-phasing our fields. The complete argument requires showing that one \textit{cannot} make this phase rotation, and in fact reduces to the fact that we've already shown from parameter counting that there is a physical phase left over. Experiments have verified that in the Standard Model this phase is non-zero. The key statement is this:
\begin{center}
	\textit{A physical complex parameter that is measured to be non-trivial implies $CP$ violation.}
\end{center}

So far we haven't mentioned flavor at all. Charge and parity are discrete symmetries of our field theory that come from spacetime symmetry. In a standard model, however, $CP$ violation \textit{always} comes with flavor violation. We could see this in our parameter counting by noting that we really \textit{needed} to have $N\geq 3$ flavors (i.e.\ $N\times N$ Yukawa matrices) in order to have a physical phase. 

Let's go back to the Wolfenstein parameterization (\ref{eq:CKM:Wolfenstein}). The interesting feature we noted at the end of the last subsection was that to leading order the only complex parameters show up in the 1--3 and 3--1 elements. In other words, to very good approximation, $CP$ violation \textit{only} occurs in interactions between the first and third generations. One can go to higher order in the expansion in $\lambda$ to get $CP$ violation in other elements, but clearly these effects will be suppressed by additional powers of $\lambda$.

\subsection{The Jarlskog Invariant}

Now that we're familiar with the existence of the $CP$-violating phase, we would like to be able to quantify it in a meaningful way that is manifestly basis-independent. What we need is some kind of invariant that identifies $CP$ violation. Such an object exists and it is called the \textbf{Jarlskog invariant}, $J$~\cite{Jarlskog:PhysRevLett.55.1039}. It is defined by
\begin{align}
	\text{Im}\left[V_{ij}V_{kl}V^*_{i\ell}V^*_{kj}\right] = J\sum_{mn}\epsilon_{ikm}\epsilon_{j\ell n}\ ,
\end{align}
where there is no sum on the left-hand side. In terms of our \CKM parameterizations, this corresponds to
\begin{align}
	J&= c_{12}c_{23}c_{13}^2 s_{12} s_{23} s_{13} \sin\delta_{\text{KM}}\label{eq:Jarlskog:standard} \approx \lambda^6 A^2 \eta \ .
 \end{align}
This parameterization-independent quantity that measures the amount of $CP$ violation in our model. The most remarkable observation is that it depends on \textit{every} physical mixing angle! Thus if \textit{any} of the mixing angles are zero, there would be \textit{no} $CP$ violation. This is another manifestation that one needs $N\geq 3$ flavors to have $CP$ violation and underlines the connection between flavor and $CP$. In fact, we can see that the amount of $CP$ violation in the Standard Model is small, but it is \textit{not} small because the $CP$ phase $\delta_{\text{KM}}$ is small. Quite on the contrary, it is small because of the mixing angles. We can see this in the Wolfenstein parameterization where the Jarlskog invariant comes along with six powers of $\lambda$.

\subsection{Unitarity triangles and \textit{the} unitarity triangle}
\label{sec:unitarity:triangle}

Using the unitarity of the \CKM matrix, we can write down equations for the off-diagonal elements of $V V^\dag$. For example,
\begin{align}
	\sum_{i=1}^3V_{id}V_{is}^* = 0\ . 
\end{align}
We have 6 such relations (three for the rows, three for the columns) and can plot each relation as a triangle in the complex plane. Each leg of the triangle is one term in the sum. These are called \textbf{unitarity triangles}. 

Some of these triangles are so flat that they are almost linear. Consider the example above:
\begin{align}
	V_{ud} V_{us}^* +  V_{cd} V_{cs}^* + V_{td} V_{ts}^* = 0.
\end{align}
Remembering the Wolfenstein parameterization, we can see that the first two terms are $\mathcal O(\lambda)$ while the third term is $\mathcal O(\lambda^5)$. The last term is very small and so one ends up with two (normalized) sides that are $\mathcal O(1)$ and one that is smaller by a factor of $\sim 10^{-3}$.

Let's look at the most interesting unitarity triangle. It is so interesting that it has a `special name,' we call it \textbf{\textit{the} unitarity triangle}\footnote{By now you've noticed that we are squeezing as much physics as we can from the linguistic distinction between the definite and indefinite articles `a' and `the.'}.
\begin{align}
	\sum_{i}V_{id}V_{ib}^* = 0. 
\end{align}
Doing the same order-of-magnitude estimate with the Wolfenstein parameterization,
\begin{align}
	V_{ud} V_{ub}^* +  V_{cd} V_{cb}^* + V_{td} V_{tb}^*
\end{align}
is a sum of terms that are \textit{each} $\mathcal O(\lambda^3)$. Each side is roughly the same order, so we can expect it to look more like a triangle than a line. It is customary to normalize the sides of the triangle. Let us divide by $V_{cd}V_{cb}^*$. The resulting relation is
\begin{align}
	\frac{V_{ud} V_{ub}^*}{V_{cd}V_{cb}^*} +  1 + \frac{V_{td} V_{tb}^*}{V_{cd}V_{cb}^*}=0.
\end{align}
How do we draw this? Treat each term in the sum as a vector on the complex plane. Recall from preschool that to sum complex numbers we can think of adding vectors by attaching the tail of each vector to the head of the preceding vector. So let's start at zero, add the unit term to get to $(1,0)$, and sum either of the remaining two terms to get the third point of the triangle. The third term brings you back to zero, of course. You end up with a nice triangle:
\begin{center}
	\begin{tikzpicture}[scale=9]
		\draw[line width=2] (0,0) -- (1,0) -- (77:.404) -- cycle;
		\begin{scope}
	    	\clip (0,0) -- (1,0) -- (77:.404) -- cycle;
	    	\draw[line width=1.5] (0,0) circle (.1);
			\draw[line width=1.5] (1,0) circle (.2);
			\draw[line width=1.5] (77:.404) circle (.1);
			\draw[->,line width=1.5] (.1,0) arc (0:38:.1);
			\draw[->,line width=1.5] (77:.304) arc (-103:-52:.1);
			\draw[-<,line width=1.5] (.8,0) arc (180:165:.2);
	  	\end{scope}
		\node at (-.05,-.05) {$\displaystyle{(0,0)}$};
		\node at (1.05,-.05) {$\displaystyle{(1,0)}$};
		\node at (77:.455) {$\displaystyle{(\rho,\eta)}$};
		\node at (35:.15) {$\displaystyle{\gamma}$};
		\node at (.75,.05) {$\displaystyle{\beta}$};
		\node at (.16,.26) {$\displaystyle{\alpha}$};
		\node at (-.1,.2) {$\displaystyle{ \left|\frac{V_{ud}V^*_{ub}}{V_{cd}V^*_{cb}}\right| }$};
		\node at (.6,.3) {$\displaystyle{ \left|\frac{V_{td}V^*_{tb}}{V_{cd}V^*_{cb}}\right| }$};
	\end{tikzpicture}	
\end{center}
The great thing about \textit{the} unitarity triangle is that because each side is $\mathcal O(1)$ relative to the others, it is robust against experimental errors.

\begin{prob}
	\textbf{A geometric interpretation of the Jarlskog invariant}. Show that the area of all six unitarity triangles are all exactly equal to $J/2$. This makes sense since each of the triangles gives the \textit{same} information about $CP$ violation and we argued that the parameterization-invariant quantity that captures this information is $J$. \textsc{Note}: this holds for unitarity triangles \textbf{before} normalizing one side to unit length.
	\begin{sol}
		This is straightforward: use (area) = (base) $\times$ (height). To be somewhat pedantic, take the following steps
		\begin{enumerate}
			\item Normalize the triangle so that one side has unit length in the real direction.
			\item The height of the triangle is given by the imaginary part of one of the other (normalized) terms in the sum. 
			\item Use the formula for the area of a triangle.
			\item Return to the previous normalization by multiplying this area by the absolute value of the side that we scaled and rotated, squared (because this is an area).
		\end{enumerate}
	\end{sol}
\end{prob}

\begin{prob}\textbf{Area of unitarity triangles}. Check the claim of the previous problem by writing out the areas of any two different unitarity triangles written with respect to the standard parameterization of the \CKM matrix. 
	\begin{sol}
		You are free to pick any two unitary triangles you want, but two of them are particularly easy. Looking at the standard \CKM parameterization (\ref{eq:CKM:standard:parameterization}) we see that $V_{ud}$, $V_{us}$, $V_{cb}$, and $V_{tb}$ are purely real. Thus there are two triangles that have one leg parallel to the real axis:
		\begin{align}
			\sum_i V_{id}V^*_{is} &= 0\label{eq:hw2:1:tri:1}\\
			\sum_i V_{cj}V^*_{tj} &= 0.\label{eq:hw2:1:tri:2}
		\end{align}
		Thus the base of the triangle (\ref{eq:hw2:1:tri:1}) is
		\begin{align}
			V_{ud}V_{us} &= c_{12}c_{13}s_{12}c_{13},
		\end{align}
		while the base of the triangle (\ref{eq:hw2:1:tri:2}) is
		\begin{align}
			V_{cb}V_{tb} &= s_{23}c_{13}c_{23}c_{13}.
		\end{align}
		Great. That's the easy part. To calculate the height it is sufficient to take the imaginary part of either of the remaining terms in the sum. For triangle (\ref{eq:hw2:1:tri:1}) we'll take $V_{cd}V^*_{cs}$
		\begin{align}
			\left|\text{Im}\left(V_{cd}V^*_{cs}\right)\right| &= \left|\text{Im}\left( s_{12}^2s_{13}c_{23}s_{23} e^{-i\delta} - c_{12}^2s_{13}c_{23}s_{23}e^{i\delta} \right)\right|\\
			&= \sin\delta \, s_{13}c_{23}s_{23}.
		\end{align}
		For triangle (\ref{eq:hw2:1:tri:1}) we'll take $V_{cs}V^*_{ts}$,
		\begin{align}
			\left|\text{Im}\left(V_{cs}V^*_{ts}\right)\right| &= \left|\text{Im}\left( -c_{12}s_{12}s_{13}c^2_{23}e^{-i\delta} + c_{12}s_{12}s_{13}s^2_{23} e^{i\delta} \right)\right|\\
			&= \sin\delta\, c_{12}s_{12}s_{13}.
		\end{align}
		Comparing the (base)$\times$(height) of each triangle, we find that both triangles have area equal to one half of
		\begin{align}
			c_{12}s_{12}s_{13}c_{13}^2c_{23}s_{23}\,\sin\delta.
		\end{align}
		This is indeed our definition of the Jarlskog, (\ref{eq:Jarlskog:standard}).
	\end{sol}
\end{prob}

This is a very famous triangle. There aren't many famous triangles in the history of humankind, but this is one of them, In the diagram above we gave the angles names, $\alpha,\beta,\gamma$. Note that there is physical significance to the direction with which we measure the angles; for example, if $(\rho,\eta)$ had been in the lower quadrant the signs would differ.  According to popular legend, these were assigned by Yossi Nir while he was at \acro{SLAC}. There's another parameterization that was also invented at \acro{SLAC} by B.J.~Bjorken,
\begin{align}
	\phi_1 = \beta \qquad\qquad\quad
	\phi_2 = \alpha \qquad\qquad\quad
	\phi_3 = \gamma.
\end{align}
These appear to have been popularized in Japan by Sanda so that now the flavor physicists at Belle use different conventions from those at BaBar and \acro{LHC}\textit{b}. 

In terms of the \CKM matrix parameters, the angles are given by
\begin{align}
	\alpha &= \text{arg}\left(-\frac{V_{td}V^*_{tb}}{V_{ud}V_{ub}^*}\right)
	&
	\beta &= \text{arg}\left(\frac{V_{cd}V^*_{cb}}{V_{td}V_{tb}^*}\right)
	&
	\gamma &= \text{arg}\left(\frac{V_{ud}V^*_{ub}}{V_{cd}V_{cb}^*}\right) \ .
\end{align}
 For completeness, we can also write out the length of the sides of the unitarity triangle in terms of these parameters,
\begin{align}
	R_u &= \left|\frac{V_{ud}V^*_{ub}}{V_{cd}V_{cb}^*}\right| = \sqrt{\rho^2+\eta^2}
	&
	R_t &= \left|\frac{V_{td}V^*_{tb}}{V_{cd}V_{cb}^*}\right| = \sqrt{(1-\rho)^2+\eta^2}\ .
\end{align}
The remaining side is, by normalization, set to unit length.

\subsection{Flavor Symmetries}
\label{sec:flavor:symmetries}

The lightest quarks are the up and the down. We have already identified the approximate SU(2) symmetry between these two states as isospin. The splitting in the up and down quark masses are a measure of isospin violation. One may also include the strange quark to promote this to an approximate SU(3) \textbf{flavor symmetry}, though this is even more approximate because of the larger mass splitting between the strange and the lighter two quarks. 

For modern students of particle physics, this classification seems very strange: we're taking three quarks---two with electric charge $-1/3$ and one with electric charge $+2/3$---and saying that they're somehow symmetric. In fact, as theorists we know that we're \emph{really} talking about \emph{six} different quarks because the Standard Model is chiral: the left-chiral up $u_L$ and the right-chiral up $u_R$ have very different quantum numbers, one is in an SU(2)$_L$ doublet and the other is not. In what sense do we have a right to talk about SU(3) flavor symmetry? It's almost as if this entire classification scheme were developed in a time before we even know quarks existed!

\emph{That}, of course, is the point. The history of SU(3) flavor symmetry is rich and may hold lessons for contemporary seekers of new physics. Long before we knew that quarks are real, we had phenomenologically identified the flavor structure of what is now the Standard Model. For a history, follow the development of the eightfold way~\cite{gell2000eightfold}. The lesson here is that these approximate symmetries, even if they're hidden behind strong dynamics, leave an imprint on the composite states that we \emph{can} measure. And indeed, in these lectures we will make ample use of SU(2) isospin and SU(3) flavor symmetry. 

Why stop there, then? Why not do flavor physics with SU(4) symmetry including the charm, or SU(6) symmetry including the bottom and top quarks? The reason is that these symmetries are \emph{too} approximate. Recall that a parameter is only small if it is dimensionless, and the dimensionless parameter that quantifies the approximation is $\delta m_q/\Lambda_\text{QCD}$. For quarks heavier than the strange, this parameter is too big to be helpful---though there are other ways to deal with the heavy quarks that we explore in the later sections of these lectures.

As model builders, we may also want to better understand the origin of flavor symmetry from the modern scaffolding of the Standard Model. If left- and right-handed quarks are different, what is it that's actually being transformed under SU(3) flavor? This comes from \textbf{chiral perturbation theory}, where SU(3) comes from the spontaneous breaking of SU(3)$_L\times$SU(3)$_R$ down to its diagonal subgroup. Here SU(3)$_L$ and SU(3)$_R$ are rotations among the left-handed light quarks and the right-handed light quarks respectively. The breaking is assumed to come from a \textbf{chiral condensate}, $\langle q\bar q\rangle$, coming from strong dynamics. The Goldstones of this breaking are the pions (for isospin) and kaons (for SU(3) flavor); these have small masses relative to $\Lambda_\text{QCD}$ because the global symmetry is only approximate. For a more complete discussion, see \cite{Donoghue:DynamicsSM, Csaki:2016kln}.

\begin{framed}
	\noindent\textbf{The $\text{SU}(2)$ subgroups of $\text{SU}(3)$}. Let's pause for some group theory. We know that $\text{SU}(2)$ isospin lives inside of the bigger $\text{SU}(3)$ flavor symmetry. In fact, we know that there are other $\text{SU}(2)$s that live inside $\text{SU}(3)$. We can see these by looking at the root diagram of the octet, which we reproduce below. Each dotted line corresponds to an $\text{SU}(2)$ subgroup of $\text{SU}(3)$. We can identify these $\text{SU}(2)$s as the familiar isospin ($u \leftrightarrow d$), $v$-spin ($u\leftrightarrow s$), and $u$-spin ($d\leftrightarrow s$).
	\begin{center}
		\begin{tikzpicture}
			\draw[line width=1.5] (0:2) -- (60:2) -- (120:2) -- (180:2) -- (240:2) -- (300:2) -- (0:2);			
			\draw[dashed, color=blue] (180:2.7) -- (0:2.7);	
			\draw[dashed, color=red] (240:2.7) -- (60:2.7);	
			\draw[dashed, color=green] (300:2.7) -- (120:2.7);	
			\draw[fill=black] (0:2) circle (.09cm);
			\draw[fill=black] (60:2) circle (.09cm);
			\draw[fill=black] (120:2) circle (.09cm);
			\draw[fill=black] (180:2) circle (.09cm);
			\draw[fill=black] (240:2) circle (.09cm);
			\draw[fill=black] (300:2) circle (.09cm);
			%
			\draw[fill=black] (0:.2) circle (.09cm);
			\draw[fill=black] (180:.2) circle (.09cm);
			\node at (0:2.6) {$\pi^+$};
			\node at (60:2.6) {$K^+$};
			\node at (120:2.6) {$K^0$};
			\node at (180:2.6) {$\pi^-$};
			\node at (240:2.6) {$K^-$};
			\node at (300:2.6) {$\bar K^0$};
			%
			\node at (270:.6) {$\eta$};
			\node at (80:.6) {$\pi^0$};
		\end{tikzpicture}
	\end{center}
	\noindent These were named by Harry Lipkin---author of an excellent representation theory text for physicists \cite{Lipkin:LieGroups}---as a pun: `I-spin, $u$-spin, $v$- (we) all spin for iso-spin.' While we know that $\text{SU}(2)$ isospin is a much better symmetry than $\text{SU}(3)$ flavor, both $\text{SU}(2)$ $u$-spin and $\text{SU}(2)$ $v$-spin are as badly broken as $\text{SU}(3)$ flavor since they both suffer from a breaking on the order of  $m_s/\Lambda_{\text{QCD}}$. Technically, $u$-spin has a \textit{small} advantage over $v$-spin since it is not broken by electromagnetism, but since $\alpha \sim 1\%$, this is a small concession compared to its $m_s/\Lambda_{\text{QCD}}$ breaking. Finally let us remark on the shift in mindset that we've been espousing: as theorists we like to think about the symmetry limit in such a way that we only care about the symmetry; this is wrong. The more constructive point of view is to think as \textit{phenomenologists}, where the main aspect that we care about the symmetry limit is how the symmetry is broken.
\end{framed}

\begin{prob} \textbf{Counting parameters in low-energy QCD}. Low-energy \QCD is a theory of the lightest three quarks: $u,d,s$. Split these up into three light left-handed quarks and three light right-handed quarks. Define the model according to the criteria in Section~\ref{sec:model:building} and perform a parameter counting analogous to Section~\ref{sec:paramter:counting:SM}.
	\begin{sol}
	Let us consider the low-energy Lagrangian for the strong force in which only the light quark species ($u$, $d$, $s$) are active. This is the underlying structure of Murray Gell-Mann's eightfold way for light hadron classification. The three ingredients for our model are:
\begin{enumerate}
	\item \textsc{Gauge group}: SU$(3)_\text{c}$. We ignore U$(1)_\text{EM}$ as a small perturbation.
	\item \textsc{Matter representations}: for each flavor ($u$, $d$, $s$) we have a left-chiral fundamental representation and a right-chiral fundamental representation. Thus we have:
	
	\begin{center}
		\begin{tabular}{rlcrl}
			$u_L$ & \;$\mathbf{3}$ &\quad\quad\quad& $u_R$ & \;$\mathbf{3}$\\
			$d_L$ & \;$\mathbf{3}$ &\quad\quad\quad& $d_R$ & \;$\mathbf{3}$\\
			$s_L$ & \;$\mathbf{3}$ &\quad\quad\quad& $s_R$ & \;$\mathbf{3}$.
		\end{tabular}
	\end{center}
	
	\item \textsc{Spontaneous symmetry breaking}: at \textit{tree level} there is none. At the quantum level SU$(3)_A$ is broken spontaneously by the \QCD chiral condensate, $\langle \bar{q}_L q_R + \bar{q}_R q_L \rangle $. 
\end{enumerate}

As before we write out the most general renormalizable Lagrangian,
\begin{align}
	\mathcal L = \sum_{i=u,d,s} \left(\bar{q}^i_L i\slashed{D} q^i_L + \bar{q}^i_R i\slashed{D} q^i_R\right) + \left(m_{ij} \bar{q}_L^i q_R^j + \text{h.c.}\right).
\end{align}
Note the important difference between this low-energy \QCD Lagrangian and the Standard Model: the flavor symmetry of the kinetic term allows one to rotate up, down, and strange quarks (of a given chirality) between one another! In the Standard Model this is prohibited because this mixes different components of SU$(2)_L$ doublets or, alternately, because this mixes particles of different charge. Do not confuse this U$(3)_L\times$U$(3)_R$ flavor symmetry with the U$(3)_L\times$U$(3)_R$ flavor symmetry of the Standard Model's quark sector which mixes the three \textit{generations} between one another. There are a lot of 3's floating around, make sure you don't mix them up.

It is useful to write down this flavor symmetry in terms of vector and axial symmetries,
\begin{align}
	U(3)_L\times U(3)_R &= U(3)_V\times U(3)_A,
\end{align}
where the vector (axial) transformation corresponds to 
\begin{align}
	q_L^i \to q_L'^i = U^i_{\phantom{i}j}q_L^j \hspace{3cm} q_R^i \to q_R'^i = U^{(\dag)i}_{\phantom{(\dag)i}j}q_R^j.
\end{align}
In other words, the vector symmetry rotates the left- and right-handed quarks in the same way while the axial symmetry rotates them oppositely. 

Now let's look at the mass terms of low-energy \QCD. We may diagonalize $m_{ij}$, find the eigenvalues, and write down the three physical mass parameters. Thus for a generic mass matrix we see that flavor is a good quantum number and the U$(3)_V\times$U$(3)_A$ symmetry of the kinetic term is broken down to U$(1)_V^3$ representing the phases of each flavor mass eigenstate. If, for some reason, the mass matrix is universal, $m\propto \mathbbm{1}$, then the mass terms break U$(3)_L\times$U$(3)_R$ to U$(3)_V$. We see that the mass terms always breaks the axial part of the kinetic term's flavor symmetry.

One might now ask the following clever question:
\begin{quote}
	\textit{Why should we include the mass terms in the \QCD Lagrangian at all since we know these come from the Yukawa sector of the Standard Model?}
\end{quote}
While this is true, we must remind ourselves that our model building rules tell us that we \textit{must} include all renormalizable terms that respect our gauge symmetries. For low-energy \QCD this means that we must include these mass terms. Now one might ask an even more clever question:
\begin{quote}
	\textit{Fine, we include these terms, but then we go out and measure them and they're very small. Shouldn't we still be able to ignore these terms since they are much smaller than the relevant mass scale, $\Lambda_{\text{QCD}}$?}
\end{quote}
Indeed! We know that $m_q \ll \Lambda_{\text{QCD}}$ for all active quarks at low energies, so the limit $m_{ij}\to 0$ (or alternately taking the dimensionless parameter $m_{ij}/\Lambda_{\text{QCD}}\to 0$) should be sensible. However, the point is that in the absence of masses the chiral condensate breaks the SU$(3)_A \subset$U$(3)_A$ and does so at the scale $\Lambda_{\text{QCD}}$ (recall that $\langle \bar{q_i}q_{j}\rangle \sim \Lambda_{\text{QCD}}^3\delta_{ij}$). Thus we really should assume that this symmetry is broken. One more clever retort:
\begin{quote}
	\textit{Fine! The SU$(3)_A$ should be broken either by mass terms or by the QCD condensate, but what about the remaining U$(1)_A\subset$U$(3)_A$?}
\end{quote}
If we do our parameter counting, we have $m_{ij}$ an arbitrary complex matrix with 18 real parameters. The breaking of U$(3)_V\to $U$(1)_V^3$ gives $3^2-3=6$ broken generators. SU$(3)_A$ is broken and gives us $3^2-1 = 8$ broken generators. Thus if we pretend that U$(1)_A$ is unbroken, we would have $18-14 = 4$ physical parameters. We know that the correct answer is 3, the three Dirac quark masses. What has happened?

The answer is that U$(1)_A$ \textit{is} broken: it is anomalous, i.e.\ broken by quantum effects. The U$(1)_A$ axial anomaly is notorious in the history in particle physics. For our purposes we must make the additional rule that at the \textit{quantum level}, anomalous symmetries are no good for counting parameters. If this is the case, then there's one less broken symmetry generator and so we expect there to be one more physical parameter relative to the classical analysis. Does such a parameter exist? Yes; it is precisely the non-perturbative $\Theta_{\text{YM}}$ term which transforms as a shift with respect to U$(1)_A$! 

Thus there are two ways of counting parameters:
\begin{enumerate}
	\item \textsc{Classically}: look only at the most general, renormalizable perturbative Lagrangian (i.e.\ without $\Theta$ terms) and perform the counting using the classical symmetries without any regard as to whether or not they are broken by anomalies.
	\item \textsc{Quantum mechanically}: consider the most general, renormalizable Lagrangian \textit{including} non-perturbative terms and only consider non-anomalous symmetries.
\end{enumerate}
Both are consistent as long as you \textit{stay within the regime of the description}, in other words, if you're counting classical (tree-level) parameters, then don't include quantum effects like anomalies and $\Theta$ angles. Conversely, if you're counting quantum parameters, then you must include \textit{both} the effect of anomalies and the non-perturbative terms associated with them. 
	\end{sol}
\end{prob}

\section{Tree-level FCNCs in the Standard Model}

\begin{quote}
	What did experimentalists see in the charged current and how did this lead to the construction of the Standard Model?
\end{quote}

\subsection{Charged versus neutral currents}

We've chosen a basis for the quarks where the only flavor off-diagonal interaction comes from the $W$-boson, which we know is charged. Thus the neutral boson interactions are flavor-conserving. Now let's switch gears and look at this from the experimental side. Suppose we don't know anything about the Standard Model, all we know are the experiments whose data is summarized in the \acro{PDG}. 

Let us `go through the data' and compare some charged current and
neutral current processes. In the following we use only one or two
significant digits and do not quote the errors.
\begin{eg}\textbf{Kaon decay}. Consider the following two decays:
	\begin{align}
		\text{Br}(K^+ \to \mu^+ \nu) = 64\% \hspace{3cm}
		\text{Br}(K_L \to \mu^+ \mu^-) = 7\times 10^{-9}.
	\end{align}
	Because the left-hand side of each decay is hadronic and the right-hand side is leptonic, we can determine that $K^+ \to \mu^+ \nu$ has a charged intermediate state (\textbf{charged current}) and that $K_L \to \mu^+ \mu^-$ has a neutral intermediate state (\textbf{neutral current}).
\end{eg}

What do we learn from this? Both processes change flavor, but the
charged current flavor-changing process proceeds at a much
\textit{much} larger rate---larger by eight orders of magnitude
(recall that the lifetime of the $K^+$ and $K_L$ are roughly the same). Na\"ively we would expect both processes to come from some flavor-violating structure so that they should both be of the same order of magnitude. Do we see the same pattern in other decays?
\begin{eg}\textbf{$B$ and $D$ decays}. Making use of the \acro{PDG}, we find for two particular $B$ decays 
	\begin{align}
		\text{Br}(B^-\to D^0\ell\bar\nu) = 2.3\%\hspace{3cm}
		\text{Br}(B^-\to K^{*-}\ell^+ \ell^-) = 5 \times 10^{-7}.
	\end{align}
	Similarly, for two particular $D$ decays,
	\begin{align}
		\text{Br}(D^\pm\to K^0\mu^\pm \nu) = 9\%\hspace{3cm}
		\text{Br}(D^0\to K^\pm\pi^\mp\mu^\pm\mu^\mp) < 5\times10^{-4}.
	\end{align}
	As one can check in the \acro{PDG}, experimentalists have not
        yet measured analogous flavor-changing neutral currents in the
        charm sector, which is why we can only quote an upper bound. \FCNCs in the charm sector have been measured in $D\bar D$ mixing~\cite{Aaij:2012nva}, though as of this writing \FCNCs from decays are a work-in-progress. 
\end{eg}
We indeed find the same pattern! The charged current interaction is much larger than the neutral current. 
This pattern \emph{only} shows up in flavor-\emph{changing} processes.
For the flavor-\text{conserving} processes, the weak interaction charged and neutral currents indeed occur at roughly the same rates. We leave it to look up demonstrative examples in the \acro{PDG}.
\begin{eg}\textbf{Isolating the weak interaction, part I}. One set of flavor-conserving processes are $\nu_e \, X \to e \, X'$ and $\nu_e \, X \to \nu_e \, X'$ where $X$ and $X'$ are nuclear states. Why would it be misleading to look at the rates for $e \, X \to e \, X'$? \textsc{Answer}: This is dominated by the electric coupling and so the overall rate will not tell us about interactions mediated by the weak force.
\end{eg}

\begin{prob}\textbf{Isolating the weak interaction, part II}. 
Explain how one can experimentally isolate the weak part of the $e \, X \to e \, X'$ rate. \textbf{Hint}: a general principle is, ``we can see very small things if they break a symmetry.''
	\begin{sol}
		Look up references on `atomic parity violation.' See, for example,  \cite{Bouchiat:Parity,Haxton:APV} for reviews. The main point is that part of the $Z$ coupling is axial, which means its coupling is proportional to a fermion's helicity. This is parity-odd unlike the electric force. The goal is then to devise experiments that are sensitive to this left--right asymmetry. This is rather non-trivial, and a na\"ive estimate of the size of such an asymmetry is \cite{Bouchiat:Parity}
		\begin{align}
			A_{\text{LR}} = \frac{P_L-P_R}{P_L+P_R}
		\end{align}
		where $P_{\text{L,R}} = |A_{\text{EM}}\pm A_{W}^\text{odd}|^2$. The weak amplitude $A_W \sim g^2/(q^2 + M_Z^2)$ where the characteristic momentum scale is given by the Bohr radius,
		\begin{align}
			q\sim \frac{1}{\alpha m_e }.
		\end{align}
		This gives us
		\begin{align}
			A_{\text{LR}} \approx \alpha^2\frac{m_e^2}{M_Z^2} \approx 10^{-15},
		\end{align}
		which is hopelessly small. Fortunately, there are various enhancement mechanisms which make the measurement of this quantity experimentally tractable. See Section 2.4 of \cite{Bouchiat:Parity} for an excellent discussion.
	\end{sol}
\end{prob}

The large discrepancy between the charged and neutral flavor-changing currents and the order-of-magnitude agreement in the flavor-conserving currents is a major experimental observation about the structure of the underlying theory. In fact, flavor-changing neutral currents (\FCNCs) are always a big issue when designing models beyond the Standard Model.

If we haven't emphasized it enough so far, \textit{familiarize yourself with the \acro{PDG}}. It is a phenomenologist's best friend. When \emph{we} want to measure something, our `experiment' is to open up the \acro{PDG} and look up the actual measured value. Take time to do this; experimentalists spent tens of years extracting and thinking about this information---you can afford to spend ten minutes doing the same thing.

\subsection{The possible sources of FCNCs}

In the Standard Model we know that the charged current (the $W$ boson) can change fermion flavors. So what about flavor-changing neutral currents? As a caveat in this section we will only discuss \textit{tree-level} \FCNCs; the appearance of loop-level \FCNCs will be a major topic later in these lectures. 
The relevant Standard Model bosons are the Higgs $h$, the $Z$, the
photon $\gamma$, and the gluon $g$. Why do these particles not have
any tree-level \FCNCs? We discuss each case below.

Before we do it, however, let's make an almost trivial remark. We know that flavor-changing
effects occur due to off-diagonal flavor couplings. Thus one might
want to say that couplings do not change flavor if they are
\textbf{diagonal}. This is generally \textit{not} true since
`diagonal' is a basis-dependent property. One can shift to a different
basis and a diagonal matrix will transform into a matrix with both
diagonal and off-diagonal parts. It is only a \textbf{universal}
matrix---those which are proportional to the identity---which are
diagonal in every basis. Usually when we talk about \acro{FCNC}s being
diagonal we implicitly assume the mass basis. We learn that 
\acro{FCNC}s
effects are protected against by having couplings that are universal
or diagonal in the mass basis.

\subsection{Photon and gluon FCNCs: gauge invariance}
The reason why the couplings of the photon and gluon to fermions are
diagonal is simple: gauge invariance. Gauge symmetry forces the
kinetic terms---where the gauge couplings live---to be universal. The
three down-type quarks all have the same gauge representations,
therefore they all have the same gauge interactions. 
This argument is valid in any \acro{UV} completion of the \acro{SM}. As
long as \acro{QCD} and \acro{QED} are unbroken, the photon and gluons are safe from
tree-level \acro{FCNC}s.

\subsection{Higgs FCNCs:  Yukawa alignment}

The Higgs couplings to the fermions comes from the Yukawa couplings. These are trivially diagonal in the fermion mass basis since they are proportional to the fermion mass matrices themselves. In other words, they are \textbf{aligned} with the mass matrices. For example, consider the down-type Yukawas,
\begin{align}
	y^d_{ij}\bar Q^i H D^j.
\end{align}
Expanding the Higgs $H$ about its vev, we obtain
\begin{align}
	y^d_{ij}\bar Q^i (v+ih) D^j.
\end{align}
Thus it is clear that diagonalizing the mass matrix $v\, y^d$
simultaneously diagonalizes the $\bar QhD$ coupling. This is one of
the examples when it is sufficient for a matrix to be diagonal in the
mass basis but not universal. This is very specific to the fact that a standard model only has one Higgs. Consider adding a second Higgs to the model in the most trivial way,
\begin{align}
	y^1_{ij}\bar Q_L^i (v_1+ih_1) D_R^j + y^2_{ij}\bar Q_L^i (v_2+ih_2) D_R^j.
	\label{eq:FCNC:vs:Yukawa:second:Higgs}
\end{align}
In general the fermion mass matrix is  $(y^1v_1 + y^2v_2)$ while the Higgs couplings are governed by $y^1$ and $y^2$ separately. One can see that now the Higgs coupling matrices are, in general, \textit{not} proportional to the fermion mass matrix and so these change flavor. 

\begin{eg}\textbf{Two Higgs Doublet Models}.
	This is \textit{not} the way we usually add a second Higgs in two Higgs doublet models (2\acro{HDM}). In such models the up- and down-type sectors couple to different Higgs bosons which each get an independent vev. In this case the Higgs couplings \textit{are} diagonal because they are aligned with the mass matrices. To provide some context, the most famous 2\acro{HDM} model is the \acro{MSSM} since supersymmetry requires at least two Higgses due to constraints from holomorphy and anomaly cancellation. Note that if you have four Higgses in the \acro{MSSM} you again generally find \FCNCs through the Higgs couplings.
\end{eg}

\begin{eg}\textbf{Experimental signature of Higgs FCNCs}.
	The model (\ref{eq:FCNC:vs:Yukawa:second:Higgs}) has explicit flavor-changing neutral currents in the quark sector. Where do you expect to see trouble? We'll narrow it down to a multiple choice question: do you expect more disagreement in the kaon sector or the $B$ sector? \textbf{Answer}: In this particular model there is some subtlety because the mass matrix can be very different from the couplings, but let us assume that this effect is not particularly perverse. Because the Higgs couples according to the Yukawas, it couples more strongly to massive particles. Thus we end up with a much stronger coupling to $B$ mesons (from the $b$ quark) than kaons (which only have an $s$ quark). This is an important lesson: there are some cases where \FCNCs in the kaon sector are negligible and offer model-builders some wiggle room.
\end{eg}

\subsection{$Z$ FCNCs: broken gauge symmetry}

Now what about the $Z$? This is a gauge boson, but it is a gauge boson of a \textit{broken} gauge symmetry so there is no reason to expect gauge invariance to protect against \FCNCs. We don't have \FCNCs at tree-level in the Standard Model, so something is still protecting the $Z$. 

Because the $Z$ is neutral, it only connects fermions with the same electric charge and color. Color is trivially satisfied since $\text{SU}(3)_c$ has nothing to do with $\text{SU}(2)_L\times \text{U}(1)_Y$. Electric charge, on the other hand, is related to the $\text{SU}(2)_L$ and $\text{U}(1)_Y$ charge by $Q=T_3+Y$. In the \acro{SM}, all quarks with the same electric charge $Q$ also have the same $T_3$ and therefore also the same $Y$. Recall that the $Z$ coupling only depends on these quantities,
\begin{align}
	g^{Zff} = g \cos\theta_W T_3 - g' \sin\theta_W Y.
\end{align}
Thus particles of the same charge all have the same coupling to the $Z$, i.e.\ the $Z$ coupling is universal for each of the up-type and (separately) down-type quarks. This is why the Standard Model $Z$ doesn't give flavor-changing neutral currents. To see this explicitly, consider the terms in the Lagrangian giving the $Z$ coupling to left- and right-handed up quarks,
\begin{align}
	\mathcal L_Z = \frac{g}{\cos\theta_W}\left[\bar u_L^i\gamma_\mu\left(\frac 12 - \frac 23\sin^2\theta_W\right)u_L^i + \bar u_R^j\gamma_\mu \left(-\frac 23\sin^2\theta_W\right)u_R^j\right]Z^\mu.\label{eq:Z:coupling:to:up}
\end{align}
To go to the mass basis, we perform a unitary rotation on the external fields. Let's just write out the $u_L$ term,
\begin{align}
	\mathcal L_Z = \frac{g}{\cos\theta_W}\left[\bar u_L V_{uL}\gamma_\mu\left(\frac 12 - \frac 23\sin^2\theta_Q\right) V_{uL}^\dag u_L\right] Z^\mu \ .
\end{align}
What is the structure of this term in flavor space? Everything in the parenthesis is universal. Thus the flavor structure is trivial,
\begin{align}
	V_{uL} V_{uL}^\dag = \mathbbm{1}.
\end{align}
and we have no FCNCs.

Let us write this in terms of a general principle,
\begin{theorem}\label{thm:broken:gens:and:FCNC}
	In order to completely prevent flavor-changing neutral currents in the gauge sector, particles with the same unbroken gauge quantum numbers must also have the same quantum numbers under the broken gauge group.
\end{theorem}

In the Standard Model we have no $Z$ \FCNCs because the particles of a given electric charge also have the same $\text{SU}(2)_L$ charge (namely $T_3$).
In general this needn't have been the case. Consider a slightly different model where the $d_L$ and $s_L$ quarks had different $\text{SU}(2)_L$ representations, but still the same electromagnetic and color charge.
\begin{align}
\mathcal L \supset 
	\bar{d_L} \left[ T_3^{(d)} - Q \sin^2\theta_W\right] \gamma^\mu d_L Z_\mu
	+  \bar{s_L} \left[ T_3^{(s)} - Q \sin^2\theta_W\right] \gamma^\mu s_L Z_\mu \ .
\end{align}
If $T_3^{(d)}\neq T_3^{(s)}$ then we would not be able to avoid \FCNCs.
\begin{eg}
	For example, consider the somewhat contrived case where $Q_{d,s}=0$, $T_3^{(d)}=1$, $T_3^{(s)}=0$. Thus the $s$ doesn't see the $Z$ and the flavor structure of the relevant term in the $Z$ coupling will be
	\begin{align}
		V_{dL}
		\begin{pmatrix}
			1 &0 \\
			0 & 0
		\end{pmatrix}
		V^\dag_{dL}.
	\end{align}
	This is clearly not universal.
\end{eg}

\begin{eg}\textbf{Why the charm shouldn't have been surprising}. The
  charm quark was discovered in the $J/\psi$ particle in 1974. The
  Cabbibo angle for two-generation mixing was understood in the 1960s,
  but charm wasn't even hypothesized until the 70s. Thus through most
  of the 60s it was well understood that there was a $(u,d)$ isospin
  (really $\text{SU}(2)_L$) doublet and a weird $s$ quark with the
  same charge as the $d$ but apparently living in an iso-singlet. This
  was a missed opportunity! We now see that this structure clearly has
  flavor-changing neutral currents and physicists should have expected
  to see $K_L \to \mu^+ \mu^-$. It wasn't until Glashow, Iliopoulos,
  and Maiani that physicists realized that the non-observation of this
  mode imples that there should be a charm quark to complete the
  doublet. By the time the $b$ was discovered, physicists had learned their lesson and realized immediately that there should also be a $t$ quark. (Or course they expected a $t$ that was not much heavier than the $b$; the story of this hierarchy is one of the great unsolved problems in flavor physics.) 
\end{eg}


\begin{prob}\textbf{Exotic light quarks, part II.}\label{prob:exotic:quarks:2}
	Recall the model in Problem~\ref{prob:exotic:quarks}, where we change the Standard Model quark sector by removing the $c$, $b$, and $t$ quarks and change the representations such that the light quarks are composed of an $\text{SU}(2)_L$ doublet $Q_L=(u_L,d_L)$, and singlets $s_L$ and $s_R$. The quantum numbers are assigned to maintain the same color and electric charge as the Standard Model. Answer the following questions:
	\begin{enumerate}
		\item Write out the gauge interactions of the quarks with the $Z$ boson in both the interaction and the mass bases. (Only write the terms that change when you rotate to the mass basis.) Are there, in general, tree-level $Z$ exchange \FCNCs?
		\item Are there photon and gluon mediated \FCNCs? Support your answer by an argument based on symmetries.
		\item Are there Higgs exchange \FCNCs?
		\item Do you expect this model to be experimentally ruled out or not? Why?
	\end{enumerate}
	\begin{sol}
		Refer to the solution to Problem~\ref{prob:exotic:quarks} for background. 
		\begin{itemize}
			\item In the interaction basis the couplings to the $Z$ are given by the usual formula
			\begin{align}
				g_z = g \cos\theta_W\, T^3 - g'\sin\theta_W \, Y.
			\end{align}
			From the particle content we see that the singlets $d_R$ and $s_L$ have the same $T^3$ and $Y$ quantum numbers. The other particles all have different quantum numbers and hence different couplings to the $Z$. From (\ref{eq:sol:exotic:quarks:matrix}) we see that the bare mass terms cause the mass matrix for the $d$ and $s$ quarks to be different from the Yukawa basis. Let us say that this mass matrix $M$ is diagonalized by $\hat M = W_L M W_R^\dag$. Then the rotations
			\begin{align}
				\begin{pmatrix}
					d'_{L,R}\\
					s'_{L,R}
				\end{pmatrix}
				=
				W_{L,R}
				\begin{pmatrix}
					d_{L,R}\\
					s_{L,R}
				\end{pmatrix}\label{eq:prob:exotic:II:rotation}
			\end{align}
			shift the interaction basis fields ($d,s$) to the mass basis fields $d',s'$. In terms of these fields, the coupling to the $Z$ in the kinetic term is written as
			\begin{align}
					\begin{pmatrix}
						\bar d'_{L,R} &
						\bar s'_{L,R}
					\end{pmatrix}
					W_L^\dag
						\begin{pmatrix}
							g_{Z}^{d_{L,R}} &\\
							& g_{Z}^{s_{L,R}}
						\end{pmatrix}
						W_R
						i\gamma^\mu
						Z_\mu
						\begin{pmatrix}
							d'_{L,R}\\
							s'_{L,R}
						\end{pmatrix}.\label{eq:prob:exotic:II:coupling}
			\end{align}
			The key point is that while $g_Z^{d_R} = g_Z^{s_R}$, i.e.\ the right-chiral coupling matrix is diagonal, $g_Z^{d_L}\neq g_Z^{s_L}$ so that the left-chiral couplings become non-diagonal after the bi-unitary rotation by $W_L$ and $W_R$. Thus there \textit{are} \FCNCs in the $Z$ coupling to the left-chiral down quark sector. 
			\item There are no photon or gluon \FCNCs. The photon and gluon couplings are all universal with respect to particles within a given flavor representation. 
			\item Because the Yukawa couplings and the mass matrix (\ref{eq:sol:exotic:quarks:matrix}) are not proportional to one another (i.e.\ not aligned) there are \FCNCs from the Higgs. Note that these can be easy to `hide' since the Higgs couplings to the light quarks is small. 
			\item We now know that this model has tree-level \FCNCs. As a quick and dirty approximation we can expect that the magnitude of the flavor-changing neutral currents and charged currents should be of the same order of magnitude. However, looking at the PDG, we can consider characteristic kaon decays (kaons because the new flavor structure is only in the $s$-$d$ sector),
			\begin{align}
				\text{Br}(K^+ \to \mu^+ \nu_\mu) &= 64\%\\
				\text{Br}(K^0_L \to \mu^+\mu^-) &=  7 \times 10^{-9}.
			\end{align}
			We know that the leptons only couple to the bosons through the $\text{SU}(2)_L$ gauge bosons, so the first decay proceeds through the charged current and the second proceeds through the neutral current. It would be very difficult to explain this discrepancy if  a model has tree-level \FCNCs and hence we expect this model to be ruled out.
		\end{itemize}
	\end{sol}
\end{prob}

\begin{prob}\textbf{Exotic light quarks, part III.}\label{prob:exotic:quarks:3}
	Repeat the analysis for Problems~\ref{prob:exotic:quarks} and \ref{prob:exotic:quarks:2} for a modified exotic light quark model where the $u_R$ and $d_R$ are part of an $SU(2)_L$ doublet. 
	\begin{sol}
		The Yukawa and mass terms in the Lagrangian for this model takes the form
		\begin{align}
			\mathcal L_{\text{Yuk.}+\text{mass}} = y_L \bar Q_L \phi s_R + y_R \bar Q_R \phi s_L + m_Q \bar Q_L Q_R + m_s \bar s_L s_R + \text{h.c.}
		\end{align}
		These terms break $\text{U}(1)^4$ flavor symmetry in the kinetic terms to a $\text{U}(1)$ overall phase rotation on each field. We thus expect a total of $4-0 = 4$ physical real parameters and $4-3=1$ phases. The Dirac mass matrix takes the form
		\begin{align}
			\begin{pmatrix}
				m_Q & y_L \frac{v}{\sqrt{2}}\\
				y_R \frac{v}{\sqrt{2}} & m_s
			\end{pmatrix},
		\end{align}
		while the up-quarks have a single mass term $m_Q \bar u_L + \bar u_R + \text{h.c.}$. We may choose the real physical parameters to be the Dirac masses for each generation and the $d$--$s$ mixing angle and the physical phase to be that of the up quark mass term. The coupling to the $Z$ boson in the interaction and mass bases follow as it did in the previous problem. In particular, in the mass basis one must perform a rotation (\ref{eq:prob:exotic:II:rotation}) so that the $Z$ couplings take the form (\ref{eq:prob:exotic:II:coupling}). Now both the left- and right-handed $d$ and $s$ quarks have different $\text{SU}(2)_L\times U(1)_Y$ quantum numbers so that $g_Z^d \neq g_Z^s$ and so the rotation introduces FCNCs through the $Z$ for both chiralities. 
		
		As before the photon and gluon still do not mediate \FCNCs because their couplings are all universal with respect to particles within a given flavor representation. The Higgs again induces \FCNCs because the bare mass terms prevent the mass matrix and Yukawa matrix from being proportional to one another. The appearance of \FCNCs in the $Z$ and Higgs sectors lead us to expect this model to be ruled out experimentally.
	\end{sol}
\end{prob}

\begin{prob} \textbf{The two Higgs doublet model.} Consider the two Higgs doublet model (2\acro{HDM}) where the Standard Model is extended by an additional Higgs doublet. We label the two Higgses $H_1$ and $H_2$ with $\text{SU}(2)_L\times \text{U}(1)_Y$ representations $\mathbf{2}_{-1/2}$ and ${\mathbf{2}}_{1/2}$. For simplicity we will work with two generations. 
	\begin{itemize}
		\item Write down the most general Yukawa potential for the quarks.
		\item Carry out the diagonalization procedure for such a model. Show that the $Z$ couplings are still flavor diagonal.
		\item Show that in general the Higgs mediates \FCNCs. 
		\item Is this model experimentally viable?
	\end{itemize}
	\begin{sol}
		For a nice history and collection of references about the 2\acro{HDM}, see \cite{Haber:2000jh}.
		\begin{itemize}
			\item The most general Yukawa potential takes the form
			\begin{align}
				\mathcal L_{\text{Yuk.}} = (y^{1u}_{ij}H_1 + y^{2u}_{ij}\tilde H_2)\bar Q^i u^j + (y^{1d}_{ij}\tilde H_1+y^{2d}_{ij}\tilde H_2)\bar Q^i  d^j + \text{h.c.}
			\end{align}
			\item The diagonalization procedure proceeds as usual for the Standard Model except that the mass matrices are now
			\begin{align}
				m_{ij}^u = y^{1u}_{ij}\frac{v_1}{\sqrt{2}} + y^{2u}_{ij}\frac{v_2}{\sqrt{2}}
				\hspace{3cm}
				m_{ij}^d = y^{1d}_{ij}\frac{v_1}{\sqrt{2}} + y^{2d}_{ij}\frac{v_2}{\sqrt{2}}.
			\end{align}
			These are diagonalized by some bi-unitary transformation. As far as the kinetic terms are concerned, this rotation on the quark fields has the same effect as the rotation in the Standard Model. In particular, the $Z$ couplings are still universal and remain flavor-diagonal.
			\item It is straightforward to see that there are \FCNCs in the Higgs sector. For example, the neutral scalar Higgs components couple to the fermions by replacing $v_i/\sqrt{2}$ in the mass matrices with $h_a$ for $a=1,2$ labeling each the two Higgs doublets. It is clear that in general the mass matrix is different from the Yukawa matrix for either Higgs since, e.g.\ $m^u_{ij} \neq y^{au}_{ij}$ for either $a=1,2$. There are, of course special cases; for example if these Higgses are allowed to mix in such a way that the mass eigenstate couples to combinations of Yukawas proportional to $m^u$ and $m^d$. We can also note that in the limits $\tan \beta \equiv v_2/v_1 \to 0, \infty$ the Yukawas align with the mass matrices because one of the vevs vanishes. It is generally true in any multi-Higgs model that \FCNCs are avoided in the Higgs sector so long as all the fermions of a given electric charge only couple to a single Higgs.
			\item Higgs \FCNCs are typically stronger for heavier quarks, though in this 2\acro{HDM} one can arrange for a cancellation between large couplings in the Yukawas. A natural place to rule out this type of is in the \FCNC constraints on the $B$ mesons.
		\end{itemize} 
	\end{sol}
\end{prob}

\section{Loops and the GIM mechanism}
\label{sec:loops}

So far we've laid down the theory of tree-level flavor violation in
the Standard Model. In the mass basis the $W$ changes flavor, the other bosons do not. Pretty simple. As is often the case, the story is more subtle at loop level. 
We now move on from tree-level to loop-level processes\footnote{\acro{Y.G.} makes an obligatory reference to the circus trees in Gilroy which are grown in such a way that their trunks form loops, \url{https://www.gilroygardens.org/play/circus-trees}. \acro{Y.G.} once spent fifteen minutes looking at photos of these trees in his office saying `\textit{Wow! Come look at this!}' to any student who passed by. }. 
Loop-level contributions are usually small---they are suppressed by $\sim(\text{coupling})^2/16\pi^2$ relative to tree-level diagrams for the same processes. A loop correction to a tree-level process is hard to observe, while a small effect which is the only source of signal has at least a chance of being measured well. A natural set of processes to look are those that generate flavor-changing neutral currents since we already know that these vanish at tree-level in the Standard Model.

One of the well-known morals of flavor physics at loop level is the \textbf{GIM mechanism}, named after Glashow, Iliopoulos, and Maiani. The underlying principle is that for a unitary matrix, any pair of columns (or rows) are orthogonal. Most readers will already be familiar with the \GIM mechanism from their courses and reading in particle physics. The first non-trivial point is that loop-level \FCNCs are most sensitive to the \textit{heavy} (!) quarks running in the loops, so that we can be sensitive to the top. 

In loop-level processes, particles go off shell and we then expect the \textbf{Appelquist-Carazzone decoupling theorem} to hold \cite{Appelquist:1974tg}. This tells us that when we take the mass of the internal particle to infinity, its effects on physics at much lower scales must vanish. But now we are saying that the heavy particles in \FCNC loops do \textit{not} decouple. In fact, the result of the amplitude is a function of the form $f({m_t^2}/{m_W^2})$, which goes to a constant for large values of its argument. What's going on?

\subsection{Example: $b\to s\gamma$}

	A classic example is the loop-level diagram for $b\to s\gamma$, which was first observed at Cornell, about halfway through \acro{CLEO}'s lifetime. This was when \acro{Y.G.} was in grad school and his first paper was on this so-called \textbf{penguin diagram}. The curious etymology of this process is best explained in John Ellis' own words, as cited in Shifman's introduction to the \acro{ITEP} Lectures in Particle Physics~\cite{Shifman:1995hc}. Here's an example of a penguin diagram,
	\begin{center}
		\begin{tikzpicture}[line width=1.5 pt, scale=1]
			\draw[fermion] (-1.75,0) -- (-1,0);
			\draw[fermion] (1,0) -- (1.75,0);
			\draw[fermion] (-1,0) arc (180:135:1);
			\draw[fermion] (135:1) arc (135:45:1);
			\draw[fermion] (45:1) arc (45:0:1); 
			\draw[vector] (1,0) arc (0:-180:1);
			\draw[vector] (45:1) -- (45:2);
		%
		\begin{scope}[rotate=135]
		\begin{scope}[shift={(1,0)}] 
			\clip (0,0) circle (.175cm);
			\draw[fermionnoarrow] (-1,1) -- (1,-1);
			\draw[fermionnoarrow] (1,1) -- (-1,-1);
		\end{scope}	
		\end{scope}
		%
		\end{tikzpicture}	
	\end{center}
	The dependence of $b\to s\gamma$ on the \CKM matrix is
	\begin{align}
		\mathcal M \propto \sum_i V^*_{ib} V_{is} = 0\ ,
	\end{align}
	where we sum over the internal quarks, $i$. This product vanishes identically by unitarity. 
	This tells us that any part of the diagram that \emph{only} depends on flavor through the sum $\sum_i V^*_{ib} V_{is}$ will vanish. 
	The only other source of flavor-dependence are the masses of the quarks themselves---recall that the quark mass matrix is diagonal, but not universal. Any term in $\mathcal M$ that is independent of the internal quark mass must necessarily vanish. In the diagram above, we draw this as a chirality-flipping mass insertion. We conclude that
	\begin{align} \label{eq:GIM-loops}
		\mathcal M = \sum_i V_{ib}^* V_{is} f(m_i).
	\end{align}
We can expand $f$ in a power series. If $m_i\ll m_W$ then $f(m_i\ll m_W) \propto m_i^2/m_W^2$. The $W$ indeed satisfies the decoupling limit. When we take $m_i$ large (say $m_i=m_t$), however, the argument is not valid. Thus we know that $f(m_i^2/m_W^2)$ is linear at small values and constant at large values. The function $f$ is called the \textbf{Inami--Lim} function, because they calculated all of these diagrams in the 1980s---it wasn't until the 80s that anyone really believed that the top quark might be so heavy~\cite{Inami:1980fz}!

The reason why decoupling is violated is that the diagrams are essentially mediated by the longitudinal part of the $W$ by the Goldstone Equivalence theorem. Thus the coupling to the fermions is essentially a Yukawa coupling, which goes like the fermion mass. When you have a particle whose coupling is proportional mass, then it is clear that decoupling fails---the heavy mass is compensated by a large coupling.
Thus the \GIM mechanism tells us that one-loop diagrams carry factors of $m_i^2/m_W^2$, where $i$ is summed over the internal quarks.

\begin{eg}
\noindent\textbf{Finiteness of penguins}. 	Penguin amplitudes are
manifestly finite. One na\"ive argument for this is that any
divergences are independent of the internal quark mass (say, using
dimensional regularization) so that unitarity kills any
divergences. This, however, obfuscates deeper reasons why the
amplitude is finite. One heuristic argument that the loop level must
be finite since symmetries preventing \FCNCs prevent any tree-level
counter-term. The detailed reason for the finiteness is that there are two sources of suppression\footnote{This kind of double suppression happens all the time in model building. For example, in little Higgs theories there's a sense of \emph{collective symmetry breaking} where dangerous parameters can be protected by multiple symmetries, see~\cite{Csaki:2016kln} for a nice review. At Cornell there are two kinds of ``double protection.'' One has to do with a supersymmetric composite Higgs~\cite{Csaki:2005fc}. The other is the following quote from \acro{Y.G.}, when \acro{P.T.} asked for a copy of his hand-written lecture notes: ``Well, they probably won't be useful to you. They're written in Hebrew. Also I lost them.''} that reduce the superficial degree of divergence for these diagrams:
	\begin{enumerate}
		\item Gauge invariance, in the form of the Ward identity, requires the amplitude to depend explicitly on the external momentum. This is explained very clearly in \cite{Lavoura:2003xp}.
		\item Lorentz invariance prevents divergences which are odd in the loop momentum, $k$. In other words, $\int d^4k \slashed{k}/k^{2n} =0$. Since the leading order contribution to $b\to s\gamma$ is even in $k$, the next-to-leading term is odd and vanishes.
	\end{enumerate}
	As a final remark on this, one can also argue that the \textit{chiral structure} of this process provides a suppression mechanism. The penguin is mediated by a $\sigma^{\mu\nu}F_{\mu\nu}$ operator that requires an explicit mass insertion. This turns out to be equivalent to gauge invariance.	
\end{eg}

\subsection{History of the GIM mechanism}

In the early days of the quark model, physicists thought there were only three quarks $u,d,s$. The SU(3) flavor symmetry between these seemed to describe the light hadronic states well and there wasn't any motivation for the fourth quark. The \GIM mechanism gave a motivation: it is a way to avoid \FCNCs at tree-level. Recall that if someone with no knowledge of the Standard Model stared at the \acro{PDG} for a long enough time, that person might realize that it was very curious that neutral current processes like $K_L \to \mu^+\mu^-$ were heavily suppressed relative to the charged current processes.
In the 60s people couldn't reach the precision to see this decay while it's charged current cousin was readily measurable. \GIM gives a way to understand this: the neutral current process is suppressed by a loop factor \emph{and} by a \GIM factor,
\begin{align}
	\mathcal M \sim \frac{g^2}{16\pi^2}\frac{m_c^2}{m_W^2}.
\end{align}
%
Gaillard and T.D.~Lee calculated $K\bar K$ mixing, keeping and dropping factors of two somewhat haphazardly as theorists are wont to do. Their calculations gave a prediction the mass of the $c$, $1.5$~GeV which is remarkably close. It turns out that---like many good theorists---Gaillard and Lee had luck on their side, since if one follows their calculation somewhat more honestly, one obtains a value between $0.5$ and $10$. The charm was finally observed in 1974 at \acro{SLAC} and Brookhaven.

Which quark dominates the mixing? It seems we have already spoiled the answer since we said that heavy quarks dominate in loops, but we should be careful since there's also a \CKM suppression associated with the top so that there is a competition between the large mass and the small \CKM elements. Comparing the loops with an internal top and charm,
\begin{align}
	\mathcal M_t &\sim m_t^2 V_{td}V_{ts}
	&
	\mathcal M_c &\sim m_c^2 V_{cd}V_{cs}.
\end{align}
It turns out that the charm wins but that the top amplitude is in the same ballpark. So in this sense Gaillard and Lee got a little bit lucky once again, since back then nobody would have believed that the top was so heavy that it might challenge the charm contribution.

\begin{eg}\textbf{Up-type versus down-type quarks}.
	What can we say about it $D\to \pi \gamma$? What about $t \to c \gamma$? These should all be small. Loop-level processes with external down-type quarks (and hence internal up-type quarks) are carry factors of $m_t^2/m_W^2$ which are not small at all. For processes with external up-type quarks, on the other hand, the \GIM mechanism is very efficient because $m^2_{d_i}/m_W^2$ is small for any down-type quark $d_i$. In other words, the fact that the $s$ and $b$ are much lighter than the $c$ and $t$ means that \FCNCs are more visible in the down quark sector. 
\end{eg}

\begin{framed}
\noindent\textbf{A poor expansion parameter}. What should we
        make of the fact that $m_t/M_W$ is a poor expansion parameter?
        For box diagrams contributing to $B\to X_s \gamma$ where the top
        is indeed the dominant fermion running in the loop, it may
        seem surprising that the approximation of the Inami--Lim
        function, defined in Section~\ref{sec:loops}, {$f(m_t^2/M_W^2)\approx m_t^2/M_W^2$} is still good even though the expansion parameter is larger than one. It turns out that this result is only off by $\mathcal O(4)$. You can check the plot of the relevant function in Fig.\ 5a of the original paper~\cite{Inami:1980fz}, from which one can see that the Inami--Lim function is indeed reasonably close to the linear approximation even for large values of its argument.
\end{framed}

\begin{eg}\textbf{FCNCs in the lepton sector}.
	\GIM is \textit{most} effective in the lepton sector. Given that neutrinos have mass, we expect to have processes like $\mu \to e \gamma$ (the leptonic analog to $b\to s \gamma$). In the Standard Model, this rate goes like $m_\nu^4/m_W^4 \sim (10^{-1}/10^{11})^4 \sim 10^{-48}$. Loop factors and couplings give a few more powers, leading to around $10^{-52}$. \GIM effectively kills leptonic \FCNC at one loop.
\end{eg}

\section{Connecting to experiment: meeting the hadrons}

The \acro{PDG} may intimidate you with the many long lists of
particles and measurements. You may worry that this aspect of particle
physics reduces to memorizing the periodic table or the botanical
names of flowers. Most of the particles listed in the \acro{PDG} are
hadronic spectra. There is sometimes a more pronounced distaste for
these hadrons from theory students. Experimentalists very quickly
familiarize themselves which the particular of particles that appear
in different parts of a detector. Theorists, on the other hand, tend
to want to live in a magical universe where Lagrangians are written
not in terms of hadrons but in terms of `fundamental' constituents
like $u$ and $d$ quarks. A major theme of flavor physics is to be able
to pull one's head out of the \acro{UV} to be able to make the
transition between `fundamental' Lagrangians and the actual measurable
physical states at low energies. This transition is far from trivial
and a good theorist must be able to go back and forth between the the
Standard Model Lagrangian, low energy spectra, and---hopefully---a new
physics Lagrangian. It will be useful to have the \acro{PDG} (booklet
or website) nearby. 
A refresher of the topics in this section are the undergraduate textbooks by Griffiths \cite{Griffiths:2008fk} and Perkins \cite{Perkins:2000uq}.

As you have known since kindergarten, \acro{QCD} states at low energies are called \textbf{hadrons}. We can distinguish between \textbf{mesons}, which are bosons, and \textbf{baryons}, which are fermions. With that out of the way, we can move on to slightly less trivial things.

\subsection{What we mean by `stable'}
\label{sec:stable}

Let us highlight an important abuse of language. We divide hadrons between \textbf{stable} particle and \textbf{unstable} particles, where the latter are often called \textbf{resonances}. These do \text{not} refer to stability in an absolute sense. Rather, they refer to whether a particle is \emph{stable on detector time scales}. Recall that there are two ways to measure an intermediate particle's lifetime: (1) one can directly measure its decay width or (2) for sufficiently long-lived particles, one can measure a displaced vertex. Stable particles are those whose lifetimes are at least large enough to be measured by the latter method. 

A more practical definition is that a stable particle is one that does not decay through \acro{QCD} interactions. In this sense a kaon is a stable meson; it certainly decays, but these channels are only through the weak interaction. In fact, let's make a somewhat fancy definition that we justify later:
\begin{definition}\label{def:stable:particle}
	A \textbf{stable particle} is one that is either an eigenstate
        of the Hamiltonian or that only decays through weak
        interactions. In other words, it is an eigenstate of the
        Hamiltonian in the limit $g,g'\to 0$
\end{definition}
\noindent Resonances are defined to be those particles that are not stable. We can now connect the `intuitive' definition of stability with this semi-technical definition by looking at the characteristic lifetimes of hadrons that decay via the strong versus weak forces.

\begin{eg}
	Consider the $\rho$ meson. This is a resonance of up an down quarks. What is its approximate mass and width?  Since the $\rho$ is composed of light quarks, the dominant contribution to its mass comes from \QCD.  We thus expect the mass to be on the order of $\Lambda_{\text{QCD}}$, which is hundreds of MeV. What is its width? The couplings are order 1 so we estimate it to again be $\Lambda_{\text{QCD}}$. Now go ahead and look it up in the \acro{PDG}. Unless we specify an excited state we will mean the lowest-mass particle by that name, so in this case, we mean the $\rho(770)$. We find $m_\rho \approx 775$ MeV and $\Gamma_\rho \approx 150$ MeV. Not bad.
\end{eg}

The $\rho$ width is very large, just a factor of five smaller than its mass. In fact, some resonances have a width that is so large---on the order of their mass---that it becomes very hard to determine whether or not it is a particle at all. Recall that when we do \acro{QFT} we work with states in the asymptotic past and future. When a particle's decay width is on the order of its mass, the notion of asymptotic state becomes ill defined. Such resonances decay very quickly.

What about the stable particles? 
\begin{eg}
	[This example about useful numbers that you should memorize.] What about the $B^0$ meson? \small{\textit{We have transcribed the following as it was dictated in lecture.}}
	\begin{quote}
		When I was young, I went to \acro{TASI}, and we started hiking. I saw this sign, and the sign said, ``Mile High City'' and ``5280 feet.'' You know, I couldn't breathe for a few seconds! That's \textit{exactly} the mass of the $B$ meson in MeV! Exactly, up to four digits. Okay? It's always much easier to remember two things than it is to remember one thing. Since most of you don't even know how many feet there is in one mile, now you have two things to remember: the number of feet in one mile is exactly mass of the $B$ meson in MeV.
	\end{quote}
	What about the lifetime of the $B^0$ meson? It decays weakly
        and turns out to have a lifetime of $\tau_{B} \approx 1.5 \times 10^{-12}$ seconds (about a picosecond). A useful conversion is 
	\begin{align}
		1 \text{ GeV} \approx 1.5 \times 10^{24} \text{ sec}^{-1}.\label{eq:GeV:seconds}
	\end{align}
	From this one ends up with a decay width of $\Gamma_B \approx 4 \times 10^{-4}$ eV.
\end{eg}

\begin{prob}
	Derive (\ref{eq:GeV:seconds}) if you are not already familiar with it.
	\begin{sol}
		This is a conversion to natural units. A quick-and-dirty way to get this is to use the Heisenberg relation $\Delta E \Delta t \sim \hbar$.
	\end{sol}
\end{prob}

Now we see the difference between a resonance and a stable particle. Consider the order of magnitudes of the widths of the $\rho$ and the $B^0$:
\begin{align}
	\Gamma_B \sim&\; 10^{-4} \text{ eV}
	&
	\Gamma_\rho \sim&\; 10^{8} \text{ eV}\ .
\end{align}
\textit{This} is why we say the $B$ is stable. You should also have intuited the significance of the term resonance, since these particles tend to be \acro{QCD} excited states of a given quark content. It is worth remembering that this big difference between the weak and the strong interaction doesn't come from the $SU(2)_L$ coupling constant being much smaller than the strong coupling---it's not; it comes from the virtual $W$ that is emitted by a $b$ quark in the decay of the $B$ meson.

A nice analogy is tritium: this is a hydrogen isotope with two neutrons. The excited 2P tritium state will emit a photon to decay to the 1S ground state at a time scale on the order of a nanosecond. The 1S state will eventually decay into helium-3 with a lifetime on the order of 18 years. One can similarly imagine an excited (heavy) $B$ resonance decaying rapidly to the $B^0$ by emitting a pion and then the $B^0$ decaying via the weak interaction on a much longer time scale. 

One should be comfortable thinking about the plethora of hadrons as excited hydrogen atoms in basic quantum mechanics. In fact, let us turn to one of the basic features of hydrogen in quantum mechanics: the quantum numbers that describe a state.

\begin{framed}
	\noindent\textbf{A PDG hint}. In the \acro{PDG}, particles whose name contains their mass in parenthesis, e.g.\ $\rho(770)$, are resonances while those that do not are stable.
\end{framed}

\subsection{Hadron quantum numbers}

We consider three kinds of hadronic quantum numbers:
\begin{enumerate}
	\item Those that are exact.
	\item Those that are exact under \QCD and \QED, but not the weak interactions
	\item Those that are approximate, even under \QCD.
\end{enumerate}

There are two exact quantum numbers for hadrons: electromagnetic
\textbf{charge} $Q$ and \textbf{spin} $J$, by which we mean the sum of
the constituent particle spins and the orbital angular momenta. Since
we are interested in phenomena like particle mixing, these quantum
numbers are especially important because they give strict rules about
which particles can mix. If two particles have the same quantum
numbers then in general they mix, i.e.\ the state which commutes with
the Hamiltonian is generically a linear combination of all such
states. However, states with different exact quantum numbers \textit{cannot} mix. 
\begin{eg}
The $\rho$ meson and the photon mix. They are both electrically neutral, spin-1 states. The massless Hamiltonian eigenstate that we call the photon actually has a small component of transverse $\rho$ meson, which you can think of as a loop of the constituent quark--anti-quark pairs correcting the photon propagator.  In the early days of hadronic physics, this mixing was proposed as the reason for the electromagnetic interactions of hadrons under an ansatz called vector meson dominance~\cite{Sakurai:1960ju, OConnell:1995nse}.
\end{eg}
\begin{eg}
The axion, $a$, is a proposed pseudoscalar particle that solves the strong \acro{CP} problem: why is it that the instanton-induced $\Theta_\text{QCD}$ phase is experimentally measured to be small, even though theoretically it could be any angle in the Standard Model. The axion couples to quarks and one of the favorite search strategies is to look for its coupling to electromagnetism through the $a\gamma\gamma$ vertex. This vertex can be understood as coming from the mixing of the axion and the neutral pion. Recall that the pion famously decays to two photons from the chiral anomaly. In this point of view, the axion inherits this two photon interaction.
\end{eg}

\begin{eg}
Do the $\pi$ and $\rho$ mesons mix? No---they have different spin and spin is an \textit{exact} symmetry and \textit{must} be conserved.
\end{eg}
In order for particles to mix they needn't have exactly degenerate masses. Of course, having a small mass difference certainly helps\footnote{This is better stated as a comparison of the diagonal versus off-diagonal elements of the mass matrix. A nice related undergraduate-level quantum mechanics question is to ask the validity of degenerate perturbation theory when the states have a small mass splitting. For a given mass splitting, how big are the corrections to degenerate perturbation theory from non-degenerate perturbation theory and how do these compare to the corrections from the next order in degenerate perturbation theory?}. 

Next let us consider the quantum numbers that are not quite exact but are at least exact with respect to \QCD. These quantum numbers are \textbf{parity} and \textbf{charge conjugation}. We will be particularly interested in their combination, $CP$. These are $\mathbbm{Z}_2$ discrete symmetries. Mesons in the \acro{PDG} are denoted by their spin, parity, and charge conjugation parity by $J^{PC}$. The superscript is a mnemonic that they are only approximate---but \textit{really good}---symmetries.

\begin{eg}
	Consider the $\pi^0$ meson. The PDG tells us that it has $J^{PC} = 0^{-+}$. Does this make sense? Why should the pion have negative internal parity? We know that it is a pair of quarks. It's the lightest meson so it wants to have $J=0$ with as little total and orbital angular momentum as possible. It does this by having two antiparallel spin one-half quarks and so it must pick up a minus sign under parity. Indeed, everything makes sense.	 
\end{eg}

\begin{framed}
	\noindent\textbf{What about $T$?} We know that $CPT$ is required for any consistent Lorentz invariant quantum field theory. The violation of $CP$ by the weak interactions thus implies a (compensating) violation of $T$. In neutrino physics, the $CP$ conjugate of the process $\nu_\mu \to \nu_e$ is $\bar{\nu}_\mu \to \bar{\nu}_e$. On the other hand, the $T$ conjugate is $\nu_e \to \nu_\mu$. $CPT$ tells us that the amplitudes for the $CP$ conjugate and $T$ conjugate processes must be equal. One particularly interesting search for $T$ violation is through the electric dipole moment (\acro{EDM}) of the neutron. \acro{EDM} of an elementary particle points in the direction of its spin $\mathbf{s}$ (by Lorentz invariance there is no other possibility) and is proportional to the electric field $\mathbf{E}$. Note that 
	\begin{center}
		\begin{tabular}{rlcrl}
			$P\mathbf{s}$ &$= \phantom{+}\mathbf{s}$ & \hspace{3cm} & $P\mathbf{E}$ &$= -\mathbf{E}$\\
			$T\mathbf{s}$ &$= -\mathbf{s}$ &\hspace{3cm}& $T\mathbf{E}$ &$= \phantom{+}\mathbf{E}$
		\end{tabular}		
	\end{center}
 Thus a nonzero \acro{EDM} is a signal of both $P$ and $T$ violation. For more information about this see \cite{Perkins:2000uq,Ramsey:1982td}.
\end{framed}

One should stop to think about why we haven't mentioned \textbf{baryon number}. Is baryon number exact or approximate? Every Feynman vertex in the Standard Model respects baryon number, so we might want to say that it is exact. However, we recall that baryon number is anomalous in the Standard Model. Thus it is only an approximate symmetry once we include the electroweak sector. However, since baryons have $2J+1$ even and mesons have $2J+1$ odd, for our purposes for particle classification baryon number conservation pops out of angular momentum conservation automatically\footnote{Similarly, you may argue that $(B-L)$ looks like a good quantum number within the Standard Model, why not classify based on $(B-L)$? Since we don't have lepto-quarks in the Standard Model, this really boils down to counting either $B$ or $L$ for a given state.}. 

Finally, the approximate quantum numbers that are approximate even with respect to \QCD are those associated with flavor. An example of `flavor' for the light (i.e.\ `flavorless') mesons composed of only $u$ and $d$ quarks is \textbf{isospin}. This is because when we neglect the mass difference between the up and the down quarks ($\Delta m_{ud} \ll \Lambda_{\text{QCD}}$) then $u$ and $d$ form an isodoublet. Because the $u$ and $d$ have different electric charges, they still cannot mix. However, electrically neutral combinations like $u\bar u$ and $d\bar d$ may mix---this is why we say that the neutral pion is a mixture: $\pi^0 = (u\bar u - d\bar d)/\sqrt{2}$.  

What other approximate flavor quantum numbers might we consider? Charm and strangeness look like they could be `okay' symmetries---their masses are considerably heavier than the $u$ and $d$ masses. Actually, strangeness is the worst possible example we could have chosen since its mass is so close to $\Lambda_{\text{QCD}}$. On the other hand, beauty (bottom-ness\footnote{Perhaps `top-ness' should be called `truthiness'? Why isn't `truthiness' a useful approximate quantum number?}, if you prefer) is a \textit{good} approximate quantum number because it's so heavy. Can a $B$ meson and a charmed meson mix? Only by a very small amount since the $b$ quark is so heavy. When you see a stable meson with a mass on the order of 5 GeV you \textit{know} it contains a $b$ quark. The $s$ quark, on the other hand, is not particularly light nor heavy. We'll get to know its mesons soon.

\begin{eg}
	Look up the listing for the $\pi^\pm$ meson in the \acro{PDG}. Why does it only list $J^P$ instead of $J^{PC}$? 
	(Hint: what do we know about degenerate states versus eigenstates?)
	The $\pi^\pm$ are degenerate states that are conjugate to one another, i.e.\ a state $\pi^+$ has neither charge parity plus or minus since it's not an eigenstate of charge parity. How can this be? Recall that degenerate states needn't be states of the symmetries of the theory, all that is required is that a superposition of these states is an eigenstate.
\end{eg}
While you're looking at the $\pi^\pm$, you'll note that there are additional quantum numbers listed: $I^G$. $I$ is isospin, which we'll discuss below. What is $G$? This is $G$-parity, a generalization of charge quantum number that acts on isospin multiplets to replace the $C$ quantum number for particles like the $\pi^\pm$. The strict definition is
\begin{align}
	G = C e^{i\pi I_2},
\end{align}
where the second factor is a rotation in isospin space that takes $I_3 \to - I_3$.

The punchline of this discussion is that degenerate states with the
same quantum numbers generally mix, sometimes with large or small
mixing. When they have different \textit{exact} quantum numbers they
\textit{cannot} mix, but having different \textit{approximate} numbers
means that they can mix only via weak interaction effects.

\subsection{Binding energy}

\begin{quote}
In the days right after the July 4$^\text{th}$, 2012 discovery of the Higgs boson, many popular news outlets would say that the Higgs is responsible for giving particles mass. This is actually rather misleading---most of the mass that we care about as humans come from protons and neutrons, whose masses dominantly do \emph{not} come from the Higgs-induced masses of the valence up- and down-quarks. Most of the mass of every atom comes, instead, from \QCD.
\end{quote}

The masses of the hadrons come from two sources: the `bare masses' of the quarks coming from electroweak symmetry breaking and from the binding energy coming from \QCD interactions. We may continue our analogy with hydrogen since the mass of a hydrogen atom comes from the masses of its constituent proton and electron plus a potential energy $V=-13.6$ eV. Indeed, we can experimentally check that the mass of hydrogen is 13.6 eV less than the sum of the masses of the proton and electron. In this sense this analogy is \textit{too} good, since this statement comes from being able to physically pull a proton and electron asymptotically apart and measure their masses independently of their mutual electric field. 

In \QCD the situation is more complicated since confinement tells us that we cannot pull individual quarks apart and measure them independently. What we can do is something a bit counterintuitive and sneaky: we can probe these quantities at high energies where \QCD is weak. If we can extract the quark masses at high scales then we can \textit{hope} to run these masses down to the \acro{IR}, but it's clear that we're cheating a little. The act of running the masses down involves the inclusion of interactions, which is precisely what we wanted to separate from our mass measurement.  Of course, even the measurement of the masses in the \acro{UV} is tricky since \acro{UV} quantities don't depend strongly on low-scale masses. Another way of saying this is that we usually define masses at low energies.

This is really a problem of the light quarks since we have probed them at energies much larger than their pole masses, to the extent that a pole mass is even well defined for confined particles. The bottom line is that these quark masses are very regularization-scheme dependent. When someone tells you a quark mass, the particular value shouldn't mean much without an accompanying regularization scheme. This is like someone telling you that it's $30^\circ$ outside without specifying Fahrenheit or Centigrade\footnote{$30^\circ$ is a typical spring temperature in both Ithaca, \acro{NY} and Riverside, \acro{CA}... if you use the appropriate units.}. 

While we're on the subject of differences between \QCD and \acro{QED}, let us remark that you should have already noticed that unlike atoms, hadrons can have masses \textit{larger} than the sum of their constituent quarks. In other words, the potential energy can be positive. This, of course, is a signature of confinement.

\subsection{Light quarks, heavy quarks, and the heaviest quark}
\label{sec:light:heavy:heaviest}

So far we've discussed two kinds of quarks and hinted at properties of their associated hadrons. There are the light quarks ($u,d,s$) which have masses---even though they are renormalization-scheme dependent---much less than $\Lambda_\text{QCD}$. These guys have their relevant mass scales set by $\Lambda_\text{QCD}$ and the differences between their masses are negligible enough to talk about approximate symmetries between them. This is a bit of a stretch for the $s$ quark, but it's certainly true for the $u$ and $d$. 

On the other hand, there are the heavy quarks ($c,b,t$) with masses much greater than $\Lambda_\text{QCD}$.  Unlike the light quarks, mesons made of heavy quarks have their mass scales set by the bare masses of their valence quarks.  Again, the $c$ is a bit of a borderline case, but we don't have to be picky. This is certainly true for the stable $B$ mesons,
\begin{align}
	m_b &\approx 4500 \text{ MeV}
	&
	m_{B^0} &= 5280 \text{ MeV}.
\end{align}
What about the top quark? 

It turns out that because the top quark is \textit{so} heavy ($m_t
\ggg \Lambda_\text{QCD}$) it decays before it hadronizes. For example,
at a collider we might produce a $b$ quark that hadronizes into an
excited $B$ meson (resonance) which quickly decays into a stable $B$
meson which must then decay by the weak interaction. We understand
this story from our $\rho$ and $B$ examples above: the weak
interaction widths are much smaller than strong interaction
widths. For the top quark, on the other hand, the weak decay is
\textit{more} important than the strong force. The top  decays mainly
via $t \to b W$ with a width of about 2 GeV, so that 
$\Gamma_t \gg\Lambda_\text{QCD}$. Thus the top quark 
 decay well before it can form any kind of hadron. In fact, because $m_t > m_W$,  the top decays into an \textit{on-shell} $W$ which gives additional kinematic enhancement since it decays via a two-body decay rather than a three-body decay required by the $b$.

We sometimes say that the top ``doesn't form hadrons.'' We should be careful with how we phrase this. Because the top decays so quickly we can never identify the hadron that it forms. Arguably it is even an ill-posed question to identify the top hadron in that time scale. In principle, though, we can ``turn off'' the weak interaction and calculate a spectrum of top hadrons. This is a very strongly coupled problem that is intractable by known techniques, but \textit{in principle} one can calculate the spectrum of top hadrons when the weak force is negligible. In this sense it's not quite correct to say that there's `no such thing' as a top hadron, it's just that we don't see them. 

The reason why this linguistic issue is important is that it puts one in danger of turning poor language into poor physics. For example, one might want to try to explain that the top quark doesn't form hadrons because such hadrons are not eigenstates of the full Hamiltonian. This is \textit{wrong}! Sure they're not eigenstates of the Hamiltonian, but \textit{neither are any of the `stable' hadrons that we've been discussing}, even those made up of light quarks! Any particle which decays is not an eigenstate of the Hamiltonian. When we discuss particles like the $B$ meson, we really mean objects that are eigenvalues of the Hamiltonian in the limit when the weak interaction is turned off. This now ties back to Definition \ref{def:stable:particle} for a stable particle which is defined in the limit when the $SU(2)_L$ coupling $g \to 0$.

\subsection{Masses and mixing in mesons}\label{sec:masses:and:mixing:mesons}

\begin{quote}
	Why do some mesons have definite quark content while others mix?
\end{quote}

We've now thought a bit about the problem that \QCD ties up quarks
into strongly-coupled bound states like mesons. There's a second
problem associated with \QCD that is of utmost importance to us: not
only is the quark tied up in a confined state, but once a quark of a
given flavor is bound up in a physical-state meson, \emph{the meson
  doesn't necessarily preserve that quark's flavor}. As you know, a
$u$ will want to hadronize into a pion. When it ends up in a neutral
state, it mixes into the neutral pion, $\pi \sim u\bar u - d\bar d$
and does not stay a in bound state of $u - \bar u$.

A major theme in these lectures is to understand how flavor-mixing is induced by objects that naturally live in different bases. The most famous example is the \CKM matrix. However, we can already see this idea at work in the mixing apparent in meson states. 
If we work in a basis where we know the masses of different quarks up to renormalization scheme dependence, then we can say that a given quark--anti-quark pair describes meson state. We can say that such a combination has a definite `flavor' in the sense that the valence quark and anti-quark are well-defined. 
However, these mesons can mix so that the physical meson states are linear combinations of quark pairs of different flavor combinations. 
\begin{eg}
	A neutral pion has valence quark structure $\pi^0 = (u\bar{u} - d\bar{d})/\sqrt{2}$. We can say that it is 50$\%$ a $u\bar u$ state and 50$\%$ a $d\bar d$ state.
\end{eg}
\begin{prob}\label{prob:pion:content}\textbf{Why is the neutral pion antisymmetric in $u$ and $d$?} Explain why $\pi^0=(u\bar u - d\bar d)/\sqrt{2}$ rather than $\pi^0 = (u\bar u + d\bar d)/\sqrt{2}$.
	\begin{sol}
This comes from the representation theory of SU(2) isospin. There's a natural point of confusion: you are familiar from the addition of angular momenta that $\vec 2\otimes \vec 2 = \vec 3 \oplus 1$: that is, the combination of two doublets (spin-1/2) combine into a triplet and a singlet. The neutral ($S_z = 0$) triplet state is symmetric, $\sim\left| \uparrow\downarrow\right\rangle + \left| \downarrow\uparrow\right\rangle$. Na\"ively, we might then expect that the pion---the $I_3=0$ state of an isotriplet---should also have a plus sign: $\pi^0 \sim u\bar u + d\bar d$. 
		The plus sign misses something critical: the pion is composed of a quark and an an anti-quark. Thus we're not combining two isodoublets, $\vec 2 \otimes \vec 2$, we're combining and isodoublet with an anti-isodoublet, $\vec 2 \otimes \bar{\vec 2}$. 
		
		Fortunately, SU(2) is special because it is a pseudo-real representation. This provides us a trick to convert an anti-doublet $\bar{\vec 2}$ into a doublet. We've already used this trick when we write the Yukawa terms in the Standard Model: we had to define a conjugate Higgs field, $\tilde H = \epsilon^{ij} H_i^*$. The ``intuitive physics way'' of understanding this is that when we work with SU($N$), we are allowed to construct objects using the $N$-component anti-symmetric tensor, $\epsilon^{i_1\cdots i_N}$. For the case of SU(2), $\epsilon^{ij}$ and its inverse $\epsilon_{ij}$ can be used to raise and lower indices. These raised and lowered indices can be understood as column versus row vectors. In other words, it lets us convert a conjugate field into an object that transforms like a non-conjugate field at the cost of the minus signs that come from the $\epsilon$ tensor. This is precisely the origin of the relative minus sign between the $u\bar u$ and $d\bar d$ terms in the pion composition.
			For explicit matrices, see Halzen \& Martin's \emph{Quarks and Leptons}, Section 2.7.
			
			There is an alternative way to see this from the theory of Goldstone bosons. Since the pion triplet is identified with the Goldstones of isospin breaking, one can use the Callan--Coleman--Wess--Zumino formalism (\acro{CCWZ}) to `pull out' these fields from the matter fields charged under the broken symmetry. The details are beyond the scope of this solution---see the Appendix of~\cite{Csaki:2016kln} for an introduction---but the essence is that the relative minus sign between the $u\bar u$ and $d\bar d$ terms comes from the $\sigma^3/2$ generator of the spontaneously broken SU(2) acting on an isodoublet. It is perhaps amusing that these two pictures of understanding the relative minus sign are equivalent, but in one case it comes from $\epsilon \sim i \sigma^2$ and in the other it comes from $\sigma^3/2$.
	\end{sol}
\end{prob}

Now consider the case of a pair of heavy quarks,
the $J/\Psi = \bar c c$ and the
$\Upsilon = \bar b b$. The $J/\Psi$ (``so good they named it twice'')
led to a Nobel prize for the discovery of charm while a $b\bar b$
resonance---the $\Upsilon(4S)$---decays to nearly-at-rest $B^0$ mesons
and play a key role in `$B$-factories' like the LHC$b$, BaBar, and
Belle. In general,  mesons composed of a heavy quark and its
antiparticle are called \textbf{quarkonia}, the $J/\Psi$ and
$\Upsilon$ are examples of charmonium and bottomonium states,
respectively.

Why is it that the $\pi^0$ has a $45^\circ$ mixing while the
$\Upsilon$ is effectively a state with only $b$ and $\bar b$? Why
can't we have a physical meson that is an appreciable admixture of
bottom and charm, $(b \bar b + c \bar c)/\sqrt{2}$? No fundamental symmetries prohibits this. 

The answer is related to the origin of meson masses, which we saw above can predominantly come from either
\begin{enumerate}
	\item The \QCD binding energy, with order $\Lambda_{\text{QCD}}$, or 
	\item The valence quark masses. 
\end{enumerate}
In \QCD we know that isospin is a good symmetry in the limit where the quark masses are small compared to $\Lambda_{\text{QCD}}$. This holds for the light quarks ($u,d,s$). In this case the dominant contribution to the meson masses comes from the \QCD potential energy and the mesons have masses on the order $\Lambda_{\text{QCD}}$ plus, for higher resonances, any additional energy from internal degrees of freedom such as orbital angular momentum. Each light quark has isospin 1/2 so that the bound state $q\bar q$ (for $q=u,d,s$) has isospin $0\oplus 1$. In other words, we may have iso-singlets and iso-triplet mesons coming from light quark bound states. 
\begin{eg}\textbf{Hydrogen atom analogy}. This is just the same story as hyperfine splitting in the hydrogen atom. Just like the hydrogen atom the singlet and triplet states have different energies (masses), but within the triplet all three states are degenerate.	
\end{eg}
We say that for light quark mesons like the $\pi^0$, the \QCD potential energy gives mass according to isospin and not flavor.

What about mesons such as the $B^0$ whose masses come from a heavy bare quark mass? In this case the meson knows about flavor because the bare mass contribution from the heavy quark is larger than the $\Lambda_{\text{QCD}}$-scale potential energy contribution. It is now \textit{flavor} and not isospin that determines the meson mass.

We should think of these effects in terms of $2\times 2$ matrices. First consider the mass term generated by the \QCD potential. This is a diagonal matrix acting on a space spanned by the iso-singlet $|1\rangle$ and iso-triplet $|3\rangle$ mesons,
\begin{align}
	\mathcal L_{\text{mass, iso}} \sim 
	\begin{pmatrix}
		\langle 1| & \langle 3|
	\end{pmatrix}
	\begin{pmatrix}
		m_1 & 0 \\
		0 & m_3
	\end{pmatrix}
	\begin{pmatrix}
		|1\rangle\\
		|3\rangle
	\end{pmatrix}.\label{eq:mass:matrix:isobasis}
\end{align}
Contrast this to the mass terms coming from the valence quarks. For simplicity let's consider only the case where this contribution is dominated by only one of the quarks. We can also write this contribution as a $2\times 2$ matrix, but this time acting on the flavor-basis states defined by the presence of the heavy quark:
\begin{align}
	\mathcal L_{\text{mass, flavor}} \sim 
	\begin{pmatrix}
		\langle q| & \langle q'|
	\end{pmatrix}
	\begin{pmatrix}
		m_{q} & 0 \\
		0 & m_{q'}
	\end{pmatrix}
	\begin{pmatrix}
		|q\rangle\\
		|q'\rangle
	\end{pmatrix}.\label{eq:mass:matrix:flavorbasis}
\end{align} 
We must take the sum of these two mass matrices, but (\ref{eq:mass:matrix:isobasis}) and (\ref{eq:mass:matrix:flavorbasis}) are written in \textit{different bases}. Thus shifting one matrix to the other basis will introduce off-diagonal terms which indicate mixing. The basis of mass eigenstates is generally something in between these two bases, the relevant question is whether the mass basis is closer to the isospin basis or the flavor basis. In other words, we must ask which effect is more important: isospin splitting (in this case singlet-triplet splitting) or the difference in heavy quark masses. If isospin splitting is more important, then we can use the isospin basis and treat the valence quark masses as a perturbation that introduces off-diagonal terms. This is what happens when the mesons such as the $\pi^0$ that are made up of light quarks whose masses are all small.

On the other hand, if the quark mass differences are more important, then we should use the flavor basis and treat the isospin-dependent masses as a perturbation that introduce off-diagonal terms. This is the case with our hypothetical mixing between a $\bar b b$ and a $\bar c c$ (these take the place of the $|q\rangle$ and $|q'\rangle$ states above). A physical mass state is some linear combination
\begin{align}
	\cos \theta\; \bar b b + \sin \theta\; \bar c c\ ,
\end{align}
where $\theta$ is the mixing angle. In this case, however, $\sin\theta$ is suppressed by a factor on the order of magnitude of
\begin{align}
	\frac{\Lambda^2_{\text{QCD}}}{m^2_b - m^2_c} \ll 1 \ .
\end{align}

It is also worth observing that the mass difference $|m_d-m_u|$ also splits the triplet. You should also now realize that we were being a bit sneaky in the previous section when we wrote $2\times 2$ matrices. We explore these ideas in the following problem.
\begin{prob}\label{prob:meson:eigenstates}\textbf{
Kaon--pion mixing}. 
Why \emph{don't} the neutral kaon and pion mix? 
\begin{sol}
They do not mix because the kaon has a flavor quantum number that is conserved
under \acro{QCD} and \acro{QED}. There is a tiny mixing due to the weak interaction,
but it is so small that we can only dream of ever detect it.
\end{sol}
\end{prob}

\begin{prob}\label{prob:meson:eigenstates}\textbf{Meson eigenstates for general quark masses}. Consider a model of \QCD with only up and down quarks of unspecified masses $m_u$ and $m_d$, that is we do not know \textit{a priori} whether the isospin or flavor basis is dominant. Write down the meson mass matrix in the flavor basis. (\textbf{Hint}: the matrix is larger than $2\times 2$.)
	\begin{sol}
		A convenient basis is $\left\{|u\bar u\rangle, |u\bar d\rangle, |d \bar u\rangle, |d\bar d\rangle\right\}$. We can convert to the isospin basis using
		\begin{align}
			|0,0\rangle &= \frac{1}{\sqrt{2}}\left(|u\bar u\rangle - |d\bar d\rangle \right)\\
			|1,-1\rangle &= |d\bar u\rangle\\
			|1,0\rangle &= \frac{1}{\sqrt{2}}\left(|u\bar u\rangle + |d\bar d\rangle \right)\\
			|1,1\rangle &= |u\bar d\rangle
		\end{align}
		where we've written states according to $|I,I_3\rangle$. Note that $u$ and $\bar d$ are analogous to $|\uparrow\,\rangle$ while $\bar u$ and $d$ are analogous to $|\downarrow\,\rangle$ in the usual SU$(2)$ spin notation. (To see why just remember that the kinetic term must be isospin invariant and that the SU$(2)$ metric is the antisymmetric tensor.)
		Let us say that the singlet gets a mass $m_1$ and the triplet states get mass $m_3$. The mass matrix thus takes the form
		\begin{align}
			\begin{pmatrix}
					2m_u^2 + \frac 12 m_3^2 + \frac 12 m_1^2	&	&	& \frac 12 m_1^2 - \frac 12 m_3^2 \\
						& m_u^2+m_d^2 + \frac 12 m_3^2 	&			& \\
						&		& m_u^2+m_d^2 + \frac 12 m_3^2	& \\
			 	\frac 12 m_1^2 - \frac 12 m_3^2	&	&	& 2m_d^2 + \frac 12 m_3^2 + \frac 12 m_1^2
			\end{pmatrix}.
		\end{align}
		From here one can happily plug into \textit{Mathematica} to solve for the exact eigenstates and eigenvalues. As a sanity-check note that in the limit ${m_{u,d}\to 0}$ the $u\bar u$ and $d\bar d$ states mix with a $45^\circ$ angle giving the $\pi^0$ and $\eta$. The $u\bar d$ and $d \bar u$ states do not mix because both bases conserve $I_3$. Note further that electromagnetism also preserves the $I_3$ quantum number. 
	\end{sol}
\end{prob}

\subsection{The light, pseudoscalar mesons}
\label{sec:pseudoscalars}

Let us briefly introduce some states in the spectrum of low-mass mesons. Heavier mesons are resonances that are excitations of internal degrees of freedom or states that carry a heavy valence quark. Let's look at the lightest mesons, the pseudoscalars ($J^P = 1^-$). We know, for example, that the $u$ and $d$ have a very good approximate isospin symmetry because $m_d-m_u \ll \Lambda_{\text{QCD}}$. From the usual addition of angular momentum in quantum mechanics, we know that mesons made out of the $u$ and $d$ can be decomposed into an iso-singlet and an iso-triplet. The singlet is the $\eta$ ($m_\eta = 550$~MeV) and the triplet is the $\pi$ ($m_\pi \sim 135$~MeV). Electromagnetism provides a higher-order splitting that distinguishes the neutral pion $\pi^0$ and the charged pions $\pi^\pm$.

Now add the $s$ quark to enlarge isospin to an SU$(3)$ symmetry called SU$(3)$ \textit{flavor}, which we introduced in Section~\ref{sec:flavor:symmetries}.
We know that since $m_s$ is quite a bit heavier than $m_{u,d}$, this is slightly dubious, but the difference is still less than $\Lambda_{\text{QCD}}$ so let's see how far we can push this. The tensor product of a fundamental and anti-fundamental of SU(3) is $\mathbf{3}\otimes \bar{\mathbf{3}}=\mathbf{1}\oplus\mathbf{8}$, where the $\mathbf{8}$ is the adjoint. 

Do we find this structure in the meson spectrum? Yes! The singlet is the $\eta'$ ($m_{\eta'}=960$ MeV) and the pseudoscalar octet composed of the pions, the $\eta$, and the kaons ($m_{K}\sim 490$ MeV). We can see that the pions form a \textit{pretty-good}  SU(2) iso-triplet and the kaons which flesh out the \textit{just-okay} SU(3) octet. These indeed have mass difference on the order of the $s$ mass ($m_s\sim 150$ MeV), which breaks the SU(3) flavor symmetry. To provide a graphical mnemonic, the weight diagram for the pseudoscalar nonet (octet + singlet) is:

\begin{center}
	\begin{tikzpicture}
		\draw[line width=1.5] (0:2) -- (60:2) -- (120:2) -- (180:2) -- (240:2) -- (300:2) -- (0:2);
		\draw[fill=black] (0:2) circle (.09cm);
		\draw[fill=black] (60:2) circle (.09cm);
		\draw[fill=black] (120:2) circle (.09cm);
		\draw[fill=black] (180:2) circle (.09cm);
		\draw[fill=black] (240:2) circle (.09cm);
		\draw[fill=black] (300:2) circle (.09cm);
		\draw[fill=black] (60:.3) circle (.09cm);
		\draw[fill=black] (120:.3) circle (.09cm);
		\draw[fill=black] (270:.2) circle (.09cm);
		\node at (0:2.6) {$\pi^+$};
		\node at (60:2.6) {$K^+$};
		\node at (120:2.6) {$K^0$};
		\node at (180:2.6) {$\pi^-$};
		\node at (240:2.6) {$K^-$};
		\node at (300:2.6) {$\bar K^0$};
		\node at (60:.8) {$\eta'$};
		\node at (120:.7) {$\eta$};
		\node at (270:.6) {$\pi^0$};
	\end{tikzpicture}
\end{center} 

Group theory tells us how states of different quark content ought to mix. In the SU$(2)$ case, we saw that the interesting mixing occurred in the $I_3=0$ states ($\pi^0,\eta \propto u\bar u \pm d\bar d$). Similarly, the interesting mixing occurs in the $I_3=0$ states of SU(3) where group theory predicts that the $\eta$ (part of the octet) and $\eta'$ (the singlet) have the quark content
\begin{align}
	\eta &= \frac{1}{\sqrt{6}}\left(u\bar u + d\bar d -  2s\bar s\right)
	&
	\eta' &= \frac{1}{\sqrt{3}}\left(u\bar u + d\bar d + s\bar s\right) \ .
\label{eq:eta:content}
\end{align}
This is the analog of Problem~\ref{prob:pion:content} applied to SU(3)
flavor symmetry. It turns out the mass difference $|m_s-m_{u,d}|$
leads to a mixing angle that gives only an order 10\% correction and
we are justified in working in the isospin basis to leading order.

For completeness, we may represent these fields in and SU(3) adjoint matrix as
\begin{align}
	\begin{pmatrix}
		\frac{1}{\sqrt{2}}\pi^0 + \frac{1}{\sqrt{6}}\eta & \pi^+ & K^+\\
		\pi^- & -\frac{1}{\sqrt{2}}\pi^0+\frac{1}{\sqrt{6}}\eta & K^0\\
		K^- & \overline{K^0} & -\frac{2}{\sqrt{6}}\eta
		\end{pmatrix} \ ,
\end{align}
where this comes naturally from identifying the pseudoscalar mesons as Goldstone bosons of the breaking of [approximate] chiral symmetry. See Appendix~\ref{sec:Goldstone:current:pion} for a historical interlude on how this relates to current algebra techniques. The $\eta'$ is the subject of the so-called ``U(1)$_A$ problem'' that plagued particle physics until instanton effects were well understood.

The light vector mesons are a separate story that we explore in Appendix~\ref{sec:vector:mesons}.

\subsection{Hadron names}

\begin{quote}
	``You can know the name of a bird in all the languages of the world, but when you're finished, you'll know absolutely nothing whatever about the bird. So let's look at the bird and see what it's doing---that's what counts. I learned very early the difference between knowing the name of something and knowing something.'' --Richard Feynman\footnote{``What is Science?'', presented at the fifteenth annual meeting of the National Science Teachers Association, in New York City (1966) published in \textit{The Physics Teacher} Vol. 7, Issue 6 (1969).}
\end{quote}

It is useful familiarize yourself with the names of common hadrons. At the very least one should be motivated that experimentalists are all very familiar with different hadrons (since these are the things which fly through the detector), so it's worth knowing what they're talking about when they say things such as, ``A hyperon can punch through and mimic a muon.'' 
The \acro{PDG} has a review on hadron names, so whenever you encounter an new
hadron, you can go there. Here, we just mention some of the most
common hadrons.

We've already talked about the pseudoscalar mesons which have $J^{PC}=0^{-+}$. We have the $\pi$, $\eta$, $\eta'$, $K$, $D, B$. The first three are called `unflavored' by the \PDG. By this we really mean that it contains no \textit{heavy} flavor, i.e.\ no net strangeness, charm, or beauty. The $\eta$s contain some admixture of $s\bar s$, but this has no net strangeness quantum number. The $K$, $D$, and $B$ are all code for flavored mesons. The kaons ($K$) all have net strangeness, the $D$s have net charm, and the $B$s have net beauty. 

There are several vector mesons which have $J^{PC}=1^{--}$. The iso-triplets are the $\rho$s and in Appendix~\ref{sec:vector:mesons} we delve into the story of the $\omega$ and $\phi$, states that mix isospin representations. The flavored vector mesons are indicated by stars relative to the pseudoscalars: $K^*, D^*, B^*$. 

There are other mesons. The $J^{PC}=0^{++}$ mesons are formed by quark configurations with orbital angular momentum $\ell = 1$ but fundamental spin $s=1$ such that the net spin is zero.  These have less exotic names, $a_0, a_1, a_2$ and so forth. What about the
$1^{+-}$ states? These are formed from states with $\ell=1$ and $s=0$. Can you guess their names? $b_0, b_1, \ldots$. 
For completeness we should also discuss the names of some popular
baryons. The \textit{most} popular baryons are well-known by
everybody: the proton and neutron. In these lectures, we won't spend much time thinking about the other baryons because they are harder to produce. It is worth briefly reviewing why this is the case.

\begin{framed}
	\noindent\textbf{Why it is harder to produce baryons}. We understand how we get mesons: we smash particles together at a collider and the hard scattering processes tend to create objects like $q\bar q$ pairs. Confinement tells us that we must pull out a $q\bar q$ pair from the vacuum in order to form color-neutral final-states. If we wanted to create final-state baryons, on the other hand, we'd need to pull out \textit{two} $q\bar q$ pairs from the vacuum. An undergraduate might then say that this must be a higher-order process since we're borrowing `more' energy from the \QCD vacuum, and thus baryon production is rare.

	This is \textit{wrong}! \QCD is strongly interacting, so the cost of pulling a $q\bar q$ pair from the vacuum is order one. Indeed, we know that for high-energy collisions we end up with jets that each have dozens of $q\bar q$ pairs. What \textit{does} end up costing you something is combinatorics. When one pulls out two $q\bar q$ pairs from the vacuum, it's far more likely (counting color multiplicities) that they will produce three mesons than three baryons. Thus the suppression is not that making enough pairs to produce baryons is less likely, but rather that once you do this it is still more likely to form mesons.
\end{framed} 

For fun, let's go over some of the baryon names.
\begin{itemize}
\item For three light quarks ($u,d$ combinations), i.e.\ isospin $I=1/2$, these are called \textbf{nucleons} $N$. This is just a fancy name for `proton or neutron.' For $I=3/2$ we have the $\Delta$s ($m_\Delta =$ 1232 MeV). These come in charges $q=+2,+1,0,-1$.
\item For baryons with two light quarks we have the $\Lambda$s ($I$=0) and $\Sigma$s ($I$=1). The isospin comes only from the light quarks. These particles are named according to their heavy quark. No index implies an $s$, while a subscript $c$ or $b$ indicates charmed or beautiful baryons, for example $\Lambda_c$.
\item Baryons with just one light quark ($I=1/2$) are called $\Xi$, which is too difficult for particle physicists to pronounce consistently so we tend to call them `cascades.' These are also named according to their heavy quarks with $s$ implied for no index. Thus we would name a baryon with one light quark, a strange quark, and a charm quark the $\Xi_c$. How do we know which light quark is active? By the charge; for example the $\Xi^+$ must be a $uus$ combination. 
\item Finally, what about the baryons with no light quarks? There is one famous one, the $\Omega$ composed of three strange quarks. This is the particle that Ne'eman and Gell-Mann predicted on the basis of the decouplet of SU(3). They wrote down its charge, quark content, and mass and the $\Omega$ was discovered a few years later. We don't often have a reason to discuss baryons with multiple heavy quarks. The first such baryon, the $\Xi_{cc}$ was only tentatively observed in the past decade. Recall that while it is easy to produce light quarks from the vacuum, heavy quarks with $m\gg \Lambda_{\text{QCD}}$ are a whole different story. Such production is suppressed by $\exp(-m/\Lambda_{\text{QCD}})$. Thus, for a $\Xi_{cb}$ one can hope to produce the valence $b$ quark from a high-energy parton-parton collision but the accompanying $c$ must be produced from the vacuum at the cost of exponential suppression.
\end{itemize}
\begin{prob}\textbf{Why we don't produce tons of baryons with two and
    three heavy quarks at colliders}. Contrarian students may argue that there should be no problem producing heavy baryons at the \acro{LHC}. The \acro{LHC} has a really large center-of-mass energy. Can't it produce all of the heavy quarks we need?
\begin{sol}
Do not be confused by the fact that the \acro{LHC} is a multi-TeV collider! We would produce these heavy hadrons just as well at \acro{HERA} as we would at the \acro{LHC}. The deep principle in play here is \textit{decoupling}. The \textit{hard scattering} event which occurs at high scales is likely to produce any of the quarks since $m_q \ll $TeV. These quarks propagate away from one another for a distance scale on the order of $1/\Lambda_{\text{QCD}}$ and it is \textit{only} at that distance scale that hadronization from \QCD really becomes active. This is a fact that is very well known by anyone who works with Monte Carlo generators: you use MadGraph for the hard scattering, but Pythia for hadronization and showering. You can separate these steps because they occur at two totally different scales and are governed by different physics.
\end{sol}
\end{prob}

\begin{prob}
	\textbf{An improved Gell-Mann--Okubo formula for the baryon octet}. The masses of the baryons in the $SU(3)$ iso-octet do not fit the na\"ive isospin predictions. Apparently there is a considerable contribution from the spin-spin interactions. Fortunately, the magnetic moments of the isospin octet baryons are relatively well-measured. Use this to write down predictions for the baryon masses taking into account the constituent quark masses \textit{and} the hyperfine splitting from spin-spin interactions. See Appendix~\ref{sec:vector:mesons} for a discussion of the Gell-Mann--Okubo formula.
	\begin{sol}
		See Griffiths \textit{Introduction to Elementary Particles}, Chapter 5.6.2--5.6.3 (Chapter 5.10 in the first edition).
	\end{sol}
\end{prob}

\section{Parameterizing QCD}

In our review of hadrons we already presented the general problem that
\QCD poses to flavor physics: because of confinement, the fundamental fields in our theory are not the degrees of freedom that we observe in experiments. Our Lagrangians are written in terms of quarks, but our world is made up of hadrons. Extracting quark interactions from these hadronic interactions is tricky.

When calculating scattering in quantum field theory, we work with asymptotic states, like electrons and positrons. In \QCD, however, we can't have asymptotic quarks. Thus, even at a formal level, a lot of what we say about quark interactions in \acro{QFT} isn't strictly very well defined. We can talk about pion field theories, but then we're working with pion couplings and are still removed from our fundamental \QCD Lagrangian. 
Said simply, the fundamental problem of \QCD is that our theory is described in one language and our experiments are done in another. 
Fortunately, there are ways around this; partly due to clever parameterizations. Before moving forward it is important to introduce some language that lies halfway between theory and experiment---a sort of phenomenology Yiddish.
The vocabulary is as follows:
\begin{itemize}
	\item \textbf{matrix element}: this is basically what we mean
          by
 transition amplitude. You should be familiar with this from non-relativistic quantum mechanics.
	\item \textbf{form factor}: this should sound familiar from (non-relativistic) scattering such as Rutherford scattering. It is the shape correction to the approximation that a scattering object is point-like ($S$-wave).
	\item \textbf{decay constant}: this is a non-perturbative
          property of mesons that parametrizes their decay widths to leptons.
	\item \textbf{factorization}: this is the deep idea that we may separate different kinds of physics. Physics at different scales ought to decouple. We are interested in the factorization of the hadronic matrix elements from the leptonic matrix elements. This factorization is realized because the operators that mix hadronic and leptonic pieces are formed by integrating out an intermediate $W$ at energies much lower than $M_W$.
\end{itemize}
Let's familiarize ourselves with the concepts behind these words. We begin by remembering how we do calculations and start with a simple purely leptonic example. Consider the process $W \to \ell \nu$. The amplitude is written as
\begin{align}
	\mathcal A = \langle \ell \nu | \mathcal O | W\rangle\ ,
\end{align}
where $\mathcal O$ is some operator in the Hamiltonian (or Lagrangian up to a sign). We then write the \textbf{decay rate} for the $W$ to a lepton and its neutrino by
\begin{align}
	\Gamma \sim \int |\mathcal A|^2 \times d(\text{phase space}).
\end{align}
We know that in this example, the amplitude is trivially $\mathcal A = g/\sqrt{2}$. You can write down the relevant tree-level term for $\mathcal O$,
\begin{align}
	\mathcal O = \bar\ell \gamma_\mu W^\mu \nu \label{eq:W:ell:nu:operator}
\end{align}
and do the appropriate Wick contractions with the external states. The point here is that the external states are asymptotic so that our operator can `literally' be written in terms of them.

Let's see what happens when we consider hadronic decays. Consider $\pi \to \ell \nu$. The matrix element is simple to write,
\begin{align}
	\mathcal A = \langle \ell \nu | \mathcal O | \pi^+ \rangle.
\end{align}
The leptonic part of this matrix element is simple, precisely because we again have asymptotic external states which appear as creation and annihilation operators in $\mathcal{O}$. What about the $\pi^+$? Let's say this pion is a $u \bar d$ combination. We must then create an up--anti-down pair. Because this is a \textit{low energy} process ($m_{\pi}\ll M_W$), we can integrate out the $W$ boson from our theory. The operator $\mathcal O$ factorizes into the leptonic and hadronic currents that attached to either end of the $W$ boson:
\begin{align}
	\mathcal O = \mathcal O_\ell \frac{1}{M_W^2} \mathcal O_H = (\text{``$\bar\ell\gamma^\mu\nu$''})\frac{1}{M_W^2}(\text{``$\bar d\gamma_\mu u$''}),
\end{align}
where we've written quotes to indicate that this is a heuristic form of the operators, we'll write it out honestly below. (In particular, we've been sloppy with factors of, e.g., $\gamma^5$.) This means that our matrix element should also factorize,
\begin{align}
	\langle \ell \nu | \mathcal O | \pi^+ \rangle = \langle \ell \nu | \mathcal O_\ell | 0 \rangle \frac{1}{M_W^2} \langle 0 | \mathcal O_H | \pi^+ \rangle.
\end{align}
The object on the left-hand side is simple to calculate. Don't be confused by the form of the  $\langle \ell \nu | \mathcal O_\ell | 0 \rangle$ matrix element, it looks like some weird expectation value for, say, the leptonic current---it doesn't make sense kinematically on its own, the matrix element isn't a kinematic object. The kinematics are split with the hadronic part and are really only accounted for in the phase space integral.  The factor of $1/M_W^2$ combines with coupling constants implicitly in the $\mathcal O$s to give the famous Fermi coupling,
\begin{align}
	\frac{G_F}{\sqrt{2}} &\equiv \frac{g^2}{8M_W^2}.
\end{align}
The numerical factors are historical and are motivated below (\ref{eq:Gf:definition}). 

The $\langle 0 | \mathcal O_H | \pi^0 \rangle$ might cause some concern. It looks something like a quark current expectation value, except for the fact that the external state has \textit{no explicit quarks}---it's a pion. Sure, a pion is made of quarks, but it's also made up of glue and is an actual asymptotic state while quarks are not. The easiest way to deal with this is to just define this object to be a parameter of our effective theory. We call it the pion \textbf{decay constant}, $f_\pi$, and we can \textit{heuristically} define it as
\begin{align}
	\langle \pi | \mathcal O | 0\rangle \sim f_\pi.
\end{align}
We'll get to a more proper and technical definition below; for now just focus on the ideas. Decay constants capture all of the hard-to-calculate non-perturbative \emph{brown muck} that keeps the pion together\footnote{`Brown muck' is a term coined by Howard Georgi, \cite{Georgi:1991mr}.}. In recent times, lattice \QCD calculations have made huge steps in being able to calculate these objects to the percent level, but in principle we should treat these as additional physical parameters whose values should be fixed by experiment.

Let's now examine more complicated hadronic decays. Consider $K^+\to \pi^0\ell^+ \nu$. 
\begin{eg} Draw the Feynman diagram for this process. (This should be very basic review!)
\begin{center}
	\begin{tikzpicture}[line width=1.5 pt, scale=1.3]
		\everymath{\displaystyle}
	%
		\node[draw,circle] (id2) at (0,0em) {$u$};
		\node[draw,circle] (iu1) at (0,-1.7em) {$\bar s$};
		\draw[dashed, black!30] (0,-.85em) ellipse (1.2em and 2.2em); 	
	%
		\node[draw,circle] (ou1) at (8em,0em) {$u$};
		\node[draw,circle] (ou2) at (8em,-1.7em) {$\bar u$};
		\draw[dashed, black!30] (8em,-.85em) ellipse (1.2em and 2.2em);
	%
		\draw[fermion] (id2) -- (ou1);
		\draw[fermion] (iu1) -- (3em,-1.7em);
		\draw[fermion] (3em,-1.7em) -- (ou2);
		\draw[snake=snake, line before snake=.5em] (3em,-1.7em) -- (5em, -4.5em); 
		\draw[fermion] (5em, -4.5em) -- (8em, -4em);
		\draw[fermionbar] (5em, -4.5em) -- (8em, -5.5em);
		\node at (3em,-4em) {$W^+$};
		\node at (8.6em,-4em) {$\nu_\ell$};
		\node at (8.5em,-5.5em) {$\ell$};
		\node at (-2.5em ,-.85em) {$K^+$};
		\node at (10.3em,-.85em) {$\pi^0$};
	\end{tikzpicture}
\end{center}
\noindent We've only drawn the $u\bar u$ content of the $\pi^0$, but we know by now that the $\pi^0$ is really the linear combination $(u\bar u - d\bar d)/\sqrt{2}$. Get used to drawing Feynman diagrams quickly in your head. Experimentalists and flavor physicists deal with processes whose diagrams all have similar topologies and no longer bother to explicitly draw diagrams for each other.
\end{eg}
The matrix element is again simple to write,
\begin{align}
	\mathcal A = \langle \pi^0 \ell \nu | \mathcal O | K^+ \rangle.
\end{align}
The leptonic part of this matrix element is simple, precisely because we again have asymptotic external states which appear as creation and annihilation operators in $\mathcal{O}$. What about the $K^+$ and the $\pi^0$? We know that $K=\bar s u$ and for this decay we only care about the $u\bar u$ part of the $\pi^0$. The operator that we need is $\bar u \gamma_\mu s$, since we want it to annihilate an anti-$s$ quark and create a $u$ quark. Of course, what we \textit{really} want is something that annihilates a \emph{kaon} and creates a \emph{pion}. This is where our \QCD `language problem' manifests itself: the quark operator coming from the Lagrangian is not directly related to the hadronic external states coming out of an experiment.

We certainly have creation and annihilation operators at the quark level, but just because you annihilate an $\bar s$ and produce a $u$ it does \textit{not} mean that you have automatically taken a $K^+$ and turned it into a $\pi$. This quark-level operator could generate any number of hadronic processes with different external states, say $\Lambda \to p \mu \nu $. So how do we calculate this matrix element?
Once again we appeal to factorization. In particular, we know that heuristically we can write
\begin{align}
	\mathcal O = \mathcal O_\ell \frac{1}{M_W^2} \mathcal O_H = \left(``\bar\ell_L\gamma^\mu\nu_L\text{''}\right)\frac{1}{M_W^2}\left(``\bar u_L\gamma_\mu s_L\text{''}\right)\ ,
\end{align}
where we've explicitly separated the leptonic and hadronic sides of the integrated-out $W$ boson diagram. The quotation marks remind us not to take the actual explicit form of the $\mathcal{O}_{\ell,H}$ operators that we wrote above too seriously yet, we'll shortly discuss their actual form. The matrix element thus factorizes,
\begin{align}
	\langle \pi^0 \ell \nu | \mathcal O | K^+ \rangle = \langle \ell \nu | \mathcal O_\ell | 0 \rangle \frac{1}{M_W^2}	\langle \pi | \mathcal O_H | K^+\rangle.
\end{align}
As we know from above, we're perfectly happy calculating the leptonic piece, $\langle \ell \nu | \mathcal O | 0 \rangle$. The hard part is the hadronic piece, $\langle \pi | \mathcal O | K\rangle$, which is deeply entrenched in how \QCD converts quarks into bound states. In the case of a single hadron we ended up packaging this non-perturbative garbage into a single parameter, the decay constant. What about when we have two hadrons, $H_1$ and $H_2$? We can define another parameter, 
\begin{align}
	\langle H_1 | \mathcal O | H_2\rangle \sim F.
\end{align}
We call this a \textbf{form factor}, where we again save a technical definition for later. We choose this name because this looks like scattering: one state $H_1$ coming in and a different state $H_2$ coming out. Usually the physical process includes leptons coming out, so form factors are closely related to semileptonic decays, e.g.\ $B\to D\ell \nu$. Form factors are scale dependent, $F({q}^2)$, encoding the extent of hadronic substructure a given scattering can probe. If you hit a hadron with a very high-energy probe---say, an intermediate boson with large ${q}^2$---then you start to probe the partonic configuration of the hadron. In the right scenarios, it is sufficient to understand the form factor at zero momentum transfer, $F(0)$.

\begin{prob} \label{prob:range:of:q2} \textbf{Range of $q^2$.} For a physical decay of the form $M_1 \to M_2 \ell \nu$, what is the range of possible values of $q^2$? Is $q^2$ time-like or space-like? \textsc{Hint}: $q$ is the momentum from the hadronic current to the leptonic current.
	\begin{sol}
		Because $q = p_\ell + p_\nu$, $q^2$ must equal the invariant mass of this system and must be positive definite. As a crutch, you can imagine that $p_\ell$ and $p_\nu$ is a bound state with some definite $m_{\text{bound}}^2$. 
	\end{sol}
\end{prob}


Being able to calculate these decay constants and form factors would already be a big step. While there have been many achievements on this front, such as correctly predicting some values before they are experimentally measured, there's still a long way to go before the \QCD `language problem' is practically solved. For example, three-body hadronic matrix elements are unlikely to be determined in the near future. We need to find a way to get around this so that we can take one non-perturbative result as a parameter and apply it to make predictions of other processes. A more recent approach involves the hope that the \acro{A}d\acro{S}/\acro{CFT} correspondence might be able to provide a perturbative method to calculate these strongly-coupled objects; this is often referred to as \acro{A}d\acro{S}/\QCD. Recent progress has led to promising results in meson spectroscopy, but predictions of actual matrix elements are still `undelivered'~\cite{Csaki:2008dt}.

To reiterate the big picture: progress with holographic and lattice techniques notwithstanding, we cannot calculate hadronic matrix elements from first principles. The best that we can do is to \textit{parameterize our ignorance}. But a bit of cleverness can go a long way. Sometimes we can measure the matrix elements in one process and apply the result to other processes---this is the best case, one doesn't even need theory to do this! When this fails, however, we have to be a bit more clever and use approximate symmetries. The program is as follows
\begin{enumerate}
	\item Use symmetries to determine which matrix elements are relevant. 
	\item For an unknown hadronic matrix element, write out the most general linear combination of dynamical variables (e.g.\ momenta) which match the Lorentz structure of the matrix element. (Do not forget the discrete symmetries!)
	\item Write the coefficients of these terms as parameters (decay constants and form factors) that have to be determined.
	\item Use every trick you can to determine which of these terms are relevant.
\end{enumerate}

\subsection{The decay constant}
Let's take a closer look at $\langle 0 |\mathcal O | \pi^+ \rangle$. 
The operator $\mathcal O$ should annihilate the valence quarks of a pion and so must have the form
\begin{align}
	\mathcal O \sim \bar u \Gamma d
\end{align}
for some Dirac structure $\Gamma$. We can always simplify the Dirac structure into a basis of $S,P,V,A,T$ (scalar, pseudoscalar, vector, axial vector, tensor).  Of these five operators, symmetries force some of them to vanish.
\begin{eg}
	\textbf{Which operators vanish?} Suppose we knew that the main contribution come from either the $V = \gamma^\mu$ or the $A=\gamma^\mu \gamma^5$ operator, while the other vanishes. Which one is zero? \textbf{Solution:} We know that $\pi$ is a pseudoscalar and \QCD conservs parity. The vacuum $|0\rangle$ is parity-even, thus $\mathcal O$ must be parity-odd, i.e.\ the vector must vanish and the main contribution must come from the axial operator. Similarly, $S=T=0$.
	\label{eg:vector:vanishes:in:decay:constant}
\end{eg}

Let's look at the the axial matrix element $\langle 0 |\bar u \gamma^\mu\gamma_5 d|\pi\rangle$ a bit more closely. 
We don't know how to calculate this object, but we saw above that we can parameterize our ignorance in terms of a \textbf{decay constant} $f_\pi$, which we now officially define to be
\begin{align}
	\langle 0 | A^\mu|\pi\rangle \equiv -ip^\mu f_\pi \ .
\end{align}
There is nothing surprising about the structure of the right-hand side.  Lorentz structure forces us to have a $p^\mu$ on the right-hand side since this is the only vectorial quantity available. The rest of the right-hand side is an overall coefficient, which we define to be $-if_\pi$ with mass-dimension 1. In principle this \textit{could} have had some dependence on Lorenz scalar quantities, of which we only have $m_\pi$, but this is not a \textit{dynamic} variable---it is some fixed value that does not change when we change the kinematic configuration of the decay. 

It turns out that $f_\pi$ can be measured (see below) and is
\begin{align}
	f_\pi \sim 131 \text{ MeV} \ ,
\end{align}
the order of magnitude here is something you should commit to memory.
Sometimes this is be written in terms of self-adjoint isospin currents and fields as\footnote{See chapter 4.3 of Georgi's \emph{Weak Interactions}~\cite{Georgi:2009vn}.},
\begin{align}
	\langle 0 |j_{5a}^\mu|\pi_b\rangle = i\delta_{ab}F_\pi p^\mu \hspace{3cm} F_\pi \equiv \frac{f_\pi}{\sqrt{2}} \approx 93 \text{ MeV}.
\end{align}
The theoretical \emph{meaning} of the decay constant comes from chiral perturbation theory; see~\cite{Csaki:2016kln} for a review and connections to models of new physics. For our present purposes, it is most useful to remember that the decay constants parameterize  incalculable strong dynamics.

How do we measure $f_\pi$? We should look at the dominant pion decay, $\pi^+ \to \mu^+ \nu$, which has a branching ratio of about 99.99\%. The amplitude for this decay is given by the simple $s$-channel diagram:
\begin{center}
	\begin{tikzpicture}[line width=1.5 pt, scale=1.3]
		\draw[fermionbar] (-140:1)--(0,0);
		\draw[fermion] (140:1)--(0,0);
		\draw[vector] (0:1)--(0,0);
		\node at (-140:1.2) {$d$};
		\node at (140:1.2) {$u$};
		\node at (.5,.3) {$W$};	
	\begin{scope}[shift={(1,0)}]
		\draw[fermion] (-40:1)--(0,0);
		\draw[fermionbar] (40:1)--(0,0);
		\node at (-40:1.2) {$\mu$};
		\node at (40:1.2) {$\nu$};	
	\end{scope}
	\end{tikzpicture}
\end{center}

%
It is straightforward to calculate the amplitude for this diagram,
\begin{align}
	\mathcal M(\pi \to \mu \nu) &= -\frac{g^2}{4 M_W^2} f_\pi V_{ud} \,m_\mu \bar\mu_R \nu_L.\label{eq:pi:to:mu:nu}
\end{align}
\begin{prob}\label{prob:pi:mu:nu:amp}\textbf{A basic Feynman diagram calculation}. Derive (\ref{eq:pi:to:mu:nu}). This is a very important exercise in \acro{QFT}. Explain the physical significance of the $m_\mu$ factor. 
	\begin{sol}
		For this calculation one needs to write in explicit projection operators. The amplitude takes the form
		\begin{align}
			\mathcal M = \frac{g^2}2 V_{ud} \bar v_d \gamma^\mu P_L u_u \frac{-i}{M_W^2} \bar u_\mu \gamma_\mu P_L v_\nu.
		\end{align}
		The first spinor contraction is just $\frac 12 \langle \pi^+ |A^\mu | 0\rangle$, where $A^\mu$ is the axial current. We have used the result example~\ref{eg:vector:vanishes:in:decay:constant} to neglect the vector piece, which vanishes by parity. Note the factor of 1/2 coming from the fact that we're only taking the axial part of the projection operator, $\gamma^\mu P_L = \frac 12\gamma^\mu (1-\gamma^5)$. Plugging in the expression for the pion decay constant,
		\begin{align}
			\mathcal M &= -\frac{g^2}{4 M_W^2}V_{ud}f_\pi \bar u \slashed{p}P_L v\\
			&= -\frac{g^2}{4 M_W^2}V_{ud}f_\pi m_\mu \bar u P_L v,
		\end{align}
		where in the second line we used the equation of motion for the outgoing charged lepton\footnote{We've been a little sloppy here, but there is some elegance to being able to do a sloppy-but-accurate calculation. The pion momentum $p$ should really be written in terms of $p=p_\mu + p_\nu$ and for each term the equation of motion for the appropriate lepton should be used. We already know, however, that $\slashed{p}_\nu v_\nu = 0$ since the neutrino is massless in the Standard Model.}. The benefit of this calculation is that one can use all of the usual Feynman rules from Peskin. However, things become a bit more clunky with factors of $\gamma^5$. These can become distracting for more complicated problems.

		The significance of the mass insertion is the chirality flip  in this decay. The pion is spin zero and the $W$ mediating the decay only couples to left-handed particles. In order to preserve angular momentum, one requires an explicit mass insertion on the final state fermions. Note that the use of the equation of motion is equivalent to a mass insertion on an external leg when using massless chiral fermions. 
	\end{sol}
\end{prob}

One can now convert this to a decay width (hint: the \PDG has a review section for two- and three-body decay kinematics),
\begin{align}
	\Gamma(\pi \to \mu\nu) = \frac{G_F^2}{8\pi}|V_{ud}|^2 f_\pi^2 m_\mu^2 m_\pi \left(1-\frac{m_{\mu}^2}{m_\pi^2}\right)^2.\label{eq:rate:pi:mu:nu}
\end{align}
The term in the parenthesis is the factor coming from the phase space integral; it is conventionally normalized so that in the limit $m_\ell \to 0$ the phase space factor goes to 1. What about the factor of $m_\mu^2$? This is clearly the same factor of $m_\mu$ in (\ref{eq:pi:to:mu:nu}). 

\begin{prob}\textbf{An exercise in calculating decay rates}. Derive (\ref{eq:rate:pi:mu:nu}) from (\ref{eq:pi:to:mu:nu}). \textit{\textbf{Hint}: If you're spending a lot of time with $\gamma$ matrix identities then you're doing this the hard way.}
	\begin{sol}
		The formula for the decay rate is
		\begin{align}
			d\Gamma &= \frac{1}{32\pi^2}|\mathcal M|^2 \frac{|\mathbf{p_\ell}|}{m_\pi^2}d\Omega,\label{eq:prob:pi:mu:nu:differential:decay:rate}
		\end{align}
		where we've written the muon subscript as $\ell$ to avoid confusion with Lorentz indices. In case you didn't feel inclined to re-derive this formula and your favorite quantum field theory textbook isn't nearby, then you can always look this up in the kinematics review of the pocket \PDG---which should \textit{always} be nearby. 
		
		The first step is to square the matrix element and sum over spins. This matrix element should be particularly simple because we only have one spinor bilinear. This is worked out in most \acro{QFT} textbooks so we won't belabor the four lines of work required to extract the relevant expression. Writing $p_\ell$ for the lepton four-vector and $p_n$ for the neutrino four-vector (to avoid confusion with Lorentz indices) we obtain from (\ref{eq:pi:to:mu:nu}),
		\begin{align}
			\sum_s |\mathcal M|^2 = \left(\frac{g^2}{4M_W^2}\right)^2 f_\pi^2 |V_{ud}|^2 m_\ell^2 \, 2 p_\ell\cdot p_n.\label{eq:prob:pi:mu:nu:spin:sum}
		\end{align}
		The factor of $2 p_\ell\cdot p_n$ is simply the result of summing over the spins of the spinor structure $|\bar\mu_R \nu_L|^2$. Note that this is \textit{trivial} and just comes from the completeness relations of the plane wave spinors. 
		
		\begin{framed}
			\noindent\textbf{Some strategy}. It is important to get the right answer, but it is also important to do so in a way that doesn't make your life difficult. This toy calculation is an important example. We have made use of the fact that we are only looking at the axial current so that our amplitude takes the form
			\begin{align}
				\mathcal M \propto f_\pi (p_\ell+p_n)_\alpha \bar\mu_L \gamma^\alpha \nu_L.
			\end{align}
			One could have na\"ively taken the square of the spinorial part as written. If one were to work with Dirac spinors, this would be a terrible mess,
			\begin{align}
				\bar u(p_\ell) \gamma^\alpha \frac{1}{2}(1-\gamma^5) v(p_n).
			\end{align}
			Squaring and summing over spins requires taking a trace over four $\gamma$ matrices times $(1-\gamma^5)$ and one obtains a funny relation in terms of two-index bilinears in $p_\ell$ and $p_n$ including an ugly term with an $\varepsilon^{\alpha\beta\mu\nu}$. See, e.g.\ equation (5.19) in Peskin and Schroeder \cite{Peskin}. One will eventually find that the $\varepsilon$ term cancels because it is contracted with $(p_\ell+p_n)_\mu(p_\ell+p_n)_\nu$. This is \textit{a lot of work!} 
			
			We can do a little better by using Weyl spinors. The simplification is obvious: the $\varepsilon$ term drops out from the very beginning because we never have to deal with the annoying $\gamma^5$ matrix. The relevant relations to perform the analogous calculation in Weyl space can be found in \cite{Dreiner:2008tw}; or in a pinch you can extract them from the Dirac spinor relations. This is indeed a meaningful simplification and justifies the use of Weyl spinors in `actual calculations' (rather than just abstractly to refer to chiral fields); however, we can do better!
			
			What we did above in (\ref{eq:prob:pi:mu:nu:spin:sum}) was to go a step further and get rid of the Dirac/Pauli matrix structure altogether. This was easy: we just used the fact that the $\gamma^\mu$ (alternately $\bar\sigma^\mu$) contracted a $(p_\ell+p_n)_\mu$. This means we can use the equation of motion to explicitly pull out the mass factor $m_\ell$, where have used $m_n = 0$. Now we get an expression of the form $\bar u(p_\ell) v(p_n)$ in Dirac notation, or in Weyl notation $\psi(p_\ell)\chi(p_n)$. Squaring and summing over spins is trivial for these since it is just the completeness relation for the plane wave spinors.
		\end{framed}
		
		It is trivial to determine the contraction $2p_\ell\cdot p_n$ since this is constrained by the kinematics,
		\begin{align}
			m_\pi^2=(p_\ell+p_n)^2 = m_\ell^2 + 2p_\ell\cdot p_n.
		\end{align}
		Thus we are led to (we drop the $\sum_s$ symbol)
		\begin{align}
			|\mathcal M|^2 = \left(\frac{2}{\sqrt{2}}G_F\right)^2 f_\pi^2 |V_{ud}|^2 m_\ell^2 \, m_\pi^2\left(1-\frac{m_\ell^2}{m_\pi^2}\right),
		\end{align}
		where we've also replaced the $g^2/8M_W^2$ with $G_F/\sqrt{2}$ according to the definition of the Fermi constant (\ref{eq:Gf:definition}). Let's simplify the other factors in (\ref{eq:prob:pi:mu:nu:differential:decay:rate}). The magnitude of the muon three-momentum can be derived trivially from kinematics; conservation of four-momentum gives us
		\begin{align}
			|\mathbf{p_n}|&=|\mathbf{p_\ell}|\\
			m_\pi^2&=|\mathbf{p_n}|+\sqrt{|\mathbf{p_\ell}|^2+m_\ell^2}.
		\end{align}
		From this we obtain
		\begin{align}
			|\mathbf{p_\ell}| &= \frac{m_\pi}{2}\left(1-\frac{m_\ell^2}{m_\pi^2}\right).
		\end{align}
		The angular integral is trivial
		\begin{align}
			d\Omega = d(\cos\theta)d\phi = 4\pi,
		\end{align}
		and we can now plug in all of these factors to obtain the result we wanted
		\begin{align}
			\Gamma &= \frac{1}{8\pi}G_F^2f_\pi^2 |V_{ud}|^2 m_\ell^2m_\pi \left(1-\frac{m_\ell^2}{m_\pi^2}\right)^2.
		\end{align}
	\end{sol}
\end{prob}
 
At first sight this factor seems puzzling: this means that the $\pi\to
\mu\nu$ decay is dominant over the $\pi\to e\nu$ decay, even though
the latter has a much bigger phase space. How did this factor of the
charged lepton mass appear? 
This \textbf{chiral suppression} comes from the fact that the pion is a spin-0 state whereas the operator, mediated by a $W$ boson, is spin-1. Because spin-1 operators connect fermions of the same chirality, one requires a mass insertion to flip the spin of one of the final state leptons so that the $\ell \nu$ final state conserves angular momentum compared to the initial pion. 


\begin{framed}
	\noindent\textbf{Chirality versus helicity}. It is interesting that experimentalists often refer to this as \textit{helicity} suppression. These are two different---but closely related---ideas that every particle physicist should be comfortable with. Chirality is a property of a \textit{field}, it is the difference between the $\text{SU}(2)_L$ doublet versus singlet fields. Helicity, on the other hand, is a property of a \textit{particle} which one can measure---it is an \textit{experimental} quantity. For massless fields chirality and helicity are the same. For massive fields one can flip chirality though mass insertions. 
\end{framed}

Everything in (\ref{eq:rate:pi:mu:nu}) is known except for $f_\pi$; we
can thus measure this decay to determine $f_\pi$. The \textit{physics}
behind this quantity is also very interesting to us. For example, how does this quantity scale? Can we understand it heuristically?
\begin{eg} \textbf{The positronium `decay constant.'}\label{eg:positronium:decay:constant}
	It is often useful to think of a toy example as a crutch for physical intuition. In this case, an instructive example is positronium. This is an $e^+e^-$ bound state and should have a decay constant, just like the pion. What is it?
	
	The relevant question is how does positronium decay. Semiclassically this occurs when the electron and positron touch, i.e.\ the probability that they are in the same place. The lifetime of positronium clearly depends on the overlap of the $e^+$ and $e^-$ wavefunctions. Assuming that the positron is at the origin, the `decay constant' of positronium can thus be heuristically interpreted as something like $|\Psi_{e}(r=0)|^2$. 
	\textit{Caveat emptor!} This is just a toy model to build physical intuition, it is not meant to be an actual representation of the physics of the pion!
\end{eg}


We've now explored the meaning of the axial matrix element and have parameterized the \QCD brown muck into a single pion decay constant. Armed with this, what can we say about the pseudoscalar matrix element? It turns out that we don't need to do any more calculations or introduce any more parameters: we can use a slick trick to get the pseudoscalar matrix element from the axial matrix element. This trick falls under the umbrella of the current algebra, something which has fallen out of favor in modern textbooks. For a more formal introduction, see \cite{ChengLi:Gauge, Treiman:Current:Algebra, Gasser:1982ap}. Here we sketch what we need.

Consider the pseudoscalar kaon matrix element,
\begin{align}
	P = \langle K(p) | \bar s \gamma_5 u |0 \rangle. 
\end{align}
This is manifestly a Lorentz scalar quantity. The result that we would like to show is
\begin{align}
	P = i f_K \frac{m_K^2}{m_u + m_s}\ .\label{eq:pseudoscalar:decay:constant:K}
\end{align}
The trick to derive this result is to take the divergence of the axial current,
\begin{align}
	i\partial_\mu (\bar s \gamma_\mu \gamma_5 u) &= i (\partial_\mu \bar s)\gamma_\mu\gamma_5 u + \bar s \gamma_\mu\gamma_5(i\partial_\mu u)\\
	&= (i {\slashed{\partial}}\bar s )\gamma_5 u - \bar s\gamma_5(i\slashed{\partial} u)\\
	&= -(m_s+m_u) \bar s \gamma_5 u.
\end{align}
Now take this and put it between the kaon and the vacuum. The derivative is just a momentum ($\partial_\mu=-ip_\mu$), so we get
\begin{align}
	\langle K | i\partial_\mu(\bar s \gamma_\mu \gamma_5 u)|  0\rangle = p_\mu (-if_Kp^\mu) = -if_K m_K^2,
\end{align}
from this we obtain (\ref{eq:pseudoscalar:decay:constant:K}) straightforwardly.

\begin{eg}\textbf{Scalar dominance}. Pop quiz: which matrix element is
  bigger, the pseudoscalar or the axial vector? The difference is a
  factor of $m_K^2/(m_u+m_s)$. Since the kaon is made up of light
  quarks, the dominant mass scale is set by $\Lambda_{\text{QCD}}$
  which is greater than the sum of the valence quark masses. Thus the
  (pseudo) scalar is enhanced relative to the (axial) vector
  operator. This is a generic feature for the light quark mesons. Note
  that this is a purely \QCD effect; electroweak theory only gives
  axial and vector contributions from the $W$. For heavy quarks the
  scalar dominance is much less significant.
\end{eg}

\subsection{Remarks on the vector mesons}

So far we've discussed the decay constants for pseudoscalars---these are the lightest mesons and can be identified as approximate Goldstone bosons from the breaking of approximate flavor symmetries\footnote{We refer to Section 4 of~\cite{Csaki:2016kln} for a discussion.}.  What about the vector mesons? Consider analogous matrix element with a $\rho$ meson,
\begin{align}
	\langle 0 | \mathcal O | \rho \rangle  = ?
\end{align}
Can we measure this object directly? What we are really asking is whether or not the $\rho$ has a nonzero matrix element with the \QCD vacuum: does it have leptonic decays via the weak interaction that would cause this matrix element to be nonzero? For the $\rho$ the answer is practically\textit{no}, it only decays via the strong force\footnote{The \PDG shows a small branching ratio of the neutral $\rho$ to $\ell^+\ell^-$, but this is due to mixing with the $\omega$.}, in other words, the $\rho$ is a resonance; it decays into two pions. The same is also true for the $\phi$. These particles do not decay weakly.
What about the (heavy) flavored vectors? Both the $J/\psi$ and the
$\Upsilon$ can decay to leptons, so we can measure  their decay
constants directly. 

We now move to the way we parametrize the decay constant.
Na\"ively we might want to write
\begin{align}
	\langle 0 | V^\mu | \rho \rangle \stackrel{?}{\propto}  f'_\rho p^\mu,
\end{align}
but this turns out to be \textit{too} na\"ive. We know from the \acro{LSZ} theorem in quantum field theory that amplitudes with external vectors are proportional to the polarization vector, $\epsilon^\mu(k)$. 
Thus the correct thing to write is
\begin{align}
	\langle 0 | V^\mu | \rho \rangle \sim f_\rho \epsilon^\mu,
\end{align}
where now it is clear that both sides transform the same way under parity.

\begin{framed}
	\noindent\textbf{Dynamical quantities}. In the example of the vector mesons we saw that there are other dynamical quantities other than the external particle momenta. This might lead one to think a bit more about similar quantities that exist in our system. A natural question is whether we should also include the spin vectors $s^\mu$ of the quarks. It is a somewhat subtle point that these should \textit{not} be considered to be real physical dynamical quantities! The spin index just identifies the chirality of the external state spinors.
\end{framed}

\subsection{Form factors}

Let us introduce \textbf{form factors} with a simple example that will be useful later on: the $\beta$ decay of a neutron into a proton, $n \to p^+ e \bar\nu$. We focus only on the vector current. The relevant matrix element that we cannot calculate from first principles is
\begin{align}
	\langle p^+(p') | \bar d \gamma^\mu u | \bar n(p) \rangle.
\end{align}
This is a `brown muck' object that we'd like to parameterize. To do so we must write out the most general linear combination of kinematic variables and products of those variables, \begin{align}
	\langle p^+(p') | \bar d \gamma^\mu u | \bar n(p) \rangle \sim a p^\mu  + b p'^\mu,
\end{align}
where the coefficients $a$ and $b$ are our \textbf{form factors}. We know that these can only depend on Lorentz scalars; there are three of these available: $p^2$, $p'^2$, and $p\cdot p'$. The first two are just masses and are not dynamical, they don't change when we change the momenta. The third one is a bona fide dynamical Lorentz scalar. It is conventional to write this in a different momentum basis.
Define the momentum $q$ by
\begin{align}
	q &\equiv p-p'.\\
	q^2 &= p^2+p'^2 - 2p\cdot p'.
\end{align}
This should be a familiar quantity from deep inelastic scattering calculations. We use  $(p+p')$ and $q$ as our basis of momenta in this problem. We write our form factors as $f_\pm$ that are functions of $q^2$, and thus write our matrix element as
\begin{align}
	\langle p^+(p') | \bar d \gamma^\mu u | \bar n(p) \rangle &= f_+(q^2)(p+p')^\mu + f_-(q^2)(p-p')^\mu. \label{eq:vector:form:factors}
\end{align}

For a more non-trivial example of how to parameterize a form factor with more dynamical variables, see the following problem. 

\begin{prob} \textbf{The $\langle D^{*+}(p_D,\epsilon)|V^\mu|\bar B(p_B)\rangle$ form factor}. 
	Consider the semi-leptonic decay $B \to D^* \ell \nu$. Let us again focus on calculating the hadronic vector matrix element.
	\begin{enumerate}
		\item Before diving into this particular decay, consider $B \to D^{+} \ell \nu$. Convince yourself that the relevant hadronic vector matrix element takes the form of (\ref{eq:vector:form:factors}). 
		\item Prove that
		\begin{align}
			\langle D^{*+}(p_D,\epsilon)|V^\mu|\bar B(p_B)\rangle = ig(q^2)\varepsilon^{\mu\nu\alpha\beta} \epsilon^*_{\nu} (p_D+p_B)_\alpha q_\beta,
		\end{align}
		where $\varepsilon$ is the totally antisymmetric tensor.
		\item Using the fact that $T$ acts as a complex conjugate, show that $g(q^2)$ is real. 
		\item Parameterize the axial matrix element, $\langle D^*(p_D,\epsilon)|V^\mu|\bar B(p_B)\rangle$.
	\end{enumerate}
	\begin{sol}
		This problem is an excellent example of a form factor multiplying a non-trivial combination of dynamical variables. 
		\begin{enumerate}
			\item The key difference between the $D$ and $D^*$ is that the latter has a polarization vector. Without this the parameterization of the matrix element reduces to (\ref{eq:vector:form:factors}).
			\item The dynamical variables available to us are the $D^*$ polarization $\epsilon$ and the two meson momenta $p_B$ and $p_D$. Eventually we will repackage the momenta into $(p_B+p_D)$ and $q\equiv (p_B-p_D)$. Now we would like to use spacetime symmetries to determine the structure of the parameterization. The $V^\mu$ operator contains a Lorentz index, so we know the right-hand side must be a vector or axial vector. The matrix element on the left-hand side is necessarily parity-even since \QCD respects $P$, so the right-hand side must be vectorial. In the next part we will use $T$ to determine that $g(q^2)$ is real, but note that we \textit{never} need to consider $C$ since QCD also respects charge conjugation and the right-hand side is composed of dynamical variables which are all trivially even under $C$. Thus all we have to worry about is constructing a Lorentz vector out of $\epsilon$, $P_B$, and $p_D$. 
			
			There is a subtlety in the parity of vectorial objects. The parity of a pure vector $V^\mu$ can be written as $P[V^\mu]=(-)^{\mu}$, by which we mean the parity is even (+) for $\mu=0$ and odd ($-$) for $\mu=1,2,3$. Note that the $\mu$ on $(-)^\mu$ isn't a Lorentz index, it's just there to tell us whether $(-)=+$ or $-$. Thus a $J^P = 1^-$ meson has parity $-(-)^\mu$. We see that the parity of the terms in the matrix element are
			\begin{align}
				P[V^\mu] = (-)^\mu \qquad\qquad
				P[D^*] = - \qquad\qquad
				P[B] = -.
			\end{align}
			We confirm that the matrix element is $P$-even with respect to `overall parities' $-$ and has the correct left-over `vector parity' $(-)^\mu$ that is required for a single-index object. The dynamical objects that we have, however, do not seem to have the correct parities:
			\begin{align}
				P[\epsilon^\mu] = -(-)^\mu \qquad\qquad\qquad P[p_{B,D}] = (-)^\mu.
			\end{align}
			In fact, we can see that the only non-zero vector bilinear $\epsilon_\mu p_B^\mu$ also does not have the correct parity. One might be led to believe that this matrix element must vanish. However, we have one more trick up our sleeves. In $d$-dimensional space we have a $d$-index totally antisymmetric $\varepsilon^{\mu_1\cdots \mu_d}$ tensor with which we can construct Lorentz contractions. In more formal language, we have the additional operation of taking a Hodge dual to convert $p$-forms into ($d-p$) forms. Our our present case this enables us to consider triple products of $\epsilon$, $p_D$, and $p_B$ to construct a Lorentz vector. The natural object to write down is 
			\begin{align}
				\varepsilon^{\mu\nu\alpha\beta} \epsilon^*_{\nu} p_{D\alpha} p_{B\beta}.
			\end{align}
			The key thing to recall is that $\varepsilon$ has parity
			\begin{align}
				P[\varepsilon^{\mu\nu\alpha\beta}]=-(-)^{\mu}(-)^{\nu}(-)^{\alpha}(-)^{\beta},
			\end{align}
			where we've written out the usual `vector parity' for a four-tensor-like object. Because of the additional overall minus sign, we see that $\varepsilon$ is a \textbf{pseudotensor}. One should already be familiar with this in the coordinate definition of the volume form in which the $\varepsilon$ tensor carries information about orientation. Armed with this we now see that the $(\varepsilon\epsilon^*p_Dp_B)^\mu$ contraction is indeed a parity-even vector to which the hadronic matrix element $\langle D^{*+}(p_D,\epsilon)|V^\mu|\bar B(p_B)\rangle$ may be proportional. 
		To complete the analysis, we remark that we are free to change momentum variables to $(p_D+p_B)$ and $q\equiv (p_D-p_B)$. It is clear that we are free to make the replacement
		\begin{align}
			p_{D\alpha} p_{B\beta} \longrightarrow (p_D+p_B)_\alpha q_\beta
		\end{align}
		since these are contracted with the $\varepsilon$ antisymmetric tensor so that only the $p_{D\alpha}p_{B\beta}-p_{D\beta}p_{B\alpha}$ term is picked out. Finally, the overall coefficient $g(q^2)$ can only be a function of $q^2$ since this is the only dynamical Lorentz scalar quantity. 
		\item The matrix element transforms as a complex conjugate under time reversal. The only complex element from our dynamical variables on the right-hand side is $\epsilon^*$, which contains an imaginary element. The sign of this element represents the transverse polarization of the $D^*$ and we expect it to flip under complex conjugation. Otherwise, the rest of the element is real and so $g(q^2)$ must also be real. 
		\item The axial current is not pure \QCD and comes from the chiral nature of the weak interactions and hence we are allowed to have $\langle D^{*+}(p_D,\epsilon)|A^\mu|\bar B(p_B)\rangle$ which is `overall parity' odd (i.e. $P = -(-)^\mu$). One immediate choice is a term proportional to $\epsilon^{*\mu}$ since this has precisely the correct parity and Lorentz structure. In addition to this, we can also form objects out of the three dynamical variables since $\epsilon^* (p_D+p_B) q$ also has the same correct parity and Lorenz structure; we just have to insert the Lorentz index at each place. The $\epsilon^{*\mu} (p_D+p_B)\cdot q$ term is the same as the $\epsilon^{*\mu}$ since $(p_D+p_B\cdot q)$ is not dynamical (it's a sum of masses). We are left with terms of the form
		\begin{align}
			\langle D^{*+}(p_D,\epsilon)|A^\mu|\bar B(p_B)\rangle =  g_{A1}(q^2)\epsilon^{*\mu} + g_{A2} \epsilon^*\cdot (p_D+p_B) q^\mu +  g_{A3}(\epsilon^* \cdot q)(p_D+p_B)^\mu.
		\end{align}
		
		\end{enumerate}
	\end{sol}
\end{prob}

\section{The CKM: Light quarks}\label{sec:CKM:I:Flavor}

Experimental physics has robust flavor structure that is explained very well by the Standard Model. What we mean by this is that we make several measurements and perform global fits to determine the  Standard Model parameters. When there are more measurements than parameters, the global fit becomes statistically meaningful. 
In order to `really probe' the Standard Model's flavor sector, then, we must measure and re-measure the four parameters of the \CKM matrix. To get to these four parameters, we will need to make additional measurements to deal with the incalculable hadronic physics of the previous section.

The motivation here is very important, since it is easy to feel like we're making a big fuss about measuring each particular element of the \CKM matrix. Nobody really cares about any particular element, say $V_{cb}$, of the \CKM. The insight from determining its value is to include it to the global fit of the Standard Model parameters to actually \emph{test} the theory. One might complain that as we go through each element of the \CKM matrix we're only measuring quantities which we could have extracted from the other elements; this is, in fact, the entire point! Each measurement only becomes meaningful when the parameter it is measuring has been measured independently in many other ways.

\subsection{Measuring $|V_{ud}|$}

Let's start with $V_{ud}$. The key process to measure $d\to u$ transitions is already very familiar: $\beta$ decay. This, however, comes in different forms:
\begin{itemize}
	\item \textbf{Nuclear} $\beta$ decay, for example $^3$H$\to ^3$He or $^{14}$C $\to ^{14}$N. 
	\item \textbf{Neutron} $\beta$ decay $n \to p e\nu$ involving a \textit{free} neutron decaying to a \textit{free} proton. 
	\item \textbf{Pion} $\beta$ decay, $\pi^+ \to \pi^0 e \nu$ or $\pi^+ \to \mu\nu$.
\end{itemize}
Each of these probe the $d\to u$ transition differently. Even though these are all mediated by the $W$ boson and hence the same \textit{operators}, these processes differ in their initial and final states. In other words, they differ according to how the quarks are dressed by \QCD. 

The basic approach for determining $V_{ud}$ in each of these processes is the same. We take the amplitude and compare it to muon decay. 
\begin{center}
	\begin{tikzpicture}[line width=1.5 pt, scale=1.3]
		\draw[fermionbar] (-140:1)--(0,0);
		\draw[fermion] (140:1)--(0,0);
		\draw[vector] (0:1)--(0,0);
		\node at (-140:1.2) {$d$};
		\node at (140:1.2) {$u$};
		\node at (.5,.3) {$W$};	
	\begin{scope}[shift={(1,0)}]
		\draw[fermion] (-40:1)--(0,0);
		\draw[fermionbar] (40:1)--(0,0);
		\node at (-40:1.2) {$e$};
		\node at (40:1.2) {$\nu_e$};	
	\end{scope}
	\begin{scope}[shift={(5,0)}]
			\draw[fermionbar] (-140:1)--(0,0);
			\draw[fermion] (140:1)--(0,0);
			\draw[vector] (0:1)--(0,0);
			\node at (-140:1.2) {$\nu_\mu$};
			\node at (140:1.2) {$\mu$};
			\node at (.5,.3) {$W$};	
		\begin{scope}[shift={(1,0)}]
			\draw[fermion] (-40:1)--(0,0);
			\draw[fermionbar] (40:1)--(0,0);
			\node at (-40:1.2) {$e$};
			\node at (40:1.2) {$\nu_e$};	
		\end{scope}
	\end{scope}
	\end{tikzpicture}
\end{center}
How do these diagrams differ? The external states and the phase space integrals are different, but the actual Feynman rules differ only by a factor of $V_{ud}$.

\subsubsection{Nuclear $\beta$ decay}

Let's start with nuclear $\beta$ decay. As a particle physicist, you
may be nervous---nuclei are complicated objects! Fortunately,
symmetries make our lives much easier. We want to determine the matrix element of an operator $\mathcal O$ sandwiched between two nucleon states $N$ and $N'$,
\begin{align}
	\langle N | \mathcal O_\beta | N'\rangle \ .
\end{align}
In the Standard Model, we know that $\mathcal O_\beta$ comes from $W$ exchange and so can be vector or axial. In fact, we can pare this down further in the non-relativistic limit. The energy emitted by the electron in nuclear $\beta$ decay is much smaller than the rest energy of the nucleus. For example, a tritium nucleus at rest hardly moves after undergoing $\beta$ decay to $^3$He. In this limit, the axial current differs from the vector by a \textbf{spin flip}. This can be seen, for example, from the non-relativistic Dirac equation or by considering the effect of the $\gamma^5$ in the Dirac fermion basis. 

We use this observation to further simplify our analysis. Independent of the non-relativistic limit, we know that we can appeal to cases where either the $A$ or $V$ operator vanishes. For example, $A=0$ when the initial and final states are spin zero since a spin-less particle vanishes when acted upon by a spin-changing operator. Said differently, the vector and axial vector are parity odd and even respectively. The external states have a definite parity and since \QCD respects parity, only one of the axial and vector operators may be non-zero. Fewer matrix elements mean fewer calculations, which makes us happy.

In nuclear physics these $N \to N'$ transitions between $J=0$ states are called a \textbf{superallowed} $\beta$ decays. In such a transition one of the neutrons within the nucleus $N$ converts into a proton to produce nucleus $N'$. Nothing else has changed, certainly not spin since both $N$ and $N'$ have $J=0$. Now invoke an approximate symmetry: in the limit of exact isospin symmetry, \textit{nothing changes} when we make this $n\to p$ transition in the nucleus. The $n$ and $p$ are just states of isospin $+1/2$ and $-1/2$ respectively. This limit ignores the mass differences and electric charges of these particles, but as far as \QCD is concerned, it is as if nothing changed. The \QCD transition matrix element should thus be unity: $\langle N'|\mathcal O_\beta|N\rangle = 1$. 

This is a deep and useful tool for us: we almost have no right to be
able to say anything about this matrix element, but since we have an
approximate symmetry that becomes exact in a well defined limit, we
can sneakily exert some control over what's going on. In this case, we 
expect isospin breaking to give one or two percent corrections. 
Before going into detail of how isospin helps us do this calculation, the result turns out to be 
\begin{align}
	\left|V_{ud}\right| = 0.97425(22) \qquad \Longrightarrow\qquad \lambda = 0.2255(10). \label{eq:Vus}
\end{align}
Look at how many significant digits we have! This is certainly an experimental triumph that we have such accuracy despite \QCD. Let's sketch how we can get this result.

The decay amplitude $A\to B$ in the Standard Model takes the form:
\begin{align}
	\mathcal M^\mu &= \langle N'(p') | A^\mu + V^\mu | N(p) \rangle.
\end{align}
We choose spin-less initial and final states, $J_N = J_{N'} = 0$, so that $\langle N' | A^\mu | N \rangle = 0$. Thus this reduces to the form factors that we introduced in (\ref{eq:vector:form:factors}),
\begin{align}
	M^\mu = \langle N'(p') | V^\mu |N(p)\rangle = f_+(q^2) (p+p')^\mu + f_-(q^2) q^\mu. 
\end{align}
Now we invoke the isospin limit where $m_N = m_{N'}$. In this limit, the decay cannot occur kinematically! This is okay. We use the isospin limit to calculate the hadronic matrix element, and \emph{the matrix element doesn't care about kinematics}. Since the real world is very close to the symmetry limit, the matrix element at $q^2 = 0$ should be very close to its true value. 
In other words, $q^2$ is very small relative to $\Lambda^2_{\text{QCD}}$ so that
\begin{align}
	f_\pm (q^2) \approx f_\pm (0).
\end{align}
\begin{eg}
	We can test this assumption by looking at the spectrum of this decay with respect to $q^2$. If $f_\pm$ is constant then the spectrum is a straight line. This is precisely the dominant feature in the so-called \textbf{Kurie plot} of tritium $\beta$ decay.
\end{eg}

In the isospin limit, we have yet another very helpful simplification: $f_-(q^2\ll \Lambda^2_{\text{QCD}}) = 0$. To see this we invoke the Ward identity,
\begin{align}
	0=q_\mu \mathcal M^\mu = f_+(q^2)(p^2-p'^2) + f_-(q^2)q^2 = f_+(q^2)(m_{N}^2-m_{N'}^2) + f_-(q^2)q^2.\label{eq:Vud:fminus:zero}
\end{align}
The exact isospin limit implies $m_N = m_{N'}$ so that the $f_+$ term vanishes. This, in turn, means that $f_-(q^2)q^2 = 0$. Note that we are \textit{not} saying that $q^2=0$, even though we use the  $q^2\to 0$ limit to get to this equation. Thus we conclude that $f_-(q^2)=0$.

We have reduced everything to a single form factor, $f_+(0)$. We should be very proud of ourselves. But, like any good infomercial,  ``\textit{wait, there's more!}'' It turns out that while we're working in the isospin limit, $f_+(0)$ doesn't even need to be determined experimentally---\emph{we can figure it out simply using group theory}!

The vector operator for the non-relativistic contribution to this matrix element changes $I_3$ by one unit; this is only the difference between $N$ and $N'$ as far as isospin and \QCD is concerned. In terms of quarks, this is because $V^\mu = \bar u \gamma^\mu d$ and each of the quarks changes isospin by 1/2. Since this is a $\Delta I_3 = 1$ operator, we know that is is proportional to the usual $\text{SU}(2)$ raising operator acting on isospin space,
\begin{align}
	 V | j,m\rangle \propto \sqrt{(j-m)(j+m+1)} |j,m+1\rangle.
\end{align}
The overall coefficient can be determined, but \textit{we don't care} because this cancels in the ratio with $\mu\to e\nu\bar \nu$. The coefficient above is just a Clebsch--Gordan coefficient and tells us that, finally,
\begin{align}
	\langle N'(p') | V^\mu | N(p) \rangle \propto \sqrt{(j-m)(j+m+1)}(p+p')^\mu,
\end{align}
where we stress that the $j$ and $m$ are \textit{isospin} numbers, not angular momentum. This is an incredible result: we started with some hadronic matrix element that we knew nothing about. Using Lorentz invariance, specific decay channels, and approximate isospin symmetry we were able to get rid of \textit{all} \QCD unknowns and end up expressing everything in terms of a single number that comes from group theory. After taking into account corrections to the approximations that we've made, one can extract the value for $V_{us}$ in (\ref{eq:Vus}).

\subsubsection{Neutron $\beta$ decay}

Neutron offer another laboratory to measure $V_{ud}$. Consider the free neutron decay process $n \to p e \bar\nu$. While this na\"ively appears to be very similar to the nucleon calculation, the external states are different---the hadronic states are now necessarily spin-1/2 so we can't get away with using the $J_N = J_{N'}=0$ trick to get rid of the axial contribution. This ends up giving six matrix elements to calculate which, then (nontrivially!) reduce to calculating
\begin{align}
	G_V &= \langle n|V^\mu|p\rangle(p+p')_\mu\\
	G_A &= \langle n|A^\mu|p\rangle(p+p')_\mu.
\end{align}
The relevant quantity turns out to be the ratio of these matrix elements, $g_A \equiv G_V/G_A$. This is a ratio between two hadronic matrix elements with no symmetries forcing any obvious hierarchies between the two terms so that we expect $g_A \sim \mathcal O(1)$. Indeed, it turns out that $g_A = 1.27$. 
\begin{eg}
	How does one measure $g_A$? We can distinguish the $V$ and $A$ matrix elements by their angular distributions so that the angular dependence of $\beta$ decay is a probe of $g_A$.
\end{eg}
$V_{ud}$ depends on the neutron lifetime and $g_A$ as
\begin{align}
	|V_{ud}|^2 \sim \tau_n \times (1+3g_A^2) (1+ \text{radiative corrections}).
\end{align}
We can  calculate the right-hand side to determine $V_{ud}$; it turns out to have roughly the same value and precision as the nuclear matrix element method. This should be very reassuring since this method carries different hadronic assumptions. For two recent reviews, see \cite{Hocker:2006xb} and \cite{Blucher:2005dc}.

\subsubsection{Pion $\beta$ decay}

Finally, for pion $\beta$ decay, $\pi^+ \to \pi^0 e \nu$, we see immediately that the hadronic matrix element is between two spin-0 states, $\langle	\pi^0 | \mathcal O | \pi^{\pm} \rangle$, which is a very nice feature.
The pions live in an isospin $I=1$ isotriplet and are much simpler objects than nuclei. It even turns out that the calculation of the corrections to the isospin limit are much easier for pions. In fact, this is the \emph{cleanest} way to measure $V_{ud}$. Why, then, are we only mentioning this decay as a remark after the nucleon and neutron decays? Why isn't this the \textit{best} way to measure $|V_{ud}|$? It is suppressed by phase space! The $\pi^\pm$ and $\pi^0$ are nearly degenerate so that the branching ratio is on the order of $10^{-8}$. Experimentally we are limited by the total number of these decays that we measure.
\begin{eg}
	The neutral pions are \textit{not} difficult to detect experimentally. Due to the axial anomaly $\pi^0 \to \gamma\gamma$ is the dominant decay (98.8\%) and it's easy to tag two photons with an invariant mass of $m_\pi = 135$ MeV. 
\end{eg}

%
%
%
%

\subsection{Measuring $|V_{us}|$}

The determination of $|V_{ud}|$ is a template for measuring all of the \CKM elements: identify the right decays, use a symmetry limit where you can appeal to tricks to simplify calculations, and then (the part that we've not done explicitly) calculate corrections to the symmetric limit. We now apply this to the up--strange \CKM element.

\textbf{Kaons} are the lightest mesons that contain a strange quark; like the pions. The the charged kaons are like pions with $d\to s$, whereas the neutral kaon is a $d\bar s$. The decays of kaons to states that do not contain a strange are thus an obvious place to measure $V_{us}$.

\subsubsection{$K\to \pi \ell \nu$}
\label{sec:Kl3}

The dominant semileptonic decay is $K \to \pi \ell\nu$. We would like
to apply the same slick maneuvers to determine $|V_{us}|$ that we used
for $|V_{ud}|$. To do this we need an approximate symmetry that
simplifies the analysis. Here, again, we have an obvious choice: the
generalization of isospin to include the strange quark is
$\text{SU}(3)$ flavor as mentioned in
Section~\ref{sec:flavor:symmetries}. In fact, if we stare at this long
enough we see that this process with respect to the $v$-spin subgroup of
$\text{SU}(3)$ flavor symmetry is really just like pion $\beta$ decay, $\pi^+ \to \pi^0 \ell^+ \nu$, with respect to isospin. 

However, we do have reasons to be very skeptical. For example, we know
that isospin is broken by effects of the order $\alpha$ and
$(m_d-m_u)/\Lambda_{\text{QCD}}$ both are of order $1\%$. Thus, with some exceptions---such as the difference in the proton versus neutron lifetimes---we expect isospin breaking effects to be small corrections. For $\text{SU}(3)$ flavor, however, the symmetry breaking is on the order of $m_s/\Lambda_{\text{QCD}} \sim 20\%$. While this is still a valid symmetry limit to expand about, we now expect the precision to be rather poor compared to the precision of our experiments. There is a second reason why we should be skeptical that our previous bag of tricks will work: in our measurement of $|V_{ud}|$ we assumed that $f(q^2)\approx f(0)$. We justified this because $q^2\ll \Lambda^2_{\text{QCD}}$. In this case, however, because the kaon and pion masses are appreciably different, the momentum transfer is not small. It seems like this approximation should also be very poor.

Having aired our objections to $\text{SU}(3)$ flavor due to the size of its breaking and the small $q^2/\Lambda^2_{\text{QCD}}$ limit, we  proceed with the calculation guilt-free and see what happens. We are interested in the two decays,
\begin{align}
	K^0 &\to \pi^-e^+ \nu_e 
	&
	K^+ &\to \pi^0 e^+\nu_e 
	\ .
	\label{eq:K:pi:nu:e}
\end{align}
Actually, we should be more honest: we don't actually see $K^0$s. The neutral kaon mass eigenstates are denoted by $K_{S,L}$ (``K-short'' and ``K-long'') due to mixing. This is an entire rich and fascinating story in itself, but we leave this for later. If you are very serious, then you should replace $K^0$ with $K_L$, but for our current purposes we can be na\"ive.

What is the difference between the two decays in (\ref{eq:K:pi:nu:e})? They have different charges. This comes from the fact that the $K^0$ decay has a spectator $\bar d$ quark while the $K^+$ decay has a spectator $\bar u$. This tells us that two decays are related by \textit{isospin}. Thus to the extent that isospin is approximately true, it is sufficient to determine the form factor for only one of these process since the other will be related by symmetry.

We write out the matrix element:
\begin{align}
	\langle \pi (p_\pi)| \bar s\gamma^\mu u| K(p_K)\rangle = f_+(q^2) (p_\pi+p_K)^\mu + f_-(q^2)q^\mu \ ,
\end{align}
where we haven't yet specified $K^0$ or $K^+$ decay.
Isospin tells us that we can relate the two form factors. We know that $K^0$ is an isospin $|\frac 12, -\frac 12\rangle$ state. This is related to $K^+$ by 
	\begin{align}
		J_-|j,m\rangle = \frac{1}{\sqrt{2}}\sqrt{(j+m)(j-m+1)}|j,m-1\rangle.
	\end{align}
	Thus we see that
	\begin{align}
		\frac{f^{K^+}_+(0)}{f_+^{K^0}(0)}= \frac{1}{\sqrt{2}}\ .
	\end{align}	
What about the other form factor, $f_-(q^2)$? In (\ref{eq:Vud:fminus:zero}) we invoked the isospin symmetry limit to explain why this term should vanish in nuclear $\beta$ decay. In the case of kaon decay, we replace isospin with SU$(3)$ flavor and this argument becomes rather fishy: after normalizing $f_+$ to unity, we would at best be able to say that $f_-$ is on the order of 20\%. Fortunately, there's another argument available to allow us to neglect $f_-$. The contributions of the $f_-$ form factor to semileptonic kaon decays are proportional to the lepton mass. Thus for decays with electrons, these decays are suppressed by factors of $m_e/m_K$. This is essentially the same chiral suppression we encountered in Problem \ref{prob:pi:mu:nu:amp}, though here we have a three body decay. The amplitude associated with this term goes like
\begin{align}
	\mathcal M(f_-) \sim f_-(q^2)\, (p_e+p_\nu)_\mu \bar u_e \gamma^\mu u_\nu \ ,
\end{align}
where we've used $p_\pi - p_K = p_e+p_\nu$. Contracting the Lorentz indices gives
\begin{align}
	\mathcal M(f_-) &\sim f_-(q^2)\, \bar u_e (\slashed{p}_e+\slashed{p}_\nu) u_\nu\\
					&\sim f_-(q^2)\,m_e\, \bar u_e u_\nu \ ,
\end{align}
where we've used the equations of motion for the leptons, $\slashed{p}_e u(p_e)=m_e u(p_e)$ and $\slashed{p}_\nu u(p_\nu)=0$.

\begin{eg}\textbf{That's a nice argument, why didn't we use it before?}
	Let's recall what just happened: in pion $\beta$ decay we had a nice symmetry argument for why $f_-(q^2)=0$. We then moved to kaon decay, and found that this argument is no longer so nice since $SU(3)$ flavor isn't as precise as $SU(2)$ isospin. So instead, we cooked up a different reason why the contribution to the amplitude mediated by this term is suppressed anyway. \emph{If this `dynamical' reason was there all along, why didn't we apply it to pion decay?}
	At first glance, one might say that pions are lighter than kaons, and so the mass suppression shouldn't be as dramatic. However, pions are only a factor of about 4 lighter than kaons, which should not be a big difference compared to the size of $m_e$. This line of thought, however, is a red herring! The point isn't that $m_e \ll m_K$, but rather that $m_e\ll (m_K-m_\pi)$. The reason why this argument doesn't hold for pion decay is that $m_e$ is not \textit{much} smaller than $(m_{\pi^+}-m_{\pi^0})$.
\end{eg}

\begin{eg}\textbf{What about muons?}
	The next obvious question is whether $f_-(q^2)$ can still be neglected when considering semileptonic decays into muons, e.g.\ $K\to \pi \mu \nu$. The answer is, as you would expect, \emph{no}, since $m_\mu \approx 100$~MeV is certainly not very small compared to the meson mass difference. This is why we were very specific in (\ref{eq:K:pi:nu:e}) to specify the outgoing electron rather than a general lepton. For semileptonic kaon decays into muons, we say that we are ``sensitive to $f_-(q^2)$.'' This is a useful process not so much for measuring $|V_{us}|$, but for checking predictions of the form factors from lattice calculations or chiral Lagrangians.
\end{eg}

Now it's time for us to face the music. To get to this point we assumed that $q^2 \ll \Lambda_{\text{QCD}}$ and that SU$(3)$ breaking is small. These manifest themselves into the statements
\begin{align}
	f(q^2) &= f(0)
	&
	f(0) &= 1 \ , 
\end{align}
respectively. While the analogous assumptions were completely sensible for pion decay, we're now well outside the regime where we would expect these to give reasonable results. It turns out, somewhat magically, that these \textit{a priori} poor approximations work out much better than we ever had a right to expect. 

First consider $f(q^2)=f(0)$. One can check this by plotting the decay spectrum with respect to energy and observing that that it is a straight line. It `turns out' that the expansion in $q^2$ takes the form 
\begin{align}
	f(q^2) = f(0) + \lambda q^2 + \cdots. 
\end{align}
where $\lambda \sim 0.04$ fm$^2$ and the higher order terms are not measured. Why should the coefficient be so small? A not-so-satisfactory hand-waving answer is that to good approximation the kaon and pion are point-line. Since the form factors really probe the structure of the particles, the small coefficient is telling us that the process is not sensitive to any meson substructure. As long as $q^2 \ll \Lambda_{\text{QCD}}^2$ this is what we expect.

Next consider the approximation $f(0)=1$. It turns out that the value we find is
\begin{align}
	f(0) = 0.961 \pm .008.
\end{align}
Na\"ively we would have expected $f(0)= 1\pm 20\%$, how did we get something which is more like $1\pm 4\%$? There's a clue here if we play with some numerology: how is 4\% related to 20\%? It is the square!
\begin{align}
	(20\%)^2 = 4\%.
\end{align}
Besides writing a trivial numerical identity, this is a big hint: it seems like the actual correction to our symmetry-limit isn't linear ($20\%$), but rather second-order. Thus our expansion about the symmetry limit is somehow slightly different from the na\"ive expansion that we would have written. Somehow $f(0)=1$ up to \textit{second-order} in SU$(3)$ breaking. This has a fancy name:
\begin{theorem}\label{thm:Ademollo:Gato} (\textbf{Ademollo--Gatto}) The form factor $f(0)=1$ up to second order in the symmetry breaking parameter~\cite{Ademollo:1964sr}.
\end{theorem}
The argument doesn't quite work for $f_-(q^2)$, but, as we argued
above, we don't care since we already have a nice $m_e/m_K$
suppression on this term. Here we present a heuristic argument\footnote{We thank Shmuel Nussinov for explaining this to us over dinner.} for the Ademollo--Gatto result. The $f_+(q^2)$ form factor of the $(p+p')^\mu$ term in the matrix element should be unity for $q^2=0$ and in the limit where $SU(3)$ is restored. This is because $f_+(0)$ should be understood to be proportional the overlap of the $s$ and $u$ wavefunctions which approaches unity in the unbroken $SU(3)$ limit. Clearly this is the maximum value of such a quantity. The heart of the Ademollo--Gatto theorem---and the reason why $f_+$ ends up being so well behaved despite the breaking of $SU(3)$---is that at the deviation of a function expanded about a critical point (such as a maximum) is always \textit{second} order, tautologically.  Note that the observation that the form factor should be interpreted as a wave function overlap is already something that we met with the `decay constant' of positronium in Example \ref{eg:positronium:decay:constant}. A similar theorem, Luke's theorem \cite{Luke:1990eg, Boyd:1990vy}, exists for the limit of large symmetry breaking where the critical point is a minimum.

\begin{prob} \textbf{Ademollo--Gato counter-argument}. In our heuristic proof of the theorem, we argued that $f(q^2)=1+\mathcal O(q^2)$ because $q^2 = 0$ is a maximum of $f(q^2)$ so that corrections are quadratic. However, a contrarian may argue that you can also have a maximum at the endpoint of a range, and because $q^2\geq 0$ (Problem~\ref{prob:range:of:q2}), our proof has holes in it. Counter this counter-argument.
	\begin{sol}
		The argument that we want to support is that $q^2=0$ is a maximum of the overlap between the constituent quarks in the kaon. This argument has \emph{nothing} to do with kinematics or the requirement that $q^2$ takes a certain range for physical processes. We can consider this overlap ``before'' we do any kinematics: $q^2$ may even be space-like ($q^2 <0$). When this is the case, we know that the overlap is still maximal at $q^2=0$, so that our heuristic argument for $f(q^2) = 1 + \mathcal O(q^2)$ still holds. One may prove this more rigorously, but the hard evidence is that it is true experimentally.
	\end{sol}
\end{prob}


Returning to our main task, we can use the above techniques to obtain 
\begin{align}
	V_{us} = 0.2257(21).
\end{align}
The precision here includes corrections from the symmetry limit that are beyond the scope of these lectures.
Back in ancient times---the 1960s---this decay was known as $K_{\ell3}$. We present a similar observable called $K_{\ell 4}$ in Appendix~\ref{sec:Kl4}.

\subsubsection{Hyperon decay}

\begin{quote}
	\textit{Thirty years ago when you say `$D$ term' and `$F$ term,' everybody knew what you were referring to. You were referring to the SU(3) invariants in hyperon decay.	Now if you say `$D$ term' or `$F$ term,' everybody thinks you're talking about supersymmetry. 
	}
\end{quote}

Moving on from the mesons, another way to measure $|V_{us}|$ is to look at the decays of baryons containing strange quarks and no heavier quarks, the \textbf{hyperons}. These include the $\Lambda$, $\Sigma$, $\Xi$, and $\Omega$ baryons. Just as we looked at neutron $\beta$ decay $n \to p e \nu$ to measure $|V_{ud}|$, we can also measure $\Lambda \to p e \nu$ in the analogous way to measure $|V_{us}|$. 
\begin{eg}\textbf{$\Lambda$ lifetime versus $n$ lifetime}. Why does
  the $\Lambda$ decay much faster than the neutron? If you answer that
  $|V_{us}|\ll |V_{ud}|$, then \textit{wake up}! This effect goes
  in the opposite direction. The correct answer is that the $n$ decay is very phase space suppressed compared to $\Lambda$ decay.
\end{eg}

The upshot of looking at baryons is that there are many more baryons to work with and that we can measure all of them, at least in principle. The SU(3) baryon analysis depends on parameters that encode the two irreducible matrix elements between two SU(3) flavor octet states, $B_n$ and $B_m$, with respect to an octet operator $\mathcal O_k$~\cite{Cabibbo:2003cu}, 
\begin{align}
	\langle B_n |\mathcal O_k | B_m\rangle = F_{\mathcal{O}} f_{knm} + D_{\mathcal{O}} d_{knm}\ ,
\end{align}
where $f_{knm}$ is the SU(3) structure constant and $d_{knm}$ is a group theory factor defined by
\begin{align}
	\left\{ \lambda^k,\lambda^n \right\} = 2\delta_{kn} + 2d_{knm}\ .
\end{align}
This just the same program of defining form factors to parameterize our ignorance of \QCD matrix elements. Note that this expansion implicitly includes the assumption of SU(3) flavor, thus we neglect the mass difference between the light quarks.

This is a pretty good deal: we have two form factors in the SU(3) limit and several baryons with which we may make measurements. We again turn the crank by following through the calculations and then doing the experiments. There are lots of tricks one can further use to keep a handle on the breaking terms, and at the end of the day one finds that one can measure $|V_{us}|$ with the same precision as the kaons. For a nice review of the current state-of-the-art for semileptonic hyperon decays, see the review by Cabibbo, Swallow, and Winston \cite{Cabibbo:2003cu}.


\subsubsection{$\tau \to K^-\nu$}

Yet another measurement of $|V_{us}|$ comes from the decay of the $\tau$. Recall that the $\tau$ is so heavy that its quark decays hadronize. In this sense, the $\tau$ is \textit{almost} a hadron. Our strategy is to divide out the hadronic uncertainties by comparing the ratio of two decay rats,
\begin{align}
	\tau^- &\to K^- \nu_\tau 
	&\text{and}&&
	\tau^- &\to \pi^- \nu\ .
\end{align}
In the limit of SU(3) flavor, the ratio is simply
\begin{align}
	R = \frac{\Gamma(\tau^- \to K^- \nu)}{\Gamma(\tau^- \to \pi^- \nu)} \quad\stackrel{SU(3)}{=}\quad \frac{|V_{us}|^2}{|V_{ud}|^2}.
\end{align}
This is too rough for us. Let us at least quantify the dominant corrections from the $SU(3)$ limit. One correction is phase space, since the $K$ has mass approximately equal to 500 MeV while the pions are around 140 MeV. Recall that the $\tau$ mass is 1777~MeV.
On top of this, we also have a correction from hadronic effects that we know to high precision. What is the hadronic matrix element here? It's just $K$/$\pi$ going to vacuum, thus they simply decay constants! Hence another breaking effect is $f_K$ versus $f_\pi$. We end up with a revised ratio,
\begin{align}
	R = \frac{\Gamma(\tau^- \to K^- \nu)}{\Gamma(\tau^- \to \pi^- \nu)} = \frac{|V_{us}|^2}{|V_{ud}|^2}\times(\text{Phase Space})\left|\frac{f_K}{f_\pi}\right|^2.
\end{align}
The powers of two should be clear since we are comparing branching ratios, not amplitudes.  

Let us make a remark about the measurement of $f_K$ and $f_\pi$. To use the above method to determine $|V_{us}|$, we need a measurement of $f_K$ that is somehow `orthogonal' to the product $f_K |V_{us}|$. We could measure $f_K$ through leptonic decays such as $K\to \mu\nu$, but this gives something proportional to the same combination $f_K |V_{us}|$. This is also true for the pions, which is why we focused on $\beta$ decay rather than $\pi \to \ell \nu$ to measure $|V_{ud}|$. It turns out that for $f_K$ we have to depend on a lattice \QCD calculation of $f_K$ as an independent input.

The best way to use the $\tau$ is to look at \textbf{inclusive} decay. The main idea is that the $\tau$ is heavy enough that at $m_\tau$ one can really treat \QCD perturbatively, $\alpha_s(m_\tau)\sim 0.3$. You can calculate how many light and heavy quarks you produce and trust your perturbative calculation.
The inclusive decay has no form factors, which makes it theoretically very nice, but currently our limiting factor is the number of experimentally observed $\tau$ decays.

\section{The CKM: Heavy quarks}\label{sec:CKM:I:Flavor-heavy}

We now move to discuss processes that involve heavy quakrs. They
sensitive to \CKM matrix elements with $c$ and $b$ quarks. The
ideas, while similar in spirit to what we seen so far, are different
in the details of how to apply the symmetries.

\subsection{Measuring $|V_{cs}|$}

The values for $|V_{cs}|$ and $|V_{cd}|$ are not measured very
well. The errors are much much larger than those for $|V_{ud}|$ and
$|V_{us}|$ by factors of ten. The reason is clear: the charm quark
mass lives in the dangerous in-between regime where it's too heavy to
be considered light and thus extent the $SU(3)$ flavor symmetry, but not heavy enough to treat in the heavy quark limit. Or, in the timeless words of Britney Spears,
\begin{quote}
	\textit{Feels like I'm caught in the middle\\
	That's when I realized---\\
	I'm not a girl,
	Not yet a woman.}
\end{quote}
Later in the song Spears sings the line ``\textit{I'm not a girl, there's no need to protect me...},'' from, we can only assume, the large corrections associated with quarks which cannot be reasonably treated in either the SU(3) flavor or heavy quark limit.

It turns out that lattice \QCD is ideal for effects that are on the
order of $\Lambda_{\text{QCD}}$ and the charm mass is about a factor
of few larger than this. Recall that the approach of lattice \QCD is to discretize spacetime so that one immediately has \acro{UV} and \acro{IR} cutoffs associated with the lattice spacing and the total lattice size, respectively. One expects that such techniques should not be able to capture effects that are much smaller or larger than $\Lambda_{\text{QCD}}$ due to these cutoffs. In other words, the \QCD effects that are precisely \textit{at} $\Lambda_{\text{QCD}}$ are those which should be best captured by lattice techniques.

\subsubsection{$D\to K e\nu$}

The obvious process to probe $|V_{cs}|$ is $D\to K e\nu$. If this is not `obvious' go back and look at our favorite processes for $|V_{ud}|$ and $|V_{us}|$. Just as we did for $K\to \pi e \nu$, we pick the electron final state in favor of the muon final state since this saves us one less form factor to worry about: $f_-$ gives an effect that is suppressed by $m_\ell/m_{D}$. Following Section~\ref{sec:Kl3}, the remaining problem is that we have no idea what $f_+(0)$ is. The lattice gives us $f_+(0)=0.7$, but we even don't know if $q^2=0$ is a good approximation. Note that this has indeed shifted quite a bit from the previous symmetry limit where $f_+(0)=1$. Going through the same rigamarole, we end up with a measurement
\begin{align}
	V_{cs} =0.98 \pm 0.01_\text{exp} \pm 0.10_\text{thy} \ .
\end{align}
This is indeed approximately equal to one. There's a pretty small experimental error, but by comparison the theory error is huge. Compare this to the case for $|V_{ud}|$ where the error was in the fourth digit! Here we give our best effort with lattice \QCD, but we're still left with 10\% theory error. 

\subsubsection{$D_s^+ \to \ell^+ \nu$}

Perhaps alternative measurements of $|V_{cs}|$ can give more insights. In that spirit, we look at the pure leptonic decays of the $D_s$. It should be clear why we use $D_s$, since this is the meson which will give a factor of $|V_{cs}|$ when it decays. We are thus interested in the decays
	$D_s^+ \to \ell^+ \nu$,
where the $\ell$ can be any of $e$, $\mu$, $\tau$. Each choice has its own pros and cons. The electron has the smallest branching ratio due to chirality suppression, so we can forget about it. Now we have to choose between the $\mu$ and $\tau$. On the one hand, chirality suppression prefers the $\tau$ decay, but on the other hand the $\tau$ is constrained by phase space since $m_\tau \approx M_{D_s}$, where $M_{D_s}=1970$ MeV. The muon is more chirality suppressed, but enjoys a much bigger phase space. Theoretically these effects balance out and both are useful. Experimentally, on the other hand, the $\tau$ is a pain to measure since it hadronizes while the muon is a very clean signal. Enterprising experimentalists, however, have been able to use both modes with success. 

Before we can extract $|V_{cs}|$, we need to know $f_{D_s}$. This is
comforting, since lattice people tell us that they really know how to
calculate 
$f_{D_s}$ accurately: the $c$ is not too heavy, and the $s$ is not too light relative to $\Lambda_{\text{QCD}}$. Indeed, the simplest things for the lattice to calculate are masses and decay constants, and $f_{D_s}$ should be ideal---it's a number that lattice techniques were \textit{born} to predict. Because there's no such thing as an `easy' lattice calculation, there was of course a $3\sigma$ disagreement. Like nearly all $3\sigma$ effects, it ended up being pushed down to $1\sigma$ and forgotten\footnote{In this case the reconciliation was only done one month prior to the original version of these lectures and was performed by the then-dean of Cornell's College of Arts and Sciences~\cite{Na:2010uf}.}. 

The lattice predicts the ratio of $f_{D_s}/f_D$, from which we can reconstruct and even better measurement for $|V_{cs}|$ than we had for the semileptonic decay:
\begin{align}
	|V_{cs}| = 1.030 \pm 0.038 \ ,
\end{align}
where the error is basically theoretical. Does this look strange to you? $|V_{cs}|>1$, but is at least within error\footnote{Even if it were not within error, we would have said that someone underestimated the error.}. Such a thing happens all the time---we measure something which is supposed to be the sine of an angle, but it ends up also being larger than unity. It boils down to how we extract parameters from measures;  we rarely are able directly measure quantities like this.

\subsubsection{$W\to c\bar{s}$}

There is yet another way to measure $|V_{cs}|$: the decay of a $W$ boson, $W\to c\bar s$. This is the most direct measurement of $|V_{cs}|$. Can we actually measure this?
\begin{eg}\label{eg:W:csbar:branching}\textbf{Quick calculation: the branching ratio of $W\to c\bar s$}. Neglecting higher order terms in the Wolfenstein expansion parameter $\lambda$, a $W$ in its rest frame can decay into any of the three lepton--neutrino pairs or any of the two \textit{active} generations of quarks, $u$--$d$ or $c$--$s$. Note that it does not have the rest energy to decay into third generation quarks. The quarks have an additional factor of three due to color so that there are nine possible decays. We can approximate each as having the same rate so that the second generation quark decays occur 1/3 of the time. This is an appreciable rate and so we expect it to be tractable experimentally.
\end{eg}

There's an immediate problem: How are we going to actually \textit{see} $W\to c\bar{s}$? What we really see is $W$ going to two jets. And as we know from West Side Story,
\begin{quote}
	\textit{When you're a Jet,\\
	you're a Jet all the way\\
	From your first cigarette\\
	To your last dying day.
	}
\end{quote}
The problem is that unlike charmed---and especially bottom---jets, we cannot tag a strange jet\footnote{Tony was a strange jet because he fell in love with Maria, the sister of the leader of the rival gang, the Sharks. Further, \emph{West Side Story} is based on \emph{Romeo and Juliet}: Romeo was charming, and Juliet even sounds like `jet.'}.
We can identify jets containing a $b$ or $c$ quark by measuring displaced vertices. This boils down to the lifetime of the $b$ and $c$ mesons.
	\begin{eg}\label{eg:B:D:lifetime} \textbf{The lifetime of the $B^0$ and $D^0$ mesons}. Which has a longer lifetime, the charmed or the bottom mesons? We can see that the $D^0$ decay goes like $|V_{\text{cs}}|^2$ while the bottom goes like $|V_{\text{cb}}|^2$. We know from our Wolfenstein parameterization that the latter is suppressed by $(\lambda^2)^2$. On the other hand, we should also account for the phase space. From dimensional analysis we write
		\begin{align}
			\Gamma \sim m^n G_F^2 \ ,
		\end{align}
		where $m$ is the meson mass, $n$ is some integer, and $G_F$ contains the necessary $M_W^{-2}$ factors from the $W$ propagator. In order to obtain the correct overall dimension we must have $n=5$, so that the ratio of the two branching fractions scale like
		\begin{align}
			R =\frac{\Gamma(b)}{\Gamma(c)} =\left(\frac{m_B}{m_D}\right)^5 \lambda^4.
		\end{align}
		It turns out that phase space factors nearly cancel the Wolfenstein suppression leaving $R\sim 0.3$, so that they lifetimes are actually comparable. Note that since $\tau \sim 1/\Gamma$, this means that the $B$ has a longer lifetime by a factor of about three.
	\end{eg}
	What we end up with are $B$ mesons which give a nice, observable displaced vertex while the $D$ meson gives a slightly harder-to-measure displaced vertex. For $W$ decays, the $b$ quark isn't an active generation, so \textit{any} displaced vertex is a signal for charm. 
The $W\to c\bar{s}$ measurement was done by the Delphi experiment, one
of the four \acro{LEP} experiments that ran at \acro{CERN}. 
Delphi found
\begin{align}
	V_{cs} &= 0.94 \pm 0.29_{\text{stat}} \pm 0.13_{\text{sys}} \ .
\end{align}
This number is constrained by the number of $W$ bosons that they measured.

\subsubsection{$W\to $ hadrons}

Another $|V_{cs}|$ measurement which all of the \acro{LEP} experiments were able to do is $W\to $ hadrons. We can normalize this to the rate for $W\to$ leptons,
\begin{align}
	R = \frac{\Gamma(W\to h)}{\Gamma(W\to \ell)} = N_c^\text{eff} \sum_{u,c,d,s,b}(V_{ij})^2 \times (\text{Phase Space})
\end{align}
Using the same approximation that we used in Example \ref{eg:W:csbar:branching}, we find that $R\approx 2$. We can write out the branching ratio of the $W$ to leptons as
\begin{align}
	\frac{1}{\text{Br}(W\to \ell)} = \frac{\Gamma(W\to h)+\Gamma(W \to \ell)}{\Gamma(W\to  \ell)} = R+1.
\end{align}
Thus measuring the branching ratio to leptons very precisely gives $R$ so that the theoretical situation is very nice. In order to measure this precisely we need to include \QCD corrections, which we are able to do since we are in a regime where we are able to do perturbative \QCD. 

This measurement ultimately gives the sum
\begin{align}
	|V_{ud}|^2 + |V_{us}|^2 + |V_{ub}|^2 + |V_{cd}|^2 +|V_{cs}|^2 +|V_{cb}|^2 .
\end{align}
Of these, the only terms that are not small are $|V_{ud}|$ and $|V_{cs}|$. Since the former value is well known to very good precision, we can take this to be a measurement of $|V_{cs}|$. Experimentally, the sum is known rather precisely,
\begin{align}
	\sum_{u,c,d,s,b}|V_{ij}|^2 \approx 2.002 \pm 0.027\ .
\end{align}
If you take it as a measurement of $V_{cs}$, then this measurement has the smallest error, but it is clearly an indirect measurement since it really probes the above sum.

\subsection{Measuring $|V_{cd}|$}

Let's quickly see how we measure $|V_{cd}|$.

\subsubsection{$D^\pm \to \ell \nu$}

A natural process to start with is $D^- \to \ell \nu$. This follows  the \textit{same} semileptonic pseudoscalar meson decay pattern that we've used many times now. We again ask which particular lepton final state is best suited for this search. We know that there is a chiral suppression, so the electron branching ratio is very small. This leaves us with the muon and $\tau$. The $\tau$, however, is hard to measure experimentally and suffers from phase space suppression. So let's go with the muon.
The $D^\pm$ are made up of a $d$ and a $c$ quark so that this probes $V_{cd}$. The problem, of course, is that the factor of $|V_{cd}|$ always comes along with the decay constant, which we have to get from the lattice. 
Combining the phase space and chirality ($m_\mu$) suppression, this process is very small. 

Another way to say this is that as the heavy quark mass gets heavier, it becomes easier for the heavy quark to decay \emph{by itself} without the assistance of light quark. The only way for a pion to decay is through an interaction between its valence quarks and the $W$, but the charm quark in the $D$ can decay by itself with the other quark simply acting as a spectator. The spectator decays are enhanced since the rate goes like $\Gamma\sim m^5$. In particular, instead of decays of the form
\begin{center}
	\begin{tikzpicture}[line width=1.5 pt, scale=1.3]
				\draw[fermion] (-140:1)--(0,0);
				\draw[fermionbar] (140:1)--(0,0);
				\draw[vector] (0:1)--(0,0);
				\node at (-140:1.2) {$c$};
				\node at (140:1.2) {$d$};
				\node at (0:1.2) {$W$};	
	\end{tikzpicture}
\end{center}
one has to consider
\begin{center}
	\begin{tikzpicture}[line width=1.5 pt, scale=1.3]
				\draw[fermion] (-1,0)--(0,0);
				\draw[fermion] (0,0)--(1,0);
				\draw[fermionbar] (-1,-.25)--(1,-.25);
				\draw[vector] (45:1)--(0,0);
				\node at (-1.2,.1) {$c$};
				\node at (1.2,0) {$s$};
				\node at (-1.2,-.2) {$d$};
				\node at (45:1.2) {$W$};	
	\end{tikzpicture}
\end{center}
Even a modest increase in the heavy quark mass leads to a dominance of the spectator decay rate over the two-body decay rate. Thus we should find a better decay mode.

\subsubsection{$D\to \pi \ell \nu$}

The best way to determine $V_{cd}$ is through the semileptonic decays,
\begin{align}
	D^\pm &\to \pi^0 \ell^\pm \nu
	&
	D^0 &\to \pi^\pm \ell^\mp \nu\ .
\end{align}
The two decays are related by isospin and so give the same information. Isospin breaking effects are small; we note that for the $J=1$ charm mesons there is a phase space `miracle' that we discuss in Appendix~\ref{sec:D:star}. There is still a problem: while the $D^\pm$ and $D^0$ are related by isospin, the $D$s and $\pi$s are not related by any symmetry. This is troubling since it takes away one symmetry limit that we'd have liked to use to simplify our analysis. Compare this to the kaon decay $K\to \pi \ell \nu$ where we had $SU(3)$ flavor to relate the left- and right-hand side of the decay equation. 

Again drawing from our bag of tricks, we can still write down a ratio of independent decay rates in which the final states are related by SU(3) flavor,
\begin{align}
	\frac{\Gamma(D\to K\ell \nu)}{\Gamma(D\to \pi \ell \nu)} \quad\stackrel{SU(3)}{=}\quad \frac{|V_{cs}|^2}{|V_{cd}|^2}. 
\end{align}
We estimate the error from phase space in the usual way. The correction for the form factor $f_+$ is measured from the lattice~\footnote{The important point about the lattice measurement is that this is done by plotting $f_+(0)$ over an unphysical strange quark mass $\tilde{m}_s$ and then extrapolating to the physical value. There are many papers on whether one should use linear versus quadratic extrapolations (giving a systematic error), but nice thing is that the \textit{ratio} of these values between $\tilde{m}_s = m_s$ and $\tilde{m}_s = m_c$ is much less sensitive to this error. Of course, one could have avoided using the ratio altogether by directly calculating $D\to \pi \ell \nu$ using the lattice from the very beginning. In this case one has to extrapolate the mass of the kaon down to the pion mass scale, but this is within the regime in which lattice techniques are effective.}.

\subsubsection{Neutrino scattering}

A surprising method for measuring $|V_{cd}|$ which claims to have the
smallest uncertainty is neutrino scattering against a fixed
target. The fixed target is effectively a bunch of $u$ and $d$
quarks. However, it is also composed of \emph{sea
  quarks}---quark--anti-quark pairs that pop in and out of the
vacuum. In particular, we expect the light ($u,d,s$) quarks to pop out
of the vacuum with some significant distribution. (The heavy quark
populations are suppressed by roughly $\exp(-m_Q/\Lambda_{\text{QCD}})$.)  This is a huge hadronic uncertainty.

Let's be n\"aive and only assume $u$ and $d$ quarks in the target. We can tag a charm final state by looking for a pair of muons since this must come from the process
\begin{align}
	\nu + d \to \mu^- + c \to \mu^- + \mu^+ + \nu.
\end{align}
The cross section goes like $\sigma \propto |V_{cd}|^2$. We take the data as a function of $q^2$ and cuts for a hard muon to isolate the charm decays against backgrounds with $\nu + u \to \mu^+ + s$.  What is nice about this is that we look at \textit{inclusive} decays, unlike our previous exclusive decays. We don't care about how the remnant of the charm hadronizes, we only tag on the muons. This gives us a different set of hadronic uncertainties which we will discuss later.

The problem is $\nu + s \to c + \mu$ coming from interactions with strange sea quarks. The ratio of these cross sections is
\begin{align}
	\frac{|V_{cs}|^2}{|V_{cd}|^2}\sim 20 \ ,
\end{align}
which we know because the Cabbibo angle squared is 5\%. The background rate is twenty times larger, but is suppressed by the density of $s$ versus $d$ in the target. This latter suppression is only on the order of hundreds and so limits our accuracy. 

One thing we can do is to also look at anti-neutrino beams while tagging on a single muon,
\begin{align}
	\bar\nu + u \to \mu + d.
\end{align}
The sea quark background here comes from a sea $c$. Fortunately, the $c$ quark is not much of a sea quark because it's heavy. (Can you imagine what it was like transcribing this sentence? If not, try reading it out loud.) Thus this anti-neutrino process effectively only measures the valence quarks so that this cross section can be used to subtract the sea $s$ contribution in the $\nu$ scattering. 

The hadronic physics involved in this measurement is completely orthogonal to that of the semileptonic $D\to \pi$ decays. Having these two processes agree gives much more confidence in the result.

\subsection{Measuring $|V_{cb}|$}

By now you're tired of using the same SU(3) flavor tricks to find clever ways to relate meson decay rates so that one can extract \CKM matrix elements. That's fine, because we have reached an impasse: for the $V_{cb}$ element we are stuck working with quarks that are \emph{not light} and therefore not subject to $SU(3)$ flavor. We have to appeal to a different bag of tricks, though these mostly fall outside the scope of these lectures. We introduce one of these tricks, heavy quark effective field theory, in Appendix~\ref{sec:HQS}.

\subsubsection{$B\to D$ decays}
\label{sec:Vcb:BtoD}

The natural decays to consider are
\begin{align}
	B &\to D \ell \bar\nu 
&
	B &\to D^* \ell \bar\nu \ . 
\end{align}

If we didn't know anything about heavy quark symmetry, we would say that $B \to D \ell \bar\nu$ contains two form factors, while $B \to D^* \ell \bar\nu$ contains four (one for the vector and three for the axial), giving a total of six form factors between these two decays. \emph{Yuck!} You may want to consider the heavy baryon decay,
\begin{align}
	\Lambda_b \to \Lambda_b \ell \bar\nu,
\end{align}
which is essentially the same as neutron decay. This also has six form factors, giving a grand total of twelve form factors between the three decays. \emph{Double yuck!}

Now for a big surprise: In the heavy quark limit, \textit{all} of
these form factors are either zero or the \textit{same} non-zero
function, called the  \textbf{Isgur--Wise function}.
This allows us to write everything in terms of a single form factor. The reason boils down to this: you can think of a $B$ or $D$ as a heavy quark surrounded by the \emph{brown muck} of \QCD. What happens if we replace these pseudoscalar mesons with their vector excited states, the $B^*$ and $D^*$? Nothing---for the most part, this physics factorizes from \QCD. Ditto for whether the heavy quark is locked up in a meson like the $B$ or a hadron like the $\Lambda_b$. In fact, the physics that differentiates the $B$ and the $D$ is simply whether or not there is a $b$ or $c$ quark inside them. If we were able to magically swap the $b$ quark with a $c$ quark---thus transforming the $B$ into a $D$---then \QCD would barely notice\footnote{There's a better pop culture reference: the fictional \acro{DC} superhero Swamp Thing is a sentient creature composed of swamp muck. The creature absorbed the consciousness of Alex Holland, who died unjustly in the swamp. If you replace Alex Holland in this story with any other ordinary person in the \acro{DC} universe, then Swamp Thing would have absorbed a different consciousness, but the \emph{magical swamp muck would be the same swamp muck}.}! All of this boils down to the observation that the twelve form factors we alluded to boil down to an \emph{single} function in the heavy quark limit. We say more about this limit appealing to heavy quark effective theory in Appendix~\ref{sec:HQS:Vcb}; but otherwise leave the discussion at that.

\begin{framed}
\noindent \textbf{What about $B^* \to D\ell \bar\nu$?}
Due to heavy quark symmetry, the
rates for this process and $B \to D \ell \bar\nu$ are the same
since the weak decay of the heavy quark does not care about the brown muck.
However, $B^* \to D\ell \bar\nu $ is a lost cause because $B^*$ decays
via $B^*\to B\gamma$ electromagnetically, which is much stronger than
the weak decay, see Example \ref{eg:Bst:decay}.
\end{framed}

\subsubsection{Inclusive decays}

There is another way to get information about $|V_{cb}|$. Consider the
decay $B\to X_c \ell\nu$, where $X_c$ means \textit{any} state
containing a $c$ quark. This is called an \textbf{inclusive
  decay}. How can we relate this process to the 
underlying calculation with respect to quarks and gluons?

There is a principle called \textbf{quark--hadron
duality}\footnote{This is from the pre-string theory days, when
  phenomenologists had more dualities than formal theorists.}, for a
review see~\cite{Shifman:2000jv}. 
The duality says that inclusive hadronic decays, when integrated over
large enough part of phase space, are described by
the underlying parton-level processes.  

This duality has been checked in many cases with much success.
Despite this success, the notion of the quark--hadron duality is vague. There are no formal criteria for when it can be used. Moreover, it is not clear how we can estimate deviations from the duality-based results. Yet, the intuition is clear: \acro{QCD} is perturbative at short distances. The strong interaction results in energy shifts of $O(\Lambda_{\rm QCD})$ between the various partons. By integrating over a large enough interval, that is, much larger than $\Lambda_{\rm QCD}$, we average over these hadronic processes and recover the underlying results.

It is important to stress that despite its seemingly innocuous appearance, this quark--hadron duality is far from trivial. Consider the measurement of the famous `$R$ ratio', 
\begin{align}
	R = \frac{\sigma(e^+e^-\to \text{hadron})}{\sigma(e^+e^-\to \mu^+\mu^-)}.
	\label{eq:R:ratio}
\end{align}
Quark--hadron duality tells us that this plot should look like a series of step functions with a step at each quark's mass threshold, with some smoothing due to phase space. 

\begin{figure}[htbp]
	\centering
		\includegraphics[width=.6\textwidth]{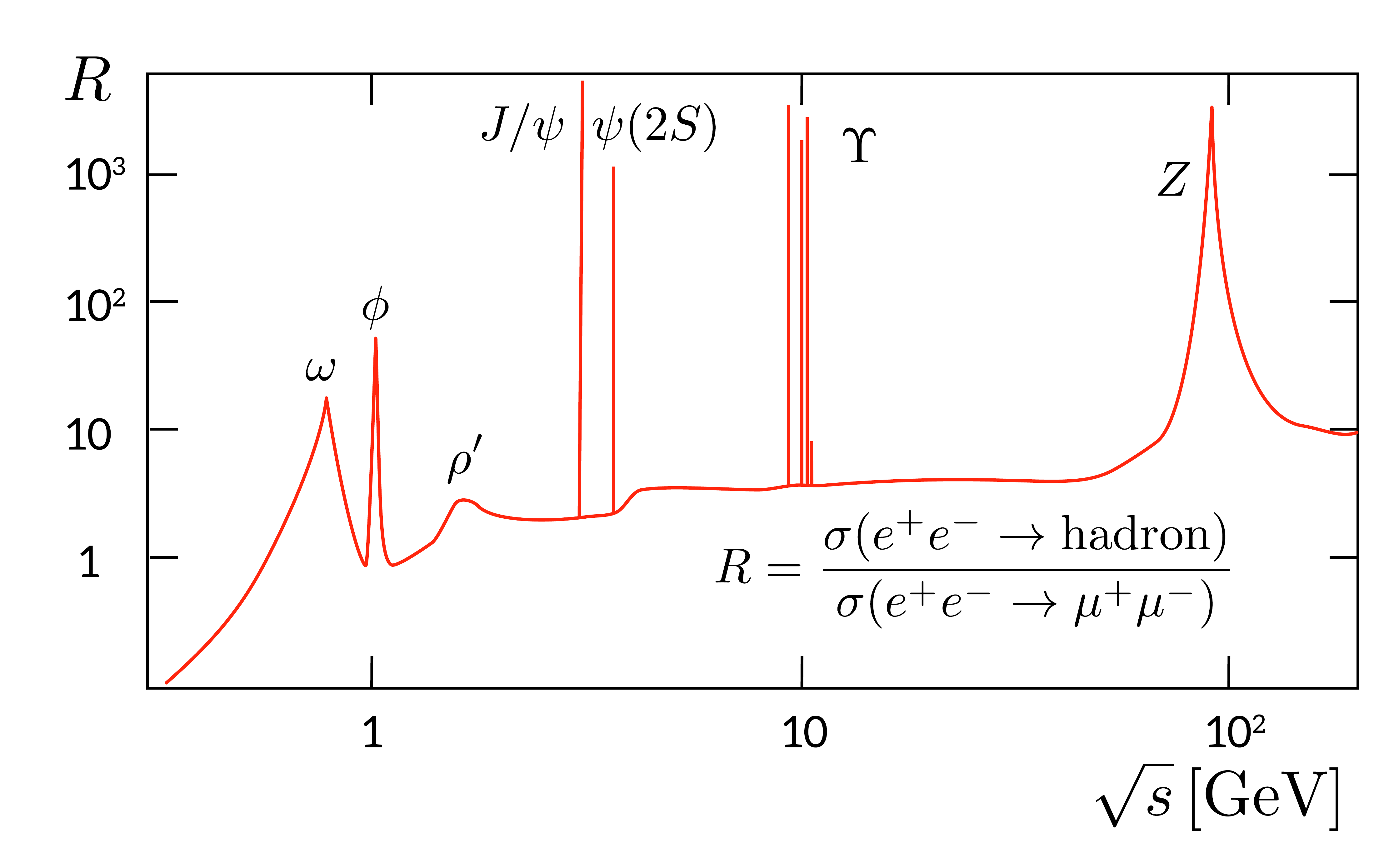}
	\caption{Sketch of the $R$ ratio, (\ref{eq:R:ratio}), one of the famous checks for the existence of quarks.}
	\label{fig:Rsmooth}
\end{figure}

If we compare this to experimental plots of the $R$ ratio, sketched in
Fig.~\ref{fig:Rsmooth}, we can immediately see a problem. There are
lots of peaks associated with hadronic resonances which clearly do not
appear in the quark-level analysis. For example, at the $\phi$
resonance, $e^+e^-\to \phi\to K^+K^-$ has a huge cross section---much larger than what one would predict from the naive quark-level diagrams. 
Clearly there's a subtlety in the quark-hadron duality principle. The subtlety is that we must \textit{smear} out the data. What is the scale of the smoothing? $\Lambda_{\text{QCD}}$, of course! When we smear out features on this order and smaller---when we integrate over these features---we begin to follow the quark-level $b\to c\ell\nu$ plot.

We are now ready to apply the principle of quark hadron duality to $B$
decays. For $B\to X_c$, for example, this duality tells us that
\begin{align}
	B\to X_c \ell \nu \approx b\to c \ell\nu.
\end{align}
The important symbol here is the `$\approx$.' The $B$ decays into something charmed and potentially a lot of other stuff. When we sum over all of this \emph{other stuff}, we say that this amplitude should be approximately the same as that of the quark-level $b\to c$ amplitude.
Said in another way, one can predict the rates for inclusive hadronic
processes by calculating the quark process.

To measure $|V_{cb}|$, we just plot the spectrum of $B\to X_c \ell
\nu$, integrate over some region of $q^2$, and pretend that we're
looking at a plot of $b\to c \ell \nu$. In the heavy quark limit it's
clear that the $B$ decay really is the same as the $b$ decay, so that
to leading order we're done. 

Quark--hadron duality tells us that we need to smear. In the case of $B
\to X_c \ell \nu$ we can smear over the final state, but not over the
initial one. Here we need to use the heavy quark symmetry argument
that tells us that in the infinite $b$ quark mass limit, the fact that the $b$
is inside a hadron is irrelevant and we can treat the $b$ as a free
quark. There are well defined corrections to this statement that we
discuss in  Appendix~\ref{sec:HQS:Vcb}.

%

\subsection{Measuring $|V_{ub}|$}

We only briefly touch on $|V_{ub}|$. The natural decays to consider are
\begin{align}
	B &\to \pi\ell\nu 
&\text{and}&&
	B &\to \rho\ell\nu \ .
\end{align}
One problem is that the $B$ certainly has a heavy quark, which may
appeal to heavy quark effective theory as introduced in
Appendix~\ref{sec:HQS}, but the $\pi$ and $\rho$ only contain light
quarks, which are well beyond the regime of heavy quark
symmetry. Phenomenologically, there is a further problem with
statistics because this process is \acro{CKM} suppressed, $|V_{ub}|
\ll |V_{cb}|$. We do not dive into the tools that can be used to solve
this problem, but only mention that some tools are available for those who are curious enough to go and dive into the literature.

One of the avenues is to look at inclusive
$	B\to X_u \ell \nu$ decays.
The extra problem here is that $X_u$ is not a well-defined signal experimentally---this would just be `hadronic junk.' For example, an experimentalist can't tell it $X_u$ apart from $X_c$: if you see a charm then you know it goes in $X_c$, but if you don't see a charm then it's not necessarily true that there wasn't one. The $X_u$ is a small theoretical error on the $X_c$ inclusive decay.

\begin{figure}[htbp]
	\centering
		\includegraphics[width=.6\textwidth]{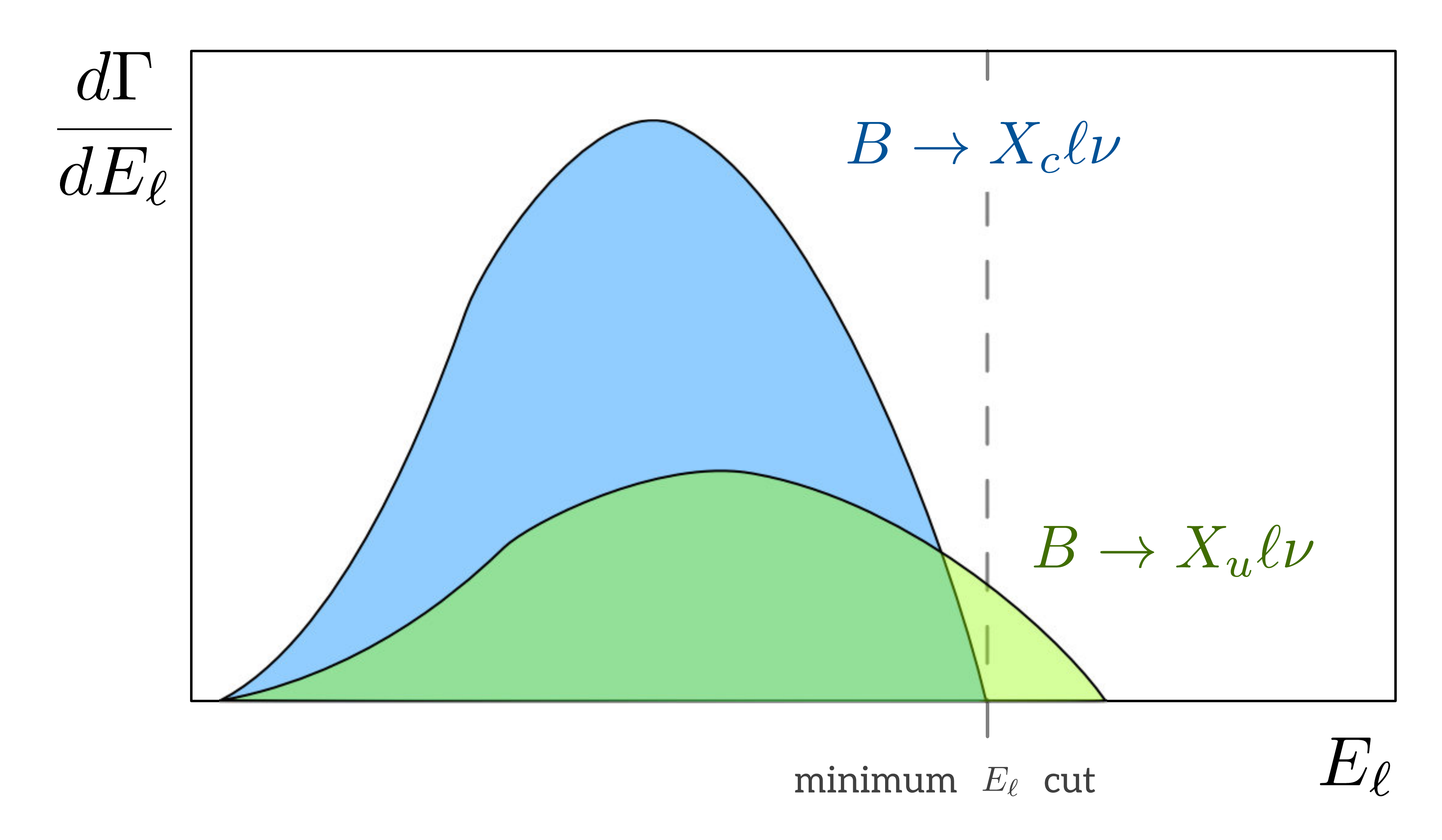}
	\caption{Sketch of the distribution of $B\to X_c\ell\nu$ and
          $B\to X_u\ell\nu$ decays as a function of the charged lepton
          energy. Note that the sketched spectra are not to scale; the difference has been exaggerated to show the effect.}
	\label{fig:BXlv}
\end{figure}

Fortunately, there is a trick. We can use phase space. There's a kinematic endpoint in the energy of the electron which is different for the $X_c$ and $X_u$ decays. The $X_u$ decays permit higher-energy electrons. 
As sketched in Fig.~\ref{fig:BXlv}, there is a little region where the up overcomes the charm in the endpoint region. Experimentally this turns out to be do-able, but it's still very hard. 

\section{The trouble with top}
\label{sec:top}

In the previous sections, we ran through how one can experimentally determine the magnitudes of six elements of the \CKM matrix. The remaining three elements brings us to the top quark. As we remarked in Section~\ref{sec:light:heavy:heaviest}, the top is special among quarks. It decays before it hadronizes. You can check this by comparing its lifetime to the characteristic hadronization timescale, $\Lambda_\text{QCD}^{-1}$. The fact that tops never get to hadronize is actually \textit{good} thing from the point of view of measuring \CKM elements since it suggests that we can avoid  worrying about dreaded form factors that encode strong dynamics. On the other hand, we also know that tops are hard to produce. Before the \acro{LHC}, physicists had only produced about a hundred or so tops in labs. Compare this to billions of $b$s and trillions of $c$s. 

Experimentally, tops are identified by tagging the associated
$b$---this involves measuring invariant mass of the $W$ and $b$
jet. The decay to $Wb$ basically happens for every on-shell top, so it
seems hard to measure anything from top decays other than
$V_{tb}$. Even if we could get around this---for example by looking at
the masses of decay products---it would still be practically
impossible to tell $V_{td}$ from $V_{ts}$.

What a mess. There are still a few things we can do. For example, we
can compare measurements of the top with and without $b$ tagging, to get
\begin{align}
	\frac{|V_{tb}|^2}{\sum |V_{ti}|^2} = 0.94 \pm 0.30.
\end{align}
Another way to do a \CKM measurement is through \textbf{single top production}. Tops are more likely to be produced from gluons than $W$s, but gluon vertices don't carry \CKM elements. Tops produced from a $W$ can be distinguished from those produced by a gluon since gluon-borne tops come in particle--anti-particle pairs. The relevant diagram is 
\begin{center}
	\begin{tikzpicture}[line width=1.5 pt, scale=1.3]
		\draw[fermionbar] (-140:1)--(0,0);
		\draw[fermion] (140:1)--(0,0);
		\draw[vector] (0:1)--(0,0);
		\node at (-140:1.2) {$d$};
		\node at (140:1.2) {$u$};
		\node at (.5,.3) {$W$};	
	\begin{scope}[shift={(1,0)}]
		\draw[fermion] (-40:1)--(0,0);
		\draw[fermionbar] (40:1)--(0,0);
		\node at (-40:1.2) {$b$};
		\node at (40:1.2) {$t$};	
	\end{scope}
	\end{tikzpicture}
\end{center}
\begin{eg}
	\textbf{Tevatron versus LHC}. Given the initial states for the above diagram, what can we say about what kind of collider we should use to probe this? The Tevatron is a proton anti-proton collider so that there are \textit{actual} up and anti-down valence quarks. Indeed, the Tevatron has produced tens of single tops to date. The \acro{LHC}, which is a proton--proton collider, requires the anti-down to appear as a sea quark. Actually, this was a trick question: this process occurs at small $x$ where the parton distribution functions become large so that there is actually an additional enhancement that makes the \acro{LHC} not-so-bad at single top production.
\end{eg}
This is all we have to say about top decay at tree-level. Unsatisfied? You should be. Fortunately, there's a sneaky way to get to the top  \CKM matrix elements: we can look at loop-level processes.

\subsection{Tops in loops}

Many \FCNCs diagrams are very sensitive to the \CKM matrix elements involving
the top. The reason, as we discussed in Section~\label{sec:loops} is
that the amplitude is proportional to the heaviest quark in the
loop and comes with the associated \CKM matrix
elements. We refer to these as \textbf{indirect probes} of the top \CKM elements.

Recall the $b \to s$ loop
decay in (\ref{eq:GIM-loops}). Taking into account importance top, we may write
\begin{align} \label{eq:GIM-loops-again}
\mathcal M (b \to s)= \sum_i V_{ib}^* V_{is} f(m_i/m_w) \approx  V_{tb}^* V_{ts} f(m_t/m_W)
\end{align}
where $f(m_t/m_W)$ is a known function. We conclude that 
measuring the magnitude of this amplitude gives information about 
$| V_{tb}^* V_{ts}|$.
Similarly, $b \to d$ \FCNCs are sensitive
 to $| V_{tb}^* V_{td}|$.

There are many processes involving $B$ mesons that can be then used
to probe the top \CKM matrices: meson oscillations,
leptonic, semi-leptonic and radiative $B$ decays. 
We postpone a discussion of oscillations to the
next section, and here we only briefly discuss the radiative case.

\subsection{Measuring the $b\to s\gamma$ penguin}

Something useful to measure in $b\to s\gamma$ is the energy spectrum of the emitted photon. In $\mu \to e\gamma$, the spectrum is trivial, it is just a line at the only kinematically allowed energy. For $b\to s\gamma$, on the other hand, hadronization allows an actual shape since the actual measurement is $B\to X_s \gamma$. Quark--hadron duality tells us that to good approximation, we can just say that we have an $s$ quark and the photon spectrum is just a line. But this approximation relies on smearing. When you cannot smear then the approximation is not good. In fact, the actual spectrum is far from a $\delta$-function. The spectrum is called a \textbf{shape function}. This is a truly non-perturbative object that we do not know from first principles. Even the lattice doesn't come close right now---it's inclusive, and the lattice at best can do one or two hadrons. The one thing that \textit{is} good---remember $V_{ub}$ which relied on the endpoint of a spectrum---is that the shape function in $b\to s\gamma$ can be related to $V_{ub}$. These hadronic unknowns can at least be related to one another so that one measurement can be used as input into other measurements. 

Given the distribution in photon energies, the reality of experiments requires us to put a cut on the photon energy, bounding it from below. These cuts introduce problems. If we want to measure the $b\to s\gamma$ branching ratio and we can only measure part of the spectrum, then we have to extrapolate to get the rest of the spectrum. Since the spectrum is pure \acro{QCD}, this is very hard to do with any confidence. Thus the calculation must also be done with a cut; and the lower the cut the more we trust the calculation. Current lower limits  from BaBar and Belle go down to about 1.6~GeV. As we push the cut lower we increase the statistics and reduce theoretical error, but we also pick up more background and lose purity. 
Eventually we get a number for the inclusive decay $B\to X_s \gamma$ which is published with an explanation for how this cut is taken. 

\subsection{Measuring $b\to s\gamma$ versus $b\to d \gamma$}

How can we distinguish $b\to s \gamma$ from $b\to d \gamma$? Consider the ratio
\begin{align}
	\frac{\text{Br}(B\to X_d \gamma)}{\text{Br}(B\to X_s \gamma)}\sim\left|\frac{V_{td}}{V_{ts}}\right|^2 \sim \lambda^2.
\end{align}
It would be very nice if we could measure this ratio of \CKM elements. However, we cannot experimentally distinguish between $X_s$ and $X_d$. This is like some corollary to the Heisenberg uncertainty principle: \emph{Things which are very nice theoretically tend to be hard to do experimentally, and if you can measure something experimentally, you don't really know what you're doing theoretically.}

Another approach is to make exclusive measurements. We could, for example, look specifically for
$	B \to K^* \gamma$.
Why not $B\to K\gamma$? The decay to a pseudoscalar meson is suppressed by angular momentum conservation. The problem with $B\to K^* \gamma$ is that the branching ratio is much smaller: it turns out to that only roughly 20\% of the $s$ quarks hadronize into $K^*$s. At least we once again have a definite $\gamma$ energy from kinematics.

The theoretical problem is calculating the hadronic part. We need form factors for
	$\langle B |\mathcal O |K^* \rangle$.
We can compare this to $B\to \rho \gamma$, allowing us to determine the ratio 
\begin{align}
	\left|\frac{V_{td}}{V_{ts}}\right|^2 \sim \lambda^2.
\end{align}
Comparing $B\to \rho \gamma$ and $B\to K^* \gamma$ gives this ratio up to SU(3) flavor symmetry breaking effects since the form factors---and hence a big chunk of the hadronic uncertainties---cancel in the SU(3) limit. Ultimately, small statistics limit the usefulness of $B\to\rho \gamma$ because the $\rho$ has such a large width that it is miserable to identify, but at least it's an independent measurement. 

Some say that $b\to s \gamma$ is one of the hardest calculations that's ever been done using pure perturbation theory, right up there with $(g-2)$. Na\"ively all you have to do is calculate these a couple of penguin diagrams. This only gives a coefficient for the dipole operator, which has a fancy name $\mathcal O_7$. This, however, is a calculation done in the electroweak where you have heavy particles like $W$s, tops, and perhaps new physics running around in loops. When you want to ask about physics at the $b$ scale, $\mathcal O_7$ \emph{mixes} with other operators through renormalization group flow. The effect is huge because the operator mixing is mediated by gluon loops. You'll be off by a factor of six if you neglect this mixing. At the end of the day, precise calculations require a 10$\times$10 matrix of anomalous dimensions. It takes a long time to do the calculation, and you still end up with some irreducible uncertainty associated with setting the matching scale. Brings you down to 5\% error and we probably cannot do much better theoretically. This is a whole industry, and physicists can spend their entire careers doing this\footnote{The places where people spend their careers on these precision calculations seems to correlate highly with places that export high quality chocolate.}.

\section{Meson Mixing and Oscillation}

Mixing and oscillation are two of the key ideas in flavor physics and always come hand-in-hand. As a point of nomenclature, it is useful to distinguish between the two very closely related ideas. \textbf{Mixing} is a theoretical statement that tells us that the flavor eigenstates are different from the mass eigenstates. For example, in \acro{SUSY} we know that gauge-eigenstate bino, neutral wino, and higgsinos mix into the neutralinos. In mesons, we have the flavor and the mass eigenstates.
\textbf{Oscillation}, on the other hand, is the time evolution of a flavor state into another flavor state because they do not match the energy eigenstates. 

Given a superposition of two Hamiltonian eigenstates that mix, the oscillation frequency is $\omega = \Delta E$ with $\hbar\equiv 1$; in the rest frame this is just $\Delta E = \Delta m$. Thus oscillations are a way to use quantum mechanics to measure the mass splitting $\Delta m$, even when the absolute mass scales are hard to probe directly.

The most famous example is kaon mixing between the $K=(\bar s d)$ and $\bar K = (\bar d s)$. In the absence of the weak interactions, these states have degenerate masses. 
The same thing is true about the $K^+(\bar s u)$ and $K^-=(\bar u s)$. In \QCD the $K$ and $K^\pm$ states are protected by $U(1)$ flavor symmetries, a $K$ does not mix with $\bar K$ nor does a $K^+$ mix with a $K^-$.
The story changes once we turn on the weak interactions. The charged kaons still cannot mix since $U(1)_\text{EM}$ steps in place of the broken flavor symmetry to protect against the mixing. However, the $K$ and $\bar K$ are left unprotected and they \textit{do} mix into one another. Are you perturbed by this? While there's no explicit flavor symmetry protecting against $K$--$\bar K$ mixing, we also know that the Standard Model Lagrangian doesn't have flavor-changing neutral currents. Since both the $K$ and $\bar K$ are neutral states, shouldn't this flavor-changing neutral effect be prohibited? 

No, though it is suppressed. At his \acro{TASI} 2008 lectures, Herbi Dreiner said that one key rule of model-building is that \emph{anything that isn't explicitly forbidden is tacitly allowed.} This is also true for flavor-changing effects: $K$--$\bar K$ mixing isn't \textit{explicitly} forbidden by symmetries, and so we should expect them to happen. In this case, however, \FCNCs are \textit{so} scandalous that it is not practiced openly---that is, it isn't advertised in the Lagrangian and is hidden at tree-level. The flavor-changing $K$--$\bar K$ mixing occurs only at loop level due to box diagrams of the form
\begin{center}
	\begin{tikzpicture}[line width=1.5 pt, scale=1]
		\draw[fermion] (-1,2.3) -- (0,2);
		\draw[fermion] (0,2) -- (2,2);
		\draw[fermion] (2,2) -- (3,2.3);
		\draw[fermionbar] (-1,-.3) -- (0,0);
		\draw[fermionbar] (0,0) -- (2,0);
		\draw[fermionbar] (2,0) -- (3,-.3);
		\draw[vector] (0,2) -- (0,0);
		\draw[vector] (2,2) -- (2,0);
	\end{tikzpicture}
\end{center}

What does this mean? Is it mysterious or magical that such an oscillation occurs? No, not at all. This is clear if we forget about the quark constituents and just treat the kaons themselves as quantum fields in their own right. The $K$ and $\bar K$ are just states of a single complex scalar field. The box diagram above is then a two point amplitude between the field and its conjugate---but we know exactly what this is, this is just a loop-induced mass term. Once again, you can admonish such a thing and have it hidden from tree level, but unless you \textit{explicitly} prohibit it, you can expect it to still happen.  


\subsection{Open system Hamiltonian}

The game that we must now play is to start with a state which is a superposition of the Hamiltonian eigenstates and let it evolve in time. There are a lot of subtleties when doing this in quantum field theory which are well-described in the literature. Following Einstein's famous adage that physics should be as simple as possible---\textit{but no simpler}---we'll sidestep many of the subtleties and present the most straightforward treatment which accurately represents the physics.

It is convenient to distinguish between states which are \textbf{asymptotic} and those which are \textbf{resonances}. In \acro{QFT} we usually work with asymptotic states, which to zeroth order in perturbative interactions have don't decay and so go off to infinity. All decays of these asymptotic states come from interactions which we treat in perturbation theory. Alternately, we can describe the \textit{same} physics with states which have finite lifetime even at zeroth order in perturbation theory. In other words, we include some of the interactions into our definition of the field. A very rough analogy is the difference between the Schr\"odinger, Heisenberg, and interaction picture in quantum mechanics, where different amounts of time evolution are packaged in the definition of an operator. As we shall see, for meson oscillations it is much easier to work with resonances rather than asymptotic states.

In regular quantum field theory, the Hamiltonian is basically the mass of the particle, $H=M$. For a resonance, the Hamiltonian is
\begin{align}
	H = M-\frac{i\Gamma}{2}.\label{eq:H:resonance:non:Hermitian}
\end{align}
This is not Hermitian! Are you worried? We never worked with non-Hermitian matrices in quantum mechanics because probabilities are conserved, everything we measure is real, and Hermitian matrices have real eigenvalues. 
Of course (\ref{eq:H:resonance:non:Hermitian}) is perfectly fine. If a particle  can decay then we shouldn't expect probability to be conserved. For example in addition to all of the diagrams that contribute to an $S$ matrix element for given incoming and outgoing states, we must also include the processes where those specified incoming and outgoing states decay during their free evolution. In other words, we are working with an \textbf{open system} in which the `interaction' with the system allows the particle to decay \emph{out of the Hilbert space}. We have taken all of the decay processes out of the system so that the `effective' system that we are work with is no longer Hermitian. 

\begin{prob} \textbf{Deriving the effective kaon Hamiltonian}. In the limit where the weak interactions are turned off, the effective kaon Hamiltonian is diagonal with respect to the $K$ and $\bar{K}$ states. By treating the weak Hamiltonian $H_W$ as a perturbation, show that the effective Hamiltonian takes the form in (\ref{eq:H:resonance:non:Hermitian}). \textit{Hint:} Start by writing out the $S$-matrix as $S = T\exp\left[-i\int dt\, H_W(t)\right]$, where $H_W(t)=e^{iHT}H_We^{-iHt}$ is the interaction picture weak Hamiltonian. 
	\begin{sol}
	This question and solutions come from G.~Ridolfi's \acro{CP} notes\footnote{\url{http://www.ge.infn.it/~ridolfi/notes/cpnew.ps}}. Labeling the kaon states with lowercase Greek indices, we have
	\begin{align}
		S_{\beta\alpha} =& S^{(0)}_{\beta\alpha} + S^{(1)}_{\beta\alpha} + S^{(2)}_{\beta\alpha}\\
		S^{(0)}_{\beta\alpha} =& \langle \beta | \alpha\rangle \\
		S^{(1)}_{\beta\alpha} =& -i \langle \beta | \int dt\, e^{iHt}H_We^{-iHt} | \alpha\rangle\\
		S^{(2)}_{\beta\alpha} =&
			-\frac 12 \langle \beta | \int dt\, e^{iHt}H_We^{-iHt} \int dt'\, e^{iHt'}H_W e^{-iHt'} \Theta(t-t') | \alpha\rangle + (t' \leftrightarrow t).
	\end{align}
	The first two terms are simple:
	\begin{align}
	S^{(0)}_{\beta\alpha} =& \delta_{\beta\alpha}\\
	S^{(1)}_{\beta\alpha} =& -2\pi i \delta(E_\beta-E_\alpha)\langle \beta| H_W|\alpha\rangle.
	\intertext{
	The second-order term requires a bit more work. The two terms are identical due to $(t\leftrightarrow t')$. By inserting a complete set of unperturbed Hamiltonian states $I=\sum_\lambda |\lambda\rangle\langle \lambda|$ one obtains
	}
		S^{(2)}_{\beta\alpha} =& -2\pi\delta(E_\beta-E_\alpha)\sum_\lambda 
			\langle\beta | H_W|\lambda\rangle
			\langle\lambda | H_W|\alpha\rangle
			\int_0^\infty d\tau\, e^{i(E_\beta-E_\lambda)\tau}.
	\end{align}
	To evaluate the last integral, perform the infinitesimal shift $(E_\beta-E_\lambda)\to (E_\beta - E_\lambda + i\epsilon)$ with $\epsilon>0$. The resulting integral is straightforward,
	\begin{align}
		\int_0^\infty d\tau\, e^{i(E_\beta-E_\lambda+i\epsilon)\tau} = \frac{i}{E_\beta-E_\lambda+i\epsilon},
	\end{align}
	so that finally we have
	\begin{align}
			S^{(2)}_{\beta\alpha} =& -2\pi i\delta(E_\beta-E_\alpha)\sum_\lambda 
				\frac{\langle\beta | H_W|\lambda\rangle
				\langle\lambda | H_W|\alpha\rangle}{E_\beta-E_\lambda+i\epsilon}.
	\end{align}
	Now we would like to define the effective weak Hamiltonian $H_W^\text{eff}$ so that $H = m_K \delta_{\beta\alpha} + H_{W\, \beta\alpha}^\text{eff}$. The point is that $H_W^\text{eff}$ should give the same contribution to the $S$-matrix as the perturbation analysis above. We thus have
	\begin{align}
		H^\text{eff}_{W\, \beta\alpha} =
		\langle \beta H_W | \alpha \rangle
		+ 
		\sum_\lambda 
			\frac{\langle\beta | H_W|\lambda\rangle
			\langle\lambda | H_W|\alpha\rangle}{E_\beta-E_\lambda+i\epsilon}.
	\end{align}
	This effective Hamiltonian is \textit{not} Hermitian. Using the handy identity,
	\begin{align}
		\frac{1}{x+i\epsilon} = P\left(\frac 1x\right) - i\pi \delta(x),
	\end{align}
	where $P$ denotes the principal value, we have $H = M -\frac i2 \Gamma$ with
	\begin{align}
		M_{\beta\alpha} =&
		m_K \delta_{\beta\alpha} + \langle | H_W | \alpha\rangle
			+ P\sum_\lambda \frac{
			\langle \beta | H_W | \lambda \rangle
			\langle \lambda | H_W | \alpha \rangle
			}{m_K-E_\lambda}\\
		\Gamma_{\beta\alpha} =& 2\pi\sum_\lambda 
			\langle \beta | H_W | \lambda \rangle
			\langle \langle | H_W | \alpha \rangle
			\delta(m_K-E_\lambda).
	\end{align}
	\end{sol}
\end{prob}

When we have a system with more than one eigenstate then the Hamiltonian is a matrix. Let us thus consider the two state $K$--$\bar K$ system where the Hamiltonian is a $2\times 2$ matrix that, in the flavor basis, takes the form
\begin{align}
	H =M - \frac{i\Gamma}{2}
	\label{eq:meson:mixing:H:form}
\end{align}
where $M$ and $\Gamma$ are $2 \times 2$ matrices.
Of course, in the limit where the weak interactions are turned off, $H_{12}=H_{21}=0$.
Our aim is to probe $H_{12}$. We will see later how it can be used to probe some
combination of \CKM parameters. 

We can diagonalize this matrix to obtain the masses and the mass
eigenstates, which we call the $K_L$ (`$K$-long') and $K_S$
(`$K$-short'). The names describe the different lifetimes, so you can
already guess that the two states not only have different masses, but
different widths. We could have alternately named the states based on
their masses, which is what is done for the $B$ system: $B_H$
(`$B$-heavy') and $B_L$ (`$B$-light'). Interestingly, for the $B$
meson the mass splitting has not yet been measured. For the $B_s$ and
neutral kaons, it was found that the heavy states have
longer lifetimes. 


\begin{framed}
	\noindent\textbf{Long and short or heavy and light}? Note that there are two mass eigenstates which we may label according to their lifetime or their mass. Typically the mass is the natural choice, though for the kaon system it turns out that the lifetimes are very different. As we'll see below, the $K_S$ is mostly \acro{CP} even and wants to decay to the \acro{CP} even state $\pi\pi$, whereas the $K_L$ is mostly \acro{CP} odd and wants to decay to the \acro{CP} odd state $\pi\pi\pi$. However, the masses of the kaons an pions are such that $K_L\to 3\pi$ is severely phase space suppressed and so $K_L$ has a much longer lifetime.
\end{framed}

\begin{prob} \textbf{CPT and the diagonal entries of $M$ and
    $\Gamma$}. Prove that \acro{CPT} invariance implies that $M_{11} =
  M_{22} \equiv m_K$ and $\Gamma_{11} = \Gamma_{22} \equiv \Gamma$
  where $m_K$ is the avarge kaon mass and $\Gamma$ is the avarge width.
	\begin{sol}
		First recall that \acro{CP} takes $K \leftrightarrow \bar K$. This means that in this basis, the \acro{CP} operator can be written as a matrix
		\begin{align}
			U_{\text{CP}} = \begin{pmatrix}
				0 & 1\\
				1 & 0
			\end{pmatrix}.
		\end{align}
		Further, time reversal acts as a complex conjugate, $M\to M*$ and $\Gamma \to \Gamma^*$. From this we conclude that \acro{CPT} acts as
		\begin{align}
			M = \begin{pmatrix}
				M_{11} & M_{12}\\
				M^*_{12} & M_{22}
			\end{pmatrix}
			\to \begin{pmatrix}
				M_{22} & M_{12}\\
				M^*_{12} & M_{11}
			\end{pmatrix}\ ,
		\end{align}
		and similarly for $\Gamma$. Thus \acro{CPT} requires $M_{11} = M_{22}$ and $\Gamma_{11}=\Gamma_{22}$.
	\end{sol}
\end{prob}

We've seen that weak interactions generate a mass term between the $K$ and $\bar K$. \acro{CPT} tells us that the flavor-conserving Hamiltonian must be the same: $H_{11} = H_{22}$. Hermiticity tells us that
\begin{align}
	M_{12} &= M_{21}^*\qquad	\Gamma_{12} &= \Gamma_{21}^*.
\end{align}
Diagonalizing $H$ gives the two mass eigenvalues and the two widths. Before doing any work, lets estimate the important quantities, $\Delta m$ and $\Delta \Gamma$. One should be able too look at the matrix and very intuitively identify
\begin{align}
	\Delta M &\sim 2 M_{12}.
\end{align}
Further, you should see right away that the mixing is $45^\circ$.
\begin{eg}
	\textbf{Diagonalizing $2\times 2$ matrices}.
	Given a Hermitian $2\times 2$ matrix,
	\begin{align}
		\begin{pmatrix}
			a & b\\
			b^* & c
		\end{pmatrix}
	\end{align}
	one should remember that the mixing angle $\theta$ is given by
	\begin{align}
		\tan 2\theta = \frac{2b}{a-c}. 
	\end{align}
\end{eg}
In the case where $a=c$,
we get a $45^\circ$ mixing even when the off-diagonal term is very
small. 
Performing a full calculation in the non-Hermitian case gives, e.g.\ for the $B$ system,
\begin{align}
	|B_{L,H}\rangle = p|B^0\rangle \pm q |\bar B^0\rangle\label{eq:mixing:BLH}
\end{align} 
subject to the constraint $|p|^2+ |q|^2 =1$. 
The eigenvalues are
\begin{align}
	\mu_\alpha = M_\alpha + \frac{i}{2}\Gamma_\alpha
\end{align}
where $\alpha = 1,2$. Note that we must diagonalize the entire
Hamiltonian; this is \textit{not} generally the same as separately
diagonalizing $M$ and $\Gamma$. The diagonalizations are equivalent
only in the \acro{CP}-conserving limit as discuss below.

Look very carefully at (\ref{eq:mixing:BLH}). Usually in a rotation between states one has 
\begin{align}
	|A_{1}\rangle &= \phantom{+}\cos\theta |B_{1}\rangle + \sin\theta|B_2\rangle\\
	|A_{2}\rangle &= -\sin\theta |B_{1}\rangle + \cos\theta|B_2\rangle.
\end{align}
Our expression for $|B_{L,H}\rangle$ is very different! A rotation from one basis to another typically preserves the vector length. This is no longer true for $|B_{L,H}\rangle$ because the Hamiltonian is not Hermitian. 
Another consequence of working in an open system is that the overlap of the two eigenstates is no longer zero, but
\begin{align}
	\langle B_L | B_H \rangle = |p|^2 - |q|^2.
\end{align}
The two eigenstates are not orthogonal! Remember linear algebra from
kindergarden: we know that we only need \textit{independent} vectors
to span a space, not necessarily orthogonal. 

\begin{framed}
\noindent
This non-orthogonality is
related to a phenomena called \textbf{vacuum regeneration}: a
\textit{mass} eigenstate can be measured to be another mass eigenstate! This is different from \emph{oscillation}, where you start with something that is not a mass eigenstate. This is a direct result of the fact that the mass eigenstates are not orthogonal.
\end{framed}

\begin{prob}\textbf{Diagonalizing the kaon Hamiltonian}. It's a useful exercise to actually diagonalize the open system Hamiltonian (\ref{eq:H:resonance:non:Hermitian}), that way you'll believe the following results. Determine the eigenvalues $\mu_\alpha$ and eigenvectors ($K_L$ and $K_S$) of the kaon Hamiltonian. In particular, show that the eigenvalues are given by
	\begin{align}
		\mu_{L,S} = M_{11} - \frac i2 \Gamma_{11} \pm \sqrt{
			\left(M_{12} - \frac{i}{2}\Gamma_{12}\right)
			\left(M_{12}^* - \frac{i}{2}\Gamma_{12}^*\right)
		}
	\end{align}
	and that eigenvectors are given by
	\begin{align}
		K_{L,S} \sim |K^0\rangle \pm \sqrt{\frac{M_{12}^* - \frac{i}{2}\Gamma_{12}^*}{M_{12} - \frac{i}{2}\Gamma_{12}}}|\bar K^0\rangle.
	\end{align} 
	Note especially that 
	\begin{align}
		\frac{p}{q} = \sqrt{
		 \frac{M_{12} - \frac i2 \Gamma_{12}}{M_{12}^*- \frac i2 \Gamma^*_{12}}
		}.
	\end{align}
	\begin{sol}
		For simplicity, let us write
		\begin{align}
			H =
			\begin{pmatrix}
				H & H_{12}\\
				H_{21} & H
			\end{pmatrix}
			=
			\begin{pmatrix}
				M_{11} - \frac i2\Gamma_{11} & M_{12} - \frac{i}{2}\Gamma_{12}\\
				M_{12}^* - \frac{i}{2}\Gamma_{12}^* & M_{11} - \frac i2\Gamma_{11}
			\end{pmatrix}.
		\end{align}
		The eigenvalue equation is
		\begin{align}
			(H-\mu)^2 - H_{12}H_{21} = 0 \Rightarrow \mu = H \pm \sqrt{H_{12} H_{21}}.
		\end{align}
		The eigenvectors are also easy to find. Calling these states $(p,q)^T$, the eigenvector equation becomes
		\begin{align}
			H p + H_{12} q &= \mu p\\
			H_{12} q &= \pm \sqrt{H_{12}H_{21}} p.
		\end{align}
		We thus have eigenvectors given by $q = \pm \sqrt{H_{21}/H_{12}} p$, which one can normalize appropriately.
	\end{sol}
\end{prob}

Let us now consider the case where \acro{CP} is conserved. \acro{CP} exchanges $B^0 \leftrightarrow \bar B^0$ and so guarantees that $|p|=|q|$. In this case $\langle B_L| B_H \rangle = 0$. The Hamiltonian, however, is \textit{still} not Hermitian. 
%

The difference in the eigenvectors is
\begin{align}
	\mu_H-\mu_L = 2\sqrt{
		\left(M_{12} - \frac{i}{2}\Gamma_{12}\right)
		\left(M_{12}^* - \frac{i}{2}\Gamma_{12}^*\right)
	} \quad\stackrel{\text{CP}}{=}\quad 2\text{Re } M_{12} - i \text{Re }\Gamma_{12},
\end{align}
where $\stackrel{\text{CP}}{=}$ means ``in the limit of \acro{CP} conservation.''
We can thus identify the differences in masses and widths between the long and short states:
\begin{align}
	\Delta m &\quad\stackrel{\text{CP}}{=}\quad 2 |M_{12}|
&
	\Delta \Gamma &\quad\stackrel{\text{CP}}{=}\quad 2|\Gamma_{12}|.
\end{align}
Writing $2\Gamma = \Gamma_1+\Gamma_2$,
we can define three physical parameters:
\begin{align}
	x&= \frac{\Delta m}{\Gamma}
	&
	y&= \frac{\Delta \Gamma}{2\Gamma}
	&
	\phi &= \text{arg}\left(\Gamma^*_{12}M_{12}\right)\ .
\end{align}
Why are these particular ratios important? We show below that $x$ tells us about how many oscillations the meson undergoes before decaying, $y$ measures the relevance of the decay width differences between the flavor eigenstates, and $\phi$ measures \acro{CP} violation.
The range of values for the parameters is $x\in[0,\infty)$, $y\in
[-1,1]$, and $\phi\in[0,\pi)$. In the \acro{CP}-conserving limit, $M$ and $\Gamma$ are relatively real so that $\phi=0$.

In the limit where \CP is conserved, $|p|=|q|$ and the $K_S$ is a \CP-even state while the $K_L$ is a \CP-odd state. The $K_S$ can decay to the \CP-even final state $\pi\pi$ whereas the $K_L$ must decay into  $\pi\pi\pi$. Because the kaons are so light, the phase space for the latter decay is rather small and the $K_S$ has a much shorter lifetime. 

Because \CP is not conserved in Nature, the $K_S$ and $K_L$ are not \CP eigenstates. Define the \CP-even and \CP-odd neutral kaon states by $K_E \sim |K^0\rangle + |\bar K^0\rangle$ and $K_O \sim |K^0\rangle - |\bar K^0\rangle$. The difference between the open system Hamiltonian eigenstates and the \CP eigenstates is parameterized by $\epsilon$,
\begin{align}
	|K_{S,L}\rangle = \frac{1}{\sqrt{1+\epsilon^2}}\left(|K_{E,O}\rangle + \epsilon |K_{O,E}\rangle\right).
\end{align}
Recalling that $K_{S,L} \sim p|K^0\rangle \mp q |\bar K^0\rangle$, we have
\begin{align}
	\frac{p}{q} = \frac{1+\epsilon}{1-\epsilon}.
\quad\Rightarrow\quad
	\epsilon = \frac{p-q}{p+q} \approx \frac{i}{2}
	\frac{\text{Im }M_{12} - i\text{Im } \Gamma_{12}}{\text{Re }M_{12} - i\text{Re } \Gamma_{12}}.
\end{align}

This is the reason why $K_L$ has been observed to decay into two pions.  The $\epsilon$ effect comes from the $K_L$ having some admixture of the $K_E$ state which may decay into two pions.
Constraints on $\epsilon_K$ provide strong bounds on new physics contributing to the flavor sector. 

\subsection{Time evolution}

Now that we know that $K$ and $\bar K$ can mix, e.g.\ from processes like $K^0 \leftrightarrow \pi\pi \leftrightarrow \bar K$, we want to understand their oscillation. And since kaons are the usual example in old textbooks, we choose to be fancy and modern by instead using the completely analogous $B$ system. The time evolution of the neutral $B$ meson is given by
\begin{align}
	|B^0(t)\rangle=a(t)|B\rangle + b(t)|\bar B\rangle,
\end{align}
subject to the Schr\"odinger equation with the open system Hamiltonian. At time $t=0$, the wave function is some linear combination of flavor states $B$ and $\bar B$. To evolve this in time we write $B$ and $\bar B$ in terms of energy eigenstates $B_H$ and $B_L$ and independently evolve each. We can then project back into the $B$ and $\bar B$ basis at any later time.

Take $b(0)=0$ to so that $B^0(t)$ a state which is a pure $B$ at $t=0$. At $t\neq 0$ it is a mixture of $B$ and $\bar B$. Assuming \CP for simplicity, we find
\begin{align}
	B^0(t) = \cos\left(\frac{\Delta E \,t}{2}\right)|B\rangle + i\sin\left(\frac{\Delta E \,t}{2}\right)|\bar B\rangle.
\end{align}
We generalize this to the \CP violating case later. $\Delta E$ is
complex when we have an open system, but let us further assume that
$\Delta\Gamma=0$ so that we have oscillations with the decay factorized.
The probability of $B^0(t)$ to remain a $B$ at time $t$ is
\begin{align}
	P\left[B^0(t)=B\right] &= |\langle B^0(t)|B^0\rangle|^2
	=\frac{1+\cos(\Delta E\, t)}{2} e^{-\Gamma t}\ . \label{eq:B:decay:probability}
\end{align}
You'll derive the general formula in Problem~\ref{prob:meson:oscillation:formula}.
Similarly, $P[B^0(t)=\bar B]$ is the same with $+ \to -$ before the cosine.
We see that the $B$ system is indeed oscillating---it's right there in
the cosine! The frequency is $\Delta E$, as advertised.
In the meson rest frame, $\Delta E = \Delta m$ and $t = \tau$, the
proper time. Thus the flavor oscillations are controlled by the mass
splittings of the physical states. Further, by measuring the
oscillation frequency we're learning something about the mass
splitting. In order to make such a measurement one would have to
determine the meson flavor at both production and decay, which is
experimentally tricky. We discuss it a bit later.

\subsection{Time scales}

The richest physical phenomena are based on the fact that there are
different scales. In this case, there are different time scales for
oscillations. The dimensionless parameters controlling the oscillation
time scales are $x$ and $y$. We start by considering the case of $y=0$.

Rewriting (\ref{eq:B:decay:probability}) in terms of $x \equiv \Delta m/\Gamma$, 
\begin{align}
	P[B^0(t)=B] &= \frac 12 \left(1+\cos \Delta m t\right) =  \frac 12 \left(1+\cos x\Gamma t\right) \ .
\end{align}
What is $t?$ In neutrino or atomic experiments, you can change the
propagation time $t$ by moving the detector. In our case, 
on the other hand, $t$ is given to you by nature: it's particle's decay time, $\langle t\rangle\sim 1/\Gamma$. This is, of course, a statement about the average $t$; each individual meson decays probabilistically. 

As always, we can learn about the system by considering asymptotic states.
\begin{itemize}
	\item First consider $x\ll 1$. In this case there are effectively no oscillations. Once the state is produced it simply doesn't have enough time to oscillate. It is produced as a flavor state and---for its short lifetime---it remains a flavor state. An example of this kind of state is the $D$ meson. It makes sense talk about flavor states as long as $x\ll 1$.
	\item Now consider the opposite extreme, $x\gg 1$. In this case the oscillations are washed out since the cosine averages out to zero. What is happening physically? Any measurement of the meson system is done over a small characteristic time scale. In the limit $x\gg 1$, the meson is able to oscillate many times over this measurement period so that the effect of the oscillation is washed out to zero. The natural objects to discuss in this scenario are thus mass eigenstates. This is the case when we produce kaons; we say that the kaon we produce is $50\%$ $K_L$ and $50\%$ $K_S$.
	\item Finally, consider the case $x\sim 1$; this is the most interesting case since  oscillations actually observable. The state is neither a flavor nor a mass eigenstate, but we really a state which evolves with time. We can go and measure the oscillation of $B$ and $K$ mesons experimentally and find that
\begin{align}
	x_B \sim x_K \sim 1.
\end{align}
In fact, $x_B\approx .73$ and $x_K\approx .94$. As we mentioned before, this is complete luck since $x$ depends on the \CKM and the masses in some non-trivial way. Other similar mesons are not so fortunate. For example,
\begin{align}
	x_s &\sim 25
	&
	x_D &\sim 10^{-2}\ ,
\end{align}
where $x_s$ is standard notation for $x_{B_s}$.
For these particles the width of the meson's weak decay is much smaller than its mass,
\begin{align}
	\Gamma \sim G_F^2 m^5 \sim 10^{-14}m \ .
\end{align}
Again, the `practical' use of these oscillations is to make precise measurements of $\Delta m$ using quantum mechanics, similar to the M\"ossbauer effect.
\end{itemize}

\begin{figure}[htbp]
	\centering
		\includegraphics[width=\textwidth]{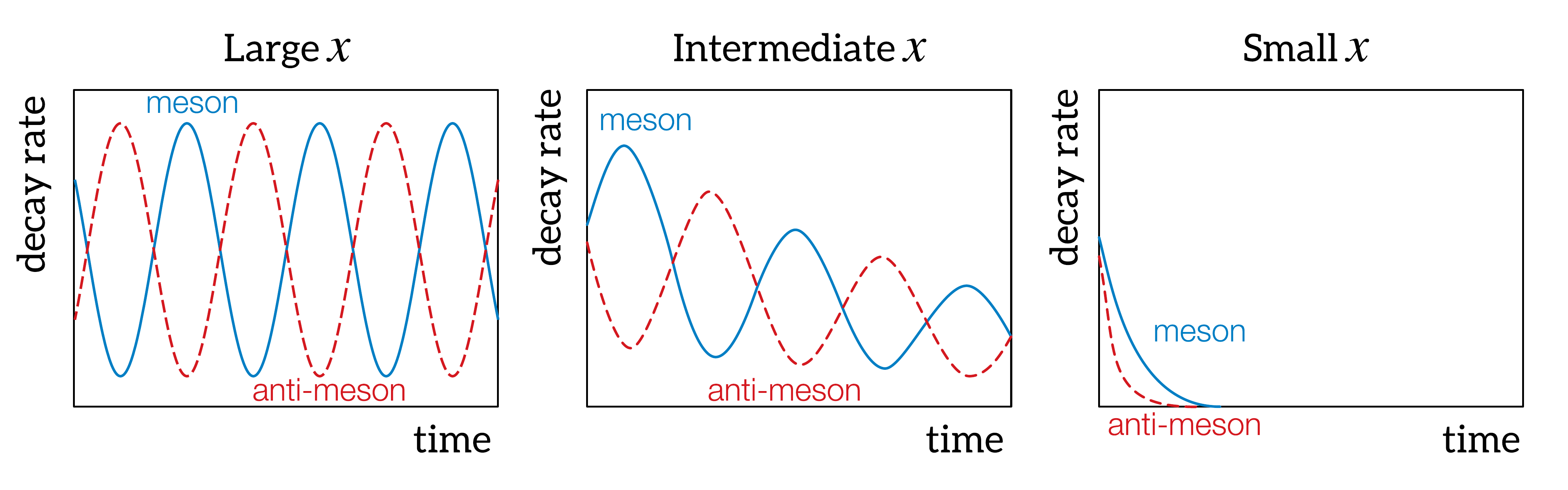}
	\caption{Illustration of how the ratio $x=\Delta m/\Gamma$ allows us to probe the mass splittings between the \CP eigenstates.
	\label{fig:mesondecayoscillate}}
\end{figure}

Fig.~\ref{fig:mesondecayoscillate} illustrates\footnote{Adapted from from David Kirky via Pat Burchat, \url{http://www.stanford.edu/dept/physics/people/faculty/docs/burchatCaltech2003.pdf}.} the
different cases discussed above, where we have the case of $x \ll 1$
relevant for $D$, $ x\sim 1$ relavant for $B$ and $K$ and $x \gg 1$
relevant ofr $B_s$.

\begin{framed}
	\noindent\textbf{We are used to $x\gg 1$.} This scenario is what we are most familiar with because it is often an implicit assumption. In supersymmetry we can produce a bino and say that it is really some combination of neutralino mass eigenstates. This statement already contains the assumption that $x\gg 1$ so that we integrate over the quick oscillation time. Even in the Standard Model, we can talk about $W\to cs$ versus $W\to cd$ decays. The probability of one decay versus another is the square of the relative \CKM matrix element, but this is also an assumption about the the down-type oscillation frequency.
\end{framed}

Let us emphasize that $\Delta m$ and $\Gamma$ come from different physics. The width of the particle $\Gamma$ has to do with the three-body decay of the $B$ meson with some \CKM matrix elements. The lifetime $\Delta m$ has to do with some box diagram with some completely different structure. For $B$ mesons, we're very lucky the lifetime is of the same order as the width so that $x\approx 0.7$. This is \textit{completely} luck and this is one reason why $B$ mesons are so interesting to study experimentally.
\begin{eg}
	If the $B$ mass were doubled, then its lifetime would decrease
        by about $2^5$ (recall that $\Gamma \sim m^5$) and we end up with $x \ll 1$.
\end{eg}

\begin{prob} \textbf{General meson oscillation formula}. \label{prob:meson:oscillation:formula}
	In (\ref{eq:B:decay:probability}) we derived the probability for a $B$ flavor state to remain a $B$ after time $t$ in the limit of \CP conservation. Derive the following general expressions:
	\begin{align}
		|K^0(t)\rangle &= \phantom{\frac{p}{q}} g_+(t) |K^0\rangle + \frac qp g_-(t)|\bar K^0\rangle\\
		|\bar K^0(t)\rangle &= \frac{p}{q} g_-(t) |K^0\rangle + \phantom{\frac qp} g_+(t)|\bar K^0\rangle\\
		g_{\pm}(t) &= \frac 12 \left(e^{im_L t}e^{-\frac 12 \Gamma_L t} \pm e^{im_S t}e^{-\frac 12 \Gamma_S t}\right).\label{eq:general:meson:oscillation}
	\end{align}
	\begin{sol}
		We start with
		\begin{align}
			|K_{L,S}\rangle = \frac{1}{\sqrt{|p|^2+|q|^2}} \left( p|K^0\rangle \pm q|\bar K^0\rangle\right),
		\end{align}
		which we can invert simply to obtain
		\begin{align}
			|K^0\rangle &= \frac{\sqrt{|p|^2+|q|^2}}{2p} (|K_L \rangle + |K_S\rangle)\\
			|K^0\rangle &= \frac{\sqrt{|p|^2+|q|^2}}{2q} (|K_L \rangle - |K_S\rangle).
		\end{align}
		Now we simply time evolve the effective Hamiltonian eigenstates. Let us write $H_L$ and $H_S$ for the long and short eigenvalues, for example $H_L = M_L - \frac i2 \Gamma_L$. For $|K^0(t)\rangle$ we have
		\begin{align}
			|K^0(t)\rangle &= \frac{1}{2p} \left[
			e^{-iH_Lt} (p|K^0\rangle + q|\bar K^0\rangle)
			+
			e^{-iH_St} (p|K^0\rangle - q|\bar K^0\rangle)
			\right]\\
			&= 
			\frac 12 \left[
				\left(e^{-iH_Lt}  + e^{-iH_St}\right)|K^0(t)\rangle +
				\frac qp \left(e^{-iH_Lt}  - e^{-iH_St}\right)|\bar K^0(t)\rangle
			\right].
		\end{align}
		The $|\bar K^0(t)\rangle$ oscillation comes from making the appropriate substitutions.
	\end{sol}
\end{prob}

Did you do the previous problem? Good. We now want to take into account the other dimensionless oscillation parameter, $y$. The generalization of (\ref{eq:B:decay:probability}) is $P[B^0(t)=B] = |g_+(t)|^2$. With a a few lines of algebra, it is easy to show that 
\begin{align}
	P\left(B^0(t)= B\right) = \frac{e^{-\Gamma t}}{2}\left[\cosh (y\Gamma t) + \cos(x\Gamma t)\right] \ .
\end{align}

Let us consider the some limits with respect to $x$ and $y$. 
\begin{itemize}
	\item $|y| \ll 1$ and $y\ll x$. In this case, the effects of $y$ are irrelevant. This is the case in the $B$ system.
	\item $y\sim x$. Both $x$ and $y$ are important. For the kaon system $x,y\sim 1$  so that we can probe this scale experimentally. The $D$ system is in the regime where $x,y \ll 1$, while the $K$ system is in the regime $x,y\sim 1$.
	\item $y\sim 1$ and $y\ll x$. Here there are two mass eigenstates with different decay rates (lifetimes) and we can forget about oscillations. We expect this to be the case for the $B_s$ system.
\end{itemize}
Other limits, like $y\gg x$, are not realized in the four meson systems.
The following table shows the order of magnitude for the $x$ and $y$ values of some our favorite mesons,
\begin{center}
	\begin{tabular}{lll}
		& $x$ & $y$\\
		\hline
		$K$ & 1 & 1\\
		$B$ & 1 & $10^{-2}$ (*)\\
		$B_s$ & 10 & $10^{-1}$ (*)\\
		$D$ & $10^{-2}$ & $10^{-2}$
	\end{tabular}
\end{center}
The stars indicate when a quantity is not well measured. 

\subsection{Flavor tagging}  

We want to measure $\Delta m$. We have to produce an initial state which we \textit{know} is a $B$ at $t=0$ and then determine the flavor of that state at some later detection time $t$. In other words, we need to \textbf{flavor tag} at both the production and decay times. 
\begin{framed}
	\noindent\textbf{Nomenclature}. The phrase `$b$-tagging' is
        used differently in different experimental communities. At the
        \acro{LHC} this refers to whether or not a jet contains a $b$
        quark, while in flavor experiments this refers to
        distinguishing between $B$ and $\bar B$ mesons. By the way, the $B$ meson nomenclature is very silly: a $B$ meson contains an \textit{anti}-$b$ quark.
\end{framed}
How do we perform these tags? Let's discuss the easier tag: once we have some unknown state propagating, how can we determine what it is at time $t$, when it decays? The easiest way to tag a $b$ quark decay is through the semileptonic decays,
\begin{align}
	b &\to c \mu^- \bar\nu
&
	\bar b &\to \bar c \mu^+ \nu \ .
\end{align}
Sometimes a $B$ will decay into final states that are also accessible to the $\bar B$, making it useless for tagging whether it contained a $b$ or $\bar b$, for example $B \to \pi^+\pi^-$. In this case the decaying object is really a superposition of both $B$ and $\bar B$, something we'll get into when we look more closely at \CP violation. 
\begin{eg}
	Can you use $B\to D^+ \pi^-$ for $b$ tagging? Na\"ively, it's easy to draw a the diagram showing that this final state tages a $\bar B$.
	\begin{center}
		\begin{tikzpicture}[line width=1.5 pt, scale=1.3]
			\draw[fermionbar] (40:1.2)--(0,0);
			\draw[fermion] (180:1)--(0,0);
			\draw[vector] (-40:1)--(0,0);
			\node at (40:1.4) {$c$};
			\node at (180:1.2) {$b$};
			\node at (0.0,-.5) {$W^-$};
		\begin{scope}[shift={(0,.3)}]
			\draw[fermionbar] (180:1) to [out=0,in=220] (40:1); 	
			\node at (40:1.4) {$d$};
			\node at (180:1.2) {$d$};
		\end{scope}
		\begin{scope}[shift={(-40:1)}]
			\draw[fermion] (-20:1)--(0,0);
			\draw[fermionbar] (20:1)--(0,0);
			\node at (-20:1.2) {$u$};
			\node at (20:1.2) {$d$};
		\end{scope}
		\end{tikzpicture}
	\end{center}
	 Can we now conclude that the parent of the $D^+\pi^-$ state was a $\bar B$ meson? No. We could also draw a diagram with an initial $B$ meson,
	\begin{center}
		\begin{tikzpicture}[line width=1.5 pt, scale=1.3]
			\draw[fermion] (40:1.2)--(0,0);
			\draw[fermionbar] (180:1)--(0,0);
			\draw[vector] (-40:1)--(0,0);
			\node at (40:1.4) {$u$};
			\node at (180:1.2) {$b$};
			\node at (0,-.5) {$W^+$};
		\begin{scope}[shift={(0,.3)}]
			\draw[fermion] (180:1) to [out=0,in=220] (40:1); 	
			\node at (40:1.4) {$d$};
			\node at (180:1.2) {$d$};
		\end{scope}
		\begin{scope}[shift={(-40:1)}]
			\draw[fermion] (-20:1)--(0,0);
			\draw[fermionbar] (20:1)--(0,0);
			\node at (-20:1.2) {$d$};
			\node at (20:1.2) {$c$};
		\end{scope}
		\end{tikzpicture}
	\end{center}
	Note the arrows. This second diagram mediates $\bar B \to D^+ \pi^-$. The two diagrams differ in how much \CKM suppression each receives, but the $D^+\pi^-$ final state doesn't come from a pure $\bar B$. Because of the \CKM suppression, it's an \textit{almost}-pure $\bar B$ since the interference is only at the percent level.
\end{eg}

Sticking to the semileptonic decays where the relatively
easy-to-measure muon versus anti-muon $b$ tags the final state, we
need a clever way to $b$-tag the initial state. The way this is done
depends crucially on the production mechanisms. We discuss them in
turn below.

We first consider the situation at 
the $B$-factories, where the trick is to use  quantum entanglement. 
Start with an electron--positron collider that operates at the $\Upsilon(4S)$ resonance. These mesons  have \textit{just} the right rest energy to decay into a $B\bar B$ pair nearly at rest. Since these particles are entangled upon production, we may use the decay of one of them as an initial-state $b$ tag for the other.
\begin{eg}
	Without looking at the \PDG, what is the spin of the $\Upsilon(4S)$? It has to be a spin-1 (vector) meson since it is produced through $e^+e^-\to \gamma\to \Upsilon(4S)$.
\end{eg}
As soon as one $B$ decays semileptonically, then at that instant we know precisely the $b$ content of its sibling. That is the moment that we have been calling $t=0$, the moment that we know the exact flavor content of the propagating state. From that moment on, the remaining state oscillates and identifying its decay allows us to probe the oscillation. If, for example, we tag a state that is initially a $B$ and we subsequently measure a semileptonic decay that could only have come from a $\bar B$, then we know that the system oscillated. The smoking gun signal for oscillations are same-sign leptons. 

Argus and \acro{CLEO} were \textit{symmetric} $B$ factories followed by
BaBar and Belle that were \textit{asymmetric} $B$ factories. 
In the symmetric case we cannot trace the time evolution and we can
only measure the time integrated oscillation probability.
In the asymmetric case, however, the center of mass frame is boosted relative to the lab frame so that the $B\bar B$ pair is not produced at rest. Suppose that at some time, $t=0$, one meson decays into a $\mu^+$, telling us that it was a $B$ meson. Thus at $t=0$ we know that the other meson is a $\bar B$. Then we can flavor tag decay final state and then use the decay position---through vertex detectors---as a measurement of the decay time. Note that the boosted system is important for being able to determine the decay time since it allows experimentalists to measure the physical distance between the vertices. 

The last two decades have been fantastic for $B$ physics. In addition
to the $B$ factories, \acro{LEP}, the Tevatron and the \acro{LHC} 
also made significant contributions to $B$ physics. Flavor tagging at
high energies is a different story because the $B\bar B$ pairs are
uncorrelated. At the $B$ factories, it is crucial that we knew the
angular momentum of the state; the fact that the $\Upsilon(4S)$ has
$L=1$ gave us the correlation. At high energies, we can have many
different angular momenta that mix; this causes decoherence. Thus, we
need a way to flavor
tag 
the $b$ at $t=0$. There are several ways to do this. For example, we may use
the fact that charged $B$ mesons do not oscillate: if we determine the
charge of the ``other-side'' $B$, we know what the flavor of
``our-side'' $B$ at $t=0$.
The decay of the oscillating $B$ can then be subsequently tagged by looking at our favorite semi-leptonic decays.
While such flavor tagging is not very efficient, hadron machines produce so many $b$s that we can afford to be very picky about using only nice events.

In the case where the $B$ is produced with large momentum and flavor tagged, it is clear we can follow the time evolution to extract $x$. Historically, however, $x$ was
first measured at Argus, where the $B$ is produced at rest and one can't measure the decay time. All you can do is flavor tag for particle production. Effectively such an experiment integrates over time. For simplicity, let's set $y=0$ and integrate from $t=0$ to $t=\infty$. In this limit we can easily perform the integral over the oscillation probability---by `easily' we mean it is easy to plug into \textit{Mathematica}---to obtain
\begin{align}
	\int_0^\infty P(B\to B)\, dt = \frac 12 \frac{2+x^2}{1+x^2}.
\end{align}
This gives us a function that we may invert to determine $x$. As good
theorists, we should check that this expression makes sense by looking
at limiting cases. Indeed, our above discussion we know that in the
limit $x\to \infty$ we expect $\int P\, dt =1/2$ while in the limit
$x\to 0$ we expect $\int P \, dt= 1$. 

\subsection{Calculating $\Delta m$ and $\Delta \Gamma$}

Now we should describe how to actually calculate $\Delta m$ and
$\Delta \Gamma$. This boils down to determining $M_{12}$ and
$\Gamma_{12}$. $M_{11}$ and $\Gamma_{11}$ are directly measured
numbers that we don't care too much about since they don't contribute to the mass or width difference.

We know that in the Standard Model, $M_{12}$ is related to some $(V-A)$ weak current operator $\mathcal O \sim (b\bar d)(b \bar d)$, 
\begin{align}
	M_{12} \propto \frac{1}{2m_B} \langle B | \mathcal O | \bar B\rangle,
\end{align}
But wait a moment---does $\mathcal O \sim (b\bar d)(b \bar d)$ look
strange to you? Indeed, one might have wanted to write down $(b\bar
d)(\bar b d)$ since, after all, we're used to operators formed out of
a field and its conjugate. Clearly, $ (b\bar d)(b \bar d)$ is not an
ordinary mass term. It represents a flavor-changing coupling between the $B$ and $\bar B$, with $\Delta B=2$ and $\Delta D=2$. This is because the mass term connects a particle and an anti-particle, but now because we're mixing $B$ and $\bar B$s, the `particle' is the $B$ and the `anti-particle' is an anti-$\bar B$ which is also a $B$.

\begin{eg}\textbf{A toy example.}
	Before going to the Standard Model, let's work with a simpler model with vector-like quarks that induce tree-level \FCNC through a $Z$ boson. Call this coupling $\kappa$. The diagram for $M_{12}$ is just the usual $s$-channel tree diagram and goes like ${\kappa^2}/{m_Z^2}$. Stop and check this again: should we have written $\kappa^\dag\kappa$ instead of $\kappa^2$? No! Check the diagram to make sure that \textit{both} vertices are given by the \textit{same} term in the Lagrangian, not the term and its Hermitian conjugate.
\end{eg}

In the Standard Model, the $B$--$\bar B$ mixing diagram occurs at one-loop level as a box diagram. At two-loop order one may also have `twin-penguin' diagrams which are subdominant. The two box diagrams are:
\begin{center}
	\begin{tikzpicture}[line width=1.5 pt, scale=1]
		\draw[fermion] (-1,2.3) -- (0,2);
		\draw[fermion] (0,2) -- (2,2);
		\draw[fermion] (2,2) -- (3,2.3);
		\draw[fermionbar] (-1,-.3) -- (0,0);
		\draw[fermionbar] (0,0) -- (2,0);
		\draw[fermionbar] (2,0) -- (3,-.3);
		\draw[vector] (0,2) -- (0,0);
		\draw[vector] (2,2) -- (2,0);
	\begin{scope}[shift={(5,0)}]
 		\draw[fermion] (-1,2.3) -- (0,2);
		\draw[vector] (0,2) -- (2,2);
		\draw[fermion] (2,2) -- (3,2.3);
		\draw[fermionbar] (-1,-.3) -- (0,0);
		\draw[vector] (0,0) -- (2,0);
		\draw[fermionbar] (2,0) -- (3,-.3);
		\draw[fermion] (0,2) -- (0,0);
		\draw[fermionbar] (2,2) -- (2,0);
	\end{scope}
	\end{tikzpicture}
\end{center}
The calculation of each diagram is identical in the limit where all internal particles are heavy. They're independent diagrams, but practically we can just calculate one of them and multiply by two. These diagrams are manifestly finite by power counting. 
We can parameterize the amplitude straightforwardly just by using power-counting and the \acro{GIM} mechanism,
\begin{align}
	\mathcal M\sim \frac{g^4}{m_W^2} \langle B|\mathcal O_{v-A}|B\rangle \sum V_{id}V_{ib}^*V_{jd}V_{jb}^* \, F(x_i, x_j)
\end{align}
where
\begin{align}
	x_i = \left(\frac{m_{u_i}}{m_W}\right)^2\ .
\end{align}

For the $B$ meson calculation, the dominant contribution comes from an internal top quark. For the kaons, there's a competition between the \CKM matrix elements and the mass insertions required by chirality and the \GIM mechanism. It turns out that for the real part of the amplitude, the internal charm is the winner, though the imaginary part is dominated by the top.

Now all we need is our old friend, the hadronic matrix element. We are faced with what seems like a rather intractable hadronic object, $\langle B | (b\bar d)(b\bar d) |B \rangle$. We have a trick up our sleeve---but like all tricks up someone sleeve, it's somewhat suspicious. We start by inserting a sum over a complete set of states,
\begin{align}
	\langle B | (b\bar d)(b\bar d) |B \rangle = \sum_n \langle B b\bar d | n\rangle \langle n | (b\bar d)|\bar B \rangle.
\end{align}
The trick is to assume that the vacuum is the most important contribution in the sum and hence approximate the sum with only the vacuum state, $\sim |n\rangle \langle n | \approx |0\rangle \langle 0$. Thus in the so-called \textbf{vacuum insertion approximation} (\acro{VIA}) we have
\begin{align}
	\langle B | (b\bar d)(b\bar d) |B \rangle \approx \langle B| (b\bar d) | 0\rangle \langle 0 | (b\bar d)|\bar B \rangle.
\end{align}
The great thing about this approximation is that we already know what $\langle 0 | (b\bar d)|\bar B\rangle$ is: it is the $B$ decay constant,
\begin{align}
	\langle B | (b\bar d)(b\bar d) |B \rangle  \approx f_B^2.
\end{align}
We've been characteristically hand-wavy since we neglected the Dirac structure of the operator. To do this properly one should invoke Fierz identities which give us some additional factors,
\begin{align}
	\langle B | (b\bar d)(b\bar d) |B \rangle = \frac{8}{3}f_B^2 B_B.
\end{align}
The prefactor includes a factor of two from the sum over the two box
diagrams and an additional $1 + \frac{1}{3}$ color factor. We've
included a parameterization of the correction to the vacuum insertion
approximation, $B_B$. Based on lattice calculations, it turns out, however, that for the $B$ meson this correction is only on the order of a few percent. To be clear, there is no \textit{a priori} reason for this approximation to work very well and the agreement is rather surprising.
Now that we are armed with this, we can proceed to cross-check the \CKM matrix.

\begin{framed}
	\noindent\textbf{The unbelievably heavy top quark.} Carlos Wagner once wrote a paper in the 80s that assumed the top mass to be around 50 GeV, for which it was promptly rejected by the journal editor as being unreasonably heavy. When you put the 50 GeV top into the above calculation you predict that $B$ mixing is very small. In the early 80s, flavor physicists found that $B$ mixing is, in fact, order one. The natural explanation was that the top was heavy, and indeed, flavor measurements in 1981 suggested $m_t\sim 150$ GeV. People didn't believe this because it was so ridiculously large. It wasn't until much later that electroweak precision tests predicted the same value. Historically people often say that electroweak precision experiments predicted a heavy top, but it was in fact $B$--$\bar B$ mixing that was the \emph{first} avatar of a heavy top---we just weren't ready to believe it!	
\end{framed}

The next object to calculate is $\Gamma_{12}$. The off-diagonal mass $M_{12}$ came from the the \textit{off shell} contribution of $\langle B | \mathcal O | \bar B \rangle$. When this matrix element goes on shell it gives a contribution to the width. This is because the real part of the two point function tells us about the particle's mass while the imaginary part tells us about its decay. Intuitively, the Hermitian part of the Hamiltonian is off shell while the anti-Hermitian part is on shell. For example, $K\to \pi\pi \to \bar K$ contributes to the width while $K\to B\to K$ with an off-shell $B$ contributes to the mass.

We propose a quark-level calculation. Can we trust such a calculation? There's something fishy.  In the calculation of $M_{12}$, we tacked on hadronic effects in $f_B$. We were off-shell with large $q^2$ so that our interactions are local and we could use effective operators. In the case of the decay width, the hadronic final states make a difference: for kaons, actual processes we care about are $K\to \text{hadrons}$ and $\bar K \to \text{hadrons}$, not the quark-level processes that are subsequently dressed by \QCD. These processes occur on shell, the quarks \textit{really} hadronize and we have to take this into account in our kinematics. By doing the quark level calculation, we implicitly invoke quark-hadron duality. However, the condition for quark-hadron duality is that we can average over a large $q^2$, i.e.\ we can smear over resonances. Here we have $q^2\to 0$; we only have \textit{one} kinematic point---the mass of the decaying particle. There is no smearing. \textit{A priori}, it is hard to predict the validity of our calculation given these conditions. We can hope that it might still be applicable to the $B$, but we already know that it is not so good for the kaons. 

Consider the kaon box diagram where only the $u$ can go on-shell. Phase space tells us that we can produce no more than three pions. It's now trivial to see that quark-hadron duality doesn't work since a phase space which permits `up to three pions' is clearly \textit{not} and inclusive decay. In fact, the $K_S$ mass eigenstate can go only to two pions, while the $K_L$ can go to three. So $y$ is basically one from phase space. $\Delta \Gamma$ is the difference between these rates. The kaon calculation is off by a large factor. In the $B$ case it is probably correct within factor of 2. 

What about the $D$ box diagram: which internal quark dominates? A quick back-of-the-envelope calculation invoking the ratios of quark masses and \CKM elements in the Wolfenstein parameterization should convince you that it is the strange quark. We could proceed   to calculate this diagram just as we did for the $B$ and $K$, but it turns out that this would be too na\"ive~\cite{Falk:2001hx}. In the $B$ box we had a heavy $t$ quark running in the loop. For the $D$ box we have an $s$ quark in the loop, which is \textit{not} heavy and so it is really \textit{not} okay to integrate it out into a local operator. If one does this calculation as if the $s$ were much heavier than the $D$, then one obtains a result which is four orders of magnitude too small. It is \textit{completely} wrong to make it a local operator; you can never integrate out an $s$ when you're talking about $c$ decay.

The correct thing to calculate is a diagram where only the $W$s are integrated out, giving the so-called \textbf{fish diagram},
\begin{center}
	\begin{tikzpicture}[line width=1.5]
		\clip (-3,-2) rectangle (3,2);
		\draw (0,3) circle (3.75);
		\draw (0,-3) circle (3.75);
	\end{tikzpicture}
\end{center}
From this one obtains $x\sim 10^{-2}$. 

\section{CP violation}
\label{sec:CP}

\emph{A priori}, \CP violation seems to have nothing to do with flavor physics. \CP is a discrete subgroup of spacetime symmetry while flavor has to do with internal symmetries. 
However, it turns out that in nature, all observations of
\CP violation happen to come along with flavor violation. We already know from these lectures that the only source of known \CP violation is the phase of the \CKM matrix. 

\subsection{General aspects of CPV}

When we look for \CP violation (\acro{CPV} or \cancel{CP}), we search for reactions where probability for one process is different from the probability of its \CP conjugate process,
\begin{align}
	P(A\to B) \neq P (\bar A \to \bar B).
\end{align}
You should distinguish this from T and CPT violation,
\begin{align}
	\cancel{\text{T}}: \quad P(A\to B) &\neq P (B \to A)\\
	\cancel{\text{CPT}}: \quad P(A\to B) &\neq P (\bar B \to \bar A).
\end{align}
However, since we believe CPT is a good symmetry of nature, \cancel{T} and \cancel{CP} are effectively equivalent.


\begin{prob}\label{prob:unitarity:CPT} \textbf{Unitarity and CPT}. This problem is slightly more formal but is a good exercise in quantum field theory.
	\begin{enumerate}
	\item  Show that in a time-reversal symmetric theory of scalar particles $A_{fi}=A_{if}$. Does this imply that the cross section for $A+B\to C+D$ equals that of $C+D\to A+B$?

	 \item Show that \acro{CPT} implies $A_{fi}=A_{\bar{i}\bar{f}}$ where $U_{\text{CPT}} |i\rangle=|\bar{i}\rangle$ and $U_{\text{CPT}}|f\rangle=|\bar{f}\rangle$.

	 \item Show that unitarity and \acro{CPT} imply $\Gamma (i\to \text{all}) = \Gamma (\bar{i}\to \text{all})$. Does this imply $\Gamma (i\to j) = \Gamma (\bar{i}\to \bar{j})$? 

	 \item Argue that unitarity and \acro{CPT} implies that $\Gamma (i\to j) = \Gamma (\bar{i}\to \bar{j})$ at \textit{lowest} order in perturbation theory. Further, argue that any CP violation in rates must come from loop effects. 

	 \end{enumerate}
	\begin{sol}
	For more background, see Preskill's lecture
        notes\footnote{\url{http://www.theory.caltech.edu/~preskill/notes.html},
          or the Sidney Coleman lecture notes upon which these are
          loosely based. The latter material is neatly encapsulated in
          the QFT textbook by Ticciati.}. 
\begin{enumerate}
\item Suppose a theory of scalar particles is $T$ symmetric. This means that $[T,H]=0$, where $H$ is the full Hamiltonian including interactions. (The free Hamiltonian is $T$-symmetric.) The unitary time evolution operator is transformed according to
		\begin{align}
		T U(t,t_0) T^{-1} =& T e^{iH_0(t-t_0)}e^{-iH(t-t_0)}T^{-1}	\\
		=& e^{-iH(t-t_0)}e^{iH(t-t_0)} = U(t_0,t) = U(t,t_0)^\dag.
		\end{align}
		Since $S=U(\infty,-\infty)$, $TST^{-1}=S^\dag$. We want a relation between amplitudes, but we must be careful with matrix elements with respect to the antiunitary $T$ operation. $\langle f|Si\rangle = \langle TSi|Tf\rangle = \langle S^\dag Ti | Tf\rangle$. For scalars we have $Ti=i$ and $Tf = f$ since they have no spin to go in the opposite direction. Thus we have the relation $\langle f|Si\rangle = \langle i|Sf\rangle$, or $A_{fi} = A_{if}$. Note that this is \textit{not} the same as saying $\sigma(i\to f) = \sigma(f\to i)$ since the cross section has phase space factors that are not the same even if the amplitudes are.
		\item \acro{CPT} gives a weaker form of the detailed balance condition. $CPT|i\rangle = \bar i\rangle$ and $CPT|f\rangle = \bar f\rangle$. Following the same steps as the previous sub-problem (because both $T$ and $CPT$ and anti-unitary), we arrive at $A_{fi} = A_{\bar i \bar f}$.
		\item Unitarity tells us that $S^\dag S = S S^\dag = 1$, or in terms of the $T$-matrix, $T^\dag T = TT^\dag$. (Do not confuse this $T$ with the time inversion operator.) This implies then that $\sum_m |A_{im}|^2 = \sum_m |A_{mi}|^2$. Acting on both sides with $CPT$ gives $\sum_{\bar m} |A_{\bar m\bar i}|^2 = \sum_m |A_{m\bar i}|^2$ since $\sum_m = \sum_{\bar m}$ when going over all states. This then implies that $\Gamma(i\to \text{ all}) = \Gamma(\bar i \to \text{ all})$. Alternately, one can use the optical theorem, $\text{Im } A_{ii}= \text{Im }A_{\bar i\bar i}$, along with the equality of the phase space factors. Note that only the \textit{inclusive} rates $i\to \text{ all}$ and ${\bar i} \to \text{ all}$ match; in general $i\to j$ and $\bar i to \bar j$ have different rates. This is a manifestation of \CP violation.
		\item For simplicity, we will use the $T$-matrix in place of the $S$ matrix. The optical theorem tells us that $T_{ij}-T^*_{ji} = i\sum_n T_{in}T^*_{jn}$. This then implies that $|T_{ij}|^2 = |i(\sum TT^\dag)_{ij}+T_{ji}^*|^2$. If we now compare $\Gamma(i\to j)$ with $\Gamma(\bar i\to \bar j)$ and not that $\Gamma \sim |T|^2$, we find that
		\begin{align}
			|T_{ij}|^2 - |T_{\bar i\bar j}|^2  =&
			|i(\sum TT^\dag)_{ij}+T^*_{ji}|^2 - |T_{ji}|^2
			& -2\text{Im }(\sum TT^\dag)_{ij}T^*_{ji} + |(\sum TT^\dag)_{ij}|^2,
		\end{align}
where in the first line we used \acro{CPT} to subtract $|T_{\bar i \bar j}|^2 = |T_{ji}|^2$ from both sides. The right-hand side of the second line is at least $\mathcal O(T^3)$ whereas the left-hand side is $\mathcal O(T^2)$. This means that the right-hand side is higher order in perturbation theory. For fixed external states, higher order diagrams come from loops so that CP violation vanishes at the lowest order of perturbation theory and can only appear in loop effects.
	\end{enumerate}
	\end{sol}
\end{prob}

If \CP were violated at tree level, then it should have shown up all over the place. However, \CP was only first observed experimentally in the 1960s in the kaon system, and not until 2000 in the $B$ system. To date, we only have a handful of experimental examples that demonstrate \CP violation. One should compare this to parity violation, which does show up at tree level and can indeed be seen all over the place. 
%

\begin{framed}
\noindent \textbf{Open question} (a free project to anyone who wants
to do it): what is the \CP violation in the Hydrogen atom? Given a
hydrogen and an anti-hydrogen, what is the relative difference in the
width of the 21 cm line? The lifetime of the $J=1$ state must be the
same from \acro{CPT}, but this does not imply that the 21 cm width is the
same; there are sub-leading decays of the $J=1$ state.
We suspect that it is something like $10^{-100}$. 
\end{framed}

\CP violation ultimately appears as the result of interference between two diagrams  with a difference in the \textbf{strong} and \textbf{weak} phases. The reason why \CP effects are hard to see is that usually for a given process there is only a single leading diagram and the higher-loop corrections are extremely small. 
\begin{itemize}
	\item The \textbf{weak phase} is a \CP-odd phase; it changes
          sign for a \CP conjugate process. This phase shows up in the
          Lagrangian. Given a coupling $V_{ub}W^+\bar u b$, then the
          Hermitian conjugate coupling is $V_{ub}^*W^- \bar b u$. It is
          called ``weak'' because in the Standard Model the only source of this phase is the weak interaction
	\item The \textbf{strong phase}, on the other hand, is a
          \CP-even phase; it does not change sign between a process
          and its \CP conjugate. Since all the phases in the
          Lagrangian are \CP odd, this phase cannot directly come from
          coupling constants. In fact, strong phases come from time
          evolution. It is called ``strong'' because in the Standard Model, the strong
          interaction generates a non-trivial \CP-even phase.
For example, a free particle has a trivial time evolution, $\exp(iEt)$, independent of whether it is a particle or anti-particle. A more complicated strongly-interacting system has much less trivial time evolution that yields a strong phase from non-perturbative dynamics.
One way this effect is manifested is when one goes near a resonance; for example, two pion scattering near the $\rho$ resonance. There's a strong phase across in the Breit--Wigner resonance in classical scattering, exactly the same phenomena as a forced oscillation with friction.  
\end{itemize}

In mesons, there are two processes of primary interest in flavor physics: decays and oscillations. Suppose there are two diagrams that allow decay of a meson $B$ into a final state $f$. We may write the amplitudes for $B\to f$ and its \CP conjugate as
\begin{align}
	A_f = A(B\to f) &= a_1 e^{i(\delta_1+\phi_1)} + a_2 e^{i(\delta_2+\phi_2)} 
	\label{eq:CP:Af:phases} \\
	\bar A_f = A(\bar B\to \bar f) &= a_1 e^{i(\delta_1-\phi_1)} + a_2 e^{i(\delta_2-\phi_2)},
	\label{eq:CP:Afbar:phases}
\end{align}
where we write $\delta$ as the strong phase and $\phi$ as the weak
phase\footnote{Admittedly this is very bad notation because the
  $\delta$ in our \CKM parameterization is the weak phase. Unfortunately, this is the standard convention. We didn't invent it.}. 
These terms, for example, typically correspond to a tree-level diagram
and a penguin. It is trivial to generalize this to the case where there are multiple diagrams with different phases.
The partitioning (\ref{eq:CP:Af:phases}--\ref{eq:CP:Afbar:phases}) into two terms is required by enforcing that the terms have the same coefficients and that the amplitude and \CP-conjugate process is related by simply flipping the sign of the weak phase. 

It should be trivial that if $a_2 =0$ or $\delta_1 = \delta_2$ or $\phi_1=\phi_2$, then $|A_f| = |\bar A_f|$. Thus the conditions to see \CP violation are that you have two terms such that there are different weak phase and strong phase.

Our goal experimentally is to find processes where $a_1\approx a_2$ and where there are different $\delta$ and $\phi$. Note that $a$s and $\delta$s are related to the strong interaction; only $\phi$ is related to the weak interaction. We return to the usual story: we have to deal with hadronic matrix elements. Here there are two good cases: either the hadronic matrix elements cancel or they can be directly measured. 

\begin{eg}
	We know that we have \CP violation if $|A_f| \neq |\bar A_f|$. Is the converse true? Can we show that there is \CP violation without actually directly measuring it? The analog is measuring the angles of a triangle without directly measuring any particular angle. In principle one can measure \CP violation just by measuring the lengths of the sides of the unitarity triangle. By measuring a few decay rates and a few complex conjugate decay rates---such that they are all the same---and we can still demonstrate that there is a phase in the Lagrangian. 
\end{eg}

When oscillations are involved we also use the following definitions
\begin{align}
	M_{12} &= |M_{12}| e^{i\phi_M} 
	&
	\Gamma_{12} &= |\Gamma_{12}|e^{i\phi_\Gamma}.
\end{align}
where the phases are \CP-odd phases.

We've talked very generally about \CP violation in mesons. Let us
briefly mention two other measurements of \CP violation outside of
flavor physics. The first is the \textbf{triple product}, where in
some decay one takes three vectors (e.g. $p_1$, $p_2$, and spin in a
three-body decay) and generate some kind of triple product
$(\mathbf{p}_a\times \mathbf{p}_2)\cdot \mathbf{s}$. These types of
quantities turn out to be \CP odd and sensitive to the sum of the
strong and weak phase. At the \acro{LHC} one can study \CP violation
with these sorts of triple products. 

The other possible measurements for \CP violation are \textbf{electric
  dipole moments} (\acro{EDM}). These are \CP violating for elementary
particles. When you take the dipole moment of such a particle, then
the \CP conjugate of such a state must change because the dipole is an
odd angular momentum and the transformation is \CP-odd. (Note that the \emph{magnetic} moment is \CP-even.) The argument only holds for elementary particles. When you have molecules you can have an \acro{EDM} and move to different states. Measurement of the \acro{EDM} for the electron is a measurement of \cancel{CP}. What about the neutron? People say that the neutron \acro{EDM} is the best bound on \acro{SUSY} \CP violation. Certainly the neutron is not elementary, but there is no excited neutron state. Thus the full statement is that the \acro{EDM} of any particle that doesn't have an excited state is a signal of \cancel{CP}. 

So far \acro{CPV} has only been unambiguously observed in $K_L$,
$B^0$, and $B^\pm$ decays. 
\CP violation is typically measured through asymmetries. These are \emph{ratios of branching ratios} of the form
\begin{align}
	A_\text{CP} \equiv 
	\frac{
		\Gamma(B\to f) - \Gamma(\bar B \to \bar f)
	}{
		\Gamma(B\to f) + \Gamma(\bar B \to \bar f)
	}.
\end{align}
There are basically two processes which are sensitive to \CP: decay and oscillation. To really be sensitive to the \CP phases, we need two amplitudes to interfere, either two decays, two mixings, or between a decay and a mixing. Thus there are three options:
\begin{enumerate}
	\item \textbf{CPV in decay}. This is also known as \textbf{direct CP violation}. This is an interference between decay amplitudes. For example the $B^\pm$ doesn't oscillate, so \acro{CPV} in this system comes its decay. This type of \acro{CPV} is characterized by $|\bar A/A| \neq 1$ where $A$ ($\bar A$) is the amplitude for a decay (\CP conjugate decay).
	\item \textbf{CPV in mixing}. This is also called \textbf{indirect CP violation}. This is an interference between $M_{12}$ and $\Gamma_{12}$, the two ways to mix form $B$ into a $\bar B$. This is the type of \CP violation seen in charged-current semileptonic neutral meson decays. This type of \acro{CPV} is defined by $|q/p| \neq 1$ where $q$ and $p$ are the mixing coefficients between the mass and flavor eigenstaets in (\ref{eq:mixing:BLH}).
	\item \textbf{CPV in interference between decays with and without mixing}. This is \CP violation coming from the interference between a decay $B\to f$ and a decay $B\to \bar B \to f$. This is defined by $\text{Im}\, \lambda_f \neq 0$, where
	\begin{align}
		\lambda_f \equiv \frac qp \frac{\bar A_f}{A_f}.
	\end{align}
\end{enumerate}

In the following we  briefly discuss the first and last options. All
we mention about the second option is that \CP in mixing measures the deviation of $|q/p|$ from one. You are encouraged to study each of these cases in more detail.

\subsection{CP Violation in decay}

Let's start with \CP violation in decay, or \textbf{direct CP violation}. We define the observable
\begin{align}
	a_\text{CP}[f] &\equiv \frac{\Gamma(\bar B\to \bar f)-\Gamma(B\to f)}{\Gamma(\bar B\to \bar f)+\Gamma(B\to f)}
	= \frac{\left|\bar A/A\right|^2 -1}{\left| \bar A/A\right|^2 +1} \\ ,
\end{align}
where we define the shorthand
\begin{align}
	A&\equiv A(B\to f)
	&
	\bar A&\equiv A(\bar B\to \bar f) \ .
\end{align}
We assume that these amplitudes get contributions from at least two
diagrams with a weak and strong phase,
(\ref{eq:CP:Af:phases}--\ref{eq:CP:Afbar:phases}) and we order
the amplitudes such that $a_2 \le a_1$. We can then parameterize the
amplitude with respect to the diagrams' relative magnitudes
$r=a_2/a_1$ and relative phases 
$\Delta\phi \equiv \phi_1-\phi_2$ and $\Delta\delta \equiv \delta_1-\delta_2$,
\begin{align}
	A(B\to f) &= a_1 e^{i(\delta_1+\phi_1)}\left(1+ re^{i(\Delta\phi+\Delta\delta)}\right)\\
	A(\bar B\to \bar f) &= a_1e^{i(\delta_1+\phi_1)} \left(1+ re^{i(-\Delta\phi+\Delta\delta)}\right).
\end{align}
Working for first order in $r$ we get the expression
\begin{align}
	a_\text{CP} = r\sin\Delta\phi\sin\Delta\delta \ .
\end{align}
Note that this indeed only depends on the physical difference in the phases, $\Delta \phi$ and $\Delta\delta$, not the unphysical value of the phases themselves. Further, this expression reiterates that we need two different diagrams with different weak \emph{and} string phases.
Up to now we haven't explained how these values are related to our
fundamental parameters. The weak phases $\phi$ is be related to the
\CKM matrix elements. We can use symmetries to get information about
the hadronic matrix element which show up in $r$. For the strong \CP
phase, $\delta$, however, the whole effect has to do with symmetry
breaking so the leading effect is zero. This is why extracting
information from \acro{CPV} measurements is not easy.  
Practically, we want ways to measure \CP asymmetries where the $\delta$ dependence is replaced by other measurements.
This is best illustrated by diving into some examples.

\subsubsection{$B\to K \pi$}

An excellent place to observe \CPV are the \CP asymmetries of the charged meson decays. We examine the interference between tree-level decays and penguin-mediated decays (\textbf{penguin pollution}). For example, consider $B^+\to \pi^0 K^+$. The two diagrams are:

	\begin{center}
		\begin{tikzpicture}[line width=1.5 pt, scale=1.3]
			\draw[fermion] (40:1.2)--(0,0);
			\draw[fermionbar] (180:1)--(0,0);
			\draw[vector] (-40:1)--(0,0);
			\node at (40:1.4) {$u$};
			\node at (180:1.2) {$s$};
			\node at (0.0,-.5) {$W$};
		\begin{scope}[shift={(0,.3)}]
			\draw[fermion] (180:1) to [out=0,in=220] (40:1); 
			\node at (40:1.4) {$u$};
			\node at (180:1.2) {$u$};
		\end{scope}
		\begin{scope}[shift={(-40:1)}]
			\draw[fermion] (-20:1)--(0,0);
			\draw[fermionbar] (20:1)--(0,0);
			\node at (-20:1.2) {$s$};
			\node at (20:1.2) {$u$};
		\end{scope}
		\draw [gray,decorate,decoration={brace,amplitude=5pt},xshift=-4pt]
	   (-1.3,-.2)  -- (-1.3,.4) 
	   node [black,midway,left=4pt,xshift=-2pt] {$B^+$};
	   \draw [gray,decorate,decoration={brace,amplitude=5pt}]
	   (46:1.8)  -- (30:1.5) 
	   node [black,midway,xshift=20pt, yshift=5pt] {$\pi^0$};
	   \draw [gray,decorate,decoration={brace,amplitude=5pt}]
	   (2.2,-.1)  -- (2.2,-1.2) 
	   node [black,midway,xshift=20pt, yshift=0] {$K^+$};
		\end{tikzpicture}
		\qquad\qquad
		\begin{tikzpicture}[line width=1.5 pt, scale=1.3]
			\draw[fermion] (36:.6)--(0,0);
			\draw[fermion] (36:1.5)--(36:.6);
			\draw[fermionbar] (-.5,0)--(0,0);
			\draw[fermionbar] (-1,0)--(-.5,0);
			\draw[vector] (-.5,0) arc (180:36:.5);
			\draw[gluon] (-40:1)--(0,0);
			\node at (36:1.7) {$s$};
			\node at (180:1.2) {$b$};
		\begin{scope}[shift={(0,-.3)}]
			\draw[fermion] (180:1) to [out=0,in=175] (1.7,-1); 
			\node at (1.8,-1) {$u$};
			\node at (180:1.2) {$u$};
		\end{scope}
		\begin{scope}[shift={(-40:1)}]
			\draw[fermion] (-20:1)--(0,0);
			\draw[fermionbar] (60:1.5)--(0,0);
			\node at (-19:1.2) {$u$};
			\node at (60:1.7) {$u$};
		\end{scope}
		\draw [gray,decorate,decoration={brace,amplitude=5pt},xshift=-4pt]
	   (-1.3,-.5)  -- (-1.3,.2) 
	   node [black,midway,left=4pt,xshift=-2pt] {$B^+$};
	   \draw [gray,decorate,decoration={brace,amplitude=5pt}]
	   (43:1.8)  -- (25:2) 
	   node [black,midway,xshift=15pt, yshift=15pt] {$K^+$};
	   \draw [gray,decorate,decoration={brace,amplitude=5pt}]
	   (2.2,-.7)  -- (2.2,-1.5) 
	   node [black,midway,xshift=20pt, yshift=0] {$\pi^0$};
		\end{tikzpicture}
	\end{center}
	
\noindent Note that we consider the \emph{gluon} penguin since this
doesn't cost us much in terms of small coupling constants. Both
diagrams are $\mathcal O(G_F)$ up to loop factors which can be
partially made up for by the enhancement from an internal top
quark. Further, note that the two processes have different weak
phases. These weak phases come from the \CKM matrix elements. Moreover,
the penguins have a different Lorentz structure and hence different
hadronic matrix elements. While this makes it hard to calculate the
ratio $r$, it also make it possible to have a \CPV signal.

We can easily estimate the size of each of these amplitudes,
\begin{align}
	\mathcal M_{\text{penguin}} &\sim \alpha_w\alpha_s \frac{\lambda^2}{16\pi^2}
&
	\mathcal M_{\text{tree}} &\sim  \alpha_w \lambda^4.
\end{align}
The tree is larger by a factor of five by our na\"ive analysis,
however experimentally it turns out to be the other way around: the
penguin is larger by a factor of about 3. It is very important that
these turn out to be roughly the same size so we can have a sizable
\CPV asymmetry.

Now that we have the two amplitudes, we must identify the phases. We
assume that the $b\to s$ penguin is dominated by the internal top,
though the charm also contributes.In the standard paramerization of
the \CKM, what's the weak phase of this transition? None! You could have seen this---the from our Wolfenstein parameterization, the phase  only shows up in the elements linking the first and third generations.  What about the tree diagram? Looking back at our unitarity triangle, the $b\bar u$ vertex comes with the $\gamma$ angle. Finally, we note that we cannot say anything about the strong phase---we just don't know.
%

There are four different $B\to K\pi$ decays. We can use isospin and
SU(3) to relate these. We only mention the important fact that the $B^+$ was the first system where we saw \CPV without flavor mixing.

%
%
%
%


\subsubsection{$B\to DK$}

Now let's do the `best case of all,' $B\to DK$ decay. See, $DK$ even sounds like `decay.' This is by far the cleanest measurement that we can do of any flavor parameter. It's more precise at the theory level than any measurement we've presented so far. This should be exciting if you're a theorist or an experimentalist. If you're a theorist you say \emph{wow}. If you're an experimentalist you say \emph{wow, I want to measure that.}

The question is not so much how do we see the \CP violation, but we
want to measure the angle $\gamma$ of the unitarity triangle. 
Let us
consider the following three decays
\begin{align}
B^+ \to D K^+ \qquad
B^+ \to \bar D K^+ \qquad
B^+ \to D_\text{CP} K^+ 
\end{align}
where $D_\text{CP}$ is a state that decays into a \CP eigenstate.
We can tell the three apart using the final state of the $D$ decay. 
$D$ and $\bar D$ can be tagged with semi-leptonic decays,
while $D_\text{CP}$ is tagged when the final state is a \CP eigenstate,
for example, $\pi^+ \pi^-$.
To good approximation we can neglect $D\bar D$ oscillation. 
Note that $D_\text{CP}$ is a $45^\circ$ rotation compared to the $D$ and $\bar D$,
\begin{align}
	D_\text{CP}= \frac{D\pm \bar D}{\sqrt{2}}.
\end{align}
The
crucial thing is that you don't know what you produce, not a $D$ or
$\bar D$ or a $D_\text{CP}$. All you know is that it's some coherent
combination of $D$ and $\bar D$. 

There are two amplitudes that give the $D_\text{CP}$ final state and
thus we can hope to utilize their interference. Usually there's a tree and a loop, but in this case it's a tree and a tree\footnote{This is doubly rare: if two amplitudes for \CP violation are not tree and loop, usually it's loop and loop!} In terms of mesons the two amplitudes are
\begin{align}
	B^+ &\to D K^+ 
	&
	B^+ &\to \bar D K^+\ .
\end{align}
At quark level, the decays are
%
\begin{center}
	\begin{tikzpicture}[line width=1.5 pt, scale=1.3]
		\draw[fermion] (15:1.5)--(0,0);
		\draw[fermionbar] (180:1)--(0,0);
		\draw[vector] (-40:1)--(0,0);
		\node at (15:1.7) {$c$};
		\node at (180:1.2) {$b$};
		\coordinate (uin) at (-1,-.3);
		\coordinate (umid) at (-80:.5);
	\begin{scope}[shift={(-40:1)}]
		\draw[fermion] (-20:1)--(0,0);
		\draw[fermionbar] (40:1)--(0,0);
		\node at (40:1.2) {$u$};
		\node at (-20:1.2) {$s$};
		\coordinate (uend) at (-40:1.2);
		\node at (-40:1.4) {$u$};
	\end{scope}
	\draw[fermion] (uin) to [out=0, in=140] (umid) to [out=-35, in=160] (uend);
	\node at (-1.2,-.3) {$u$};
	\draw [gray,decorate,decoration={brace,amplitude=5pt},xshift=-4pt]
	   (-1.2,-.5)  -- (-1.2,.2) 
	   node [black,midway,left=4pt,xshift=-2pt] {$B^+$};
	\draw [gray,decorate,decoration={brace,amplitude=5pt},xshift=-4pt]
	   (2,.6)  -- (2,-.1) 
	   node [black,midway,right=4pt] {$\bar{D}$};
	\draw [gray,decorate,decoration={brace,amplitude=5pt},xshift=-4pt]
	   (2.2,-.8)  -- (2.2,-1.7) 
	   node [black,midway,right=4pt] {$K^+$};
	\end{tikzpicture}
	\qquad\qquad
	\begin{tikzpicture}[line width=1.5 pt, scale=1.3]
		\draw[fermion] (15:1.5)--(0,0);
		\draw[fermionbar] (180:1)--(0,0);
		\draw[vector] (-40:1)--(0,0);
		\node at (15:1.7) {$u$};
		\node at (180:1.2) {$b$};
		\coordinate (uin) at (-1,-.3);
		\coordinate (umid) at (-80:.5);
	\begin{scope}[shift={(-40:1)}]
		\draw[fermion] (-20:1)--(0,0);
		\draw[fermionbar] (40:1)--(0,0);
		\node at (40:1.2) {$c$};
		\node at (-20:1.2) {$s$};
		\coordinate (uend) at (-40:1.2);
		\node at (-40:1.4) {$u$};
	\end{scope}
	\draw[fermion] (uin) to [out=0, in=140] (umid) to [out=-35, in=160] (uend);
	\node at (-1.2,-.3) {$u$};
	\draw [gray,decorate,decoration={brace,amplitude=5pt},xshift=-4pt]
	   (-1.2,-.5)  -- (-1.2,.2) 
	   node [black,midway,left=4pt,xshift=-2pt] {$B^+$};
	\draw [gray,decorate,decoration={brace,amplitude=5pt},xshift=-4pt]
	   (2,.6)  -- (2,-.1) 
	   node [black,midway,right=4pt] {$D$};
	\draw [gray,decorate,decoration={brace,amplitude=5pt},xshift=-4pt]
	   (2.2,-.8)  -- (2.2,-1.7) 
	   node [black,midway,right=4pt] {$K^+$};
	\end{tikzpicture}
\end{center}
These appear to have different final states, but we only look at
decays where the $D$ or $\bar D$ go into a \CP eigenstate, so that the
actual experimentally measured final state is, for example, $K^+
(\pi^+\pi^-)_D$ for both decays. Thus there really is interference. 
(We assume that there is no \CP violation in the $D$ decay
as we work within the Standard Model.)

When you see two diagrams, what do you do? Some people want to
calculate. We will estimate. In terms of the Wolfenstein expansion
parameter, the ratio of these diagrams is of order unity. We thus expect these two diagrams to be of the same order. However, once you look into the details, the ratio is actually more like 7. Part of this difference comes from the not-small Wolfenstein parameters. More importantly, there is a color suppression. The outgoing up quarks in the first diagram can swap color assignments. The other diagram doesn't have a factor of three coming from the sum over colors that appears in the first `color-allowed' diagram, so we say that the second diagram is `color-suppressed.'

So we have established the first condition for \CP violation: there
are two diagrams. We are happy that they're roughly the same
order. What about the weak phase? In the Wolfenstein parameterization,
the weak phase difference is $\gamma$ in the second diagram coming from $V_{ub}$. 
Now the third ingredient: what is the strong phase? Do we expect a
strong phase? Yes we expect one, but we do even \emph{need} a strong phase? It turns out that in this case, we do \textit{not} need a strong phase to measure $\gamma$! 
Let us define the amplitudes 
\begin{align}
	A^+_1&\equiv A(B^+\to DK^+)\\
	A^-_1&\equiv A(B^-\to DK^-)\\
	A^+_2&\equiv A(B^+\to \bar DK^+)\\
	A^-_2&\equiv A(B^-\to \bar DK^-)\\
	A^+_\text{CP}&\equiv A(B^+\to \bar D_\text{CP}K^+)\\
	A^-_\text{CP}&\equiv A(B^-\to \bar D_\text{CP}K^-)\ .
\end{align}
Given the definition of $D_{\text{CP}}$ in terms of the $D$ and $\bar D$ states, it is clear that
\begin{align}
	A^\pm_\text{CP} = \frac{A^\pm_1 + A^\pm_2}{\sqrt{2}}\ .
\end{align}
If the sum of three complex numbers is zero then it's a triangle. The sum of the $A^+$ and $A^-$ amplitudes (with a relative minus sign on the $A_\text{CP}$) are thus two independent triangles in the complex plan. 

Write the amplitude of the color-allowed diagram as $A_1^+ = A$.
The amplitude which is color suppressed can be parameterized as
\begin{align}
	A_2^+ = A r e^{i(\gamma+\delta)}.
\end{align}
Thus the way that we parameterized the amplitude tells us that
\begin{align}
	A_1^- = A,
\end{align}
since this is just the \CP conjugate of an amplitude that we defined to be real. Similarly,
\begin{align}
	A_2^- = A r e^{i(\delta-\gamma)}.
\end{align}
Finally, we can write the width
\begin{align}
	\Gamma(B^+\to D_\text{CP}K^+) = (A^+_\text{CP})^2 = \frac {A^2}{2}\left(1+r^2 + 2r\cos (\gamma+\delta)\right).
\end{align}
Do you see why this is so nice? The sensitivity to the phase shows up right where we expect, in the interference term and proportional to the ratio of the amplitudes. The expression for the $B^-$ decay is the same thing with a relative sign in the argument of the cosine,
\begin{align}
	\Gamma(B^-\to D_\text{CP}K^-) = (A^-_\text{CP})^2 = \frac {A^2}{2}\left(1+r^2 + 2r\cos (\gamma-\delta)\right).
\end{align}
When we take the difference the non-cosine terms cancel. We end up with $\sim r\sin(\gamma-\delta)$. Direct \CP violation! The situation becomes more interesting because we can actually measure $A$ and $r$ and extract $\delta$. 
If we have a decay that measures the flavor, i.e.\ that tags the $D$ flavor---for example our favorite semileptonic decay, then we know the magnitude of $A$ and $r$. Then we can get $\delta$ and $\gamma$ up to a discrete ambiguity. 

What is the difference between this and the $K\pi$ process? In the $K\pi$ case we didn't know the strong phase or $r$. For $B\to D K$ we can measure $r$ and the strong phase. We can actually measure the magnitude of the things which are interfering, this is what makes this decay unique---there are no theoretical uncertainties. 

If $\delta=0$, then both the $B^+$ and $B^-$ have the same expression. However, since we know $r$ we can still measure $\cos\gamma$! We can present this in terms of triangles. 
\begin{center}
	\begin{tikzpicture}[scale=1.5]
		\draw (0,0) -- (45:2) -- (-2,0);
		\draw (0,0) -- (100:2) -- (-2,0);
		\draw (-2,0) -- (0,0);
		\node at (-1.5,1.5) {$D_{\text{CP}}$};
		\node at (60:1.75) {$D_{\text{CP}}$};
		\node at (93:1.75) {$r$};
		\node at (35:1.5) {$r$};
		\draw (180:.35) arc (180:45:.35);
		\draw (100:.6) arc (100:180:.6);
		\draw (100:.65) arc (100:180:.65);
		\node at (.65,.0) {\footnotesize $(\gamma+\delta)$};
		\node at (-.7,-.3) {\footnotesize $(\gamma-\delta)$};
	\end{tikzpicture}
\end{center}
You measure six rates, from which you make two triangles and then you
can get the two angles $\gamma$ and $\delta$.

The example above can be generalized by including any state which can come from a $D$ and $\bar D$. For example, $B\to DK$ and $B\to \bar D K$ where $D\to f$ or $D\to \bar f$. If $f$ is not a \CP eigenstate, then you don't need to have the $D_\text{CP}$ to be a 50--50 composition of $D$ and $\bar D$ (it wouldn't be a \CP eigenstate anymore). Even when you do this you don't need to put any theory into the game, you still have all the information to measure everything. 
We could also look at $B\to D^* K$ with $D^*\to D\pi\to f \pi$. We can do the
same thing with three body $D$ decays. It is the combination
of all of these decays that one use in a global fit to get $\gamma$.

\subsection{CP violation from mixing}

Now we want to develop the formalism of oscillation when the decay does not go into a flavor state. The situation is much more complicated than the nice oscillations we first met. We consider some decays $B\to f$ and $\bar B \to f$ with amplitudes,
\begin{align}
	A_f &\equiv A (B\to f)
	&
	\bar A_f &\equiv A (\bar B\to f) \ ,
\end{align}
where, in general, $f$ needn't be a \CP eigenstate. For example, for semileptonic decays $\bar A_f=0$. Define a parameter
\begin{align}
	\lambda_f = \frac{q}{p}\frac{\bar A_f}{A_f},
\end{align}
where $p$ and $q$ are the mixing parameters of the $B$ energy eigenstates: $B_{L,H} = p|B\rangle \pm q|\bar B\rangle$. We  state without proof that $\lambda_f$ is a \textit{physical} combination of parameters.  The first fraction has to do with mixing while the second fraction has to do with decay. This product is a natural basis-independent object to write that depends on both the mixing and the decay. What is the most general evolution of a $B$ \textit{without} assuming \CP? Given a state that started as a $|B\rangle$ at $t=0$, the time evolution is
\begin{align}
	B(t) &= g_+(t)|B\rangle - \frac qp g_-(t)|\bar B\rangle
	&
	\bar B(t) &= g_+(t)|B\rangle - \frac pq g_-(t)|\bar B\rangle
\end{align}
where
\begin{align}
	g_\pm(t) = \frac 12\left(
		e^{
			im_H t - \frac 12 \Gamma_H t 
			\pm im_L t \mp \frac 12 \Gamma_L t}
	\right).
\end{align}
This is just the Schr\"odinger time evolution. Details of this derivation are presented in Appendix~\ref{app:meson:mixing:and:CP}. What is the probability for this to decay to some final state $f$? Writing a dimensionless time parameter $\tau = \Gamma t$, we have
\begin{align}
	\Gamma(B\to f)
	&= 
|A_f|^2 e^{-\tau} \left[\cosh (y\tau)+ \cos(x\tau)\right] 
\nonumber\\&
+|\lambda_f|^2 \left[\cosh(y\tau) + \cos(x\tau)\right]
\nonumber\\  &
-2\text{Re}\left[\lambda_f(\sinh(y \tau)-i\sin(x\tau))\right] \ .
\end{align}
Recall that $x \sim \Delta m$ and $y\sim \Delta \Gamma$. The first term is the decay \emph{before} oscillation. The second term has to do with the decay \emph{after} an oscillation. The third term encodes interference. This interference is proportional to $\lambda_f$, which is what we're after. We will see what $\lambda_f$ has to do with the \CKM. There is a similar expression for $\bar B$, and we can define the \CP asymmetry,
\begin{align}
	\mathcal A_f(t) \equiv \frac{\Gamma[\bar B(t)\to \bar f] - \Gamma[B(t)\to f)]}{\Gamma[\bar B(t)\to \bar f] + \Gamma[B(t)\to f)]}\ .
\end{align}
This is the oscillatory, time-dependent version of $a_\text{CP}$. In principle you just plug in the time-dependent decay rates listed above. We won't go through the long and messy general expression. Instead, we consider specific final states that simplify everything.

Start by assuming that $\Delta \Gamma = 0$ ($y=0$) since this sets all
the hyperbolic cosines to one. What's a good meson for this? The
$B$. The other approximation that we'll make is $|q/p|=1$. Do not be
confused---we said that $|q/p|\neq 1$ implies \CP violation, but we
can also have $|q/p|=1$ even when there is \CP violation in the
theory. This is a very good approximation for the $B$ system. The last
thing to assume is that $|A_f| = |\bar A_f|$. This is the case when
the final state is a \CP eigenstate and there is one dominant decay amplitude, or if the weak or strong phase
difference vanish. Our conditions are
\begin{enumerate}
	\item $\Delta \Gamma=0$
	\item No \CP violation in the decay
	\item No \CP violation in the mixing
	\item The final state $f$ is a \CP eigenstate with one decay amplitude.
\end{enumerate}
Where is the \CP violation? We want to find \CP violation in the interference. Under these assumptions, $|\lambda_f| = 1$ and
\begin{align}
	\mathcal A_f(t) = \text{Im}(\lambda_f) \sin (\Delta m t)\ .
\end{align}
Note that $\lambda_f$ is thus a pure phase and $\text{Im}(\Lambda_f)$ picks out information about this phase. $\Delta m$ is a parameter we measure, and the decay itself gives us the time $t$. All we have to do is plot the amplitude with respect to the range of $t$s that we get experimentally. There was a twenty year difference between the measurement of $\Delta m$ and the measurement of $\text{Im}(\lambda_f)$; during this time we got a very good measurement of $\Delta m$ so that all we had to think about was measuring $\text{Im}(\lambda_f)$. 

This measurement gives us the phase of $\lambda_f$. Now we want to connect it to a calculation of $\lambda_f$. Consider the \emph{golden mode}, $B\to \Psi K_S$:
\begin{align}
	\lambda_f = e^{-i\phi_B }\frac{\bar A}{A}\ .
\end{align}
No matter what decay we have, $q/p$ is the same. In the standard parameterization of the \CKM, 
\begin{align}
	e^{-i\phi_B} = \frac{V_{tb}^* V_{td}}{V_{tb}V_{td}^*}.
\end{align}
You can see this just by looking at the box diagram phase.
What would differ if we look at $B_s$? We just replace $d$ by $s$ to good approximation. 

Now consider $\bar A/A$. The diagram is:
\begin{center}
	\begin{tikzpicture}[line width=1.5 pt, scale=1.3]
		\draw[fermion] (15:1.5)--(0,0);
		\draw[fermionbar] (180:1)--(0,0);
		\draw[vector] (-40:1)--(0,0);
		\node at (15:1.7) {$c$};
		\node at (180:1.2) {$b$};
		\coordinate (uin) at (-1,-.3);
		\coordinate (umid) at (-80:.5);
	\begin{scope}[shift={(-40:1)}]
		\draw[fermion] (-20:1)--(0,0);
		\draw[fermionbar] (40:1)--(0,0);
		\node at (40:1.2) {$c$};
		\node at (-20:1.2) {$s$};
		\coordinate (uend) at (-40:1.2);
		\node at (-40:1.4) {$d$};
	\end{scope}
	%
	\draw[fermion] (uin) to [out=0, in=140] (umid) to [out=-35, in=160] (uend);
	\node at (-1.2,-.3) {$d$};
	\draw [gray,decorate,decoration={brace,amplitude=5pt},xshift=-4pt]
	   (-1.2,-.5)  -- (-1.2,.2) 
	   node [black,midway,left=4pt,xshift=-2pt] {$B$};
	\draw [gray,decorate,decoration={brace,amplitude=5pt},xshift=-4pt]
	   (2,.6)  -- (2,-.1) 
	   node [black,midway,right=4pt] {$\Psi$};
	\draw [gray,decorate,decoration={brace,amplitude=5pt},xshift=-4pt]
	   (2.2,-.8)  -- (2.2,-1.7) 
	   node [black,midway,right=4pt] {$K_S$};
	\end{tikzpicture}
\end{center}
The hadronic part of the ratio vanishes in the ratio, the remaining piece goes like
\begin{align}
	\frac{V_{cb}V^*_{cs}}{V_{cb}^* V_{cs}}\ .
\end{align}
We have:
\begin{align}
	\lambda_{\psi K_S} \equiv \frac{ V_{tb}^* V_{td} V_{cb} V_{cs}^*}{ V_{tb} V_{td}^* V_{cb}^* V_{cs}} = e^{2i\beta}\ .
\end{align}
We can see this from looking at the unitarity triangle,
\begin{center}
	\begin{tikzpicture}[scale=9]
		\draw[line width=2] (0,0) -- (1,0) -- (77:.404) -- cycle;
		\begin{scope}
	    	\clip (0,0) -- (1,0) -- (77:.404) -- cycle;
	    	\draw[line width=1.5] (0,0) circle (.1);
			\draw[line width=1.5] (1,0) circle (.2);
			\draw[line width=1.5] (77:.404) circle (.1);
			\draw[->,line width=1.5] (.1,0) arc (0:38:.1);
			\draw[->,line width=1.5] (77:.304) arc (-103:-52:.1);
			\draw[-<,line width=1.5] (.8,0) arc (180:165:.2);
	  	\end{scope}
		\node at (-.05,-.05) {$\displaystyle{(0,0)}$};
		\node at (1.05,-.05) {$\displaystyle{(1,0)}$};
		\node at (77:.455) {$\displaystyle{(\rho,\eta)}$};
		\node at (35:.15) {$\displaystyle{\gamma}$};
		\node at (.75,.05) {$\displaystyle{\beta}$};
		\node at (.16,.26) {$\displaystyle{\alpha}$};
		\node at (-.1,.2) {$\displaystyle{ \lambda_{u}= V_{ud}V^*_{ub} }$};
		\node at (.6,.3) {$\displaystyle{ \lambda_{t}= V_{td}V^*_{tb} }$};
		\node at (.5,-.07) {$\displaystyle{ \lambda_{c}= V_{cd}V^*_{cb} }$};
	\end{tikzpicture}	
\end{center}
What we find is
\begin{align}
	\text{Im}(\lambda) = \sin 2\beta.
\end{align}
What about the loop diagrams? The point is that to leading order they
have the same weak phase as the tree diagram, and thus they do not
affect the above result.

We give one more example. What about $B\to D^+ D^-$? Again we can look primarily at the tree diagram. 
\begin{center}
	\begin{tikzpicture}[line width=1.5 pt, scale=1.3]
		\draw[fermion] (40:1.2)--(0,0);
		\draw[fermionbar] (180:1)--(0,0);
		\draw[vector] (-40:1)--(0,0);
		\node at (40:1.4) {$c$};
		\node at (180:1.2) {$b$};
	\begin{scope}[shift={(0,.3)}]
		\draw[fermion] (180:1) to [out=0,in=220] (40:1); 
		\node at (40:1.4) {$d$};
		\node at (180:1.2) {$d$};
	\end{scope}
	\begin{scope}[shift={(-40:1)}]
		\draw[fermion] (-10:1)--(0,0);
		\draw[fermionbar] (10:1)--(0,0);
		\node at (-10:1.2) {$d$};
		\node at (10:1.2) {$c$};
	\end{scope}
	\draw [gray,decorate,decoration={brace,amplitude=5pt},xshift=-4pt]
	   (-1.2,-.2)  -- (-1.2,.5) 
	   node [black,midway,left=4pt,xshift=-2pt] {$B$};
	\draw [gray,decorate,decoration={brace,amplitude=5pt},xshift=-4pt]
	   (1.4,1.3)  -- (1.4,.7) 
	   node [black,midway,right=4pt] {$D^-$};
	\draw [gray,decorate,decoration={brace,amplitude=5pt},xshift=-4pt]
	   (2.2,-0.3)  -- (2.2,-1.1) 
	   node [black,midway,right=4pt] {$D^+$};
	\end{tikzpicture}
\end{center}
What angle do you get? Just plot the diagram and see what appears in
the expression for $\lambda$. The only difference is $s\to d$. Thus
the phase is still $\beta$. Yet, here the loop diagram does have a
different \CKM phase and thus the extraction of the  phase does suffer
from some hadronic uncertainties.

\begin{prob}\textbf{Time dependent CP asymmetry in $B\to \pi^+ \pi^-$}.
Show that to leading order (that is, when considering  only the tree-level
diagram of the decay) the time dependent \CP asymmetry in $B\to \pi^+
\pi^-$  is sensitive to $\alpha$.
\begin{sol}
The decay amplitude is proportional to $V_{ub}^* V_{ud}$ and thus to
the $u$ side of the unitarity triangle, while the mixing is
proportional to the $t$ side. The asymmetry is sensitive to the angle
between these two two, that is, we pick up the angle $\alpha$.
\end{sol}
\end{prob}

%
%
%
%
%

\section*{Acknowledgements}

%
\textsc{y.g.} thanks Rouven Essig and Ian Low for inviting him to give these lectures at \acro{TASI} 2016.
\textsc{p.t.}\ hopes to one day be invited to give \acro{TASI} lectures. We thank the students of the Cornell University Physics 7661 course in 2010 for their feedback on an early version of these notes; especially Panagiotis Athanasopoulos, Hardik Panjwani, Yuhsin Tsai, and Yariv Yanay.
We thank the Aspen Center for Physics (\acro{NSF} grant \#1066293) and the 2017 \acro{SLAC} Summer Institute for its hospitality during a period where part of this work was completed. We acknowledge the Just A Taste tapas restaurant in Ithaca, \acro{NY} for borrowing their name for these lecture notes and for their chipotle aioli (David Curtin's favorite).

 \appendix

\section{Some Useful Facts}

\begin{itemize}
	\item The Fermi constant in terms of the $SU(2)_L$ coupling,
	\begin{align}
		\frac{G_F}{\sqrt{2}}&=\frac{g^2}{8M_W^2.}\label{eq:Gf:definition}
	\end{align}
The factor of 8 on the right-hand side comes from the $W$ coupling constant being $g/\sqrt{2}$ and the factors of 1/2 that come from, for example, only picking up the axial part of a chiral interaction (e.g.\ only the second term of $\frac 12 \gamma^\mu(1-\gamma^5)$).
	\item The \textbf{Wolfenstein parameterization} of the \CKM matrix:
	\begin{align}
		V = 
		\begin{pmatrix}
			1-\lambda^2/2 & \lambda & A\lambda^3(\rho-i\eta)\\
			-\lambda & 1-\lambda^2/2 & A\lambda^2\\
			A\lambda^3(1-\rho-i\eta) & -A\lambda^2 & 1
		\end{pmatrix} + \mathcal O(\lambda^4),
	\end{align}
	where one should note that $\lambda \approx 0.2$ is a good expansion parameter and all other parameters are $\mathcal O(1)$. Note that the top-left $2\times 2$ has the structure of the Cabbibo mixing matrix and that CP violation \textit{only} appears with a $\lambda^3$ in the 3-1 mixing terms.
	\item The unitarity triangle:
	\begin{center}
		\begin{tikzpicture}[scale=9]
			\draw[line width=2] (0,0) -- (1,0) -- (77:.404) -- cycle;
			\begin{scope}
		    	\clip (0,0) -- (1,0) -- (77:.404) -- cycle;
		    	\draw[line width=1.5] (0,0) circle (.1);
				\draw[line width=1.5] (1,0) circle (.2);
				\draw[line width=1.5] (77:.404) circle (.1);
				\draw[->,line width=1.5] (.1,0) arc (0:38:.1);
				\draw[->,line width=1.5] (77:.304) arc (-103:-52:.1);
				\draw[-<,line width=1.5] (.8,0) arc (180:165:.2);
		  	\end{scope}
			\node at (-.05,-.05) {$\displaystyle{(0,0)}$};
			\node at (1.05,-.05) {$\displaystyle{(1,0)}$};
			\node at (77:.455) {$\displaystyle{(\rho,\eta)}$};
			\node at (35:.15) {$\displaystyle{\gamma}$};
			\node at (.75,.05) {$\displaystyle{\beta}$};
			\node at (.16,.26) {$\displaystyle{\alpha}$};
			\node at (-.1,.2) {$\displaystyle{ \left|\frac{V_{ud}V^*_{ub}}{V_{cd}V^*_{cb}}\right| }$};
			\node at (.6,.3) {$\displaystyle{ \left|\frac{V_{td}V^*_{tb}}{V_{cd}V^*_{cb}}\right| }$};
		\end{tikzpicture}	
	\end{center}
	\item 1 GeV = $1.5 \times 10^{24}$ sec$^{-1}$
	\item 1 GeV = $5 \times 10^{13}$ cm$^{-1}$
	\item The mass of the $B$ meson is 5280 MeV, which is precisely the number of feet in a mile. See also the \textit{Spiked Math} comic \cite{SpikedMath:5280}.
	\item The pion decay constant is $f_\pi \sim 131 \text{ MeV}$.	 
	\item The $\tau$ mass is 1777 MeV. From the original version of these lectures:
	\textit{What is the mass of the $\tau$? 1777 MeV. That's a 1 and then 777, like you see in casinos.}
\end{itemize}

\section{Goldstones, currents, and pions}\label{sec:Goldstone:current:pion}

	There is a somewhat antiquated way of looking at Goldstone's theorem based on what is called the \textbf{current algebra}. This presentation comes from Cheng and Li Section 5.3 \cite{ChengLi:Gauge} and Coleman's lecture on soft pions \cite{Coleman:Aspects}.
	Given a conserved current $\partial_\mu J^\mu(x)=0$ we may define a conserved charge $Q(t) = \int d^3x J^0(x)$ that acts as the generator of the symmetry. When the symmetry is not spontaneously broken, this charge is also conserved. In the case of spontaneous symmetry breaking, the charge is no longer very well defined since the integral does not converge---but for our purposes this is a technical detail since we will only care about commutators. Current conservation trivially tells us that
	\begin{align}
		0 &= \int d^3x[\partial_\mu J^\mu(x),\phi(0)]\\
		&= -\partial_0\int d^3x[J^0(x),\phi(0)] + \int d\mathbf{S}\cdot[\mathbf J(x),\phi(0)]\ .
	\end{align}
	We identify the conserved charge associated with the symmetry current, 
	\begin{align}
		Q = \int d^3x \, J^0(x),\label{eq:Goldstone:Q:from:J}
	\end{align}
	this also plays the role of the generator of the symmetry group with respect to the fields in the Lagrangian.
	For a sufficiently large volume the surface integral vanishes and we thus find that for a symmetry-preserving vacuum configuration,
	\begin{align}
		\frac{d}{dt}[Q(t),\phi(0)] &= 0.\label{eq:NLSM:commutatorconservation}
	\end{align} 
	Spontaneous symmetry breaking can thus be \textit{defined} by the condition that this commutator does not vanish on the vacuum,
	\begin{align}
		\langle 0 | [Q(t),\phi(0)] | 0 \rangle &\neq 0.
	\end{align}
	This is intuitively clear since it tells us that the generator of the symmetry enacts a non-trivial transformation of the vacuum configuration of field $\phi(0)$. Note that (\ref{eq:NLSM:commutatorconservation}) \textit{still holds} since this came from $\partial_\mu J^\mu(x)=0$, which is a statement about the Lagrangian \textit{independent} of the vacuum.  
	
	To properly understand this spontaneous breaking we may insert a complete set of orthonormal momentum states to write the left-hand side as (dropping an overall normalization),
	\begin{align}
		\sum_n \delta^{(3)}(\mathbf{p}_n)\left(\langle 0| J_0(0)|n\rangle\langle n | \phi(0)|0 \rangle e^{-iE_n t} - \langle 0 | \phi(0)|n \rangle \langle n| J_0(0)|0 \rangle e^{iE_n t}\right).
	\end{align}
	Spontaneous symmetry breaking is the condition where this expression is non-vanishing. Further, (\ref{eq:NLSM:commutatorconservation}) tells us that this expression must be time-independent so that spontaneous symmetry breaking requires that there exists a state $|n'\rangle$ such that $E_{n'} = 0$ for $\mathbf{p}_{n'}=0$, i.e. the state is massless. Further, we have the properties that $\langle n' |\phi(0)|0\rangle \neq 0$ and $\langle 0 | J_0(0) | n\rangle \neq 0$.

This turns out to be closely related to the \acro{LSZ} reduction formula, which gives us a rigorous prescription for extracting off-shell matrix elements from the path integral by extracting the poles in the external particles. Going back to the definition of the pion form factor,
\begin{align}
\langle 0 | A^\mu|\pi\rangle \equiv -ip^\mu f_\pi,	
\end{align}
we find an expression very similar (\textit{un}surprisingly!) to our manipulations above. Taking a divergence, we find that
\begin{align}
	\langle 0 |\partial_\mu A^\mu|\pi\rangle \equiv  m_\pi^2 f_\pi.\label{eq:div:A:pion:mass}
\end{align}
This is an interesting statement in light of the \acro{LSZ} formula, which one can interpret as a statement about what it means to assign a field to a particle---this, of course, is related the whole point of parameterizing our ignorance about \QCD matrix elements. In particular, any number of local operators can used as field for a particle. If $\phi$ is a `good' field, then so are objects made up of powers and derivatives of $\phi$ that maintain its canonical normalization. This is fixed from the \acro{LSZ} formula and Lorentz invariance,
\begin{align}
	\langle p | \phi(x) | 0 \rangle = (2\pi)^{-3/2}(2E)^{-1/2}e^{ip\cdot x}\label{eq:LSZ:scalar:normalized}
\end{align}
for a scalar field.
\begin{eg} \textbf{A sketch of the proof}. The \acro{LSZ} reduction formula tells us that to obtain off-shell matrix elements we should take the $n$-point Green's function
	\begin{align}
		G^{(n)}(k_1,\ldots, k_n) =  \int \prod_k^i d^4x_i  \, e^{i\sum_j k_j\cdot x_j} \langle 0 | \hat T \left[\hat\phi(x_1)\cdots \hat\phi(x_n)\right]| 0\rangle
	\end{align}
	where $\hat \phi(x_n)$ is the interaction picture field operator, and multiply it by inverse propagators to identity the on-shell pole in the external particles, e.g.,
	\begin{align}
		\mathcal M(\text{init}\to\text{final}) \propto \int \prod_i^n d^4x_i \prod_{j}^n \mathcal{O}_je^{ik_j\cdot x_j} \, \langle 0 | \hat T \left[\hat\phi(x_1)\cdots \hat\phi(x_n)\right]| 0\rangle,
	\end{align}
	where $\mathcal O_j$ is the appropriate operator in the free Lagrangian which cancels the external state propagators, in this case it is just the Klein-Gordon operator. We've neglected overall prefactors of the form $(2\pi)^{-3/2}(2E)^{-1/2}$ as well as renormalization factors. Here we have taken all external legs off shell by Fourier transforming in each of them; let us consider the simplified case where we do this for only one field, $\tilde \phi$ ,which may need not be a `fundamental' field but is canonically normalized according to (\ref{eq:LSZ:scalar:normalized}). The \acro{LSZ} prescription tells us that to obtain an on-shell matrix element with $(n-1)$ external $\phi$ particles and one external off-shell) $\tilde\phi$ field, one calculates the necessary $n$-point functions, multiplies by $(p^2-m^2)$ for the $\tilde\phi$ external line, and then takes the $\tilde\phi$ on shell. In other words, one identifies the pole in the $\tilde\phi$ squared momentum. The only diagrams with such poles are those with one particle intermediate $\tilde\phi$ states on this leg, and these are normalized by (\ref{eq:LSZ:scalar:normalized}) to contribute a factor of unity. Thus we indeed find that we generate the \textit{same} matrix element (i.e.\ combinations of couplings and internal propagators) that we would have generated for \textit{any} canonically  $\tilde\phi$ field. This is further explained in Section 2.1 of Coleman's lectures \cite{Coleman:Aspects}.
\end{eg}

From all of this we see that in light of (\ref{eq:LSZ:scalar:normalized}), what (\ref{eq:div:A:pion:mass}) is \textit{really} trying to tell us is that
\begin{align}
	\phi_\pi=\frac{\partial_\mu A^\mu}{f_\pi m_\pi^2}\label{eq:LSZ:Goldstone:pion}
\end{align}
 is a perfectly good field for the pions. Purists can attach SU(2) indices as appropriate, e.g.\ $f_\pi \to \delta^{ab}f_\pi$, but---like minus signs---if you complain about it, then it's \textit{your} homework to do it thoroughly\footnote{This policy is attributed to Csaba Cs\'aki and his graduate electrodynamics course.}. This form of the pion field should not be surprising. After all, we know that in the limit where it is a Goldstone boson, it should have a shift symmetry. This is made explicit here by the derivative form of the field which guarantees the `derivative coupling' that one finds in chiral perturbation theory. The relation of the divergence of the axial current to the pion is called the \textbf{partially-conserved axial current} (\acro{PCAC}) hypothesis.

In fact, more generally, the \acro{LSZ} statement about writing down an appropriate field to represent a given particle is at the heart of chiral perturbation theory. This was first put on solid footing by Callan, Coleman, Wess, and Zumino in what they called `phenomenological Lagrangians' \cite{Coleman:1969sm, Callan:1969sn, Coleman:1968wj}. See Chapter 4 of Donoghue \cite{Donoghue:DynamicsSM} for a more modern treatment with explicit demonstrations of the representation independence of such models.

\subsection{Goldberger--Treiman Relation}

Armed with the insight of (\ref{eq:LSZ:Goldstone:pion}), we can start to get a feel for the current algebra techniques once practiced by ancient physicists. One quick result is called the \textbf{Goldberger--Treiman} relation. 
\begin{prob}\textbf{The Goldberger--Treiman relation.}
	Derive the Goldberger--Treiman relation,
	\begin{align}
		f_\pi g_{\pi NN} = m_N g_A(0),
	\end{align}
	where $g_{\pi NN}$ is the pion-nucleon vertex function evaluated at $q^2=0$,
	\begin{align}
		\langle p(k_p) | \partial_\mu A^\mu |n(k_n)\rangle  
		&= \frac{2f_\pi m_\pi^2}{q^2-m_\pi^2} g_{\pi NN}(q^2) i\bar u_p(k_p) \gamma_5 u_n(k_n),
	\end{align}
	and $g_A(0)$ is the the nucleon axial vector coupling in
	\begin{align}
		\langle p(k_p) | A_\mu |n(k_n)\rangle = \bar u_p(k_p)\left[g_A(q^2)\gamma_\mu\gamma_5 + q_\mu h_A(q^2)\gamma_5 \right] u_n(k_n).
	\end{align}
	\begin{sol}
		Let us write the hadronic matrix element for neutron $\beta$ decay $n\to p$ through the axial current,
		\begin{align}
			\langle p(k_p) | A_\mu |n(k_n)\rangle = \bar u_p(k_p)\left[g_A(q^2)\gamma_\mu\gamma_5 + q_\mu h_A(q^2)\gamma_5 \right] u_n(k_n).
		\end{align}
		Here $g_A$ and $h_A$ are functions which depend on the momentum transfer $q\equiv k_n - k_p$. Note that we are \textit{not} `parameterizing our ignorance' about \QCD (we still have cumbersome spinor structure on the right-hand side); instead, we are saying that the neutron decays into the proton \textit{literally} through the tree-level coupling to the axial current. If you want to be precise about isospin indices, one ought to write $A_\mu \to A^+_\mu = A_\mu^1 + i A_\mu^2$, but we'll try to keep the equations as simple as possible. We can now take the divergence of this equation to obtain 
		\begin{align}
			\langle p(k_p) | \partial_\mu A^\mu |n(k_n)\rangle = i \left[g_A(q^2) m_N^2 + q^2 h_A(q^2)\right] \bar u_p(k_p) \gamma_5 u_n(k_n),\label{eq:Goldberger:Treiman:axial}
		\end{align}
		where $m_N$ is the nucleon mass. The eft-hand side of this equation, however, is simply what we called the pion field in (\ref{eq:LSZ:Goldstone:pion}), up to overall factors:
		\begin{align}
			\langle p(k_p) | \partial_\mu A^\mu |n(k_n)\rangle &= f_\pi m_\pi^2 \langle p(k_p) | \phi_\pi |n(k_n)\rangle\\
			&= \frac{2f_\pi m_\pi^2}{q^2-m_\pi^2} g_{\pi NN}(q^2) i\bar u_p(k_p) \gamma_5 u_n(k_n), \label{eq:Goldberger:Treiman:pion}
		\end{align}
		where in the second line we have defined the pion-nucleon vertex function $g_{\pi NN}(q^2)$ and pulled out the pion pole. This gives the $\pi NN$ coupling for $q^2 = m_\pi^2$. Comparing (\ref{eq:Goldberger:Treiman:axial}) and (\ref{eq:Goldberger:Treiman:axial}) we obtain,
		\begin{align}
			f_\pi g_{\pi NN}(0) = m_N g_A(0).
		\end{align}
		This is nearly the Goldberger-Treiman relation, except that the pion-nucleon vertex function is off-mass shell. We must make the assumption that $g_{\pi NN}(q^2)$ is slowly varying such that $g_{\pi NN}(0) \approx g_{\pi NN}(m_\pi^2)$. This gives us a relation which holds within 10\%. Note that this relation is actually rather trivial from the point of view of chiral perturbation theory, see e.g.\ Donoghue Section 12.3 \cite{Donoghue:DynamicsSM}.
	\end{sol}
\end{prob}

We can make more powerful use of the connection between the pion and the divergence of the axial current by invoking the algebra of the symmetry group. Recall that from a Lagrangian one can write down the current associated with a transformation
\begin{align}
	\phi_i \to \phi'_i = \phi_i + i\epsilon^a t^a_{ij}\phi_j
\end{align}
in the usual way, 
\begin{align}
	J^a_\mu = -i \frac{\delta\mathcal L}{\delta(\partial^\mu \phi_i)}t^a_{ij}\phi_j.
\end{align}
From this we can define the generator/charge of the transformation using (\ref{eq:Goldstone:Q:from:J}). For a symmetry of the Lagrangian, these generators satisfy the commutation relations of the symmetry group itself,
\begin{align}
	[Q^a(t),Q^b(t)]=if^{abc}Q^c(t),
\end{align}
where we note that the charges are time-dependent and the algebra is satisfied at equal times. Charges are nice, but even better quantities are currents, since currents actually contain local (i.e.\ not integrated) products of the actual fields we care about. From the rigorous derivation of the charge commutation relations above one can also simply write out the commutation relation for charges with the time component of currents,
\begin{align}
	[Q^a(t),J^b_0(t,\mathbf{x})] = if^{abc}J^c_0(t,\mathbf{x}).
\end{align}
By Lorentz invariance we can promote this tho
\begin{align}
	[Q^a(t),J^b_\mu(t,\mathbf{x})] = if^{abc}J^c_\mu(t,\mathbf{x}).
\end{align}
In fact, it turns out we may go even further and write out current-current commutations relations, yielding the \textbf{current algebra},
\begin{align}
	[J^a_0(t,\mathbf{x}),J^b_0(t,\mathbf{y})] = if^{abc}J^c_0(t,\mathbf{x})\delta^{(3)}(\mathbf{x}-\mathbf{y}).
\end{align}
We neglect the spatial $J^a_i$ commutators since they turn out to include extra terms which usually end up canceling unphysical local singularities ($x\to y$). For the SU$(N)_L\times$SU$(N)_R$ ($N=2,3$) flavor symmetries of interest to us, the ancient (\acro{B.C.}, ``before chiral perturbation theory'') current algebra between the vector and axial currents were used to produce relations between form factors and pion scattering amplitudes. Commutation relations naturally arose when pulling out partial derivatives from time ordered products, for example
\begin{align}
	\partial^\mu \langle f | T\left( A_\mu(x)\mathcal O(0)\right)| i\rangle 
	=
	\langle f | T\left(\partial^\mu A_\mu(x)\mathcal O(0)\right)| i\rangle  
	+
	\delta(x_0) \langle f | T\left[ A_0(x),\mathcal O(0)\right]| i\rangle,
\end{align}
where typically $\mathcal O$ is another current. Thus the current algebra provided a way to simplify the matrix elements in hadronic physics before more sophisticated techniques were available.
You may learn more about these ancient techniques in Coleman's lecture on soft pions \cite{Coleman:1968wj}, Cheng \& Li \cite{ChengLi:Gauge}, the review by Scadron \cite{Scadron:1981dn}, or the book by Treiman, Jackiw, and Gross \cite{Treiman:Current:Algebra}.

\subsection{Ademollo--Gatto from current algebra}
 
	
	Recall the Theorem~\ref{thm:Ademollo:Gato} . A cute and slightly more rigorous way of proving this theorem for the vector form factors involves the current algebra. A very rough sketch of the procedure is to note the commutator of the SU(3) generators
	\begin{align}
		\left[Q^{4+i5},Q^{4-i5}\right] = Q^3+\sqrt{3}Q^8 = Q^{\text{EM}}+Y,
	\end{align}
	where $Q^{4\pm i5}= (Q^4\pm iQ^5)/2$. One can then consider taking the expectation value of both sides with respect to an octet state, from which we obtain
	\begin{align}
		(Q^{\text{EM}}+Y)_H = -|\langle H'| Q^{4+i5}|H\rangle|^2 + \sum_m | \langle m | Q^{4-i5}|H\rangle |^2 - \sum_n | \langle n | Q^{4+i5}|H\rangle|^2,
	\end{align}
	where $H'$ is another state in the same octet. For example, for $H=\Sigma^-$, $H'=\Lambda$ or $\Sigma^0$. The states $\langle m|$ and $\langle n|$ are not in the same octet. In the appropriate limit, it turns out that $\langle H'| Q^{4+i5}|B\rangle \to f_1(0)$. Taking such a limit and rearranging terms we obtain
	\begin{align}
		f_1(0)^2 = -(Q^{\text{EM}}+Y)_B + \mathcal O(\delta^2),
	\end{align}
	where $\delta=(M_{H}-M_{H'})/M_H$ is a parameter for $SU(3)$ breaking. This is further explained in the review article by Cabbibo et al.\ \cite{Cabibbo:2003cu}. 

\subsection{$K_{\ell 4}$}
\label{sec:Kl4}

In Section~\ref{sec:Kl3} we looked at the $K_{\ell 3}$ semi-leptonic decay as a way to measure $|V_{us}|$. Another way to probe this is to use the four-body decay
	$K^+ \to \pi^+ \pi^- e\nu$.
This is creatively named $K_{\ell 4}$.
This story is more complicated than $K_{\ell 3}$ because this is a one-to-two hadron decay with two additional leptons. On the plus side, we may find some consolation in the observation that the two charged pions are at least easy to identify in a detector.

We parameterize the matrix elements in terms of three \textit{independent} momenta. Chose these to be the momenta $p_\pm$ of the charged pions and $k$ of the kaon so that the vector and axial matrix elements are
\begin{align}
	\langle \pi^+ \pi^- | A^\mu | K \rangle &= C \left[ f_1(p_+ +  p_-)^\mu + f_2 (p_+-p_-)^\mu + f_3(k-p_+ - p_-)^\mu \right]\\
	\langle \pi^+ \pi^- | V^\mu | K \rangle &= C' g \epsilon^{\mu\alpha\beta\gamma} k_\alpha p_{+\beta} p_{-\gamma}.
\end{align}
The axial matrix element is a sum of three terms that are linear in linearly independent combinations of the basis momenta. The vector matrix element, on the other hand, is composed of a contraction of all three momenta with the totally antisymmetric $\epsilon^{\mu\alpha\beta\gamma}$ tensor as required by parity. This is clear since the vector matrix element has the opposite parity of the axial matrix element and the only way to construct quantity with a single Lorentz index out of four-vectors necessarily requires the intrinsic parity of the $\epsilon$ tensor. 
\begin{prob}\textbf{We don't need $f_3$}. Argue that $f_3$ is negligible.
\begin{sol} By the same argument that we've seen several times now, we know that the $f_3$ form factor is proportional to the electron mass upon contraction with the leptonic part of the full matrix element.
\end{sol}	
\end{prob}

The prefactors $C$ and $C'$ are overall normalizations 
\begin{align}
	C&= -\frac{i}{f_\pi}
	&
	C' &= \frac{1}{4\pi^2f_\pi^3}
\end{align}
which we pull out to normalize $f_1=f_2=g=1$. The factor of $4\pi^2$ is very familiar---in fact, it's \textit{almost} `dimensionful' since it carries dimension of loop number. Indeed, this really is a loop effect since the vector matrix element is related by SU(3) to the matrix element for $\gamma \to 3\pi$ by an SU(3) flavor rotation. Of course, $\gamma\to 3\pi$ is nonsense as a physical process, but it has a nonzero \emph{matrix element} coming form the axial anomaly. 
The normalization of $f_1$ and $f_2$ is more subtle and comes from the current algebra techniques mentioned in Appendix~\ref{sec:Goldstone:current:pion}; they are \textit{not} simply Clebsch-Gordan coefficients. These come from soft pion techniques where one uses \acro{LSZ} reduction and current commutation relations to relate the amplitude for a soft pion emission to an amplitude without the pion. In the case of $K\to \pi\pi e\nu$, there are \textit{two} pions which one may take to be soft. One limit gives $f_1=f_2$ while the other gives $f_1 + f_2 = 2C$. Note that these relations hold in the limit of \textit{soft} pions, i.e.\ the limit $p_\pm \to 0$. In principle $f_1$ and $f_2$ are functions of \textit{three} dynamical scalars\footnote{To be precise, in the soft pion manipulations one takes the pion off shell so that $p_\pm^2$ becomes a `dynamical' variable as well.}: $k\cdot p_+$, $k\cdot p_-$, and $p_+\cdot p_i$. However, in the soft pion limit, these are all small. Experiments confirm that there is no strong dependence on the external momenta.
\begin{framed}
	\noindent\textbf{The mystery of $f_3$.} One of the mysteries of the ancient world  was the failure of na\"ive current algebra techniques to sensibly predict the $f_3$ form factor (even though we already know $f_3$ is $m_e/m_K$ suppressed). While the two constraints on $f_1$ and $f_2$ were consistent, the $f_3$ constraints coming from taking each pion to its soft limit gave what appeared to be contradictory predictions: 
	\begin{align}
		f_3 = 0 = f_+ + f_-,
	\end{align}
	where the $f_\pm$ come from the $K\to \pi e\nu$ decays. In other words, $f_3$ seems to vary strongly with the momenta. In 1966 Weinberg solved the puzzle by identifying the $p_\pm$ dependence of $f_3$ as coming from a kaon pole coming from the following diagram, \cite{Weinberg:1966}
	\begin{center}
	\begin{tikzpicture}[line width=1.5 pt]
		\node at (-2.5,0) {$K^+$};
		\node at (1.8,1.8) {$\pi^+$};
		\node at (2.4,1.8) {$\pi^-$};
		\node at (4.7,1.2) {$e$};
		\node at (4.7,-1.2) {$\nu$};
		\node at (3,-.6) {$A_\mu$};
		\node at (1.5,-.3) {$K^+$};
		\draw (-2,0) -- (0,0) -- (3,0);
		\draw[fermion] (4.5,1) -- (3,0);
		\draw[fermion] (3,0) -- (4.5,-1);
		\draw[scalarnoarrow] (-.2,0) -- (1.8,1.5);
		\draw[scalarnoarrow] (.2,0) -- (2.2,1.5);
		\draw[fill=black] (0,0) circle (.3cm);
		\draw[fill=white] (0,0) circle (.29cm);
		\begin{scope}
	    	\clip (0,0) circle (.3cm);
	    	\foreach \x in {-1.0,-.8,...,.4}
				\draw[line width=1 pt] (\x,-.3) -- (\x+.6,.3);
	  	\end{scope}
		\begin{scope}[shift={(3,0)}]
			\draw[fill=black] (0,0) circle (.3cm);
			\draw[fill=white] (0,0) circle (.29cm);
			\begin{scope}
		    	\clip (0,0) circle (.3cm);
		    	\foreach \x in {-1.0,-.8,...,.4}
					\draw[line width=1 pt] (\x,-.3) -- (\x+.6,.3);
		  	\end{scope}
		\end{scope}
	 \end{tikzpicture}
	\end{center}
	This diagram \textit{only} contributes to $f_3$ so that one ultimately finds
	\begin{align}
		f_3 = (f_+ + f_-) \frac{k\cdot p_-}{k\cdot(p_-+p_+)},
	\end{align}
	which indeed gives the correct interpolating behavior between the two single-soft pion limits.
\end{framed}

%
%

%

Having set up the problem thusly, one can go on to do the same analysis that we presented for three body semileptonic kaon decay ($K_{\ell 3}$). The procedure is the same, though there are more coefficents and the entire ordeal is generally more complicated and not as fun. The punchline is that one can indeed make an independent measurement to confirm the value of the previous method.

\section{The vector mesons}
\label{sec:vector:mesons}

In Section~\ref{sec:pseudoscalars} we introduce the light pseudoscalar mesons and explored their properties.
The vector mesons $J^P=1^-$, however, are a bit of a different story\footnote{This section borrows from Chapter 4.4 of \cite{ChengLi:Gauge}.}.  We can draw the $\mathbf{8}\oplus\mathbf{1}$ weight diagram to compare with the pseudoscalars:
\begin{center}
	\begin{tikzpicture}
		\draw[line width=1.5] (0:2) -- (60:2) -- (120:2) -- (180:2) -- (240:2) -- (300:2) -- (0:2);
		\draw[fill=black] (0:2) circle (.09cm);
		\draw[fill=black] (60:2) circle (.09cm);
		\draw[fill=black] (120:2) circle (.09cm);
		\draw[fill=black] (180:2) circle (.09cm);
		\draw[fill=black] (240:2) circle (.09cm);
		\draw[fill=black] (300:2) circle (.09cm);
		\draw[fill=black] (60:.3) circle (.09cm);
		\draw[fill=black] (120:.3) circle (.09cm);
		\draw[fill=black] (270:.2) circle (.09cm);
		\node at (0:2.6) {$\rho^+$};
		\node at (60:2.6) {$K^{*+}$};
		\node at (120:2.6) {$K^{*0}$};
		\node at (180:2.6) {$\rho^{-}$};
		\node at (240:2.6) {$K^{*-}$};
		\node at (300:2.6) {$\bar K^{*0}$};
		\node at (60:.75) {$\omega$};
		\node at (120:.7) {$\phi$};
		\node at (270:.6) {$\rho^0$};
	\end{tikzpicture}
\end{center}
 The $\rho$s ($m_\rho = 770$ MeV) form an SU$(2)$ iso-triplet, the analog of the pions. The SU$(2)$ iso-singlet is the $\omega$ ($m_\omega$ = 783 MeV), which is the analog of the $\eta$. Including the strange quark and enlarging to SU$(3)$ isospin symmetry the $K^*$s ($m_{K^*} = $ 892 MeV) are the analogs of the spin-0 kaons and the $\phi$ ($m_\phi =$ 1020 MeV) is the $I_3$ analog of the $\eta'$. 
Do you see the problem? Here's a hint.

\begin{prob}\textbf{The pseudoscalar Gell-Mann--Okubo mass formula.} In the early 1960s, Gell-Mann and Okubo independently were working on predictions of the quark model on the spectrum of mesons. Write down the masses of the different particles ($\pi$, $K$, $\eta$) in the pseudoscalar octet$\oplus$singlet by summing some universal (flavor-independent) \QCD-induced mass with the valence quark masses. Assume $m_u = m_d$ but leave $m_s$ independent. Assume that the quark masses and the universal mass are unknown. \textbf{Hint}: Write the \textit{squared} meson masses as the sum of the universal part and terms which are \textit{linear} in the quark masses times some overall dimensionful constant. Show that
	\begin{align}
		4m_K^2 = m_\pi^2 + 3m_\eta^2.\label{eq:Gell:Mann:Okubo:pseudoscalar}
	\end{align}
\textbf{Remark}: Writing out the \textit{squared} mesons masses \textit{not} the natural thing to do; in general one should write the \textit{linear} meson masses as a universal part plus a linear combination quark masses. We discuss below why the pseudoscalar octet is special.
	\begin{sol}
		Writing the universal mass contribution as $m_\Lambda$ and using the quark content predicted by SU$(3)$, e.g.\ by using (\ref{eq:eta:content}), we find
		\begin{align}
			m_\pi^2 &= m_\Lambda^2 + 2\mu m_u\\
			m_K^2 &= m_\Lambda^2 + \mu\left(m_u + m_s\right)\\
			m_\eta^2 &= m_\Lambda^2 + \frac{2\mu}{3}\left(m_u+2m_s\right).
		\end{align}
		We can `solve' these equations by eliminating the unknown quantities $m_\Lambda$, $m_u$, and $m_s$ to obtain the \textbf{Gell-Mann--Okubo mass formula} for pseudoscalar mesons,
		\begin{align}
			4m_K^2 = m_\pi^2 + 3m_\eta^2.
		\end{align}
		Note that the meson masses in this expression are summed in quadrature. For general SU$(3)$ octets (vector mesons) the meson masses would be summed linearly. 
	\end{sol}	
\end{prob}

Na\"ively we expect the vector meson states to have the analogous quark content as the pseudoscalars. The $I_3$ states ought to have the same mixing of quark flavor states so that 
\begin{align}
	\omega &\stackrel{?}{=} \frac{1}{\sqrt{6}}\left(u\bar u + d\bar d -  2s\bar s\right)
&
	\phi &\stackrel{?}{=} \frac{1}{\sqrt{3}}\left(u\bar u + d\bar d + s\bar s\right)\ .
	\label{eq:phi:omega:content}
\end{align}
In particular, we expect the Gell-Mann--Okubo formula to hold since the additional spin-spin term in the mass is flavor-universal:
\begin{align}
	4m_{K^*}^2 \stackrel{?}{=} 3m_\omega^2 + m_\rho^2.
\end{align}
Plugging in the values above we get a prediction that $m_\omega = 926.5$ MeV, which is very different from the experimentally observed $m_\omega^{\text{exp.}} = 783$ MeV. What went wrong?

\begin{quote}
	\textit{Ah, this seems like it's on the order of $m_s$. Shouldn't this just be a correction to our SU$(3)$ isospin approximation that we already expected?}
\end{quote}
\noindent No! The Gell-Mann--Okubo formula \textit{already} treats the valence quark masses as being different and so should \textit{already} account for the SU$(3)$ isospin breaking from $m_s \neq m_{u,d}$. We're still missing something. 

\begin{quote}
	\textit{Well, we remarked that the quadratic sum of meson masses should only hold for the pseudoscalar mesons. [Why?] The correct form of the Gell-Mann--Okubo formula for the vector mesons, then, should be \textit{linear} in the meson masses.}
\end{quote}
\noindent Good! This is absolutely true. We'll discuss this in the next section. But meanwhile, have a revised Gell-Mann--Okubo formula for vector (and any higher spin) mesons:
\begin{align}
	4m_{K^*} \stackrel{?}{=} 3m_\omega + m_\rho.\label{eq:linear:Gell:Mann:Okubo:for:vectors}
\end{align}
Unfortunately this predicts $m_\omega = 1446$, which is an even \textit{worse} fit! We're really in trouble now!

\begin{prob}
	If you haven't figured you the mystery yet, \textit{stop}. Just stop right there. Don't read beyond this problem. Go to the \PDG and look up the quark content of the $\omega$ and $\rho$ mesons.
	\begin{sol} This is actually a bit of a trick question because it's hard to find the quark content from the summary table. Textbooks list the quark content as being approximately
		\begin{align}
			\omega &= \frac{1}{\sqrt{2}}\left(u\bar u + d\bar d\right)\\
			\phi &= s\bar s.
		\end{align}
	\end{sol}
\end{prob}

It turns out that our assumption (\ref{eq:phi:omega:content}) about the quark content of the $I_3$ states is wrong. Given the theme of these lectures, you should already know why: these are two particles with the same quantum numbers, so that---in general---\textit{\textbf{they mix}}! We can thus make the ansatz that the $\omega$ and $\phi$ are not actually the $I_3$ components of the iso-octet and iso-singlet as we had initially assumed (and that was true for the $\eta$s). Instead, these states are assumed to be linear combinations of the \textit{actual} $I_3$ components of the iso-octet and iso-singlet, which we shall call $\psi_8$ and $\psi_1$ with masses $m_8$ and $m_1$. The quark content is given by
\begin{align}
	\psi_8 &= \frac{1}{\sqrt{6}}\left(u\bar u + d\bar d- 2s \bar s\right)
&
	\psi_1 &= \frac{1}{\sqrt{3}}\left(u\bar u + d\bar d + s\bar s\right)
	\label{eq:psi18}.
\end{align}
In other words, the mass matrix for $\psi_8$ and $\psi_1$ contains off-diagonal elements $m_{81}=m_{18}$,
\begin{align}
	M = \begin{pmatrix}
		m_8 & m_{81}\\
		m_{18} & m_{1}
	\end{pmatrix}.\label{eq:octet:singlet:mass:matrix}
\end{align} 
It may not seem like this helps much since it looks like we have three free parameters ($m_8$, $m_1$, and the mixing angle $\theta$) to fit two experimentally determined values $m_\omega$ and $m_\phi$. However, we should remember that the Gell-Mann--Okubo formula \textit{must work} for the $\psi_8$ and $\psi_1$ states. Thus (\ref{eq:linear:Gell:Mann:Okubo:for:vectors}) should be modified to
	\begin{align}
		4m_{K^*} = 3m_8 + m_\rho.\label{eq:linear:Gell:Mann:Okubo:general}
	\end{align}
where $m_8=1446$ MeV is the mass of the $I_3=0$ member of the octet. One might also wonder about the off-diagonal elements of the $\psi_{8,1}$ mass matrix; these masses are also fixed by the same valence quark analysis---see the following problem. (In what follows we will always be careful to take differences of meson masses so that the universal mass contributions from angular momentum and the \QCD potential can be ignored.)

%

\begin{prob}\label{prob:omega:rho:mixing:1}\textbf{The Schwinger meson mass relation}. Shortly after Gell-Mann and Okubo, Schwinger published his own mass sum rule for mesons for $SU(3)$ nonets (octet+singlet) that is robust against mixing \cite{Schwinger:1964zza}.  In this problem we will prove Schwinger's relation and see that it works for the vector mesons. First determine the elements of the mass matrix (\ref{eq:octet:singlet:mass:matrix}) in terms of the measurable masses of the other octet mesons ($m_{K}^*$, $m_\rho$). Note that these mesons don't mix (why?) and so their masses are good parameters. Next use the invariance of the trace and determinant to prove the following relations
	\begin{align}
		m_\omega +  m_\phi &= 2 m_{K^*}\\
		m_\omega\, m_\phi &= \frac 19\left[ (4m_{K^*}-m_\rho)(2m_{K^*}+m_\rho) - 8(m_\rho-m_{K^*})^2 \right]
	\end{align}
	Finally, combine these equations to write down the \textbf{Schwinger meson mass relation},
		\begin{align}
			-(m_8-m_\omega)(m_8-m_\phi) = \frac 89(m_\rho-m_{K^*})^2,\label{eq:Schwinger:meson:mass}
		\end{align}
		where $m_8$ is expressed in terms of measurable masses by the Gell-Mann--Okubo relation (\ref{eq:linear:Gell:Mann:Okubo:general}). Check that the vectors satisfy this relation. (Please do this question \textit{especially} if you are confused about what's going on!)
		
	\begin{sol}
		Representation theory for SU(3) tells us that the $I_3=0$ octet and singlet states have the quark content
		\begin{align}
			\psi_8 &= \frac{1}{\sqrt{6}}\left(u\bar u + d\bar d -2 s \bar s\right)\\
			\psi_1 &= \frac{1}{\sqrt{3}}\left(u\bar u + d\bar d + s \bar s\right).
		\end{align}
		As we mentioned before we will be working with differences of masses, so we can ignore all universal contributions to meson masses coming from angular momentum and the \QCD potential. Further, we will assume that SU(2) isospin is a good symmetry such that $m_d=m_u$. Then the mass of the $\psi_8$ comes from averaging over the probabilities that $\psi_8$ will contain a given quark pair. In other words, one should look at
		\begin{align}
			\langle\psi_8 | \psi_8\rangle = \frac 16 \left(\langle u\bar u|u\bar u\rangle + \langle d\bar d|d \bar d \rangle + 4\langle s\bar s|s \bar s\rangle\right).
		\end{align}
		to determine that 
		\begin{align}
			m_8 = \frac{1}{6}\left(2\times 2m_u + 2\times 4m_s\right) = \frac{2}{3}(m_u+2m_s),
		\end{align}
		where we remember an overall factor of two because the state $u\bar u$ has valence quark mass contribution $\Delta m = 2 m_u$. Note that we look at the square of the state $|psi\rangle$ because we want the \textit{probability} that the meson contains a given quark content; this has \textit{nothing} to do with the idea that we should be summing the masses linearly or quadratically. It is ``more obvious than obvious\footnote{This is a delightful phrase borrowed from Tony Zee.}'' that we should sum the masses linearly. (The pseudoscalars are special because they are pseudo-Goldstone bosons that need to be treated in chiral perturbation theory where the \textit{squared} meson masses are linear in the symmetry-breaking parameters, the quark masses.) Using the same analysis we may write out the valence quark mass contribution for the vector kaon and $\rho$,
		\begin{align}
			m_{K^*} &= m_u+m_s\\
			m_\rho &= 2m_u.
		\end{align}
		This gives us a basis to re-express quark masses in terms of the non-mixing octet meson masses. (These mesons don't mix because the quark masses break $SU(3)\to U(1)^3$ so that strangeness is conserved.) Thus we readily obtain
		\begin{align}
			m_8 &= \frac 13(4m_{K^*}-m_\rho)\\
			m_1 &= \frac 13(2m_{K^*}+m_\rho).
		\end{align}
		You might have worried that we do not know what the mixing term $m_{18}=m_{81}$ should look like, but the point is that the mixing term comes precisely from the fact that when $SU(3)$ is broken, $\langle \psi_8 | \psi_1\rangle \neq 0$. In particular,
		\begin{align}
			\langle \psi_8 |\psi_1\rangle = \frac{1}{3\sqrt{2}}\left(
			\langle u\bar u|u\bar u\rangle + \langle d\bar d|d \bar d \rangle -2\langle s\bar s|s \bar s\rangle
			\right),
		\end{align}
		note the minus sign! We can thus read off
		\begin{align}
			m_{81} = m_{18} = \frac{2}{3\sqrt{2}}\left(2m_u-2m_s\right) = \frac{2\sqrt{2}}{3}\left(m_\rho - m_{K^*}\right).
		\end{align}
		
		Good. Now we know that the flavor basis matrix is made up of these elements, but upon rotation to the mass basis the matrix is diagonalized to
		\begin{align}
			R \begin{pmatrix}
				m_8 & m_{81}\\
				m_{18} & m_1
			\end{pmatrix} R^T = 
			\begin{pmatrix}
				m_\phi & \\
				& m_\omega
			\end{pmatrix}.
		\end{align}
		 We can relate the physical masses $m_{\phi,\omega}$ to the expressions above by considering invariants under rotations, namely the trace and determinant. The trace relation is easy and gives us
		\begin{align}
			m_\omega + m_\phi = 2m_{K^*}.\label{eq:sol:schwinger:trace}
		\end{align}
		The determinant relation gives us
		\begin{align}
				m_\omega\, m_\phi &= \frac 19\left[ (4m_{K^*}-m_\rho)(2m_{K^*}+m_\rho) - 8(m_\rho-m_{K^*})^2 \right],
		\end{align}
		where we recognize $m_8 = 3(4m_{K^*}-m_\rho)$ in the first term on the right-hand side.
		We would like to combine these equations into the Schwinger relation. This is most easily obtained by starting with the determinant relation and, in the first term on the left-hand side, writing
		\begin{align}
			(2m_{K^*}+m_\rho) = 6m_{K^*}+m_\rho - 4m_{K^*} = 3(m_\omega + m_\phi) - 3m_8.
		\end{align}
		This leads us to 
		\begin{align}
			3m_8\left[3(m_\omega+m_\phi)-3m_8\right] - 8(m_\rho-m_{K^*}) = 9m_\omega m_\phi.
		\end{align}
		Rearranging terms and then factorizing we finally obtain
		\begin{align}
			-(m_8-m_\omega)(m_8-m_\phi) = \frac 89(m_a-m_{K^*})^2.
		\end{align}
		This is satisfied up to an error of $\mathcal{O}(160 \text{ MeV}^2)$, which is quite good considering that it is \textit{quadratic} in mass.
	\end{sol}
\end{prob}

The observation that the vector meson masses obey the Schwinger relation tells us that the model of SU(3) broken by $m_s$ is consistent and \textit{does} explain the $\omega$ and $\phi$ masses once one accounts for mixing. The next thing we need to check is to find out how much the $\psi_8$ and $\psi_1$ states actually mix. If you did the previous problem, then you already have the explicit form of the mass matrix elements $m_{8}, m_{1}, m_{18}=m_{81}$ in terms of meson masses that one can look up in the \PDG. Thus it shouldn't be any trouble for you to do the following problems. We're back to diagonalizing $2\times 2$ matrices!

\begin{prob}\textbf{The $\omega$--$\phi$ mixing angle, part I}. Parameterize the mixing between the SU(3) flavor eigenstates $\psi_8$ and $\psi_1$ into the mass eigenstates $\omega$ and $\phi$ as follows:
	\begin{align}
		\begin{pmatrix}
			\phi \\
			\omega
		\end{pmatrix}
		= 
		\begin{pmatrix}
			\cos\theta & -\sin\theta\\
			\sin\theta &\ \cos\theta
		\end{pmatrix}
		\begin{pmatrix}
			\psi_8\\
			\psi_1
		\end{pmatrix},
	\end{align}
	where the flavor eigenstates have quark content (\ref{eq:psi18}). In this problem we derive several formulae for the mixing angle. For $\tan^2\theta$ show that
	\begin{align}
		\tan^2\theta &= -\frac{m_8-m_\phi}{m_8-m_\omega} 
		= -\frac{m_1-m_\omega}{m_1-m_\phi} 
		= \frac{m_8-m_\phi}{m_1-m_\phi}
		= \frac{m_1 - m_\omega}{m_8-m_\omega}.\label{eq:mixing:tan:2:theta}
	\end{align}
	These do not give the overall sign of $\theta$. Prove the following $\tan\theta$ formulae which do:
	\begin{align}
		\tan\theta &= \frac{-m_{81}}{m_8-m_\omega} = \frac{m_{81}}{m_1-m\phi} = \frac{m_8-m_\phi}{m_{81}} = \frac{m_1 - m_\omega}{-m_{81}}.\label{eq:mixing:tan:theta}
	\end{align}
	Finally, for good measure, prove for $\cos 2\theta$ and $\sin 2\theta$,
	\begin{align}
		\cos 2\theta = \frac{m_8 - m_1}{m_\phi-m_\omega} \hspace{3cm} 
		\sin 2\theta = \frac{2m_{81}}{m_\omega-m_\phi}.\label{eq:mixing:cos:sin:2:theta}
	\end{align}
\begin{sol}
	The rotation from the $\psi_8$,$\psi_1$ basis to the physical basis can be written as
	\begin{align}
		R^T M R &= \begin{pmatrix}
			m_\phi & \\
			& m_\omega
		\end{pmatrix}, \hspace{2cm}
		R\equiv\begin{pmatrix}
			\cos\theta & \sin\theta\\
			-\sin\theta & \cos\theta
		\end{pmatrix}.
	\end{align}
	Inverting this relation one finds
	\begin{align}
		\begin{pmatrix}
			m_8 & m_{81}\\
			m_{18} & m_{1}
		\end{pmatrix}
		= 
		\begin{pmatrix}
			\text{c}^2 m_\phi + \text{s}^2 m_\omega
			& \text{c}\,\text{s}\,(m_\omega-m_\phi)\\
			\text{c}\,\text{s}\,(m_\omega-m_\phi)
			& \text{s}^2 m_\phi + \text{c}^2 m_\omega 
		\end{pmatrix},
	\end{align}
	where we've written $\sin\theta = \text{s}$ and $\cos\theta=\text{c}$ for short-hand. Now note the following relations:
	\begin{align}
		m_8 - m_\omega &= \text{c}^2(m_\phi - m_\omega)\\
		m_8 - m_\phi &= \text{s}^2(m_\omega - m_\phi)\\
		m_1 - m_\omega &= \text{s}^2(m_\phi - m_\omega)\\
		m_1 - m_\phi &= \text{c}^2(m_\omega - m_\phi),
	\end{align}
	where all we've used is $\cos^2\theta + \sin^2\theta =1$. From this it is easy to take ratios and derive the relations in (\ref{eq:mixing:tan:2:theta}). 
	For the $\tan\theta$ expressions use the observation that
	\begin{align}
		m_{81} &= \text{c}\,\text{s}(m_\omega-m_\phi)
	\end{align}
	and take the appropriate ratios with the $\sin^2\theta$ and $\cos^2\theta$ expressions above.
	Finally, for the double angle formulae use
	\begin{align}
		m_8-m_1 &= (\text{c}^2-\text{s}^2)(m_\phi-m_\omega)
	\end{align} 
	and
	\begin{align}
		\cos 2\theta &= \cos^2\theta -\sin^2\theta\\
		\sin 2\theta &= 2\cos\theta\sin\theta.
	\end{align}
\end{sol}
\end{prob}

\begin{prob}\label{prob:omega:rho:mixing:2}\textbf{The $\omega$--$\phi$ mixing angle, part II}. Now that we have the formulae above, rewrite them in terms of physical meson masses so that it is easy to go to the \PDG and plug in the appropriate numbers. In particular, prove the relations given in the ``Quark Model'' review of the \PDG:
	\begin{align}
		\tan^2\theta &= -\frac{4m_{K^*}-m_\rho - 3m_\phi}{4m_{K^*}-m_\rho - 3m_\omega}\\
		\tan\theta &= \phantom{+}\frac{4m_{K^*}-m_\rho - 3m_\phi}{2\sqrt{2}(m_\rho-m_{K^*})}.
	\end{align}
	Show that the expression for $\tan 2\theta$ obtained from (\ref{eq:mixing:cos:sin:2:theta}) is independent of physical measurements and remark on the significance of this value.
	\begin{sol}
		Writing out only the valence quark mass contribution (and taking $m_u=m_d$) we have 
		\begin{align}
			m_{K^*} &= m_u + m_s\\
			m_{\rho} &= 2m_u.
		\end{align}
		From this we may express the diagonal flavor-eigenstate masses 
		\begin{align}
			m_8 &= \frac 13(4m_{K^*}-m_\rho)\\
			m_1 &= \frac 13(2m_{K^*}+m_\rho).
		\end{align}
		Taking the inner product $\langle \phi_8 | \phi 1\rangle$ we may also extract the off-diagonal mass term 
		\begin{align}
			m_{81} = m_{18} = \frac{2\sqrt{2}}{3}\left(m_u-m_s\right) = \frac{2\sqrt{2}}{3}\left(m_\rho - m_{K^*}\right).
		\end{align}
		Armed with these it is straightforward to plug into (\ref{eq:mixing:tan:2:theta}), (\ref{eq:mixing:tan:theta}), and (\ref{eq:mixing:cos:sin:2:theta}). To give a complete list for each expression,
		\begin{align}
			\tan^2\theta &= -\frac{4m_{K^*}-m_\rho-3m_\phi}{4m_{K^*}-m_\rho -3m_\omega}\\
			&= -\frac{2m_{K^*}+m_\rho-3m_\omega}{2m_{K^*}+m_\rho-3m_\phi}\\
			&= \phantom{+}\frac{4m_{K^*}-m_\rho-3m_\phi}{2m_{K^*}+m_\rho-3m_\phi}\\
			&= \phantom{+}\frac{2m_{K^*}+m_\rho-3m_\omega}{4m_{K^*}-m_\rho-3m_\omega}.
		\end{align}
		\begin{align}
			\tan\theta &= \frac{-2\sqrt{2}(m_\rho - m_{K^*})}{4m_{K^*}-m_\rho -3m_\omega}\\
			&= \frac{-2\sqrt{2}(m_\rho - m_{K^*})}{2m_{K^*}+m_\rho -3m_\phi}\\
			&= \frac{4m_{K^*}-m_\rho -3m_\phi}{2\sqrt{2}(m_\rho - m_{K^*})}\\
			&= \frac{2m_{K^*}+m_\rho -3m_\omega}{-2\sqrt{2}(m_\rho - m_{K^*})}.
		\end{align}
		\begin{align}
			\cos 2\theta &= \frac{2}{3}\frac{m_{K^*-m_\rho}}{m_\phi - m_\omega}\\
			\sin 2\theta &= \frac{4\sqrt{2}}{3}\frac{m_\rho - m_{K^*}}{m_\omega - m_\phi}.
		\end{align}
		Note that the value of $\tan 2\theta$ obtained from the last two equations is independent of any meson masses. This is because (\ref{eq:mixing:cos:sin:2:theta}) was written such that the $m_\phi$ and $m_\omega$ dependence cancels and we're left with two quantities which are calculated using valence quark masses. The upshot is that the value of $\theta$ derived this way is precisely what is predicted by \textit{ideal mixing}.
	\end{sol}
\end{prob}

These are very important relations and we end up with a mixing angle of $\theta \approx 37^\circ$. Note that the mixing angle is determined by the ratio of mass differences! (Remind yourself that $m_{81}\propto (m_s-m_u)$.) So the reason why the Gell-Mann--Okubo relation failed for the vector mesons is that there is an appreciable mixing between the $\psi_8$ and $\psi_1$. We say `appreciable' because $37^\circ \gg 0^\circ$, but how much mixing is actually contained in $37^\circ$? To ask the question in a different way: is there a way that we could have expected this amount of mixing? Shouldn't we have expected either zero or $45^\circ$? This would be a good homework problem, but it's so important (and simple) that we'll do it together.
\begin{eg}\textbf{Ideal mixing}. We assume that SU(3) is broken only by the difference $(m_s-m_{u})$ (where $m_d=m_u$ by isospin). Thus it is \textit{intuitively} clear that the physical states in the theory should appear in a way that reflects this breaking. This is just like saying that the in the fine structure of Hydrogen one breaks along total angular momentum so that one obtains the singlet and triplet states; then in the hyperfine structure one breaks along the direction of orbital angular momentum so that one obtains the $|1,\pm 1\rangle, |1,\pm 0\rangle$ states. We can use (\ref{eq:mixing:cos:sin:2:theta}) to determine the expected angle $\theta$. A quick way to do this is to write down the valence quark masses in each component. You know from the previous problems that
	\begin{align}
		m_8 &= \frac{2}{3}(m_u+2m_s)\\
		m_1 &= \frac{2}{3}(2m_u+m_s)\\
		m_{81}&= \frac{2\sqrt{2}}{3}(m_u-m_s).
	\end{align}
	Plugging these in we obtain the \textbf{ideal mixing} angle
	\begin{align}
		\tan 2\theta_\text{ideal} = 2\sqrt{2},
	\end{align}
	which is approximately $35.5^\circ$.
\end{eg}
Aha! So we see that the value of $\theta=37^\circ$ that we found indeed is very close to the prediction of \textbf{ideal mixing}. More importantly, by now you should \textit{physically} understand what ideal mixing signifies: the $s\bar s$ component of the $I_3=0$ mesons should separate from the $u\bar u$ and $d\bar d$ components. The physical states should mirror the breaking pattern, just like the fine structure of hydrogen. Indeed, writing
\begin{align}
	\omega &=\cos\theta\, \psi_8 - \sin\theta\, \psi_1\\
	\phi &= \sin\theta\, \psi_8 + \cos\theta\, \psi_1
\end{align}
and remembering the iso-state quark content
\begin{align}
	\psi_8 &= \frac{1}{\sqrt{6}}\left(u\bar u + d\bar d -2 s \bar s\right)\\
	\psi_1 &= \frac{1}{\sqrt{3}}\left(u\bar u + d\bar d + s \bar s\right),
\end{align}
one finds that for ideal mixing the $\phi$ is a \textit{pure} $s\bar s$ state while the $\omega$ contains \textit{no} strangeness, 
\begin{align}
	\omega = \frac{1}{\sqrt{2}}\left(u\bar u + d\bar d\right) \hspace{3cm}
	\phi = s\bar s.
\end{align}
In this sense it is \textit{maximally} sensitive to how much the $s$ quark mass breaks  $SU(3)$ isospin symmetry and so we end up with the $\omega$ and $\phi$ being much heavier than their cousins.

So now we've completely turned around the `problem' that we originally proposed at the beginning of this section. There's nothing odd about the vector mesons---they behave \textit{precisely} as we would expect given the way that SU(3) is broken. Indeed, one finds that the $J=2$ tensor mesons have a mixing angle that is also very close to ideal mixing \cite{Isgur:1975ib}. The \textit{real} problem is that the \textit{pseudoscalars} are misbehaving because they somehow conspire to give zero mixing when we expect ideal mixing! It is a deep principle in physics that nothing is ever zero `by accident.'

\subsection{Why are the pseudoscalar and vector octets so different?}\label{sec:vec:vs:pseudoscalar}


Now that we've identified the pseudoscalars as the trouble-makers, why should they behave so differently from the higher-spin mesons? We already mentioned part of the answer: the pseudoscalars are secretly under-cover Goldstone bosons of $SU(3)_A$, i.e.\ they are pseudo-Goldstone bosons which obtain masses according to chiral perturbation theory. Unfortunately the formalism for chiral perturbation theory is beyond the scope of this course\footnote{For some reviews see \cite{Pich:1993uq, Pich:1995bw, Ecker:1994gg, Scherer:2002tk} or more recent reviews on modern versions of this idea \cite{Contino:2010rs, Schmaltz:2005ky}. Additionally, older-style textbooks such as Cheng \& Li \cite{ChengLi:Gauge} or Donoghue et al.\ \cite{Donoghue:DynamicsSM} give very nice introductions.}. However, the crux of the matter is that the breaking of $SU(3)$ by quark masses gives relations of the form (using soft pion techniques)
\begin{align}
	f_\pi^2m_\pi^2 &= \frac{m_u+m_d}2 \langle 0 | u\bar u + d\bar d |0\rangle\\
	f_K^2m_K^2 &= \frac{m_u+m_d}2 \langle 0 | u\bar u + s\bar s |0\rangle\\
	f_\eta^2m_\eta^2 &= \frac{m_u+m_d}6 \langle 0 | u\bar u + d\bar d |0\rangle+
	\frac{4m_s}3 \langle 0 | s\bar s |0\rangle.
\end{align}
We can assume that all of the expectation values are at the same QCD condensate scale, $\langle 0 | q\bar q |0\rangle=\Lambda^3_{\text{QCD}}$ and further that the decay constants are all equivalent $f_\pi = f_K = f_\eta = f$. What we are left with is a relation for the \textit{quadratic} meson masses that is \textit{linear} in the symmetry-breaking terms, the quark masses. This explains why the original Gell-Mann--Okubo formula for the pseudoscalar mesons (\ref{eq:Gell:Mann:Okubo:pseudoscalar}) is written with squared meson masses while the formulae for the vectors (\ref{eq:linear:Gell:Mann:Okubo:for:vectors}) is linear in the meson masses. 

So maybe this is the source of the problem. Unfortunately, one can go ahead and try to use all of the mixing angle formulae we derived in Problems \ref{prob:omega:rho:mixing:1} and \ref{prob:omega:rho:mixing:2}. You can work this out as an exercise, but it turns out that one doesn't get consistent answers! Another way to say this is that the Schwinger mass relation (\ref{eq:Schwinger:meson:mass}) doesn't hold. Clearly something is still wrong. Before moving on, though, let use remark that an independent measurement of the pseudoscalar mixing angle can be obtained by looking at various decay rates (`partial widths') as explained in the PDG's ``Quark model'' review; the angle is found to be between $-10^\circ$ and $-20^\circ$.

So what is it that is causing the pseudoscalars to mix in such an unexpected and seemingly inconsistent way? It turns out that we've already met this before when we were counting parameters: it is our old friend the $U(1)_A$ axial anomaly. The fact that $U(1)_A$ is broken means that $\psi_1$ gets an \textit{additional} mass contribution (in principle from the $F\tilde F$ term). See, for example, the discussions in the book by Donoghue et al.\ \cite{Donoghue:DynamicsSM} or the older article by Isgur \cite{Isgur:1975ib}. The reviews by Feldmann \cite{Feldmann:1999uf} and Gasser and Leutwyler \cite{Gasser:1982ap} are particularly thorough.

\begin{prob}
	\textbf{A physical interpretation of $\theta=-10^{\circ}$.} Change to the basis where the mesons have quark content following the $SU(3)$ breaking pattern preferred by the symmetry-breaking $(m_s-m_u)$ parameter,
	\begin{align}
		\psi_x &= s\bar s\\
		\psi_y &= \frac{1}{\sqrt{2}}\left(u\bar u + d\bar d\right).
	\end{align}
	Write the deviation from the ideal mixing angle $\theta_\text{ideal}$ as $\phi$ so that $\theta = \theta_{\text{ideal}}+\phi$. Show that
	\begin{align}
		\tan 2\phi &= \frac{m_{xy}}{m_x^2 - m_y^2}.
	\end{align}
	Consider the limits $m_{xy}\ll m_x^2 -m_y^2$ and $m_{xy}\gg m_x^2 - m_y^2$ and use these to explain the significance of the mixing angles $\theta = \theta_{\text{ideal}}$ and $\theta=\theta_{\text{ideal}}-45^\circ$. Comment on the stability of these values in the aforementioned limits.
	\begin{sol}
		When the mixing term $m_{xy}\ll m_x^2 -m_y^2$ then $\phi=0$ and we have the ideal mixing condition. On the other hand, if the splitting $m_x^2 -m_y^2 \ll m_{xy}$ then $\phi \to 45^{\circ}$. This value is an asymptotic and so is relatively stable in this regime. For more discussion see Isgur \cite{Isgur:1975ib}. 
	\end{sol}
\end{prob}

\begin{prob}
	\textbf{Mixing of the $I_3\neq 0$ states}. Explain why the $K^\pm$ and $\pi^\pm$ do not mix appreciably. It is not sufficient to invoke $SU(3)$ flavor symmetry since we know that $m_s\gg m_{u,d}$ so that this is not a very good symmetry. \textbf{Hint:} Assume that $SU(3)$ is only broken by the quark masses.
	\begin{sol}
		These states do not mix because the light quark masses break $SU(3)$ to $U(1)_u\times U(1)_d\times U(1)_s$ such that strangeness is conserved. 
	\end{sol}
\end{prob}

\subsubsection{Peculiarities of $D^*$ decays.}\label{sec:D:star}
	Let us remark on an interesting coincidence about the vector meson $D^*$, which has the same quark content as $D$ but is in an angular momentum $J=1$ state. How does this particle decay? It chould certainly decay to its pseudoscalar sibling, $D^*\to D + (\text{stuff})$. The `stuff' can be either a pion or a photon,
	\begin{align}
		D^* &\to D\pi^0\\
		D^*&\to D\gamma
	\end{align}
	Which one dominates? At first glance, we would say that the pion emission dominates since it is a strong process while the photon emission is electromagnetic. In other words, the $D^*$ is a resonance in the sense of Section~\ref{sec:stable}. The strong decay should then dominate by a factor of $\alpha_s /\alpha \gg 1$. This is indeed what happens the kaons, where $K^* \to K\pi$ is the dominant decay mode. 
	
	\textit{However}, there is an amazing coincidence in nature that the mass difference between the $D^*$ and $D$ is just a \textit{little} larger than the pion mass
	\begin{align}
		(m_{D^*}^2-m_D^2) \gtrsim m_\pi^2,
	\end{align}
	such that the phase space suppression of $D^* \to D\pi$ is almost the same as the ratio of coupling constants so that the rate for is only a factor of two larger than $D^* \to D\gamma$. The ratio $\alpha_s/\alpha$ is very large, but recall that phase space closes extremely quickly. Heuristically, 
	\begin{center}
		\begin{tikzpicture}
			\draw[very thick,->] (0,0) -- (0,3) node[above] {$\Gamma$};
			\draw[very thick,->] (0,0) -- (5,0) node[right] {$\Delta m_{D}^2$};
			\draw (0,2.05) -- (3,2);
			\draw (3,2) to [out=0,in=100] (4,0);
			\node at (4,-.3) {$m_\pi^2$};
		\end{tikzpicture}
	\end{center}
	so that the decay rate lies roughly halfway down the steep precipice because the mass splitting between the vector and pseudoscalar $D$ mesons, $\Delta m_D^2 \equiv m_{D^*}^2-m_D^2$, is very close to the pion mass. There's no deep reason for this, it is a complete accident like to why the angular size of the moon and the sun happen to be close enough to produce \emph{just right} solar eclipses.

	We can use isospin symmetry to relate this decay to trivially relate this process to 
	\begin{align}
		D^{*0} &\to D^+\pi^-\\
		D^{*0} &\to D^0\pi^0\\
		D^{*+} &\to D^0\pi^+\\
		D^{*+} &\to D^+\pi^0.
	\end{align}
	The relations between these processes in the isospin limit is broken by the mass difference between $D^*$ and $D^0$, which is ordinarily negligible. However, because of the delicate phase space tuning, the isospin breaking effects are order one. For example, $D^{*0} \to D^+\pi^-$ is completely forbidden because the isospin-breaking mass differences barely pushes it out of the kinematically allowed region. On the other hand, the $D^{*+}$ decays are pushed further into the kinematically allowed region and away from the steep drop so that pion decays indeed dominate over the photon decay: Br($D^{*+}\to D^0\pi^+$)$\approx 67.7\%$ and Br($D^{*+}\to D^+\pi^0$)$\approx 30.7\%$. The upshot with these decay modes is that they can teach us a lot about the strong interaction since we can reliably calculate the photon decays.

Remember that the story of isospin splitting is something that we already know very well from our physics childhood: it's simply hyperfine structure of the hydrogen atom\footnote{To zeroth order all of physics reduces to the harmonic oscillator. The first order correction to this statement is the hydrogen atom. You should know everything about the hydrogen atom.} with `hydrogen' replaced by `meson'. The mass difference between the $D^{*}$ and $D$ is just the analog of hyperfine splitting due to the magnetic interaction between the proton and electron in hydrogen. The pseudoscalar $D$ is the singlet and the vector $D^{*}$ is triplet of the spin--spin interaction. 

\begin{eg}\label{eg:Bst:decay}\textbf{$B^*$ decay}. Based on the analogy to the hyperfine splitting of hydrogen, what is the dominant $B^*$ decay? Recall in the hydrogen atom that the spin-spin coupling goes like $1/m_p$ due to the dipole moment of the proton. Thus we expect the splitting to go like $1/m_{B^*}$. Given that the $B$ meson masses are roughly a factor of three larger the $D$ masses---coming from the relative $c$ and $b$ quark masses---we expect the $\Delta m_B\equiv m_{B^*}-m_B$ splitting should be a factor of three smaller than the $\Delta m_D$ splitting. Since $\Delta m_D \sim m_\pi \sim 140$ MeV, we expect $\Delta m_B \approx 50$ MeV, which is smaller than the pion mass and hence the dominant $B^*$ decay is $B^* \to B \gamma$.
\end{eg}

\section{Heavy Quark Symmetry} 
\label{sec:HQS}

In Section \ref{sec:masses:and:mixing:mesons} we mention that the
dominant contribution to the masses of the heavy quark mesons ($D$ and $B$) comes
from the mass of the heavy quark rather than \QCD. Because of that
fact, we lose the ability to use a symmetry like SU(3) flavor symmetry
of the light quarks. Instead, we may use a different handle on these
systems, \textbf{heavy quark symmetry}. We discuss this symmetry in
this appendix.

Heavy quark symmetry is different from the symmetries that we are already used to. In the usual case for an approximate symmetry, there is a parameter in the Lagrangian that we can set to zero to yield a more symmetric theory. In heavy quark symmetry, we take the limit where the heavy quark mass goes to \emph{infinity}. This is not a symmetry that is manifest in the Lagrangian, but rather one that is specific to particular meson systems. This is why the subject only blossomed in the 1990s. There is a nice set of review literature:
\begin{itemize}
	\item A short review by Mark Wise~\cite{Wise:1994vh}
	\item Technical lectures by Benjamin Grinstein~\cite{Grinstein:1995uv}
	\item Neubert's lecture notes \cite{Neubert:1996wg} and a detailed review \cite{Neubert:1993mb}
	\item The yellow book by Manohar and Wise \cite{Manohar:Wise:HQ:phys}
	\item Finally, the insightful and idiosyncratic \acro{TASI} lectures by Howard Georgi, which appear to only be available on his website\footnote{\url{http://www.people.fas.harvard.edu/~hgeorgi/tasi.pdf}}. 
\end{itemize}

\subsection{The hydrogen atom}

All of the essential physics of heavy quark effective theory is contained in the undergraduate quantum mechanics of the hydrogen atom. Let's start with the following question:
\begin{quote}
	\textit{What is the difference between hydrogen and deuterium?}
\end{quote}
Deuterium is a hydrogen atom with an additional neutron. Chemists call these states isotopes which is a fancy way of saying \emph{basically the same.} We know that technically hydrogen and deuterium differ by their quantum numbers under the Lorentz group: they have different mass and spin. Why, then, is it that these two \emph{technically different particles} are \emph{basically the same} in chemistry? To put this on more concrete footing, if you were suddenly transported to a parallel universe where all of the oxygen on Earth were replaced by oxygen-17, we would survive just fine because for the most part chemistry doesn't care about isotopes. 

The reason why hydrogen and deuterium are chemically \emph{basically the same} is that the electrons really don't care about the nucleus. Or, in more general terms,
\begin{quote}
	\textit{The light degrees of freedom are insensitive to the heavy degrees of freedom that source the potential.}
\end{quote}
In the hydrogen atom, the nucleus just plays the role of the electromagnetic potential about which we can do non-relativistic quantum mechanics and calculate energy spectra. To leading order the potential is insensitive to the mass and spin of the actual nuclear source of the electromagnetic field. What do the corrections look like?

Recall from the hydrogen atom that we calculate not only the leading order spectra, $E_n$, but also the energy splittings coming from fine structure (e.g.\ spin-orbit coupling) and hyperfine structure (e.g.\ spin-spin coupling),
\begin{align}
	E_n &\sim m_e\frac{\alpha^2}{n^2} &
	\Delta E^{\text{fs}} &\sim m_e\alpha^4 &
	\Delta E^{\text{hf}} &\sim m_e \alpha^4 \frac{m_e}{m_p} \ .
\end{align} 
Note the extra $m_e/m_p$ suppression on the hyperfine splitting. Physically this tells us that in the limit of infinite mass, one cannot rotate the proton to reverse its dipole moment. The additional spin coming from the neutron in deuterium clearly changes the hyperfine splitting. 

The other difference coming from the larger nuclear mass leads to a
different reduced mass. Recall that the `reduced' in `reduced mass'
means we are reducing a two-body problem to a one-body problem such as
the Kepler system. In the rest frame of the Earth--Sun Kepler system,
\textit{both} the Earth \textit{and} the sun rotate about a common
center of mass,
\begin{center}
	\begin{tikzpicture} 
	\draw (0,0) circle (1cm) 
		circle (0.1cm); 
	\node [fill=black, draw, circle, inner sep=1] at (45:1) {};
	\node [fill=black, draw, circle, inner sep=1] at (-90:.1) {};
	\end{tikzpicture}
\end{center}
naturally the heavier object orbits with a much smaller radius. In other words, the whole system is at rest, but the heavy guy has some small velocity. Thus let us use suggestive language and say that deuterium also differs from hydrogen by the velocity of the nucleus.

The two differences between hydrogen and deuterium lead to energy splitting, but both effects decouple in the limit where the nucleus is taken to infinite mass since both effects scale like $1/m_\text{nucl}$.

\subsection{Heavy quark symmetry: heuristics}

The infinite nuclear mass limit is the precise analog to the heavy quark limit that we'll be using. The heavy quark mesons which we'll consider---primarily $B$ mesons---are just like hydrogen systems where the binding comes not from electromagnetism but from \QCD. In fact, we may consider a na\"ive picture where the light quark orbits the $b$. 

We might worry that this is too na\"ive. There's a very big difference between the electromagnetic potential and the \QCD potential: while the former is perturbative, the latter is hopelessly non-perturbative. Since virtual partons are $\mathcal O(1)$ effects, we shouldn't even be able to say that the $B$ meson is composed of a $b$ and a light quark, say $\bar u$.

This is precisely the issue of the \textbf{brown muck}, the dirty nonperturbative physics that should make the heavy mesons intractable. Our goal is to get around the brown muck without getting ourselves too dirty; we want to find ways to calculate \textit{the things that we care about} without having to understand the intractable physics of the muck. Should this be possible? Yes! In fact, this should now sound familiar: this is precisely what we meant when we said that chemists don't care about isotopes, and why we don't have to re-derive all of the details of deuterium from scratch. 

When the bottom quark becomes extremely heavy, the excitations of $B$ meson spectrum becomes independent of bottom mass and the spin. Mass, for us, is really an index for flavor (e.g.\ $b$ or $c$ quark). We can, of course, completely neglect electromagnetism. The calculation of the `energy levels' of the $B$---that is,the spectrum of mesons)---follows precisely as the hydrogen atom. For example, the $B$ and $B^*$ states represent the hyperfine splitting, while the higher spin resonances $B_1, B_2\,\cdots$ are the $p$-wave states and so on.

Now assume that $m_c, m_b\gg \Lambda_\text{QCD}$ and that the spectrum
of $D$ mesons have been measured in nature along with the mass of the
lightest $B$ meson. Just as we could figure out everything about
deuterium to leading order once we understood hydrogen, we can similarly determine the rest of the $B$ spectrum simply based on the $D$ spectrum and the lowest $B$ mass as a reference point. Further, we are able to do this without knowing the physics of the brown muck just as we didn't have to re-derive the deuterium spectrum from scratch. The only difference between the $D$s and $B$s are the mass of the heavy quark and, as we said before,
\begin{quote}
	\textit{the light degrees of freedom just don't care}.
\end{quote}

\begin{eg}
	For example, in the $m_b\to \infty$ limit, we have the relation
	\begin{align}
		\frac{m_{B^*} - m_B}{m_{B^*}+m_B} = 0.
	\end{align}
	The $B$ and $B^*$ are degenerate in the heavy quark limit. In
        the language of symmetry, we say that these two states form an
        SU(2) doublet with respect to the spin orientation of the
        heavy quark since these are just the singlet and triplet
        states of the hyperfine structure.
\end{eg}
\begin{eg}
	Consider the form factor between a $B$ and a $D$ that are at rest. Up to some normalization we know that
	\begin{align}
		\langle B|\mathcal O | D\rangle = 1 \ . 
	\end{align}
	For this transition---once again---\textit{the light degrees
          of freedom just don't care}. The $b\to c$ transition does
        not change the \QCD potential which the light quark feels. In
        fact, it doesn't matter what the operator $\mathcal O$ is that
        enacts the $b\to c$ decay. This is just a statement about the
        wave function overlap. We say that all the heavy flavors form
        a fundumental under a flavor group SU$(2N_f)$, where there are
        $N_f$ flavors and the factor of two comes from the spin
        orientation. In nature, where $N_f=2$, we say that the spin up
        and down $D$s and the  spin up
        and down $B$s form a $\mathbf{4}$ of $SU(4)$.
\end{eg}

\subsection{Heavy quark symmetry: specifics}

We now see that heavy quark symmetry is very different from the usual symmetries that we work with in effective field theory. In the chiral Lagrangian, for example, we can take terms to zero and see how we recover symmetries. Heavy quark symmetry is different in a very fundamental way. The $M\to\infty$ limit does \textit{not} increase the symmetry of the Lagrangian. In fact, we will only `integrate out' \textit{part} of the degrees of freedom. 

When we do a regular \acro{EFT}, we expand about the vacuum and integrate out high-frequency modes. In heavy quark effective theory, we will expand about the background of a single heavy quark, say a $b$. In such a background one \textit{assumes} that the heavy quark is there classically; we get it `for free' without having to think about how it popped out of the vacuum. If we want any \textit{additional} heavy quarks, however, we have to honestly pay the cost of introducing an additional high frequency mode to the background.

Consider the spectrum of mesons containing a heavy quark. This turns out to be very simple and can be done even before dipping into heavy quark effective theory. The main idea is to expand in powers of $1/m_Q$, where $m_Q$ is the mass of the quark. Due to the ambiguities in defining a quark mass, we should really say that it is some effective mass---a pole mass, $\bar{MS}$ mass, whatever---which is identified with a physical mass by some prescription. The mass of a hadron containing this heavy quark, $m_H$, is 
\begin{align}
	m_H = m_Q\left(1 + \frac{\bar\Lambda}{m_Q} + \frac{a}{m_Q^2} + \cdots\right)\ .\label{eq:HQS:spectrum}
\end{align}
 The leading term reminds us that in the heavy quark limit the mass of the hadron and the mass of the quark are basically the same. The sub-leading terms depend on the light degrees of freedom: $\bar\Lambda$ is the mass of the light degrees of freedom and $a$ parameterizes the interaction between the light degrees of freedom and the heavy quark. 
$\bar \Lambda$ doesn't know about the mass or spin of the heavy quark. It is a parameter that \textit{only} deals with the light degrees of freedom \textit{independently} of the heavy degree of freedom sourcing the \QCD potential. Only the interaction term takes these into account; it is analog of hyperfine splitting in hydrogen. 
\begin{eg}
	Consider the case when the hadron $H$ is a $B$ or $B^*$. One only observes a difference between the $B$ and $B^*$ mass at $\mathcal O(1/m_Q)$, i.e.\ in the $a$ term. On the other hand, the difference between the $B$ and the $B_s$ masses occurs at $\mathcal O(m_Q^0)$, i.e.\ in the $\bar\Lambda$ term.
\end{eg}
By now you should have guessed what kind of information is contained in $a$: (1) the reduced mass or kinetic energy of the heavy quark in the meson rest frame and (2) the spin. In fact, we can further parameterize $a$ as
\begin{align}
	a = -\lambda_1 + 2\left[J(J+1)-\frac{3}{2}\right]\lambda_2.\label{eq:HQET:a}
\end{align}
Here the $\lambda_1$ term is spin-independent and is related to the effect of
the reduced mass while the $\lambda_2$ term is associated with the
spin--spin interaction between the heavy quark and the brown muck. 
\begin{eg}
	Which spin does $J$ correspond to: the spin or the heavy quark or the spin of the meson? $J$ corresponds to the total spin of the meson, just as in the case of the hydrogen atom.
\end{eg}
By dimensional analysis $\lambda_{1,2}$ have mass dimension two. In fact, by the method of ``\textit{there's only one other scale lying around},'' we can guess that $\lambda_{1,2}\approx \Lambda_{\text{QCD}}^2$ and $\bar\Lambda \approx \Lambda_{\text{QCD}}$.
\begin{eg}
For example, consider the ratio of the $B_s$--$B_d$ mass splitting to the $D_s$--$D_d$ splitting. Heavy quark symmetry predicts that these splittings are the same up to effects on the order of $1/m_c$. Applying  (\ref{eq:HQS:spectrum}) gives
\begin{align}
	r_1\equiv \frac{m(B_s)-m(B_d)}{m(D_s)-m(D_u)} = \frac{\bar\Lambda_s-\bar\Lambda_d + \mathcal O(\frac{1}{m_b})}{\bar\Lambda_s-\bar\Lambda_d + \mathcal{O}(\frac{1}{m_c})} = 1 + \mathcal O(1/m_c).
\end{align}
\end{eg}
As another example, one can show that
\begin{align}
	m^2(B^*)-m^2(B) = 4\lambda_2\ .
\end{align}
It is, of course, not surprising that this splitting only comes from the non-universal part of the $a$ term. Comparing to actual measurements, we find 
\begin{align}
	\lambda_2(m_B) \approx 0.12\text{ GeV}^2.
\end{align}
We can do the same calculation for the charmed mesons and we again find 0.12 GeV$^2$. The $B$ meson gives preferable results since the error goes like $1/m_Q$. In fact, looking at the differences in the squared masses of the vector and pseudoscalar mesons for the $B$ and the $D$,
\begin{align}
	m^2(B^*)-m^2(B) &= 0.47 \text{ GeV}^2\\
	m^2(D^*)-m^2(D) &= 0.55 \text{ GeV}^2.
\end{align}
What about the kaon? Of course, we don't expect this heavy quark symmetry to hold, but it turns out that $m^2(K^*)- m^2(K) = 0.55 \text{ GeV}^2$. What do we get for the $\rho$ and pion? We can neglect the pion mass, and we recall that $m_\rho = 770$ MeV so that the difference of the squared masses is $0.57 \text{ GeV}^2$. There's no reason why these should hold, but it is a notable observation that they do.

\subsection{Heavy Quark Effective Theory}

Now let's get to the heart of heavy quark effective theory (\acro{HQET}), where we expand about the background of a classical heavy quark. We have to be a little formal. Suppose that we have a $b$ quark in an on-shell hadron $H$ with momentum 
\begin{align}
	p_H^\mu = m_H v^\mu \ .
\end{align}
We have introduced the relativistic four-velocity, $v^\mu$, in anticipation that we will eventually take a non-relativistic limit. If the hadron is on-shell and if the quark is indeed heavy, then we can say that the $b$ quark itself is also `almost on-shell' in the sense that to good approximation we know the momentum of the heavy quark,
\begin{align}
	p_Q^\mu=m_Qv^\mu + k^\mu,
\end{align}
where $k^\mu$ is a small correction. Define a four-velocity for the heavy quark,
\begin{align}
	v_Q^\mu = \frac{p_Q^\mu}{m_Q} = v^\mu + \frac{k^\mu}{m_Q}.
\end{align}
The second term is now manifestly a small correction and in the heavy quark limit, $v_Q=v$. 

We now introduce a trick. Define a projection operator
\begin{align}
	P_\pm = \frac{1\pm \slashed{v}}{2}.
\end{align}
where $\slashed{v} = v_\mu \gamma^\mu$. This projects between particle and anti-particle states. One way to see this is to write out the Dirac equation acting on plane wave fermions and anti-fermions:
\begin{align}
	m\left(\slashed{v} - 1\right)u(p) &= 0
	&
	m\left(\slashed{v} + 1\right)v(p) &= 0 \ .
\end{align} 
The projected states are,
\begin{align}
	h_v(x) &= e^{im_Qv\cdot x}P_+Q(x)\\
	H_v(x) &= e^{im_Qv\cdot x}P_-Q(x).
\end{align}
The overall phase is like the unphysical phase of the Schr\"odinger equation where the energy is arbitrary. All we're doing is pulling out the quickly oscillating part of the $Q$ time evolution. 

Write the heavy quark field as
\begin{align}
	Q(x) = e^{im_Q v\cdot x}\left(h_v(x)+H_v(x)\right).
\end{align}
We can now plug it into the Lagrangian to obtain
\begin{align}
	\mathcal L &= \bar Q(i\slashed{D }-m_Q)Q\\ 
	&= \bar h_v iv\cdot D h_v - \bar H_v (iv\cdot D +2 m_q)H_v + \bar h_v i\slashed{D}_\perp H_v + \bar H_v i\slashed{D}_\perp h_v,
\end{align}
where we define a `perpendicular' covariant derivative,
\begin{align}
	D_\perp \equiv D^\mu - v^\mu(v\cdot D).
\end{align}
In the rest frame of the particle, $v_\mu = (1,\mathbf{0})$. Thus $v\cdot D_\perp=0$ so that $D_\perp = (0, \mathbf{D})$ in this frame. Now we integrate out the heavy degrees of freedom, $H_v$.  To do this, we solve the Euler--Lagrange equation for $H_v$,
\begin{align}
	H_v &= \frac{1}{2m_Q+iv\cdot D}\cdot i D_\perp h_v. 
\end{align}
We can use this to eliminate $H_v$ from the Lagrangian,
\begin{align}
	\mathcal L = \bar h_v i v\cdot D h_v + \bar h_v i\slashed{D}_\perp \frac{1}{2m_q+iv\cdot D} i\slashed{D}_\perp h_v.
\end{align}
Great! Mission accomplished. Note, however, that the second term is rather funny-looking; it has a derivative in the denominator. Intuitively this is the remnant of the $H_v$ propagator. It is a non-local term. In order to convert this into a local term---as required for an effective Lagrangian---we do an operator product expansion in $v\cdot D/m_Q \ll 1$, since the derivative acts on the light field and picks up a momentum on the order of $\Lambda_\text{QCD}$.  Expanding we get
\begin{align}
	\mathcal L &= \mathcal{L}_\text{kin} + \frac{1}{2m_Q}\sum \bar h_v i\slashed{D}_\perp \left(\frac{i-v\cdot D}{2m_q}\right)^n i\slashed{D}_\perp j_v\\
	&= \mathcal{L}_\text{kin} + \frac{1}{2m_Q} \bar h_v\left(iD_\perp\right)^2 h_v + \frac{g_s}{4m_Q}\bar h_v \sigma_{\mu\nu}G^{\mu\nu} h_v + \cdots.
\end{align}
So now we've written out the leading-order higher-dimensional operators in our heavy quark \acro{EFT}. We should actually look at these two terms with great fondness. The first term is just a factor on the order of $p^2/m_Q \sim \Lambda_\text{QCD}^2/m_Q$ while the second term is a chromo-magnetic operator. These are precisely the $\lambda_1$ and $\lambda_2$ terms in (\ref{eq:HQET:a}), which we had identified from very physical intuition.
Let us define effective operators
\begin{align}
\mathcal{O}_1 + \mathcal{O}_2=\frac{1}{2m_Q}\bar h_v (iD_\perp)^2 h_v + \frac{g_s}{4m_Q}\bar h_v \sigma_{\mu\nu}G^{\mu\nu}h_v \ .
\end{align}
It just jumps out at you that these terms correspond to kinetic energy of the heavy quark in the rest frame of the meson and the hyperfine interaction between the heavy quark and the brown muck. This allows us to identify the physical meaning of the $\lambda_{1,2}$ parameters,
\begin{align}
	\langle H(v)|O_{1,2}|H(v)\rangle \sim \lambda_{1,2}.
\end{align}

The `perpendicular' covariant derivative $D_\perp$ is kind of annoying, but fortunately the deviation between $D_\perp$ and $D$ is third order so that when we work only to second order it can be replaced by the full covariant derivative, $D$. 
 
\subsection{HQET for $|V_{cb}|$}
\label{sec:HQS:Vcb}

In Section~\ref{sec:Vcb:BtoD} we began a discussion of the determination of the $|V_{cb}|$ matrix element of the \CKM matrix. We argued that the heavy quark limit factorizes the physics of the heavy quark from the \QCD muck so that there is only one form factor to determine.
It is easy to understand how actual \QCD differs from the heavy quark limit in terms of an object called the Isgur--Wise function which is universal and depends only on $v_b \cdot v_c$. 
Let us work in the specific limit
\begin{align}
	\frac{m_b}{m_c} = \text{const } (\sim 3)
\end{align}
while simultaneously taking $m_b \to \infty$. Imagine a $B$ meson to be composed of a $b$ quark surrounded by brown muck. Consider the velocity of the $b$ to be the same as the velocity of the entire meson, $v_b=v$ and suppose that the $b$ decays into a $c$. Let us write the velocity of the $c$ as $v_c=v'$. 
Now we would like to know what would happen if we replaced the $B$ meson with a $B^*$ or if we changed the $D$ to a $D^*$. Would the brown muck also be excited? No, nothing would change. 
We have a theory of \QCD where we treat electroweak currents as external sources. If Harry Potter waved his magic wand and turned the $b$ into a $c$ quark, the brown muck would more or less stay the same. This is just like our isotope analogy where adding another neutron to an atom doesn't lead to a big change in its chemical properties; nothing happens. And when nothing happens, the intuitive value of the form factor doesn't change.

In the limit of $v\to v'$, we don't care about $b$ versus $c$, $D$ versus $D^*$, or any of these modifications that don't significantly affect the brown muck. In fact, we don't even care \textit{how} the $b\to c$ transition occurs, as long as it is short distance. This transition could just as well have come from a vector current, an axial current, some $V\pm A$ combination, or Harry Potter's magic wand.

Now consider $v\neq v'$. In the decay $B\to D$ this corresponds to the $c$ quark picking up a non-zero velocity in the meson rest frame. We are \textit{not} allowed to assume the non-relativistic limit for the velocity of the $c$ quark, $v'$. The only thing we know is that the $b$ is non-relativistic in the meson rest frame---it is a fallacy to say that `everything' is non-relativistic in the heavy-quark limit. The $c$ can be relativistic in the rest frame of the $B$. In this case, the $b$ decays to $c$ with some generically-not-small velocity and suddenly brown muck sees the color flow moving. Then one of two things happen,
\begin{enumerate}
	\item The brown muck can follow the color flow. This corresponds to $B\to D$ decay.
	\item The brown muck can pop other things out of the vacuum to follow the color flow.
\end{enumerate}
This brings us back to the meaning of the form factor: it is the probability for $B\to D$ versus other decay processes when things are not at rest. The analogous statement in the hydrogen/deuterium atom is that not only do we take away a neutron, but we give the proton a kick. The form factor can be interpreted as the overlap between the electron wave function before and after the kick and represents the probability of finding an electron in a given shell after the proton kick. The case with the brown muck is completely analogous except that we do not know the explicit wave function.

There's one simple option for generating a Lorentz invariant out of
$v$ and $v'$. Define $w\equiv v\cdot v'$, where $w$ is just what we
usually call $\gamma$, the boost. The form factor is the universal
Isgur--Wise function of only $w$, $\xi(w)$, such that $\xi(1)=1$, i.e.\ when $v=v'$. In general $w\geq 1$. We don't know much more about $\xi(w)$, but we can see how far we can go with these properties. 
Let us choose a somewhat different normalization and we write the form factor as
\begin{align}
	\frac{1}{m_B}\langle B(v)| V^\mu | B(v')\rangle = \xi(w)\cdot (v\cdot v')^\mu.
\end{align}
We would like to calculate the form factors based on the Isgur--Wise function. Let's look at the decay of $B\to D$ so that we would like
\begin{align}
	\frac{1}{\sqrt{m_Bm_D}}\langle B(v)| V^\mu | D(v')\rangle = \xi(w)\cdot (v\cdot v')^\mu.
\end{align}
In the `usual' notation,
\begin{align}
	\langle B | V^\mu | D\rangle = F_+(q^2)(p+p')^\mu + F_-(q^2)q^\mu.
\end{align}
Our hope is to relate the form factors $F_\pm$ to the Isgur--Wise
function $\xi(w)$. First consider the case $v=v'$ where $w=1$ and $\xi(w)=1$. We find
\begin{align}
	\left.F_\pm(q^2)\right|_{w=1} = \frac{m_B\pm m_D}{2\sqrt{m_B m_D}} \ .
\end{align}
One can then restore $\xi(w)$ by tacking it onto the right-hand side.
In principle there are two form factors, but in the heavy quark limit they are related so that we only have to determine one of them before we can start making predictions.
Three remarks are in order: 
\begin{enumerate}
	\item Does this work for $B\to D^*$? Yes, it's more complicated because the form factors include more terms, e.g.\ contractions with $\epsilon$ tensors, but at the end of the day everything is indeed expressed in terms of $\xi(w)$.
	\item What kind of corrections are there to the heavy quark limit? There's a correction from the finiteness of the heavy quark mass, as well as \textit{perturbative} $\alpha_S$ corrections from soft gluons. 
	\item \textbf{Luke's theorem}. If you expand about an
          extremum, the correction is always second order. This is why
          the leading order correction to $\xi(1)$ starts at $\mathcal O(1/m_W^2)$. The point $\xi(1)$ is an extremum because we're looking at a situation with `maximum wave function overlap.' This should remind you of the Ademollo--Gatto theorem in Section~\ref{sec:Kl3}.
\end{enumerate}
 
What do we do with all of this? We would like to measure $B\to D$ and use the spectrum to obtain $|V_{cb}|$. Ideally we would like to look at the case where $w=1$, i.e.\ where the charm is at rest, but the phase space---and hence statistics---for this is very small. We must thus look at all events, plotting the total number events as a function of $w$. We can then use the data to extrapolate the curve to the $w=1$ case. Since we know $\xi(1)=1$ we can expand about this point,
\begin{align}
	\xi(w) = 1 - \rho^2(w-1),
\end{align}
and fit for the parameter $\rho$. There is a small unremovable
theoretical uncertainty associated with this. The number that we get
is something like $\rho= 0.44$. 

At the end of the day, we then get $|V_{cb}|$ up to uncertainties from
  several sources: higher order in $1/M_Q$, higher oder in
  $\alpha_S$, and the deviation from the zero recoil limit. As of now,
  these uncertainties are estimated to be at the level of few percent.
 
\section{Meson Mixing and CP formulae}
\label{app:meson:mixing:and:CP}

The meson open system Hamiltonian is given by $H=M-\frac i2 \Gamma$. The eigenstates of this Hamiltonian are
\begin{align}
	|B_{L,H}\rangle = p |B^0\rangle \pm q |\bar B^0\rangle \ ,
\end{align}
with proper normalization $|p|^2+|q|^2 = 1$.

In order to make our conventions clear we include a full derivation of some important formulae. The open system Hamiltonian describing the oscillation between flavor eigenstate mesons is
\begin{align}
	H=
	\begin{pmatrix}
		H_{11} & H_{12}\\
		H_{21} & H_{11}
	\end{pmatrix}
	=
	\begin{pmatrix}
		M_{11}-\frac i2 \Gamma_{11} & M_{12}-\frac i2 \Gamma_{12}\\
		M^*_{12}-\frac i2 \Gamma^*_{12} & M_{11}-\frac i2 \Gamma_{11}
	\end{pmatrix}.
\end{align}
Note that $H_{21} \neq H_{12}^*$. The eigenvalue equation is $(H_{11}-\lambda)^2 - H_{12}H_{21}=0$ and can be solved trivially by inspection,
\begin{align}
	\lambda_{L,H} = H_{11} \pm \sqrt{H_{12}H_{21}}.
\end{align}
This also gives a useful expression of the diagonal element of the open system Hamiltonian in terms of its eigenvalues,
\begin{align}
	H_{11} = \frac 12 (\mu_L + \mu_H) \equiv M - \frac i2 \Gamma.
\end{align}
Here we have written
\begin{align}
M = (M_H+M_L)/2,\qquad \Gamma = (\Gamma_H+\Gamma_L)/2. 
\end{align}
We also have the differences in the masses and widths,

A few parameters of particular significance are
\begin{align}
	x = \frac{\Delta m}{\Gamma}
	\qquad\qquad
	y= \frac{\Delta\Gamma}{2\Gamma}
\end{align}
Note that the factor of $2$ in $y$ is a
convention for simplifying expressions later on.

We can further use the eigenvalue equation to obtain useful relations between $\Delta M$ and $\Delta \Gamma$ in terms of the off-diagonal open Hamiltonian elements. This is useful since these are the elements that can be obtained by calculating box diagrams.
\begin{align}
	(H_{11}-\mu_{L,H})^2 = H_{12}H_{21}
	\label{eq:mixing:eigenvalue:eq}.
\end{align}
Writing out the left-hand side of (\ref{eq:mixing:eigenvalue:eq}) we have
\begin{align}
	(H_{11}-\mu_{L,H})^2 &= \left[ \frac 12(\mu_L-\mu_{L,H}) + \frac 12(\mu_H-\mu_{L,H}) \right]^2\\
	&= \frac 14 (\Delta \mu)^2\\
	&= \frac 14 \left[ (\Delta M)^2 - \frac 14 (\Delta\Gamma)^2 - i \Delta M\Delta \Gamma \right].
	\label{eq:mixing:eigenvalue:LHS}
\end{align}
Similarly, the left-hand side of (\ref{eq:mixing:eigenvalue:eq}) gives
\begin{align}
	H_{12}H_{21} &= |M_{12}|^2  - \frac i2 \Gamma_{12}^* M_{12} - \frac i2 M_{12}^* \Gamma_{12} - \frac 14 |\Gamma_{12}|^2\\
	&= \left(|M_{12}|^2 - \frac 14 |\Gamma_{12}|^2\right) - i \text{Re}(M_{12}^*\Gamma_{12}).
	\label{eq:mixing:eigenvalue:RHS}
\end{align}
Now comparing the real and imaginary parts of (\ref{eq:mixing:eigenvalue:LHS}) and (\ref{eq:mixing:eigenvalue:RHS}) we have the relations
\begin{align}
	(\Delta M)^2 - \frac 14 (\Delta \Gamma)^2 &= 4|M_{12}|^2 - |\Gamma_{12}|^2\\
	\Delta M \Delta \Gamma &= 4 \text{Re}(M_{12}^*\Gamma_{12}).
\end{align}

Moving on to the eigenvectors, we write $|B_{L,H}\rangle = p|B\rangle
\pm q|\bar B\rangle$ so that we can represent a mass eigenstate in the
flavor basis as a column vector $(p,q)^T$. Solving for the
eignevectors of the open system Hamiltonian gives
\begin{align}
	\frac qp &= \pm\sqrt{
	\frac{
		M^*_{12}-\frac i2 \Gamma^*_{12}
		}{
		M_{12}-\frac i2 \Gamma_{12}
		}
	}.
\end{align}
where the top sign corresponds to the $L$ state and the bottom sign corresponds to the $H$ state\footnote{Note that an opposite convention is used in \textit{CP Violation} by Branco, Lavoura, and Silva. This leads to different expressions for the following formulae.}. 
Now consider a mass state $|B(t)\rangle$. By this we mean a state that at $t=0$ was produced as a flavor eigenstate $|B(0)\rangle = |B\rangle$ but then oscillates as a mass eigenstate. We can determine the flavor admixture at time $t$ by writing $|B(0)\rangle$ in terms of mass eigenstates and then time evolving those:
\begin{align}
	|B(t)\rangle &= \frac{1}{2p} e^{-i\mu_Lt} |B_L\rangle
					+ \frac{1}{2p} e^{-i\mu_Ht} |B_H\rangle.
\end{align}
Then converting these back into flavor states,
\begin{align}
	|B(t)\rangle &= \frac{1}{2p} e^{-i\mu_Lt} \left(p|B\rangle +q |\bar B \rangle\right)
					+ \frac{1}{2p} e^{-i\mu_Ht} \left(p|B\rangle -q |\bar B \rangle\right)\\
		&= \frac 12 \left( e^{-i\mu_L t}+e^{-i\mu_H t}\right) |B\rangle
			+ \frac 12 \left( e^{-i\mu_L t}-e^{-i\mu_H t}\right) \frac{q}{p} |\bar B\rangle.
\end{align}
For future simplicity, we define the functions
\begin{align}
	g_\pm(t) = \frac 12 \left( e^{-i\mu_L t}\pm e^{-i\mu_H t}\right).
\end{align}

Now let us consider the decay $B(t)\to f$. The object which appears in our CP asymmetries is $\Gamma(B(t)\to f)$, where $\Gamma \sim |A|^2$, where we've written $A$ as the amplitude for the process. Let's calculate this quantity (the other factors cancel in the expression for the asymmetry). 
\begin{align}
	A &= \langle f | g_+(t) |B\rangle + \frac qp \langle f| g_-(t) |\bar B\rangle\\
	&\equiv g_+(t)A_f + \frac qp g_-(t) \bar A_f.
\end{align}
The squared amplitude is
\begin{align}
	|A|^2 & = |g_+(t)|^2 |A_f|^2 + \left| \frac{q}{p}g_-(t)\right|^2 |\bar A_f|^2 + 
	g_+(t)A_f \left(\frac qp g_-(t)\bar A_f\right)^*
	+ \left(g_+(t)A_f\right)^* \frac qp g_-(t)\bar A_f.
	\label{eq:CP:decay:derivation:Asquared}
\end{align}
It's very useful to write out the relevant products of $g_\pm(t)$ before going further. Again, note that some sources (e.g.\ Branco, Lavoura, and Silva) use different conventions so that their $g_\pm(t)$ expressions are different.
\begin{align}
	|g_\pm(t)|^2 &= \frac 14 \left| e^{-i\mu_L t} \pm e^{-i\mu_H t} \right|^2\\
	&= \frac 14 \left| e^{-iM_L t}e^{-\frac 12\Gamma_L t} \pm e^{-iM_H t}e^{-\frac 12 \Gamma_H t} \right|^2\\
 	&= \frac{e^{-\Gamma t}}{4} \left| 
			e^{-iM_L t}e^{\Delta\Gamma t/4} 
			\pm e^{-iM_H t}e^{-\Delta\Gamma t/4}
			\right|^2\\
	&= \frac{e^{-\Gamma t}}{4} \left( 
			e^{-iM_L t}e^{\Delta\Gamma t/4} 
			\pm e^{-iM_H t}e^{-\Delta\Gamma t/4}
			\right)\left( 
			e^{iM_L t}e^{\Delta\Gamma t/4} 
			\pm e^{iM_H t}e^{-\Delta\Gamma t/4}
			\right)\\
	&= \frac{e^{-\Gamma t}}{4} \left( 
		e^{\Delta\Gamma t/2}  
		\pm 
		e^{i(M_H-M_L) t}
		\pm
		e^{i(M_H-M_L) t}
		+
		e^{-\Delta\Gamma t/2} 
	\right)\\
	&= \frac{e^{-\Gamma t}}{2} \left(\cosh \frac{\Delta\Gamma t}{2}
	\pm \cos(\Delta M t) \right).
\end{align}
Similarly, 
\begin{align}
	g_\pm(t)^* g_\mp(t)&= \frac 14 
	\left(
	e^{-iM_L t}e^{-\Gamma_L t/2} 
	\pm e^{-iM_H t}e^{-\Gamma_H t/4}
	\right)^*
	\left(
	e^{-iM_L t}e^{-\Gamma_L t/2} 
	\mp e^{-iM_H t}e^{-\Gamma_H t/4}
	\right)\\
	&= 
	\frac 14 
	\left(
	e^{iM_L t}e^{-\Gamma_L t/2} 
	\pm e^{iM_H t}e^{-\Gamma_H t/4}
	\right)
	\left(
	e^{-iM_L t}e^{-\Gamma_L t/2} 
	\mp e^{-iM_H t}e^{-\Gamma_H t/4}
	\right)\\
	&=
	\frac 14 
	\left(
	e^{-\Gamma_L t} 
	\mp e^{i(M_L-M_H)t}e^{-(\Gamma_L+\Gamma_H)t/2}
	\pm e^{i(M_H-M_L)t}e^{-(\Gamma_H+\Gamma_L)t/2}
	- e^{-\Gamma_H t} 
	\right)\\
	&=
	\frac{e^{-\Gamma t}}{4} 
	\left(
	e^{\Delta\Gamma t/2} 
	\mp e^{-i\Delta M t}
	\pm e^{i\Delta M t}
	- e^{-\Delta\Gamma t/2} 
	\right).
\end{align}
This gives (recall $2i\sin\theta = e^{i\theta}-e^{-i\theta}$),
\begin{align}
	g_+(t)^*g_-(t) &= \frac{e^{-\Gamma t}}{2}\left[
		\sinh\frac{\Delta \Gamma t}{2} + i\sin(\Delta M t)
	\right]\\
	g_-(t)^*g_+(t) &= \frac{e^{-\Gamma t}}{2}\left[
		\sinh\frac{\Delta \Gamma t}{2} - i\sin(\Delta M t)
	\right].
\end{align}
Plugging this all back into (\ref{eq:CP:decay:derivation:Asquared}) we have
\begin{align}
	|A|^2 &= 
	\frac{e^{-\Gamma t}}{2} \left(\cosh \frac{\Delta\Gamma t}{2}
	+ \cos(\Delta M t) \right) |A_f|^2 
	+ \frac{e^{-\Gamma t}}{2} \left(\cosh \frac{\Delta\Gamma t}{2}
	- \cos(\Delta M t) \right)
		\left| \frac{q}{p} \bar A_f\right|^2
		\nonumber\\
	&\phantom{=}
	+ \frac{e^{-\Gamma t}}{2}\left[
		\sinh\frac{\Delta \Gamma t}{2} - i\sin(\Delta M t)
	\right]  A_f \left(\frac qp \bar A_f\right)^*
	+ 
	\frac{e^{-\Gamma t}}{2}\left[
		\sinh\frac{\Delta \Gamma t}{2} + i\sin(\Delta M t)
	\right]
	\frac qp A_f^* \bar A_f \nonumber.
\end{align}
What a mess. Let's simplify this by grouping together according to [hyperbolic] trigonomic function and also writing $\Delta \Gamma t/2 = y\Gamma$ and $\Delta M t= x \Gamma$.
\begin{align}
	2|A|^2e^{\Gamma t} \;=&
	\cosh (y\Gamma) \left( |A_f|^2 + \left|\frac qp \bar A_f \right|^2 \right)
	+\cos (x\Gamma) \left( |A_f|^2 - \left|\frac qp \bar A_f \right|^2 \right)
	\nonumber\\
	&+ \sinh (y\Gamma) \, 2\, \text{Re}\left[\frac qp A_f^* \bar A_f\right]
	- \sin (y\Gamma) \, 2\, \text{Im}\left[\frac qp A_f^* \bar A_f\right].
\end{align}
We've done all of this for $|B(t)\rangle$. For the corresponding equation where you start out as a $|\bar B\rangle$ at $t=0$, we note that $\bar B = (B_L-B_H)/q$. You can follow through the calculation to make the appropriate sign changes and $p/q \to q/p$. Anyway, it is useful to rewrite $|A|^2$ in terms of the \CP parameter $\lambda_f = (q/p)(\bar A_f/A_f)$, 
\begin{align}
	\frac{2|A|^2 e^{\Gamma t}}{|A_f|^2} \;=& 
	 \left( 1 + \left|\lambda_f\right|^2 \right)\cosh (y\Gamma)
	+ \left( 1 - \left|\lambda_f \right|^2 \right) \cos (x\Gamma)
	\nonumber\\
	&+ \sinh (y\Gamma) \, 2\, \text{Re}\, \lambda_f
	- \sin (y\Gamma) \, 2\, \text{Im}\, \lambda_f
\end{align}

\section{Solutions to Problems}\label{app:solutions}
\Closesolutionfile{answers}
\input{answers1}

\bibliographystyle{utcaps} 	
\bibliography{FlavorNotes}

\end{document}

%% file: answers1.tex
\begin{Solution}{1.1}
These are trivial to look up in the \acro{PDG}. The paper version is
slightly more satisfying to thumb through, while the online version is
kinder to the environment.
\begin{enumerate}
\item The easiest place to look is the Meson Summary Table. The $D^+$
is a $c\bar d$ bound state with mass $m_{D^+}\approx 1870$ MeV.
\item The easiest place to look is the Baryon Summary Table. The
$\Lambda$ is a $uds$ bound state with spin $1/2$.
\item The easiest place to look is the Lepton Summary Table. (Do you
see a theme?) The $\tau$ will decay to $\mu \nu\bar\nu$ 17\% of the
time.
\item The $B^+$ meson has a lifetime of $\tau_{B^\pm} = 1.6 \times
10^{-12}$ s. This translates to a width of
\begin{align}
\Gamma_{B^+} = (1.6 \times 10^{-12} \text{ s})^{-1}\times \frac{1
\text{ GeV}}{1.5 \times 10^{24} \text{ s}^{-1}} = 4.4 \times 10^{-13}
\text{ GeV} = 4.4 \times 10^{-4} \text{ eV}.
\end{align}
\item The \acro{PDG} tells us that $c\tau = 491$ $\mu$m. With a boost
factor $\gamma = 4$ this gives us $\gamma c\tau = 1.9 $ mm.
\end{enumerate}
\end{Solution}
\begin{Solution}{2.1}
		Let us review our counting for real parameters and phases. Each Yukawa matrix has $N^2$ elements which can each be written as $y_{ij} = r_{ij}e^{i\theta_{ij}}$. Thus each Yukawa matrix contains $N^2$ real parameters and $N^2$ phases. There are two Yukawas (up and down), so we have a total of $2N^2$ real parameters and $2N^2$ phases.

		We would now like to identify the number of broken generators. An element of $\text U(N)$ is an $N\times N$ complex matrix ($N^2$ real parameters and $N^2$ phases) satisfying
		\begin{align}
			UU^\dag = \mathbbm{1}.
		\end{align}
		This gives one equation of constraint for each complex component for a total of $N^2$ parameters per unitary rotation.  How many of these parameters are real? This is precisely the number of parameters of an element of $\text O(N)$, namely $\frac 12 N(N-1)$. The remaining $\frac 12 N(N+1)$ parameters are phases.

		In the quark sector of our $N$ generation Standard Model, we have $2N^2$ real parameters and $2N^2$ phases coming from the $y^u$ and $y^d$ matrices. How many of these are physical? We subtract the number of broken generators. For generic Yukawa matrices with no underlying structure, the $\text U(N)^3$ flavor symmetry of the $Q, u_R, d_R$ is broken down to $\text U(1)_B$.  Thus we have $3\times \frac{1}{2}N(N-1)$ broken generators associated with real parameters and $3\times \frac 12 N(N+1) -1$ broken generators associated with phases. This gives a total of
		\begin{align}
			2N^2 - \frac 32 N(N-1) &= \frac 12 N(N+3) \text{\; physical real parameters}\\
			2N^2 - \frac 32 N(N+1) + 1 &= \frac 12 (N-2)(N-1) \text{\; physical phases}.
		\end{align}
		Of the real parameters, $2N$ of them are Dirac masses and so $\frac 12 N(N-1)$ must be mixing angles---this is consistent with our counting of O($N$) generators. Let us remark that for the full $N$-generation Standard Model we must also add $N$ real parameters for the lepton masses, the 3 real parameters for the gauge couplings, and the two real parameters governing the Higgs sector.
	
\end{Solution}
\begin{Solution}{2.2}
		The problem already defined the three main ingredients of a model. It's up to us to now write the most general Lagrangian and identify the number of physical parameters. We will only consider the quark sector since all other sectors are the same as the Standard Model. There are five matter fields in the quark sector ($Q_L$, $u_R$, $d_R$, $s_L$, $s_R$). The only difference from the Standard Model is that the $s_L$ is an $\text{SU}(2)_L$ singlet so that it must have $Y=-1/3$ to obtain the correct electric charge. The kinetic terms have a $\text U(2)\times \text U(1)^3$ symmetry where the $\text U(2)$ corresponds to mixing the $d_R$ and $s_R$ fields. Note that the $s_L$ cannot mix with these fields even though it seems to have the same quantum numbers as the $s_R$; the $s_L$ and $s_R$ are different Lorentz representations. In other words, $s_L$ is a left-chiral spinor $\chi_\alpha$ while $s_R$ is a right-chiral spinor $\bar\psi^{\dot\alpha}$. As one can see from the indices, these transform differently under rotations. Using the counting introduced in the previous problem for the number of real parameters and phases in unitary matrices, we see that the kinetic terms have a symmetry group with 1 real parameter generator and 6 phase generators.

		Let us now identify how many total parameters are in the Lagrangian and how many of of the above symmetries are broken. The most general quark Yukawa sector is
		\begin{align}
			\mathcal L_{\text{Yuk.,} Q} &= y_u \bar{Q}_LH u_R + y_d \bar{Q}_L\tilde H d_R + y_s \bar{Q}_L\tilde H s_R + \text{h.c.},\label{eq:sol:exotic:quarks:Yuk}
		\end{align}
		where the $y_i$ are complex numbers so that we have 3 real parameters and 3 phases. In the Standard Model the Yukawa sector was the only source of additional quark interactions. In our exotic model, however, the fact that the $s_L$ is a singlet allows us to write additional bare mass terms,
		\begin{align}
			\mathcal L_{\text{mass}} &= m_s \bar{s}_L s_R + m_d \bar{s}_L d_R + \text{h.c.}\label{eq:sol:exotic:quarks:mass}
		\end{align}
		Each $m_a$ is a complex number so that this gives 2 additional real parameters and 2 additional phases. We now have a total of 5 real parameters and 5 phases. What symmetries are preserved by these Lagrangian terms? We are left with an analog of $U(1)_B$ where the left-handed and right-handed fields all transform with the same phase.  This means that of the symmetries in the kinetic sector, the one real parameter and 5 of the 6 phases are broken. The number of physical parameters is equal to the number of total parameters minus broken generators, so this gives us a total of four real parameters and no phases.

		Does this make sense? Three real parameters are associated to the Dirac masses of the particles. We have one left over real parameter, but we can see that when the Higgs gets its electroweak symmetry-breaking vev there is a mass term mixing the $s$ and $d$ quarks so that the mass term after EWSB looks like
		\begin{align}
			\mathcal L \supset
			\begin{pmatrix}
				\bar d_L & \bar s_L
			\end{pmatrix}
			\begin{pmatrix}
				y_d \frac{v}{\sqrt{2}} & y_s \frac{v}{\sqrt{2}}\\
				m_d & m_s
			\end{pmatrix}
			\begin{pmatrix}
				d_R\\
				s_R
			\end{pmatrix}
			+ \text{h.c.}\label{eq:sol:exotic:quarks:matrix}
		\end{align}
		so that we can see that the left over real parameter gives the $d-s$ mixing angle.
	
\end{Solution}
\begin{Solution}{2.3}
		A careful counting by Dimopoulos and Sutter gives 110 parameters: 30 masses, 29 mixing angles, and 41 phases \cite{Dimopoulos:1995ju}. You are referred to their paper for a very nice analysis. See also a nice presentation by Haber \cite{Haber:2000jh} for some variants of the MSSM.
	
\end{Solution}
\begin{Solution}{2.4}
		The conventional explanation for why the $\text{SU}(2)_L$ $\Theta$ angle is not physical comes from the use of the anomalous $\text{U}(1)_{B+L}$ symmetry to rotate it away. (One cannot do this in \acro{QCD} with $\text{U}(1)_A$ since this would give the quark masses an imaginary component.) A cute way to check this is to count the number of physical parameters in the $\text{SU}(2)_L$ sector at a \textit{quantum} level and check that they are all accounted for \textit{without} having to introduce a $\Theta_{\text{L}}$ term. We can cast this argument in a different way in light of our parameter counting. We saw that the $\text{U}(1)_A$ anomaly of \acro{QCD} led to the appearance of another physical parameter at the quantum level, $\Theta_{\text{QCD}}$. If we look at the $\text{SU}(2)_L$ sector of the Standard Model, we might also expect an additional parameter to appear, $\Theta{\text{weak}}$, because of the $\text{U}(1)_{B+L}$ anomaly. However, the key point is that $\text{U}(1)_{B+L}$ isn't broken by the Yukawas, and so it was never counted as a `broken symmetry' in our tree-level analysis. Thus the $\text{U}(1)_{B+L}$ anomaly doesn't lead to one less broken symmetry and so does not introduce an additional physical parameter. In other words, $\Theta_{\text{Weak}}$ is not physical.  By the way, a useful way to remember which global symmetries are broken anomalies is to look at the instanton that induces the appropriate $\Theta$ term. In this case the sphaleron violates $B+L$ while preserving $B-L$.
	
\end{Solution}
\begin{Solution}{3.1}
		See \cite{Silva:CPviolation}; one of the authors of that book attended the lecture where this problem was posed and nodded approvingly when the book was referenced.
	
\end{Solution}
\begin{Solution}{3.2}
		This is straightforward: use (area) = (base) $\times$ (height). To be somewhat pedantic, take the following steps
		\begin{enumerate}
			\item Normalize the triangle so that one side has unit length in the real direction.
			\item The height of the triangle is given by the imaginary part of one of the other (normalized) terms in the sum.
			\item Use the formula for the area of a triangle.
			\item Return to the previous normalization by multiplying this area by the absolute value of the side that we scaled and rotated, squared (because this is an area).
		\end{enumerate}
	
\end{Solution}
\begin{Solution}{3.3}
		You are free to pick any two unitary triangles you want, but two of them are particularly easy. Looking at the standard \CKM parameterization (\ref{eq:CKM:standard:parameterization}) we see that $V_{ud}$, $V_{us}$, $V_{cb}$, and $V_{tb}$ are purely real. Thus there are two triangles that have one leg parallel to the real axis:
		\begin{align}
			\sum_i V_{id}V^*_{is} &= 0\label{eq:hw2:1:tri:1}\\
			\sum_i V_{cj}V^*_{tj} &= 0.\label{eq:hw2:1:tri:2}
		\end{align}
		Thus the base of the triangle (\ref{eq:hw2:1:tri:1}) is
		\begin{align}
			V_{ud}V_{us} &= c_{12}c_{13}s_{12}c_{13},
		\end{align}
		while the base of the triangle (\ref{eq:hw2:1:tri:2}) is
		\begin{align}
			V_{cb}V_{tb} &= s_{23}c_{13}c_{23}c_{13}.
		\end{align}
		Great. That's the easy part. To calculate the height it is sufficient to take the imaginary part of either of the remaining terms in the sum. For triangle (\ref{eq:hw2:1:tri:1}) we'll take $V_{cd}V^*_{cs}$
		\begin{align}
			\left|\text{Im}\left(V_{cd}V^*_{cs}\right)\right| &= \left|\text{Im}\left( s_{12}^2s_{13}c_{23}s_{23} e^{-i\delta} - c_{12}^2s_{13}c_{23}s_{23}e^{i\delta} \right)\right|\\
			&= \sin\delta \, s_{13}c_{23}s_{23}.
		\end{align}
		For triangle (\ref{eq:hw2:1:tri:1}) we'll take $V_{cs}V^*_{ts}$,
		\begin{align}
			\left|\text{Im}\left(V_{cs}V^*_{ts}\right)\right| &= \left|\text{Im}\left( -c_{12}s_{12}s_{13}c^2_{23}e^{-i\delta} + c_{12}s_{12}s_{13}s^2_{23} e^{i\delta} \right)\right|\\
			&= \sin\delta\, c_{12}s_{12}s_{13}.
		\end{align}
		Comparing the (base)$\times$(height) of each triangle, we find that both triangles have area equal to one half of
		\begin{align}
			c_{12}s_{12}s_{13}c_{13}^2c_{23}s_{23}\,\sin\delta.
		\end{align}
		This is indeed our definition of the Jarlskog, (\ref{eq:Jarlskog:standard}).
	
\end{Solution}
\begin{Solution}{3.4}
	Let us consider the low-energy Lagrangian for the strong force in which only the light quark species ($u$, $d$, $s$) are active. This is the underlying structure of Murray Gell-Mann's eightfold way for light hadron classification. The three ingredients for our model are:
\begin{enumerate}
	\item \textsc{Gauge group}: SU$(3)_\text{c}$. We ignore U$(1)_\text{EM}$ as a small perturbation.
	\item \textsc{Matter representations}: for each flavor ($u$, $d$, $s$) we have a left-chiral fundamental representation and a right-chiral fundamental representation. Thus we have:

	\begin{center}
		\begin{tabular}{rlcrl}
			$u_L$ & \;$\mathbf{3}$ &\quad\quad\quad& $u_R$ & \;$\mathbf{3}$\\
			$d_L$ & \;$\mathbf{3}$ &\quad\quad\quad& $d_R$ & \;$\mathbf{3}$\\
			$s_L$ & \;$\mathbf{3}$ &\quad\quad\quad& $s_R$ & \;$\mathbf{3}$.
		\end{tabular}
	\end{center}

	\item \textsc{Spontaneous symmetry breaking}: at \textit{tree level} there is none. At the quantum level SU$(3)_A$ is broken spontaneously by the \QCD chiral condensate, $\langle \bar{q}_L q_R + \bar{q}_R q_L \rangle $.
\end{enumerate}

As before we write out the most general renormalizable Lagrangian,
\begin{align}
	\mathcal L = \sum_{i=u,d,s} \left(\bar{q}^i_L i\slashed{D} q^i_L + \bar{q}^i_R i\slashed{D} q^i_R\right) + \left(m_{ij} \bar{q}_L^i q_R^j + \text{h.c.}\right).
\end{align}
Note the important difference between this low-energy \QCD Lagrangian and the Standard Model: the flavor symmetry of the kinetic term allows one to rotate up, down, and strange quarks (of a given chirality) between one another! In the Standard Model this is prohibited because this mixes different components of SU$(2)_L$ doublets or, alternately, because this mixes particles of different charge. Do not confuse this U$(3)_L\times$U$(3)_R$ flavor symmetry with the U$(3)_L\times$U$(3)_R$ flavor symmetry of the Standard Model's quark sector which mixes the three \textit{generations} between one another. There are a lot of 3's floating around, make sure you don't mix them up.

It is useful to write down this flavor symmetry in terms of vector and axial symmetries,
\begin{align}
	U(3)_L\times U(3)_R &= U(3)_V\times U(3)_A,
\end{align}
where the vector (axial) transformation corresponds to
\begin{align}
	q_L^i \to q_L'^i = U^i_{\phantom{i}j}q_L^j \hspace{3cm} q_R^i \to q_R'^i = U^{(\dag)i}_{\phantom{(\dag)i}j}q_R^j.
\end{align}
In other words, the vector symmetry rotates the left- and right-handed quarks in the same way while the axial symmetry rotates them oppositely.

Now let's look at the mass terms of low-energy \QCD. We may diagonalize $m_{ij}$, find the eigenvalues, and write down the three physical mass parameters. Thus for a generic mass matrix we see that flavor is a good quantum number and the U$(3)_V\times$U$(3)_A$ symmetry of the kinetic term is broken down to U$(1)_V^3$ representing the phases of each flavor mass eigenstate. If, for some reason, the mass matrix is universal, $m\propto \mathbbm{1}$, then the mass terms break U$(3)_L\times$U$(3)_R$ to U$(3)_V$. We see that the mass terms always breaks the axial part of the kinetic term's flavor symmetry.

One might now ask the following clever question:
\begin{quote}
	\textit{Why should we include the mass terms in the \QCD Lagrangian at all since we know these come from the Yukawa sector of the Standard Model?}
\end{quote}
While this is true, we must remind ourselves that our model building rules tell us that we \textit{must} include all renormalizable terms that respect our gauge symmetries. For low-energy \QCD this means that we must include these mass terms. Now one might ask an even more clever question:
\begin{quote}
	\textit{Fine, we include these terms, but then we go out and measure them and they're very small. Shouldn't we still be able to ignore these terms since they are much smaller than the relevant mass scale, $\Lambda_{\text{QCD}}$?}
\end{quote}
Indeed! We know that $m_q \ll \Lambda_{\text{QCD}}$ for all active quarks at low energies, so the limit $m_{ij}\to 0$ (or alternately taking the dimensionless parameter $m_{ij}/\Lambda_{\text{QCD}}\to 0$) should be sensible. However, the point is that in the absence of masses the chiral condensate breaks the SU$(3)_A \subset$U$(3)_A$ and does so at the scale $\Lambda_{\text{QCD}}$ (recall that $\langle \bar{q_i}q_{j}\rangle \sim \Lambda_{\text{QCD}}^3\delta_{ij}$). Thus we really should assume that this symmetry is broken. One more clever retort:
\begin{quote}
	\textit{Fine! The SU$(3)_A$ should be broken either by mass terms or by the QCD condensate, but what about the remaining U$(1)_A\subset$U$(3)_A$?}
\end{quote}
If we do our parameter counting, we have $m_{ij}$ an arbitrary complex matrix with 18 real parameters. The breaking of U$(3)_V\to $U$(1)_V^3$ gives $3^2-3=6$ broken generators. SU$(3)_A$ is broken and gives us $3^2-1 = 8$ broken generators. Thus if we pretend that U$(1)_A$ is unbroken, we would have $18-14 = 4$ physical parameters. We know that the correct answer is 3, the three Dirac quark masses. What has happened?

The answer is that U$(1)_A$ \textit{is} broken: it is anomalous, i.e.\ broken by quantum effects. The U$(1)_A$ axial anomaly is notorious in the history in particle physics. For our purposes we must make the additional rule that at the \textit{quantum level}, anomalous symmetries are no good for counting parameters. If this is the case, then there's one less broken symmetry generator and so we expect there to be one more physical parameter relative to the classical analysis. Does such a parameter exist? Yes; it is precisely the non-perturbative $\Theta_{\text{YM}}$ term which transforms as a shift with respect to U$(1)_A$!

Thus there are two ways of counting parameters:
\begin{enumerate}
	\item \textsc{Classically}: look only at the most general, renormalizable perturbative Lagrangian (i.e.\ without $\Theta$ terms) and perform the counting using the classical symmetries without any regard as to whether or not they are broken by anomalies.
	\item \textsc{Quantum mechanically}: consider the most general, renormalizable Lagrangian \textit{including} non-perturbative terms and only consider non-anomalous symmetries.
\end{enumerate}
Both are consistent as long as you \textit{stay within the regime of the description}, in other words, if you're counting classical (tree-level) parameters, then don't include quantum effects like anomalies and $\Theta$ angles. Conversely, if you're counting quantum parameters, then you must include \textit{both} the effect of anomalies and the non-perturbative terms associated with them.
	
\end{Solution}
\begin{Solution}{4.1}
		Look up references on `atomic parity violation.' See, for example,  \cite{Bouchiat:Parity,Haxton:APV} for reviews. The main point is that part of the $Z$ coupling is axial, which means its coupling is proportional to a fermion's helicity. This is parity-odd unlike the electric force. The goal is then to devise experiments that are sensitive to this left--right asymmetry. This is rather non-trivial, and a na\"ive estimate of the size of such an asymmetry is \cite{Bouchiat:Parity}
		\begin{align}
			A_{\text{LR}} = \frac{P_L-P_R}{P_L+P_R}
		\end{align}
		where $P_{\text{L,R}} = |A_{\text{EM}}\pm A_{W}^\text{odd}|^2$. The weak amplitude $A_W \sim g^2/(q^2 + M_Z^2)$ where the characteristic momentum scale is given by the Bohr radius,
		\begin{align}
			q\sim \frac{1}{\alpha m_e }.
		\end{align}
		This gives us
		\begin{align}
			A_{\text{LR}} \approx \alpha^2\frac{m_e^2}{M_Z^2} \approx 10^{-15},
		\end{align}
		which is hopelessly small. Fortunately, there are various enhancement mechanisms which make the measurement of this quantity experimentally tractable. See Section 2.4 of \cite{Bouchiat:Parity} for an excellent discussion.
	
\end{Solution}
\begin{Solution}{4.2}
		Refer to the solution to Problem~\ref{prob:exotic:quarks} for background.
		\begin{itemize}
			\item In the interaction basis the couplings to the $Z$ are given by the usual formula
			\begin{align}
				g_z = g \cos\theta_W\, T^3 - g'\sin\theta_W \, Y.
			\end{align}
			From the particle content we see that the singlets $d_R$ and $s_L$ have the same $T^3$ and $Y$ quantum numbers. The other particles all have different quantum numbers and hence different couplings to the $Z$. From (\ref{eq:sol:exotic:quarks:matrix}) we see that the bare mass terms cause the mass matrix for the $d$ and $s$ quarks to be different from the Yukawa basis. Let us say that this mass matrix $M$ is diagonalized by $\hat M = W_L M W_R^\dag$. Then the rotations
			\begin{align}
				\begin{pmatrix}
					d'_{L,R}\\
					s'_{L,R}
				\end{pmatrix}
				=
				W_{L,R}
				\begin{pmatrix}
					d_{L,R}\\
					s_{L,R}
				\end{pmatrix}\label{eq:prob:exotic:II:rotation}
			\end{align}
			shift the interaction basis fields ($d,s$) to the mass basis fields $d',s'$. In terms of these fields, the coupling to the $Z$ in the kinetic term is written as
			\begin{align}
					\begin{pmatrix}
						\bar d'_{L,R} &
						\bar s'_{L,R}
					\end{pmatrix}
					W_L^\dag
						\begin{pmatrix}
							g_{Z}^{d_{L,R}} &\\
							& g_{Z}^{s_{L,R}}
						\end{pmatrix}
						W_R
						i\gamma^\mu
						Z_\mu
						\begin{pmatrix}
							d'_{L,R}\\
							s'_{L,R}
						\end{pmatrix}.\label{eq:prob:exotic:II:coupling}
			\end{align}
			The key point is that while $g_Z^{d_R} = g_Z^{s_R}$, i.e.\ the right-chiral coupling matrix is diagonal, $g_Z^{d_L}\neq g_Z^{s_L}$ so that the left-chiral couplings become non-diagonal after the bi-unitary rotation by $W_L$ and $W_R$. Thus there \textit{are} \FCNCs in the $Z$ coupling to the left-chiral down quark sector.
			\item There are no photon or gluon \FCNCs. The photon and gluon couplings are all universal with respect to particles within a given flavor representation.
			\item Because the Yukawa couplings and the mass matrix (\ref{eq:sol:exotic:quarks:matrix}) are not proportional to one another (i.e.\ not aligned) there are \FCNCs from the Higgs. Note that these can be easy to `hide' since the Higgs couplings to the light quarks is small.
			\item We now know that this model has tree-level \FCNCs. As a quick and dirty approximation we can expect that the magnitude of the flavor-changing neutral currents and charged currents should be of the same order of magnitude. However, looking at the PDG, we can consider characteristic kaon decays (kaons because the new flavor structure is only in the $s$-$d$ sector),
			\begin{align}
				\text{Br}(K^+ \to \mu^+ \nu_\mu) &= 64\%\\
				\text{Br}(K^0_L \to \mu^+\mu^-) &=  7 \times 10^{-9}.
			\end{align}
			We know that the leptons only couple to the bosons through the $\text{SU}(2)_L$ gauge bosons, so the first decay proceeds through the charged current and the second proceeds through the neutral current. It would be very difficult to explain this discrepancy if  a model has tree-level \FCNCs and hence we expect this model to be ruled out.
		\end{itemize}
	
\end{Solution}
\begin{Solution}{4.3}
		The Yukawa and mass terms in the Lagrangian for this model takes the form
		\begin{align}
			\mathcal L_{\text{Yuk.}+\text{mass}} = y_L \bar Q_L \phi s_R + y_R \bar Q_R \phi s_L + m_Q \bar Q_L Q_R + m_s \bar s_L s_R + \text{h.c.}
		\end{align}
		These terms break $\text{U}(1)^4$ flavor symmetry in the kinetic terms to a $\text{U}(1)$ overall phase rotation on each field. We thus expect a total of $4-0 = 4$ physical real parameters and $4-3=1$ phases. The Dirac mass matrix takes the form
		\begin{align}
			\begin{pmatrix}
				m_Q & y_L \frac{v}{\sqrt{2}}\\
				y_R \frac{v}{\sqrt{2}} & m_s
			\end{pmatrix},
		\end{align}
		while the up-quarks have a single mass term $m_Q \bar u_L + \bar u_R + \text{h.c.}$. We may choose the real physical parameters to be the Dirac masses for each generation and the $d$--$s$ mixing angle and the physical phase to be that of the up quark mass term. The coupling to the $Z$ boson in the interaction and mass bases follow as it did in the previous problem. In particular, in the mass basis one must perform a rotation (\ref{eq:prob:exotic:II:rotation}) so that the $Z$ couplings take the form (\ref{eq:prob:exotic:II:coupling}). Now both the left- and right-handed $d$ and $s$ quarks have different $\text{SU}(2)_L\times U(1)_Y$ quantum numbers so that $g_Z^d \neq g_Z^s$ and so the rotation introduces FCNCs through the $Z$ for both chiralities.

		As before the photon and gluon still do not mediate \FCNCs because their couplings are all universal with respect to particles within a given flavor representation. The Higgs again induces \FCNCs because the bare mass terms prevent the mass matrix and Yukawa matrix from being proportional to one another. The appearance of \FCNCs in the $Z$ and Higgs sectors lead us to expect this model to be ruled out experimentally.
	
\end{Solution}
\begin{Solution}{4.4}
		For a nice history and collection of references about the 2\acro{HDM}, see \cite{Haber:2000jh}.
		\begin{itemize}
			\item The most general Yukawa potential takes the form
			\begin{align}
				\mathcal L_{\text{Yuk.}} = (y^{1u}_{ij}H_1 + y^{2u}_{ij}\tilde H_2)\bar Q^i u^j + (y^{1d}_{ij}\tilde H_1+y^{2d}_{ij}\tilde H_2)\bar Q^i  d^j + \text{h.c.}
			\end{align}
			\item The diagonalization procedure proceeds as usual for the Standard Model except that the mass matrices are now
			\begin{align}
				m_{ij}^u = y^{1u}_{ij}\frac{v_1}{\sqrt{2}} + y^{2u}_{ij}\frac{v_2}{\sqrt{2}}
				\hspace{3cm}
				m_{ij}^d = y^{1d}_{ij}\frac{v_1}{\sqrt{2}} + y^{2d}_{ij}\frac{v_2}{\sqrt{2}}.
			\end{align}
			These are diagonalized by some bi-unitary transformation. As far as the kinetic terms are concerned, this rotation on the quark fields has the same effect as the rotation in the Standard Model. In particular, the $Z$ couplings are still universal and remain flavor-diagonal.
			\item It is straightforward to see that there are \FCNCs in the Higgs sector. For example, the neutral scalar Higgs components couple to the fermions by replacing $v_i/\sqrt{2}$ in the mass matrices with $h_a$ for $a=1,2$ labeling each the two Higgs doublets. It is clear that in general the mass matrix is different from the Yukawa matrix for either Higgs since, e.g.\ $m^u_{ij} \neq y^{au}_{ij}$ for either $a=1,2$. There are, of course special cases; for example if these Higgses are allowed to mix in such a way that the mass eigenstate couples to combinations of Yukawas proportional to $m^u$ and $m^d$. We can also note that in the limits $\tan \beta \equiv v_2/v_1 \to 0, \infty$ the Yukawas align with the mass matrices because one of the vevs vanishes. It is generally true in any multi-Higgs model that \FCNCs are avoided in the Higgs sector so long as all the fermions of a given electric charge only couple to a single Higgs.
			\item Higgs \FCNCs are typically stronger for heavier quarks, though in this 2\acro{HDM} one can arrange for a cancellation between large couplings in the Yukawas. A natural place to rule out this type of is in the \FCNC constraints on the $B$ mesons.
		\end{itemize}
	
\end{Solution}
\begin{Solution}{6.1}
		This is a conversion to natural units. A quick-and-dirty way to get this is to use the Heisenberg relation $\Delta E \Delta t \sim \hbar$.
	
\end{Solution}
\begin{Solution}{6.2}
This comes from the representation theory of SU(2) isospin. There's a natural point of confusion: you are familiar from the addition of angular momenta that $\vec 2\otimes \vec 2 = \vec 3 \oplus 1$: that is, the combination of two doublets (spin-1/2) combine into a triplet and a singlet. The neutral ($S_z = 0$) triplet state is symmetric, $\sim\left| \uparrow\downarrow\right\rangle + \left| \downarrow\uparrow\right\rangle$. Na\"ively, we might then expect that the pion---the $I_3=0$ state of an isotriplet---should also have a plus sign: $\pi^0 \sim u\bar u + d\bar d$.
		The plus sign misses something critical: the pion is composed of a quark and an an anti-quark. Thus we're not combining two isodoublets, $\vec 2 \otimes \vec 2$, we're combining and isodoublet with an anti-isodoublet, $\vec 2 \otimes \bar{\vec 2}$.

		Fortunately, SU(2) is special because it is a pseudo-real representation. This provides us a trick to convert an anti-doublet $\bar{\vec 2}$ into a doublet. We've already used this trick when we write the Yukawa terms in the Standard Model: we had to define a conjugate Higgs field, $\tilde H = \epsilon^{ij} H_i^*$. The ``intuitive physics way'' of understanding this is that when we work with SU($N$), we are allowed to construct objects using the $N$-component anti-symmetric tensor, $\epsilon^{i_1\cdots i_N}$. For the case of SU(2), $\epsilon^{ij}$ and its inverse $\epsilon_{ij}$ can be used to raise and lower indices. These raised and lowered indices can be understood as column versus row vectors. In other words, it lets us convert a conjugate field into an object that transforms like a non-conjugate field at the cost of the minus signs that come from the $\epsilon$ tensor. This is precisely the origin of the relative minus sign between the $u\bar u$ and $d\bar d$ terms in the pion composition.
			For explicit matrices, see Halzen \& Martin's \emph{Quarks and Leptons}, Section 2.7.

			There is an alternative way to see this from the theory of Goldstone bosons. Since the pion triplet is identified with the Goldstones of isospin breaking, one can use the Callan--Coleman--Wess--Zumino formalism (\acro{CCWZ}) to `pull out' these fields from the matter fields charged under the broken symmetry. The details are beyond the scope of this solution---see the Appendix of~\cite{Csaki:2016kln} for an introduction---but the essence is that the relative minus sign between the $u\bar u$ and $d\bar d$ terms comes from the $\sigma^3/2$ generator of the spontaneously broken SU(2) acting on an isodoublet. It is perhaps amusing that these two pictures of understanding the relative minus sign are equivalent, but in one case it comes from $\epsilon \sim i \sigma^2$ and in the other it comes from $\sigma^3/2$.
	
\end{Solution}
\begin{Solution}{6.3}
They do not mix because the kaon has a flavor quantum number that is conserved
under \acro{QCD} and \acro{QED}. There is a tiny mixing due to the weak interaction,
but it is so small that we can only dream of ever detect it.
\end{Solution}
\begin{Solution}{6.4}
		A convenient basis is $\left\{|u\bar u\rangle, |u\bar d\rangle, |d \bar u\rangle, |d\bar d\rangle\right\}$. We can convert to the isospin basis using
		\begin{align}
			|0,0\rangle &= \frac{1}{\sqrt{2}}\left(|u\bar u\rangle - |d\bar d\rangle \right)\\
			|1,-1\rangle &= |d\bar u\rangle\\
			|1,0\rangle &= \frac{1}{\sqrt{2}}\left(|u\bar u\rangle + |d\bar d\rangle \right)\\
			|1,1\rangle &= |u\bar d\rangle
		\end{align}
		where we've written states according to $|I,I_3\rangle$. Note that $u$ and $\bar d$ are analogous to $|\uparrow\,\rangle$ while $\bar u$ and $d$ are analogous to $|\downarrow\,\rangle$ in the usual SU$(2)$ spin notation. (To see why just remember that the kinetic term must be isospin invariant and that the SU$(2)$ metric is the antisymmetric tensor.)
		Let us say that the singlet gets a mass $m_1$ and the triplet states get mass $m_3$. The mass matrix thus takes the form
		\begin{align}
			\begin{pmatrix}
					2m_u^2 + \frac 12 m_3^2 + \frac 12 m_1^2	&	&	& \frac 12 m_1^2 - \frac 12 m_3^2 \\
						& m_u^2+m_d^2 + \frac 12 m_3^2 	&			& \\
						&		& m_u^2+m_d^2 + \frac 12 m_3^2	& \\
			 	\frac 12 m_1^2 - \frac 12 m_3^2	&	&	& 2m_d^2 + \frac 12 m_3^2 + \frac 12 m_1^2
			\end{pmatrix}.
		\end{align}
		From here one can happily plug into \textit{Mathematica} to solve for the exact eigenstates and eigenvalues. As a sanity-check note that in the limit ${m_{u,d}\to 0}$ the $u\bar u$ and $d\bar d$ states mix with a $45^\circ$ angle giving the $\pi^0$ and $\eta$. The $u\bar d$ and $d \bar u$ states do not mix because both bases conserve $I_3$. Note further that electromagnetism also preserves the $I_3$ quantum number.
	
\end{Solution}
\begin{Solution}{6.5}
Do not be confused by the fact that the \acro{LHC} is a multi-TeV collider! We would produce these heavy hadrons just as well at \acro{HERA} as we would at the \acro{LHC}. The deep principle in play here is \textit{decoupling}. The \textit{hard scattering} event which occurs at high scales is likely to produce any of the quarks since $m_q \ll $TeV. These quarks propagate away from one another for a distance scale on the order of $1/\Lambda_{\text{QCD}}$ and it is \textit{only} at that distance scale that hadronization from \QCD really becomes active. This is a fact that is very well known by anyone who works with Monte Carlo generators: you use MadGraph for the hard scattering, but Pythia for hadronization and showering. You can separate these steps because they occur at two totally different scales and are governed by different physics.
\end{Solution}
\begin{Solution}{6.6}
		See Griffiths \textit{Introduction to Elementary Particles}, Chapter 5.6.2--5.6.3 (Chapter 5.10 in the first edition).
	
\end{Solution}
\begin{Solution}{7.1}
		Because $q = p_\ell + p_\nu$, $q^2$ must equal the invariant mass of this system and must be positive definite. As a crutch, you can imagine that $p_\ell$ and $p_\nu$ is a bound state with some definite $m_{\text{bound}}^2$.
	
\end{Solution}
\begin{Solution}{7.2}
		For this calculation one needs to write in explicit projection operators. The amplitude takes the form
		\begin{align}
			\mathcal M = \frac{g^2}2 V_{ud} \bar v_d \gamma^\mu P_L u_u \frac{-i}{M_W^2} \bar u_\mu \gamma_\mu P_L v_\nu.
		\end{align}
		The first spinor contraction is just $\frac 12 \langle \pi^+ |A^\mu | 0\rangle$, where $A^\mu$ is the axial current. We have used the result example~\ref{eg:vector:vanishes:in:decay:constant} to neglect the vector piece, which vanishes by parity. Note the factor of 1/2 coming from the fact that we're only taking the axial part of the projection operator, $\gamma^\mu P_L = \frac 12\gamma^\mu (1-\gamma^5)$. Plugging in the expression for the pion decay constant,
		\begin{align}
			\mathcal M &= -\frac{g^2}{4 M_W^2}V_{ud}f_\pi \bar u \slashed{p}P_L v\\
			&= -\frac{g^2}{4 M_W^2}V_{ud}f_\pi m_\mu \bar u P_L v,
		\end{align}
		where in the second line we used the equation of motion for the outgoing charged lepton\footnote{We've been a little sloppy here, but there is some elegance to being able to do a sloppy-but-accurate calculation. The pion momentum $p$ should really be written in terms of $p=p_\mu + p_\nu$ and for each term the equation of motion for the appropriate lepton should be used. We already know, however, that $\slashed{p}_\nu v_\nu = 0$ since the neutrino is massless in the Standard Model.}. The benefit of this calculation is that one can use all of the usual Feynman rules from Peskin. However, things become a bit more clunky with factors of $\gamma^5$. These can become distracting for more complicated problems.

		The significance of the mass insertion is the chirality flip  in this decay. The pion is spin zero and the $W$ mediating the decay only couples to left-handed particles. In order to preserve angular momentum, one requires an explicit mass insertion on the final state fermions. Note that the use of the equation of motion is equivalent to a mass insertion on an external leg when using massless chiral fermions.
	
\end{Solution}
\begin{Solution}{7.3}
		The formula for the decay rate is
		\begin{align}
			d\Gamma &= \frac{1}{32\pi^2}|\mathcal M|^2 \frac{|\mathbf{p_\ell}|}{m_\pi^2}d\Omega,\label{eq:prob:pi:mu:nu:differential:decay:rate}
		\end{align}
		where we've written the muon subscript as $\ell$ to avoid confusion with Lorentz indices. In case you didn't feel inclined to re-derive this formula and your favorite quantum field theory textbook isn't nearby, then you can always look this up in the kinematics review of the pocket \PDG---which should \textit{always} be nearby.

		The first step is to square the matrix element and sum over spins. This matrix element should be particularly simple because we only have one spinor bilinear. This is worked out in most \acro{QFT} textbooks so we won't belabor the four lines of work required to extract the relevant expression. Writing $p_\ell$ for the lepton four-vector and $p_n$ for the neutrino four-vector (to avoid confusion with Lorentz indices) we obtain from (\ref{eq:pi:to:mu:nu}),
		\begin{align}
			\sum_s |\mathcal M|^2 = \left(\frac{g^2}{4M_W^2}\right)^2 f_\pi^2 |V_{ud}|^2 m_\ell^2 \, 2 p_\ell\cdot p_n.\label{eq:prob:pi:mu:nu:spin:sum}
		\end{align}
		The factor of $2 p_\ell\cdot p_n$ is simply the result of summing over the spins of the spinor structure $|\bar\mu_R \nu_L|^2$. Note that this is \textit{trivial} and just comes from the completeness relations of the plane wave spinors.

		\begin{framed}
			\noindent\textbf{Some strategy}. It is important to get the right answer, but it is also important to do so in a way that doesn't make your life difficult. This toy calculation is an important example. We have made use of the fact that we are only looking at the axial current so that our amplitude takes the form
			\begin{align}
				\mathcal M \propto f_\pi (p_\ell+p_n)_\alpha \bar\mu_L \gamma^\alpha \nu_L.
			\end{align}
			One could have na\"ively taken the square of the spinorial part as written. If one were to work with Dirac spinors, this would be a terrible mess,
			\begin{align}
				\bar u(p_\ell) \gamma^\alpha \frac{1}{2}(1-\gamma^5) v(p_n).
			\end{align}
			Squaring and summing over spins requires taking a trace over four $\gamma$ matrices times $(1-\gamma^5)$ and one obtains a funny relation in terms of two-index bilinears in $p_\ell$ and $p_n$ including an ugly term with an $\varepsilon^{\alpha\beta\mu\nu}$. See, e.g.\ equation (5.19) in Peskin and Schroeder \cite{Peskin}. One will eventually find that the $\varepsilon$ term cancels because it is contracted with $(p_\ell+p_n)_\mu(p_\ell+p_n)_\nu$. This is \textit{a lot of work!}

			We can do a little better by using Weyl spinors. The simplification is obvious: the $\varepsilon$ term drops out from the very beginning because we never have to deal with the annoying $\gamma^5$ matrix. The relevant relations to perform the analogous calculation in Weyl space can be found in \cite{Dreiner:2008tw}; or in a pinch you can extract them from the Dirac spinor relations. This is indeed a meaningful simplification and justifies the use of Weyl spinors in `actual calculations' (rather than just abstractly to refer to chiral fields); however, we can do better!

			What we did above in (\ref{eq:prob:pi:mu:nu:spin:sum}) was to go a step further and get rid of the Dirac/Pauli matrix structure altogether. This was easy: we just used the fact that the $\gamma^\mu$ (alternately $\bar\sigma^\mu$) contracted a $(p_\ell+p_n)_\mu$. This means we can use the equation of motion to explicitly pull out the mass factor $m_\ell$, where have used $m_n = 0$. Now we get an expression of the form $\bar u(p_\ell) v(p_n)$ in Dirac notation, or in Weyl notation $\psi(p_\ell)\chi(p_n)$. Squaring and summing over spins is trivial for these since it is just the completeness relation for the plane wave spinors.
		\end{framed}

		It is trivial to determine the contraction $2p_\ell\cdot p_n$ since this is constrained by the kinematics,
		\begin{align}
			m_\pi^2=(p_\ell+p_n)^2 = m_\ell^2 + 2p_\ell\cdot p_n.
		\end{align}
		Thus we are led to (we drop the $\sum_s$ symbol)
		\begin{align}
			|\mathcal M|^2 = \left(\frac{2}{\sqrt{2}}G_F\right)^2 f_\pi^2 |V_{ud}|^2 m_\ell^2 \, m_\pi^2\left(1-\frac{m_\ell^2}{m_\pi^2}\right),
		\end{align}
		where we've also replaced the $g^2/8M_W^2$ with $G_F/\sqrt{2}$ according to the definition of the Fermi constant (\ref{eq:Gf:definition}). Let's simplify the other factors in (\ref{eq:prob:pi:mu:nu:differential:decay:rate}). The magnitude of the muon three-momentum can be derived trivially from kinematics; conservation of four-momentum gives us
		\begin{align}
			|\mathbf{p_n}|&=|\mathbf{p_\ell}|\\
			m_\pi^2&=|\mathbf{p_n}|+\sqrt{|\mathbf{p_\ell}|^2+m_\ell^2}.
		\end{align}
		From this we obtain
		\begin{align}
			|\mathbf{p_\ell}| &= \frac{m_\pi}{2}\left(1-\frac{m_\ell^2}{m_\pi^2}\right).
		\end{align}
		The angular integral is trivial
		\begin{align}
			d\Omega = d(\cos\theta)d\phi = 4\pi,
		\end{align}
		and we can now plug in all of these factors to obtain the result we wanted
		\begin{align}
			\Gamma &= \frac{1}{8\pi}G_F^2f_\pi^2 |V_{ud}|^2 m_\ell^2m_\pi \left(1-\frac{m_\ell^2}{m_\pi^2}\right)^2.
		\end{align}
	
\end{Solution}
\begin{Solution}{7.4}
		This problem is an excellent example of a form factor multiplying a non-trivial combination of dynamical variables.
		\begin{enumerate}
			\item The key difference between the $D$ and $D^*$ is that the latter has a polarization vector. Without this the parameterization of the matrix element reduces to (\ref{eq:vector:form:factors}).
			\item The dynamical variables available to us are the $D^*$ polarization $\epsilon$ and the two meson momenta $p_B$ and $p_D$. Eventually we will repackage the momenta into $(p_B+p_D)$ and $q\equiv (p_B-p_D)$. Now we would like to use spacetime symmetries to determine the structure of the parameterization. The $V^\mu$ operator contains a Lorentz index, so we know the right-hand side must be a vector or axial vector. The matrix element on the left-hand side is necessarily parity-even since \QCD respects $P$, so the right-hand side must be vectorial. In the next part we will use $T$ to determine that $g(q^2)$ is real, but note that we \textit{never} need to consider $C$ since QCD also respects charge conjugation and the right-hand side is composed of dynamical variables which are all trivially even under $C$. Thus all we have to worry about is constructing a Lorentz vector out of $\epsilon$, $P_B$, and $p_D$.

			There is a subtlety in the parity of vectorial objects. The parity of a pure vector $V^\mu$ can be written as $P[V^\mu]=(-)^{\mu}$, by which we mean the parity is even (+) for $\mu=0$ and odd ($-$) for $\mu=1,2,3$. Note that the $\mu$ on $(-)^\mu$ isn't a Lorentz index, it's just there to tell us whether $(-)=+$ or $-$. Thus a $J^P = 1^-$ meson has parity $-(-)^\mu$. We see that the parity of the terms in the matrix element are
			\begin{align}
				P[V^\mu] = (-)^\mu \qquad\qquad
				P[D^*] = - \qquad\qquad
				P[B] = -.
			\end{align}
			We confirm that the matrix element is $P$-even with respect to `overall parities' $-$ and has the correct left-over `vector parity' $(-)^\mu$ that is required for a single-index object. The dynamical objects that we have, however, do not seem to have the correct parities:
			\begin{align}
				P[\epsilon^\mu] = -(-)^\mu \qquad\qquad\qquad P[p_{B,D}] = (-)^\mu.
			\end{align}
			In fact, we can see that the only non-zero vector bilinear $\epsilon_\mu p_B^\mu$ also does not have the correct parity. One might be led to believe that this matrix element must vanish. However, we have one more trick up our sleeves. In $d$-dimensional space we have a $d$-index totally antisymmetric $\varepsilon^{\mu_1\cdots \mu_d}$ tensor with which we can construct Lorentz contractions. In more formal language, we have the additional operation of taking a Hodge dual to convert $p$-forms into ($d-p$) forms. Our our present case this enables us to consider triple products of $\epsilon$, $p_D$, and $p_B$ to construct a Lorentz vector. The natural object to write down is
			\begin{align}
				\varepsilon^{\mu\nu\alpha\beta} \epsilon^*_{\nu} p_{D\alpha} p_{B\beta}.
			\end{align}
			The key thing to recall is that $\varepsilon$ has parity
			\begin{align}
				P[\varepsilon^{\mu\nu\alpha\beta}]=-(-)^{\mu}(-)^{\nu}(-)^{\alpha}(-)^{\beta},
			\end{align}
			where we've written out the usual `vector parity' for a four-tensor-like object. Because of the additional overall minus sign, we see that $\varepsilon$ is a \textbf{pseudotensor}. One should already be familiar with this in the coordinate definition of the volume form in which the $\varepsilon$ tensor carries information about orientation. Armed with this we now see that the $(\varepsilon\epsilon^*p_Dp_B)^\mu$ contraction is indeed a parity-even vector to which the hadronic matrix element $\langle D^{*+}(p_D,\epsilon)|V^\mu|\bar B(p_B)\rangle$ may be proportional.
		To complete the analysis, we remark that we are free to change momentum variables to $(p_D+p_B)$ and $q\equiv (p_D-p_B)$. It is clear that we are free to make the replacement
		\begin{align}
			p_{D\alpha} p_{B\beta} \longrightarrow (p_D+p_B)_\alpha q_\beta
		\end{align}
		since these are contracted with the $\varepsilon$ antisymmetric tensor so that only the $p_{D\alpha}p_{B\beta}-p_{D\beta}p_{B\alpha}$ term is picked out. Finally, the overall coefficient $g(q^2)$ can only be a function of $q^2$ since this is the only dynamical Lorentz scalar quantity. 
		\item The matrix element transforms as a complex conjugate under time reversal. The only complex element from our dynamical variables on the right-hand side is $\epsilon^*$, which contains an imaginary element. The sign of this element represents the transverse polarization of the $D^*$ and we expect it to flip under complex conjugation. Otherwise, the rest of the element is real and so $g(q^2)$ must also be real. 
		\item The axial current is not pure \QCD and comes from the chiral nature of the weak interactions and hence we are allowed to have $\langle D^{*+}(p_D,\epsilon)|A^\mu|\bar B(p_B)\rangle$ which is `overall parity' odd (i.e. $P = -(-)^\mu$). One immediate choice is a term proportional to $\epsilon^{*\mu}$ since this has precisely the correct parity and Lorentz structure. In addition to this, we can also form objects out of the three dynamical variables since $\epsilon^* (p_D+p_B) q$ also has the same correct parity and Lorenz structure; we just have to insert the Lorentz index at each place. The $\epsilon^{*\mu} (p_D+p_B)\cdot q$ term is the same as the $\epsilon^{*\mu}$ since $(p_D+p_B\cdot q)$ is not dynamical (it's a sum of masses). We are left with terms of the form
		\begin{align}
			\langle D^{*+}(p_D,\epsilon)|A^\mu|\bar B(p_B)\rangle =  g_{A1}(q^2)\epsilon^{*\mu} + g_{A2} \epsilon^*\cdot (p_D+p_B) q^\mu +  g_{A3}(\epsilon^* \cdot q)(p_D+p_B)^\mu.
		\end{align}

		\end{enumerate}
	
\end{Solution}
\begin{Solution}{8.1}
		The argument that we want to support is that $q^2=0$ is a maximum of the overlap between the constituent quarks in the kaon. This argument has \emph{nothing} to do with kinematics or the requirement that $q^2$ takes a certain range for physical processes. We can consider this overlap ``before'' we do any kinematics: $q^2$ may even be space-like ($q^2 <0$). When this is the case, we know that the overlap is still maximal at $q^2=0$, so that our heuristic argument for $f(q^2) = 1 + \mathcal O(q^2)$ still holds. One may prove this more rigorously, but the hard evidence is that it is true experimentally.
	
\end{Solution}
\begin{Solution}{11.1}
	This question and solutions come from G.~Ridolfi's \acro{CP} notes\footnote{\url{http://www.ge.infn.it/~ridolfi/notes/cpnew.ps}}. Labeling the kaon states with lowercase Greek indices, we have
	\begin{align}
		S_{\beta\alpha} =& S^{(0)}_{\beta\alpha} + S^{(1)}_{\beta\alpha} + S^{(2)}_{\beta\alpha}\\
		S^{(0)}_{\beta\alpha} =& \langle \beta | \alpha\rangle \\
		S^{(1)}_{\beta\alpha} =& -i \langle \beta | \int dt\, e^{iHt}H_We^{-iHt} | \alpha\rangle\\
		S^{(2)}_{\beta\alpha} =&
			-\frac 12 \langle \beta | \int dt\, e^{iHt}H_We^{-iHt} \int dt'\, e^{iHt'}H_W e^{-iHt'} \Theta(t-t') | \alpha\rangle + (t' \leftrightarrow t).
	\end{align}
	The first two terms are simple:
	\begin{align}
	S^{(0)}_{\beta\alpha} =& \delta_{\beta\alpha}\\
	S^{(1)}_{\beta\alpha} =& -2\pi i \delta(E_\beta-E_\alpha)\langle \beta| H_W|\alpha\rangle.
	\intertext{
	The second-order term requires a bit more work. The two terms are identical due to $(t\leftrightarrow t')$. By inserting a complete set of unperturbed Hamiltonian states $I=\sum_\lambda |\lambda\rangle\langle \lambda|$ one obtains
	}
		S^{(2)}_{\beta\alpha} =& -2\pi\delta(E_\beta-E_\alpha)\sum_\lambda
			\langle\beta | H_W|\lambda\rangle
			\langle\lambda | H_W|\alpha\rangle
			\int_0^\infty d\tau\, e^{i(E_\beta-E_\lambda)\tau}.
	\end{align}
	To evaluate the last integral, perform the infinitesimal shift $(E_\beta-E_\lambda)\to (E_\beta - E_\lambda + i\epsilon)$ with $\epsilon>0$. The resulting integral is straightforward,
	\begin{align}
		\int_0^\infty d\tau\, e^{i(E_\beta-E_\lambda+i\epsilon)\tau} = \frac{i}{E_\beta-E_\lambda+i\epsilon},
	\end{align}
	so that finally we have
	\begin{align}
			S^{(2)}_{\beta\alpha} =& -2\pi i\delta(E_\beta-E_\alpha)\sum_\lambda
				\frac{\langle\beta | H_W|\lambda\rangle
				\langle\lambda | H_W|\alpha\rangle}{E_\beta-E_\lambda+i\epsilon}.
	\end{align}
	Now we would like to define the effective weak Hamiltonian $H_W^\text{eff}$ so that $H = m_K \delta_{\beta\alpha} + H_{W\, \beta\alpha}^\text{eff}$. The point is that $H_W^\text{eff}$ should give the same contribution to the $S$-matrix as the perturbation analysis above. We thus have
	\begin{align}
		H^\text{eff}_{W\, \beta\alpha} =
		\langle \beta H_W | \alpha \rangle
		+
		\sum_\lambda
			\frac{\langle\beta | H_W|\lambda\rangle
			\langle\lambda | H_W|\alpha\rangle}{E_\beta-E_\lambda+i\epsilon}.
	\end{align}
	This effective Hamiltonian is \textit{not} Hermitian. Using the handy identity,
	\begin{align}
		\frac{1}{x+i\epsilon} = P\left(\frac 1x\right) - i\pi \delta(x),
	\end{align}
	where $P$ denotes the principal value, we have $H = M -\frac i2 \Gamma$ with
	\begin{align}
		M_{\beta\alpha} =&
		m_K \delta_{\beta\alpha} + \langle | H_W | \alpha\rangle
			+ P\sum_\lambda \frac{
			\langle \beta | H_W | \lambda \rangle
			\langle \lambda | H_W | \alpha \rangle
			}{m_K-E_\lambda}\\
		\Gamma_{\beta\alpha} =& 2\pi\sum_\lambda
			\langle \beta | H_W | \lambda \rangle
			\langle \langle | H_W | \alpha \rangle
			\delta(m_K-E_\lambda).
	\end{align}
	
\end{Solution}
\begin{Solution}{11.2}
		First recall that \acro{CP} takes $K \leftrightarrow \bar K$. This means that in this basis, the \acro{CP} operator can be written as a matrix
		\begin{align}
			U_{\text{CP}} = \begin{pmatrix}
				0 & 1\\
				1 & 0
			\end{pmatrix}.
		\end{align}
		Further, time reversal acts as a complex conjugate, $M\to M*$ and $\Gamma \to \Gamma^*$. From this we conclude that \acro{CPT} acts as
		\begin{align}
			M = \begin{pmatrix}
				M_{11} & M_{12}\\
				M^*_{12} & M_{22}
			\end{pmatrix}
			\to \begin{pmatrix}
				M_{22} & M_{12}\\
				M^*_{12} & M_{11}
			\end{pmatrix}\ ,
		\end{align}
		and similarly for $\Gamma$. Thus \acro{CPT} requires $M_{11} = M_{22}$ and $\Gamma_{11}=\Gamma_{22}$.
	
\end{Solution}
\begin{Solution}{11.3}
		For simplicity, let us write
		\begin{align}
			H =
			\begin{pmatrix}
				H & H_{12}\\
				H_{21} & H
			\end{pmatrix}
			=
			\begin{pmatrix}
				M_{11} - \frac i2\Gamma_{11} & M_{12} - \frac{i}{2}\Gamma_{12}\\
				M_{12}^* - \frac{i}{2}\Gamma_{12}^* & M_{11} - \frac i2\Gamma_{11}
			\end{pmatrix}.
		\end{align}
		The eigenvalue equation is
		\begin{align}
			(H-\mu)^2 - H_{12}H_{21} = 0 \Rightarrow \mu = H \pm \sqrt{H_{12} H_{21}}.
		\end{align}
		The eigenvectors are also easy to find. Calling these states $(p,q)^T$, the eigenvector equation becomes
		\begin{align}
			H p + H_{12} q &= \mu p\\
			H_{12} q &= \pm \sqrt{H_{12}H_{21}} p.
		\end{align}
		We thus have eigenvectors given by $q = \pm \sqrt{H_{21}/H_{12}} p$, which one can normalize appropriately.
	
\end{Solution}
\begin{Solution}{11.4}
		We start with
		\begin{align}
			|K_{L,S}\rangle = \frac{1}{\sqrt{|p|^2+|q|^2}} \left( p|K^0\rangle \pm q|\bar K^0\rangle\right),
		\end{align}
		which we can invert simply to obtain
		\begin{align}
			|K^0\rangle &= \frac{\sqrt{|p|^2+|q|^2}}{2p} (|K_L \rangle + |K_S\rangle)\\
			|K^0\rangle &= \frac{\sqrt{|p|^2+|q|^2}}{2q} (|K_L \rangle - |K_S\rangle).
		\end{align}
		Now we simply time evolve the effective Hamiltonian eigenstates. Let us write $H_L$ and $H_S$ for the long and short eigenvalues, for example $H_L = M_L - \frac i2 \Gamma_L$. For $|K^0(t)\rangle$ we have
		\begin{align}
			|K^0(t)\rangle &= \frac{1}{2p} \left[
			e^{-iH_Lt} (p|K^0\rangle + q|\bar K^0\rangle)
			+
			e^{-iH_St} (p|K^0\rangle - q|\bar K^0\rangle)
			\right]\\
			&=
			\frac 12 \left[
				\left(e^{-iH_Lt}  + e^{-iH_St}\right)|K^0(t)\rangle +
				\frac qp \left(e^{-iH_Lt}  - e^{-iH_St}\right)|\bar K^0(t)\rangle
			\right].
		\end{align}
		The $|\bar K^0(t)\rangle$ oscillation comes from making the appropriate substitutions.
	
\end{Solution}
\begin{Solution}{12.1}
	For more background, see Preskill's lecture
        notes\footnote{\url{http://www.theory.caltech.edu/~preskill/notes.html},
          or the Sidney Coleman lecture notes upon which these are
          loosely based. The latter material is neatly encapsulated in
          the QFT textbook by Ticciati.}. 
\begin{enumerate}
\item Suppose a theory of scalar particles is $T$ symmetric. This means that $[T,H]=0$, where $H$ is the full Hamiltonian including interactions. (The free Hamiltonian is $T$-symmetric.) The unitary time evolution operator is transformed according to
		\begin{align}
		T U(t,t_0) T^{-1} =& T e^{iH_0(t-t_0)}e^{-iH(t-t_0)}T^{-1}	\\
		=& e^{-iH(t-t_0)}e^{iH(t-t_0)} = U(t_0,t) = U(t,t_0)^\dag.
		\end{align}
		Since $S=U(\infty,-\infty)$, $TST^{-1}=S^\dag$. We want a relation between amplitudes, but we must be careful with matrix elements with respect to the antiunitary $T$ operation. $\langle f|Si\rangle = \langle TSi|Tf\rangle = \langle S^\dag Ti | Tf\rangle$. For scalars we have $Ti=i$ and $Tf = f$ since they have no spin to go in the opposite direction. Thus we have the relation $\langle f|Si\rangle = \langle i|Sf\rangle$, or $A_{fi} = A_{if}$. Note that this is \textit{not} the same as saying $\sigma(i\to f) = \sigma(f\to i)$ since the cross section has phase space factors that are not the same even if the amplitudes are.
		\item \acro{CPT} gives a weaker form of the detailed balance condition. $CPT|i\rangle = \bar i\rangle$ and $CPT|f\rangle = \bar f\rangle$. Following the same steps as the previous sub-problem (because both $T$ and $CPT$ and anti-unitary), we arrive at $A_{fi} = A_{\bar i \bar f}$.
		\item Unitarity tells us that $S^\dag S = S S^\dag = 1$, or in terms of the $T$-matrix, $T^\dag T = TT^\dag$. (Do not confuse this $T$ with the time inversion operator.) This implies then that $\sum_m |A_{im}|^2 = \sum_m |A_{mi}|^2$. Acting on both sides with $CPT$ gives $\sum_{\bar m} |A_{\bar m\bar i}|^2 = \sum_m |A_{m\bar i}|^2$ since $\sum_m = \sum_{\bar m}$ when going over all states. This then implies that $\Gamma(i\to \text{ all}) = \Gamma(\bar i \to \text{ all})$. Alternately, one can use the optical theorem, $\text{Im } A_{ii}= \text{Im }A_{\bar i\bar i}$, along with the equality of the phase space factors. Note that only the \textit{inclusive} rates $i\to \text{ all}$ and ${\bar i} \to \text{ all}$ match; in general $i\to j$ and $\bar i to \bar j$ have different rates. This is a manifestation of \CP violation.
		\item For simplicity, we will use the $T$-matrix in place of the $S$ matrix. The optical theorem tells us that $T_{ij}-T^*_{ji} = i\sum_n T_{in}T^*_{jn}$. This then implies that $|T_{ij}|^2 = |i(\sum TT^\dag)_{ij}+T_{ji}^*|^2$. If we now compare $\Gamma(i\to j)$ with $\Gamma(\bar i\to \bar j)$ and not that $\Gamma \sim |T|^2$, we find that
		\begin{align}
			|T_{ij}|^2 - |T_{\bar i\bar j}|^2  =&
			|i(\sum TT^\dag)_{ij}+T^*_{ji}|^2 - |T_{ji}|^2
			& -2\text{Im }(\sum TT^\dag)_{ij}T^*_{ji} + |(\sum TT^\dag)_{ij}|^2,
		\end{align}
where in the first line we used \acro{CPT} to subtract $|T_{\bar i \bar j}|^2 = |T_{ji}|^2$ from both sides. The right-hand side of the second line is at least $\mathcal O(T^3)$ whereas the left-hand side is $\mathcal O(T^2)$. This means that the right-hand side is higher order in perturbation theory. For fixed external states, higher order diagrams come from loops so that CP violation vanishes at the lowest order of perturbation theory and can only appear in loop effects.
	\end{enumerate}
	
\end{Solution}
\begin{Solution}{12.2}
The decay amplitude is proportional to $V_{ub}^* V_{ud}$ and thus to
the $u$ side of the unitarity triangle, while the mixing is
proportional to the $t$ side. The asymmetry is sensitive to the angle
between these two two, that is, we pick up the angle $\alpha$.
\end{Solution}
\begin{Solution}{B.1}
		Let us write the hadronic matrix element for neutron $\beta$ decay $n\to p$ through the axial current,
		\begin{align}
			\langle p(k_p) | A_\mu |n(k_n)\rangle = \bar u_p(k_p)\left[g_A(q^2)\gamma_\mu\gamma_5 + q_\mu h_A(q^2)\gamma_5 \right] u_n(k_n).
		\end{align}
		Here $g_A$ and $h_A$ are functions which depend on the momentum transfer $q\equiv k_n - k_p$. Note that we are \textit{not} `parameterizing our ignorance' about \QCD (we still have cumbersome spinor structure on the right-hand side); instead, we are saying that the neutron decays into the proton \textit{literally} through the tree-level coupling to the axial current. If you want to be precise about isospin indices, one ought to write $A_\mu \to A^+_\mu = A_\mu^1 + i A_\mu^2$, but we'll try to keep the equations as simple as possible. We can now take the divergence of this equation to obtain
		\begin{align}
			\langle p(k_p) | \partial_\mu A^\mu |n(k_n)\rangle = i \left[g_A(q^2) m_N^2 + q^2 h_A(q^2)\right] \bar u_p(k_p) \gamma_5 u_n(k_n),\label{eq:Goldberger:Treiman:axial}
		\end{align}
		where $m_N$ is the nucleon mass. The eft-hand side of this equation, however, is simply what we called the pion field in (\ref{eq:LSZ:Goldstone:pion}), up to overall factors:
		\begin{align}
			\langle p(k_p) | \partial_\mu A^\mu |n(k_n)\rangle &= f_\pi m_\pi^2 \langle p(k_p) | \phi_\pi |n(k_n)\rangle\\
			&= \frac{2f_\pi m_\pi^2}{q^2-m_\pi^2} g_{\pi NN}(q^2) i\bar u_p(k_p) \gamma_5 u_n(k_n), \label{eq:Goldberger:Treiman:pion}
		\end{align}
		where in the second line we have defined the pion-nucleon vertex function $g_{\pi NN}(q^2)$ and pulled out the pion pole. This gives the $\pi NN$ coupling for $q^2 = m_\pi^2$. Comparing (\ref{eq:Goldberger:Treiman:axial}) and (\ref{eq:Goldberger:Treiman:axial}) we obtain,
		\begin{align}
			f_\pi g_{\pi NN}(0) = m_N g_A(0).
		\end{align}
		This is nearly the Goldberger-Treiman relation, except that the pion-nucleon vertex function is off-mass shell. We must make the assumption that $g_{\pi NN}(q^2)$ is slowly varying such that $g_{\pi NN}(0) \approx g_{\pi NN}(m_\pi^2)$. This gives us a relation which holds within 10\%. Note that this relation is actually rather trivial from the point of view of chiral perturbation theory, see e.g.\ Donoghue Section 12.3 \cite{Donoghue:DynamicsSM}.
	
\end{Solution}
\begin{Solution}{B.2}
 By the same argument that we've seen several times now, we know that the $f_3$ form factor is proportional to the electron mass upon contraction with the leptonic part of the full matrix element.
\end{Solution}
\begin{Solution}{C.1}
		Writing the universal mass contribution as $m_\Lambda$ and using the quark content predicted by SU$(3)$, e.g.\ by using (\ref{eq:eta:content}), we find
		\begin{align}
			m_\pi^2 &= m_\Lambda^2 + 2\mu m_u\\
			m_K^2 &= m_\Lambda^2 + \mu\left(m_u + m_s\right)\\
			m_\eta^2 &= m_\Lambda^2 + \frac{2\mu}{3}\left(m_u+2m_s\right).
		\end{align}
		We can `solve' these equations by eliminating the unknown quantities $m_\Lambda$, $m_u$, and $m_s$ to obtain the \textbf{Gell-Mann--Okubo mass formula} for pseudoscalar mesons,
		\begin{align}
			4m_K^2 = m_\pi^2 + 3m_\eta^2.
		\end{align}
		Note that the meson masses in this expression are summed in quadrature. For general SU$(3)$ octets (vector mesons) the meson masses would be summed linearly.
	
\end{Solution}
\begin{Solution}{C.2}
 This is actually a bit of a trick question because it's hard to find the quark content from the summary table. Textbooks list the quark content as being approximately
		\begin{align}
			\omega &= \frac{1}{\sqrt{2}}\left(u\bar u + d\bar d\right)\\
			\phi &= s\bar s.
		\end{align}
	
\end{Solution}
\begin{Solution}{C.3}
		Representation theory for SU(3) tells us that the $I_3=0$ octet and singlet states have the quark content
		\begin{align}
			\psi_8 &= \frac{1}{\sqrt{6}}\left(u\bar u + d\bar d -2 s \bar s\right)\\
			\psi_1 &= \frac{1}{\sqrt{3}}\left(u\bar u + d\bar d + s \bar s\right).
		\end{align}
		As we mentioned before we will be working with differences of masses, so we can ignore all universal contributions to meson masses coming from angular momentum and the \QCD potential. Further, we will assume that SU(2) isospin is a good symmetry such that $m_d=m_u$. Then the mass of the $\psi_8$ comes from averaging over the probabilities that $\psi_8$ will contain a given quark pair. In other words, one should look at
		\begin{align}
			\langle\psi_8 | \psi_8\rangle = \frac 16 \left(\langle u\bar u|u\bar u\rangle + \langle d\bar d|d \bar d \rangle + 4\langle s\bar s|s \bar s\rangle\right).
		\end{align}
		to determine that
		\begin{align}
			m_8 = \frac{1}{6}\left(2\times 2m_u + 2\times 4m_s\right) = \frac{2}{3}(m_u+2m_s),
		\end{align}
		where we remember an overall factor of two because the state $u\bar u$ has valence quark mass contribution $\Delta m = 2 m_u$. Note that we look at the square of the state $|psi\rangle$ because we want the \textit{probability} that the meson contains a given quark content; this has \textit{nothing} to do with the idea that we should be summing the masses linearly or quadratically. It is ``more obvious than obvious\footnote{This is a delightful phrase borrowed from Tony Zee.}'' that we should sum the masses linearly. (The pseudoscalars are special because they are pseudo-Goldstone bosons that need to be treated in chiral perturbation theory where the \textit{squared} meson masses are linear in the symmetry-breaking parameters, the quark masses.) Using the same analysis we may write out the valence quark mass contribution for the vector kaon and $\rho$,
		\begin{align}
			m_{K^*} &= m_u+m_s\\
			m_\rho &= 2m_u.
		\end{align}
		This gives us a basis to re-express quark masses in terms of the non-mixing octet meson masses. (These mesons don't mix because the quark masses break $SU(3)\to U(1)^3$ so that strangeness is conserved.) Thus we readily obtain
		\begin{align}
			m_8 &= \frac 13(4m_{K^*}-m_\rho)\\
			m_1 &= \frac 13(2m_{K^*}+m_\rho).
		\end{align}
		You might have worried that we do not know what the mixing term $m_{18}=m_{81}$ should look like, but the point is that the mixing term comes precisely from the fact that when $SU(3)$ is broken, $\langle \psi_8 | \psi_1\rangle \neq 0$. In particular,
		\begin{align}
			\langle \psi_8 |\psi_1\rangle = \frac{1}{3\sqrt{2}}\left(
			\langle u\bar u|u\bar u\rangle + \langle d\bar d|d \bar d \rangle -2\langle s\bar s|s \bar s\rangle
			\right),
		\end{align}
		note the minus sign! We can thus read off
		\begin{align}
			m_{81} = m_{18} = \frac{2}{3\sqrt{2}}\left(2m_u-2m_s\right) = \frac{2\sqrt{2}}{3}\left(m_\rho - m_{K^*}\right).
		\end{align}

		Good. Now we know that the flavor basis matrix is made up of these elements, but upon rotation to the mass basis the matrix is diagonalized to
		\begin{align}
			R \begin{pmatrix}
				m_8 & m_{81}\\
				m_{18} & m_1
			\end{pmatrix} R^T =
			\begin{pmatrix}
				m_\phi & \\
				& m_\omega
			\end{pmatrix}.
		\end{align}
		 We can relate the physical masses $m_{\phi,\omega}$ to the expressions above by considering invariants under rotations, namely the trace and determinant. The trace relation is easy and gives us
		\begin{align}
			m_\omega + m_\phi = 2m_{K^*}.\label{eq:sol:schwinger:trace}
		\end{align}
		The determinant relation gives us
		\begin{align}
				m_\omega\, m_\phi &= \frac 19\left[ (4m_{K^*}-m_\rho)(2m_{K^*}+m_\rho) - 8(m_\rho-m_{K^*})^2 \right],
		\end{align}
		where we recognize $m_8 = 3(4m_{K^*}-m_\rho)$ in the first term on the right-hand side.
		We would like to combine these equations into the Schwinger relation. This is most easily obtained by starting with the determinant relation and, in the first term on the left-hand side, writing
		\begin{align}
			(2m_{K^*}+m_\rho) = 6m_{K^*}+m_\rho - 4m_{K^*} = 3(m_\omega + m_\phi) - 3m_8.
		\end{align}
		This leads us to
		\begin{align}
			3m_8\left[3(m_\omega+m_\phi)-3m_8\right] - 8(m_\rho-m_{K^*}) = 9m_\omega m_\phi.
		\end{align}
		Rearranging terms and then factorizing we finally obtain
		\begin{align}
			-(m_8-m_\omega)(m_8-m_\phi) = \frac 89(m_a-m_{K^*})^2.
		\end{align}
		This is satisfied up to an error of $\mathcal{O}(160 \text{ MeV}^2)$, which is quite good considering that it is \textit{quadratic} in mass.
	
\end{Solution}
\begin{Solution}{C.4}
	The rotation from the $\psi_8$,$\psi_1$ basis to the physical basis can be written as
	\begin{align}
		R^T M R &= \begin{pmatrix}
			m_\phi & \\
			& m_\omega
		\end{pmatrix}, \hspace{2cm}
		R\equiv\begin{pmatrix}
			\cos\theta & \sin\theta\\
			-\sin\theta & \cos\theta
		\end{pmatrix}.
	\end{align}
	Inverting this relation one finds
	\begin{align}
		\begin{pmatrix}
			m_8 & m_{81}\\
			m_{18} & m_{1}
		\end{pmatrix}
		=
		\begin{pmatrix}
			\text{c}^2 m_\phi + \text{s}^2 m_\omega
			& \text{c}\,\text{s}\,(m_\omega-m_\phi)\\
			\text{c}\,\text{s}\,(m_\omega-m_\phi)
			& \text{s}^2 m_\phi + \text{c}^2 m_\omega
		\end{pmatrix},
	\end{align}
	where we've written $\sin\theta = \text{s}$ and $\cos\theta=\text{c}$ for short-hand. Now note the following relations:
	\begin{align}
		m_8 - m_\omega &= \text{c}^2(m_\phi - m_\omega)\\
		m_8 - m_\phi &= \text{s}^2(m_\omega - m_\phi)\\
		m_1 - m_\omega &= \text{s}^2(m_\phi - m_\omega)\\
		m_1 - m_\phi &= \text{c}^2(m_\omega - m_\phi),
	\end{align}
	where all we've used is $\cos^2\theta + \sin^2\theta =1$. From this it is easy to take ratios and derive the relations in (\ref{eq:mixing:tan:2:theta}).
	For the $\tan\theta$ expressions use the observation that
	\begin{align}
		m_{81} &= \text{c}\,\text{s}(m_\omega-m_\phi)
	\end{align}
	and take the appropriate ratios with the $\sin^2\theta$ and $\cos^2\theta$ expressions above.
	Finally, for the double angle formulae use
	\begin{align}
		m_8-m_1 &= (\text{c}^2-\text{s}^2)(m_\phi-m_\omega)
	\end{align}
	and
	\begin{align}
		\cos 2\theta &= \cos^2\theta -\sin^2\theta\\
		\sin 2\theta &= 2\cos\theta\sin\theta.
	\end{align}
\end{Solution}
\begin{Solution}{C.5}
		Writing out only the valence quark mass contribution (and taking $m_u=m_d$) we have
		\begin{align}
			m_{K^*} &= m_u + m_s\\
			m_{\rho} &= 2m_u.
		\end{align}
		From this we may express the diagonal flavor-eigenstate masses
		\begin{align}
			m_8 &= \frac 13(4m_{K^*}-m_\rho)\\
			m_1 &= \frac 13(2m_{K^*}+m_\rho).
		\end{align}
		Taking the inner product $\langle \phi_8 | \phi 1\rangle$ we may also extract the off-diagonal mass term
		\begin{align}
			m_{81} = m_{18} = \frac{2\sqrt{2}}{3}\left(m_u-m_s\right) = \frac{2\sqrt{2}}{3}\left(m_\rho - m_{K^*}\right).
		\end{align}
		Armed with these it is straightforward to plug into (\ref{eq:mixing:tan:2:theta}), (\ref{eq:mixing:tan:theta}), and (\ref{eq:mixing:cos:sin:2:theta}). To give a complete list for each expression,
		\begin{align}
			\tan^2\theta &= -\frac{4m_{K^*}-m_\rho-3m_\phi}{4m_{K^*}-m_\rho -3m_\omega}\\
			&= -\frac{2m_{K^*}+m_\rho-3m_\omega}{2m_{K^*}+m_\rho-3m_\phi}\\
			&= \phantom{+}\frac{4m_{K^*}-m_\rho-3m_\phi}{2m_{K^*}+m_\rho-3m_\phi}\\
			&= \phantom{+}\frac{2m_{K^*}+m_\rho-3m_\omega}{4m_{K^*}-m_\rho-3m_\omega}.
		\end{align}
		\begin{align}
			\tan\theta &= \frac{-2\sqrt{2}(m_\rho - m_{K^*})}{4m_{K^*}-m_\rho -3m_\omega}\\
			&= \frac{-2\sqrt{2}(m_\rho - m_{K^*})}{2m_{K^*}+m_\rho -3m_\phi}\\
			&= \frac{4m_{K^*}-m_\rho -3m_\phi}{2\sqrt{2}(m_\rho - m_{K^*})}\\
			&= \frac{2m_{K^*}+m_\rho -3m_\omega}{-2\sqrt{2}(m_\rho - m_{K^*})}.
		\end{align}
		\begin{align}
			\cos 2\theta &= \frac{2}{3}\frac{m_{K^*-m_\rho}}{m_\phi - m_\omega}\\
			\sin 2\theta &= \frac{4\sqrt{2}}{3}\frac{m_\rho - m_{K^*}}{m_\omega - m_\phi}.
		\end{align}
		Note that the value of $\tan 2\theta$ obtained from the last two equations is independent of any meson masses. This is because (\ref{eq:mixing:cos:sin:2:theta}) was written such that the $m_\phi$ and $m_\omega$ dependence cancels and we're left with two quantities which are calculated using valence quark masses. The upshot is that the value of $\theta$ derived this way is precisely what is predicted by \textit{ideal mixing}.
	
\end{Solution}
\begin{Solution}{C.6}
		When the mixing term $m_{xy}\ll m_x^2 -m_y^2$ then $\phi=0$ and we have the ideal mixing condition. On the other hand, if the splitting $m_x^2 -m_y^2 \ll m_{xy}$ then $\phi \to 45^{\circ}$. This value is an asymptotic and so is relatively stable in this regime. For more discussion see Isgur \cite{Isgur:1975ib}.
	
\end{Solution}
\begin{Solution}{C.7}
		These states do not mix because the light quark masses break $SU(3)$ to $U(1)_u\times U(1)_d\times U(1)_s$ such that strangeness is conserved.
	
\end{Solution}